\documentclass[10pt]{article} % For LaTeX2e

\usepackage[accepted]{tmlr}
\usepackage{amsmath,amsfonts,bm}

\def\eqref#1{equation~\ref{#1}}
\def\1{\bm{1}}

\DeclareMathAlphabet{\mathsfit}{\encodingdefault}{\sfdefault}{m}{sl}
\SetMathAlphabet{\mathsfit}{bold}{\encodingdefault}{\sfdefault}{bx}{n}

\usepackage[colorlinks=true,citecolor=RoyalBlue,linkcolor=Bittersweet]{hyperref}
\usepackage{url}

\usepackage[skip=0pt]{caption}
\usepackage{subcaption}
\usepackage{graphicx}
\usepackage{amsmath}
\usepackage{amssymb}

\usepackage{booktabs}
\usepackage{algorithm}
\usepackage{algpseudocode}
\usepackage{algpseudocode}
\usepackage{threeparttable}
\usepackage{multirow}
\usepackage{tabularx}

\usepackage[dvipsnames]{xcolor}
\usepackage{soul}
\usepackage{colortbl}
\usepackage{transparent}
\usepackage{xspace}
\usepackage{paralist}
\usepackage{enumitem}
\usepackage{longtable}
\usepackage[edges]{forest}
\usepackage{url}
\usepackage{hyperref}
\usepackage{cleveref}

\crefname{figure}{Figure}{Figures}
\crefname{section}{Section}{Sections}
\crefname{equation}{Equation}{Equations}
\crefname{appendix}{Appendix}{Appendice}
\crefname{table}{Table}{Tables}

\newcommand{\interalia}{\emph{inter alia}}

\newcommand{\bert}{\textsc{BERT}}

\title{A Survey of Model Architectures in Information Retrieval}

\author{\name Zhichao Xu \email zhichao.xu@utah.edu \\
      \addr Kahlert School of Computing\\
      University of Utah
      \AND
      \name Fengran Mo \email fengran.mo@umontreal.ca \\
      \addr University of Montreal
      \AND
      \name Zhiqi Huang \email zhiqi.huang@capitalone.com \\
      \addr Capital One
      \AND
      \name Crystina Zhang \email xinyucrystina.zhang@uwaterloo.ca \\
      \addr University of Waterloo
      \AND 
      \name Puxuan Yu \email puxuan.yu@snowflake.com \\
      \addr Snowflake Inc. 
      \AND 
      \name Bei Wang \email beiwang@sci.utah.edu \\
      \addr Kahlert School of Computing \\
      Scientific Computing and Imaging (SCI) Institute \\
      University of Utah
      \AND 
      \name Jimmy Lin \email jimmylin@uwaterloo.ca \\
      \addr University of Waterloo 
      \AND 
      \name Vivek Srikumar \email svivek@cs.utah.edu \\
      \addr Kahlert School of Computing \\
      University of Utah
}

\def\month{02}  % Insert correct month for camera-ready version
\def\year{2026} % Insert correct year for camera-ready version
\def\openreview{\url{https://openreview.net/forum?id=xAIbTbHRrX}} % Insert correct link to OpenReview for camera-ready version

\begin{document}

\maketitle

\begin{abstract}
The period from 2019 to the present marks one of the most significant paradigm shifts in information retrieval (IR) and natural language processing (NLP), culminating in the emergence of powerful large language models (LLMs) from 2022 onward. Methods based on pretrained encoder-only architectures (e.g., BERT) as well as decoder-only generative LLMs have outperformed many earlier approaches, demonstrating particularly strong performance in zero-shot scenarios and complex reasoning tasks.
This survey examines the evolution of model architectures in IR, with a focus on two key aspects: backbone models for feature extraction and end-to-end system architectures for relevance estimation. To maintain analytical clarity, we deliberately separate architectural design from training methodologies, enabling a focused examination of structural innovations in IR systems. 
We trace the progression from traditional term-based retrieval models to modern neural approaches, highlighting the transformative impact of transformer-based architectures and subsequent LLM developments. The survey concludes with a forward-looking discussion of open challenges and emerging research directions, including architectural optimization for efficiency and scalability, robust handling of multimodal and multilingual data, and adaptation to novel application domains such as autonomous search agents, which may represent the next paradigm in IR.
\end{abstract}

\section{Introduction}
\label{sec:intro}

\begin{figure*}[h!]
\centering
\includegraphics[width=0.95\textwidth]{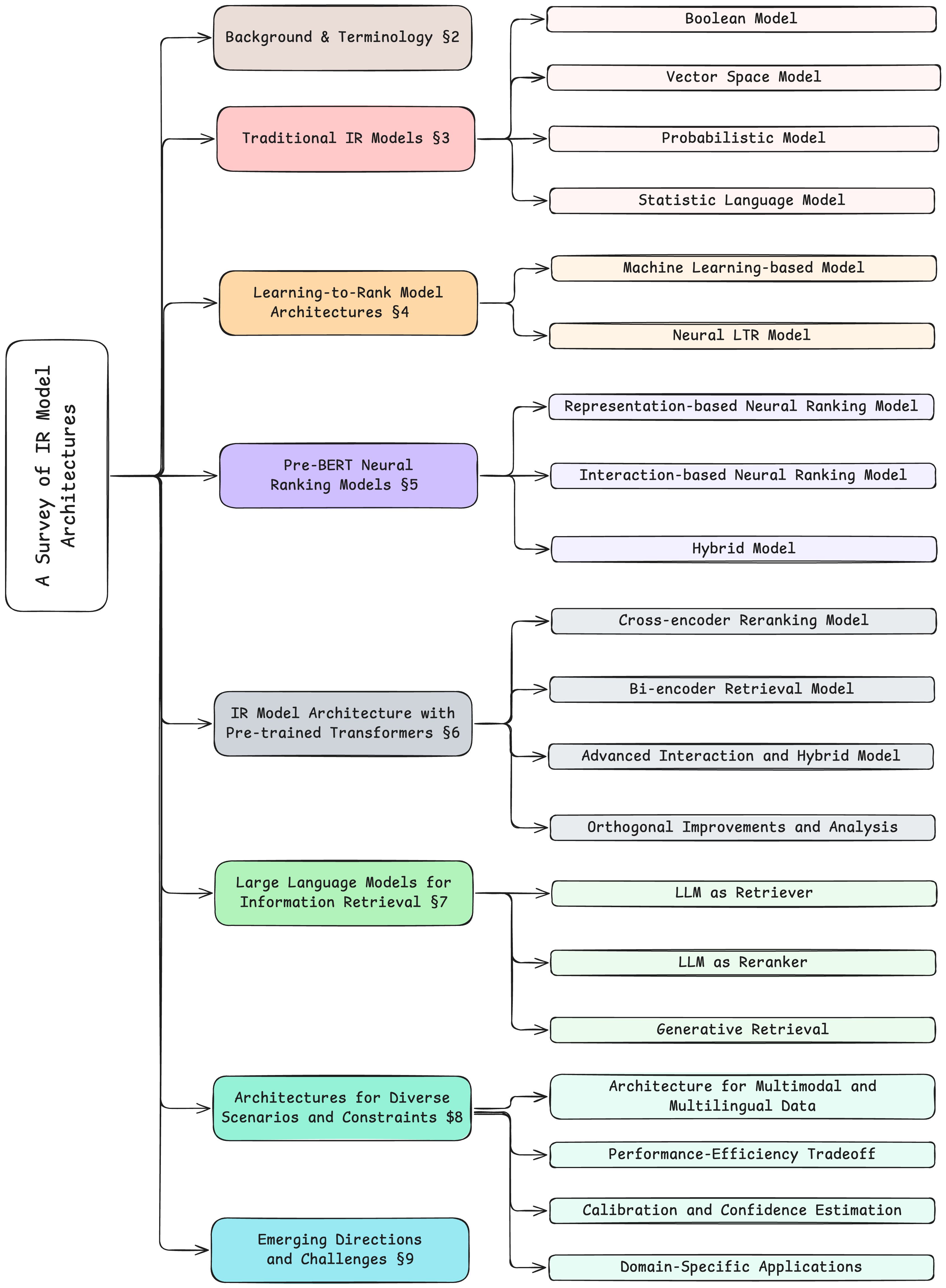}
\caption{High-level overview of the survey structure and scope.}
\label{fig:structure}
\end{figure*}

Information Retrieval (IR) aims to retrieve relevant information sources to satisfy users' information needs. In the past decades, IR has become indispensable for efficiently and effectively accessing vast amounts of information across various applications. Beyond its traditional role, IR now also plays a critical role in assisting large language models (LLMs) to generate grounded and factual responses under the generative AI era. Research in IR primarily centers on two key aspects: (1) \textit{extracting better query and document feature representations}, and (2) \textit{developing more accurate relevance estimators}. Extracting better query and document feature representations focuses on modeling textual content so that queries and documents can be compared in a meaningful space, ranging from early term-frequency vectors (e.g., TF–IDF~\citep{sparck1972statistical} and BM25~\citep{robertson2009probabilistic}) to modern contextual embeddings derived from pre-trained language models. Developing more accurate relevance estimators then builds on these representations to assess how well a document satisfies an information need, using scoring functions or learned ranking models such as BM25, neural interaction models, or later learning-to-rank frameworks that combine multiple signals. The approaches for extracting query and document features have evolved from traditional term-based methods, such as boolean logic~\citep{radecki1979fuzzy,kraft1983fuzzy} and vector space models~\citep{salton1975vector}, to modern solutions such as dense retrieval based on pre-trained language models~\citep[\interalia]{Lee2019LatentRF,Karpukhin2020DensePR,logeswaran-etal-2019-zero,lin2022pretrained}.

Relevance estimators have evolved alongside advances in feature representations. Early approaches, including probabilistic and statistical language models, computed relevance with simple similarity functions based on term-based features.
Learning-to-rank (LTR) techniques later emerged, incorporating machine learning models like support vector machines~\citep{cortes1995supportvectornetworks}, boosting methods~\citep{kearns1994cryptographiclimitations,freund1995desiciontheoreticgeneralization} as well as multi-layer neural networks for relevance estimation~\citep{li2011learning}. The success of LTR methods can be largely attributed to their extensive use of manually engineered features, derived from both statistical properties of text terms and user behavior data collected from web browsing traffic~\citep{qin2013mslr}.
In the 2010s, a vast literature explored neural rerankers in different architectures to capture the semantic similarity between queries and documents~\cite[\interalia]{pang2016matchpyramid,guo2016deep,xiong2017knrm,dai2018cknrm}.
Then pre-trained transformers, represented by \textsc{BERT}~\citep{devlin-etal-2019-bert} and its variants~\citep{liu2019roberta,sun2019ernie,lan2020albert,beltagy2020longformer},
quickly revolutionized the model design,
leading to an era where retrieval and ranking models adopt simpler architectures for relevance estimation, such as dot product operations and MLP prediction heads, which operate on learned neural representations~\citep{macavaney2019cedr,Lee2019LatentRF,Karpukhin2020DensePR,nogueira2020documentrankingpretrainedsequencetosequence,lin2022pretrained,formal21splade,formal21spladev2}.

Recent advancements in LLMs have revolutionized applied machine learning (ML) communities, including IR. One intriguing property of modern instruction-following LLMs (e.g., ChatGPT~\cite{openai2022chatgpt}) is that they can be used for feature extraction and relevance estimation, achieving strong performance without extensive training~\citep[\interalia]{ni-etal-2022-sentence,neelakantan2022text,behnamghader2024llmvec,sun2023chatgpt,qin-etal-2024-large}.
The rise of these models builds upon a rich foundation of neural architectures, including the classical Transformer architecture with multi-head attention~\citep[MHA,][]{vaswani2017attention}, Recurrent Neural Networks~\citep[RNN,][]{elman1990findingstructureintime}, attention mechanisms~\citep{bahdanau2014neural}, and pre-trained static word representations like Word2Vec~\citep{mikolov2013efficient} and GloVe~\citep{pennington-etal-2014-glove}~\citep[\interalia]{collobert2011natural,le2014distributedrepresentations}. 
 
This work reviews the evolution of model architectures in IR (with an overview in~\cref{fig:structure}). Here, the meaning of model architecture is twofold: it describes (1) backbone models for extracting query and document feature representations, and (2) end-to-end system architectures that process raw inputs, perform feature extraction, and estimate relevance.
Different from prior works and surveys~\citep{lin2022pretrained,zhu2023large}, we intentionally separate our discussion of model architectures from training methodologies and deployment best practices to deliver a focused architectural analysis. This analysis centers on the core components of AI infrastructure in the LLM era. 
The shift towards neural architectures, particularly Transformer-based models, has fundamentally transformed IR by enabling rich, contextualized representations and improved capacity for handling complex queries. While this evolution enhanced retrieval performance, it also presents new challenges, especially with the development of LLMs.
These challenges include the need for architectural innovations to optimize performance and scalability, handle multimodal and multilingual data, and understand complex instructions. Moreover, as IR systems are increasingly integrated into diverse applications\,---\,from robotics~\citep{xie2024embodiedrag}, protein structure discovery~\citep{jumper2021highly} to autonomous agents~\citep[\interalia]{wu2023autogen,chen2025learningtoreasonwithsearchforllmsviareinforcementlearning,hu2025owloptimizedworkforcelearning,deepresearchsystemcard,wu-etal-2025-webwalker} that are capable of reasoning and search\,---\,the field must evolve beyond traditional search paradigms. We conclude this survey by examining these challenges and discussing their implications for the future of IR model architectures research.

\section{Background and Terminology}
\label{sec:background}
We focus on the classical \textit{ad hoc} retrieval task, which forms the foundation for many modern IR applications. In this section, we define the core task, introduce key system architectures and evaluation paradigms, and clarify the scope of our architectural review.

\paragraph{Task Definition and Evaluation.} Given a query $\mathcal{Q}$, the task is to find a ranked list of $k$ documents, denoted as $\lbrace \mathcal{D}_1, \mathcal{D}_2, \ldots, \mathcal{D}_k \rbrace$, that exhibit the highest relevance to $\mathcal{Q}$. 
This is achieved either by \textit{retrieving} top-$k$ documents from a large collection $\mathcal{C}$ ($|\mathcal{C}| \gg k$), which typically comprises millions or billions of documents,
or by \textit{reranking} the top-$k$ candidates returned by a retriever. 
System performance is measured using standard, list-wise IR metrics. Common metrics include:
\begin{itemize}
\item \textbf{Mean Reciprocal Rank (MRR):} Measures the rank of the first relevant document. It is particularly useful for tasks where finding one correct answer is the primary goal (e.g., question answering).
\item \textbf{Recall@k:} Measures the fraction of all relevant documents that are found within the top-$k$
 results. It emphasizes the system's ability to find all relevant items.
\item \textbf{Normalized Discounted Cumulative Gain (nDCG@k):} A sophisticated metric that evaluates the quality of the ranking over the top-$k$ documents. It gives higher scores for ranking highly relevant documents at the top of the list and uses a logarithmic discount to penalize relevant documents that appear lower in the ranking. It is the de facto standard for evaluating ranked lists with graded relevance judgments.
\end{itemize}

\paragraph{The Multi-Stage ``Retrieve-then-Rerank'' Architecture.}
Modern large-scale IR systems almost universally operate on a multi-stage pipeline, commonly known as the ``retrieve-then-rerank'' architecture. This design balances the tradeoff between efficiency and effectiveness.
\begin{enumerate}
\item \textbf{Retrieval (or First-Stage Ranking):} In the first stage, a computationally efficient but less precise model scans the entire collection $\mathcal{C}$ (potentially billions of documents) to quickly identify an initial set of several hundred or thousand candidate documents. These models, often called \textit{retrievers}, must be extremely fast. Examples include traditional models like BM25 (\cref{sec:traditional}) or modern bi-encoder models (\cref{sec:transformer}).
\item \textbf{Reranking (or Second-Stage Ranking):} In the second stage, a more powerful but computationally expensive model, known as a \textit{reranker}, is applied only to the small candidate set returned by the retriever. This model can afford to perform deep, fine-grained analysis of the interaction between the query and each candidate document to produce a more accurate final ranking. Examples include Learning-to-Rank models (\cref{sec:ltr}) and modern cross-encoder transformers (\cref{sec:transformer}).
\end{enumerate}
This two-stage process is a central architectural pattern in IR, and much of the evolution discussed in this survey can be understood as developing more advanced models for each of these stages.

\paragraph{Query and Document.}
A \textit{query} expresses an information need and serves as input to the \textit{ad hoc} retrieval system. We denote \textit{document} as the atomic unit for retrieval and ranking. Our discussions are primarily based on text-based documents, but a document can also refer to a webpage or an email, depending on the actual IR application of interest.

\paragraph{Disentangling Model Architecture from Training Strategies.}
Similar to other applied ML domains, the performance of an IR system is a product of its model architecture, its training methodology (e.g., loss functions, data augmentation, optimization algorithms), and deployment best practices (e.g., indexing, quantization, parallelization, algorithm-hardware co-design). In this survey, we intentionally seek to disentangle these aspects to provide a focused analysis on the \textbf{evolution of model architecture}. This focus allows for a clearer narrative on how the core components for representation learning and relevance estimation have changed over time, from term-based logic to deep neural networks. We refer readers to dedicated surveys for in-depth reviews of training strategies and other related topics~\citep{schutze2008introduction,lin2022pretrained,song2023llm}.

\section{Traditional IR Models}
\label{sec:traditional}
In this section, we briefly review traditional Information Retrieval (IR) models prior to neural methods, with a focus on the \textbf{Boolean model}, \textbf{vector space model}, \textbf{probabilistic model}, and \textbf{statistical language model}. These models, which serve as the foundation for later developments in IR (\cref{sec:ltr,sec:neural_ranking,sec:transformer,sec:llm4ir}), are built upon the basic unit of a ``term'' in their representations~\citep{nie2010cross}.

\paragraph{Boolean Model.}
In the Boolean Model, a document $\mathcal{D}$ is represented by a set of terms it contains, i.e., $\mathcal{D}=\{t_1, t_2, \dots, t_n\}$, and a query $\mathcal{Q}$ is represented as a similar boolean expression of terms. 
A document is considered relevant to a query only if a logical implication $\mathcal{D} \rightarrow \mathcal{Q}$ holds, i.e., the document representation logically implies the query expression. 
This basic model can be extended by incorporating term weighting, allowing both queries and documents to be represented as sets of weighted terms. Consequently, the logical implication $\mathcal{D} \rightarrow \mathcal{Q}$ is also weighted. Common approaches for this include using a fuzzy set extension of Boolean logic~\citep{radecki1979fuzzy,kraft1983fuzzy} and the $p$-norm~\citep{salton1983extended}.

\paragraph{Vector Space Model.}
In Vector Space Models~\citep[VSMs,][]{salton1975vector}, the queries and documents are represented by vectors, e.g., $\mathcal{Q}=<q_1, q_2, \dots, q_n>$ and $\mathcal{D} = <d_1, d_2, \dots, d_n>$. 
The vector space is defined by a vocabulary of terms $\mathcal{V} = <t_1, t_2, \dots, t_n>$ and each element ($q_i$ or $d_i$, $1 \leq i \leq  n$) in the vectors represents the weight of the corresponding term in the query or the document. 
The weights $q_i$ or $d_i$ could be binary, representing presence or absence. 
Given the vector representations, the relevance score is estimated by a similarity function between the query $\mathcal{Q}$ and the document $\mathcal{D}$.  
The weights $q_i$ or $d_i$ can be determined by more sophisticated schema~\citep{salton1988termweighting}, such as TF-IDF~\citep{sparck1972statistical} and BM25~\citep{robertson2009probabilistic,robertson1995okapi}. This allows for more abundant features that can improve the capacity and accuracy of the models.
Besides, given the vector representations of query $\mathcal{Q}$ and document $\mathcal{D}$, the most commonly used is cosine similarity, defined as:
$$
\text{sim}(\mathcal{Q},\mathcal{D}) = \frac{\mathcal{Q}\cdot\mathcal{D}}{|\mathcal{Q}|\times|\mathcal{D}|}, 
$$
where $\mathcal{Q}\cdot\mathcal{D}$ is the dot product and $|\mathcal{Q}|,|\mathcal{D}|$ denotes the length of the vector. 

\paragraph{Probabilistic Model.}
In the probabilistic model, the relevance score of a document $\mathcal{D}$ to a query $\mathcal{Q}$ depends on a set of events $\{x_i\}_1^{n}$ representing the occurrence of term $t_i$ in this document. 
The simplest probabilistic model is the binary independence retrieval model~\citep{robertson1976relevance}, which assumes terms are independent so only $x_i=1$ and $x_i=0$ exist in the representation. 
Given a set of sample documents whose relevance is judged, the estimation of the relevance score can be derived as 
$$
\text{Score}(\mathcal{Q},\mathcal{D}) \propto \sum_{(x_i=1)\in\mathcal{D}} \log \frac{r_i(T-n_i-R+r_i)}{(R-r_i)(n_i-r_i)}
$$
where $T$ and $R$ are the total number of sampled judged documents and relevant samples, and $n_i$ and $r_i$ denote the number of samples and relevant samples containing $t_i$, respectively.

In contrast, a line of statistical retrieval functions such as TF-IDF~\citep{sparck1972statistical} move beyond binary term indicators by incorporating term frequency (TF) and inverse document frequency (IDF), allowing more nuanced term weighting while still assuming term independence. 
We illustrate the famous BM25 algorithm~\citep{robertson1995okapi}:
$$
\text{BM25}(\mathcal{Q},\mathcal{D}) = \sum_{t_i \in \mathcal{Q} \cap \mathcal{D}} \text{IDF}(t_i) \cdot \frac{f_i \cdot (k_1 + 1)}{f_i + k_1 \cdot \left(1 - b + b \cdot \frac{|\mathcal{D}|}{\text{avgdl}}\right)},
$$
where $f_i$ is the frequency of term $t_i$ in document $\mathcal{D}$, $|\mathcal{D}|$ is the length of the document, $\text{avgdl}$ is the average document length in the collection, and $k_1$ and $b$ are hyperparameters typically set between $[1.2, 2.0]$ and $[0.5, 0.8]$, respectively. The inverse document frequency term is computed as $\text{IDF}(t_i) = \log \frac{N - n_i + 0.5}{n_i + 0.5},$
where $N$ is the total number of documents in the collection and $n_i$ is the number of documents containing term $t_i$.

The smoothing mechanisms~\citep{baeza1999modern} are necessary to deal with zero occurrences of $t_i$.
Except for the binary independence retrieval model, more sophisticated probabilistic models have been proposed in the literature~\citep{wong1989probability,fuhr1992probabilistic}, such as the inter-dependency between terms~\citep{van1979information}.

\paragraph{Statistical Language Model.}
The general idea of a statistical language model is to estimate the relevance score of a document $\mathcal{D}$ to a query $\mathcal{Q}$ via $\mathcal{P}(\mathcal{D}|\mathcal{Q})$~\citep{ponte1998language}. 
Based on Bayes' Rule, $\mathcal{P}(\mathcal{D}|\mathcal{Q})$ can be derived as directly proportional to $\mathcal{P}(\mathcal{Q}|\mathcal{D})\mathcal{P}(\mathcal{D})$. 
For simplification, most studies assume a uniform distribution for $\mathcal{P}(\mathcal{D})$. 
The main focus is on modeling $\mathcal{P}(\mathcal{Q}|\mathcal{D})$ as a ranking function. 
By treating the query as a set of independent terms $\mathcal{Q}=\{t_i\}_{i=1}^n$, we have $\mathcal{P}(\mathcal{Q}|\mathcal{D})=\prod_{t_i \in \mathcal{Q}}\mathcal{P}(t_i|\mathcal{D}).$
The probability $\mathcal{P}(t_i|\mathcal{D})$ is determined using a statistical language model $\theta_{D}$ that represents the document. The relevance is then estimated by log-likelihood:
$\text{Score}(\mathcal{Q},\mathcal{D}) = \log\mathcal{P}(\mathcal{Q}|\theta_{D}) = \sum_{t_i \in \mathcal{Q}}\log\mathcal{P}(t_i|\theta_{D}),$
where the estimation of the language model $\theta_{D}$ is usually achieved by maximum likelihood.

The statistical language models for IR~\citep{miller1999hidden,berger1999information,song1999general,hiemstra1999twentyonetrec7} also encounter the problem of zero occurrences of a query term $t_i$, i.e., the probability $\mathcal{P}(\mathcal{Q}|\theta_{D})$ becomes zero if a query term $t_i$ does not appear in the document. This is too restrictive for IR, as a document can still be relevant even if it contains only some of the query terms. To address this zero-probability issue, smoothing techniques are applied, assigning small probabilities to terms that do not appear in the document. The principle behind smoothing is that any text used to model a language captures only a limited subset of its linguistic patterns (or terms, in this case).
The commonly used smoothing methods~\citep{zhai2004study,zhai2008statistical} include Jelinek-Mercer smoothing~\citep{jelinek1980interpolatedestimation}, Dirichlet smoothing~\citep{mackay1995hierarchicaldirichlet}, etc.

These foundational models, while useful, share a common characteristic: they rely on heuristic-based scoring functions derived from statistical properties of terms. Although their parameters can be tuned (e.g., $k_1$ and $b$ in BM25), the functional form of the model is fixed. A natural evolution is to frame ranking as a supervised machine learning task, where a model learns to combine various signals of relevance automatically from labeled data. This approach allows for the systematic combination of not only scores from traditional models like BM25 but also a multitude of other features describing the query, the document, and their interaction. This paradigm shift from hand-crafted formulas to learned functions is the core idea behind Learning-to-Rank models, which we discuss next.

\section{Learning-to-Rank Model Architectures}
\label{sec:ltr}
Different from traditional IR models that rely on heuristic-based scoring formulas (\cref{sec:traditional}), Learning-to-Rank (LTR) frames ranking as a supervised machine learning problem~\citep{liu2009ltr}. The core idea is to train a model that can optimally combine a wide array of signals, or ``features'', to predict the relevance of documents to a query.
For each query-document pair $(\mathcal{Q}_i, \mathcal{D}_i)$, a $k$-dimensional feature vector $\mathbf{x}_i \in \mathbb{R}^{k}$ is extracted, and a relevance label $\mathbf{y}_i$ (e.g., from human judgments) is provided. 
The goal is to learn a ranking model $f$ parameterized by $\theta$ that minimizes an empirical loss $l(\cdot)$ on a labeled training set $\Psi$: 
$$\mathcal{L} = \frac{1}{|\Psi|} \sum_{(\mathbf{x}_i, \mathbf{y}_i) \in \Psi} l(f_{\theta}(\mathbf{x}_i), \mathbf{y_i}). $$

LTR methods are typically categorized into three main approaches based on their input and loss function: pointwise, pairwise, and listwise.

\paragraph{Feature Engineering in LTR.}
A cornerstone of traditional LTR is the meticulous engineering of the feature vector $\mathbf{x}_i$. These features are designed to capture diverse aspects of relevance and can be grouped into several categories: (1) \textbf{Query-based features}, such as the number of terms in the query; (2) \textbf{Document-based features}, which are query-independent, such as document length, PageRank~\citep{brin1998anatomy}, or the number of incoming URL links; and (3) \textbf{Query-document interaction features}, which form the largest and most critical group. This category includes scores from traditional IR models like BM25 and language models, counts of matching terms, proximity features measuring how close query terms are in the document, and various TF-IDF-related statistics. The power of LTR lies in its ability to learn complex, non-linear combinations of these diverse signals, moving beyond what a single hand-tuned formula could achieve.

\begin{table*}[t!]
\centering
% \small
\caption{A list of learning-to-rank works and their model architectures.}
\label{tab:ltr_model_appendix}

\rowcolors{2}{gray!15}{white} % Alternating row colors

\begin{tabular}{lll}
\toprule
\textbf{Name} & \textbf{Backbone Architecture} & \textbf{Loss Function} \\
\midrule
\textsc{MART}~\citep{friedman2001greedy} & Boosting & Pointwise \\
\textsc{RankBoost}~\citep{freund2003efficient} & Boosting & Pairwise \\
\textsc{RankNet}~\citep{burges2005learning} & Neural Network & Pairwise \\ 
\textsc{RankSVM}~\citep{joachims2006training} & SVM & Pairwise \\
\textsc{LambdaRank}~\citep{burges2006learning} & Neural Network & Pairwise \\
\textsc{ListNet}~\citep{cao2007learning} & Neural Network & Listwise \\
\textsc{SoftRank}~\citep{taylor2008softrank} & Neural Network & Listwise  \\
\textsc{ListMLE}~\citep{xia2008listwise} & Linear & Listwise\\
\textsc{LambdaMART}~\citep{burges2010ranknet} & GBDT & Listwise \\
\textsc{ApproxNDCG}~\citep{qin2010general} & Linear & Listwise \\
\textsc{DLCM}~\citep{ai2018learning} & Neural Network & Listwise  \\
\textsc{GSF}~\citep{ai2019learning} & Neural Network & Listwise \\
\textsc{ApproxNDCG}~\citep{bruch2019revisiting} & Neural Network & Listwise \\
\textsc{SetRank}~\citep{pang2020setrank} & Self Attention Blocks & Listwise\\

\bottomrule
\end{tabular}

\end{table*}

\subsection{Pointwise, Pairwise, and Listwise Approaches with ML Models}
The pointwise approach is the simplest, treating each document independently. It frames the problem as a regression or classification task, where the model $f(\mathbf{x}_i)$ aims to predict the exact relevance label $\mathbf{y}_i$. While straightforward, this approach ignores the crucial fact that ranking is about the relative order of documents, not their absolute scores~\citep{burges2010ranknet}.

The pairwise approach addresses this by focusing on the relative order of document pairs. Given two documents $\mathcal{D}_i$ and $\mathcal{D}_j$ for the same query, the goal is to predict which one is more relevant. This transforms ranking into a binary classification problem. Seminal pairwise models include \textsc{RankSVM}~\citep{joachims2006training}, which adapts the Support Vector Machine framework to maximize the number of correctly ordered pairs, and \textsc{RankNet}~\citep{burges2005learning}, which uses a neural network and a probabilistic cost function based on pairwise logistic loss. While more aligned with the nature of ranking than the pointwise approach, pairwise methods still do not directly optimize the list-based evaluation metrics (e.g., nDCG, MRR) that are standard in IR evaluation.

The listwise approach directly tackles this issue by defining the loss function over an entire list of documents for a given query. These methods aim to directly optimize ranking metrics. A pivotal line of work began with \textsc{LambdaRank}~\citep{burges2006learning}, which observed that for gradient-based optimization, one only needs the gradients of the loss function. It introduced ``Lambda gradients'', which are derived from the change in an IR metric (like nDCG) when two documents in a ranked list are swapped. This technique was then combined with Multiple Additive Regression Trees (MART), a Gradient Boosted Decision Tree (\textsc{GBDT}) algorithm~\citep{friedman2001greedy}, to create \textsc{LambdaMART}~\citep{wu2010adapting}. Due to its strong performance and robustness, \textsc{LambdaMART} became a dominant industry standard for many years~\citep{ke2017lightgbm}.

\subsection{Neural LTR Models}
While \textsc{LambdaMART} represents a peak for GBDT-based LTR, early works also explored neural networks for the ranking function $f_{\theta}$.
\textsc{RankNet} and \textsc{LambdaRank} both parameterized the LTR model with neural networks. More recent works such as \textsc{GSF}~\citep{ai2019learning} and \textsc{ApproxNDCG}~\citep{bruch2019revisiting} have continued this trend, using multiple fully connected layers and designing differentiable approximations of IR metrics. Other architectures like \textsc{DLCM}~\citep{ai2018learning}, based on RNNs, and \textsc{SetRank}~\citep{pang2020setrank}, using self-attention, explore ways to model the entire document list jointly. A rigorous benchmark by~\citet{qin2021neural} compared the performance of these modern neural ranking models against strong GBDT-based baselines. A summary of LTR models and their backbone architectures is provided in~\cref{tab:ltr_model_appendix}.

\subsection{Orthogonal Directions}
Beyond core model architectures, LTR research has explored many other important directions. A significant portion of the literature focuses on loss functions and feature transformations~\citep{qin2021neural,bruch2019revisiting,burges2010ranknet}. Other critical areas include developing methods for unbiased relevance estimation from biased user feedback (e.g., clicks)~\citep{joachims2017unbiased,ai2018unbiased,wang2018position,hu2019unbiased} and jointly optimizing for both effectiveness and fairness in ranking systems~\citep{singh2018fairness,biega2018equity,yang2023fara,yang2023vertical,yang2023marginal}. We omit detailed discussions here and refer readers to the original papers and prior surveys~\citep{liu2009ltr,li2011learning}.

Despite their success, traditional LTR models have a fundamental ceiling. 
Their reliance on handcrafted features prevents them from operating in an end-to-end manner and, more importantly, limits their ability to bridge the \textbf{lexical gap}\,---\,the difference between the words in a query and the semantically related words in a relevant document. Their understanding is based on pre-defined statistical signals, not the underlying meaning of the text. This limitation created a clear need for a new class of models capable of learning semantic representations directly from raw text, setting the stage for the rise of neural ranking (\cref{sec:neural_ranking}).

\section{Neural Ranking Models}
\label{sec:neural_ranking}
\begin{figure*}[t]
    \begin{subfigure}{0.5\linewidth}
        \centering
        \includegraphics[width=\linewidth]{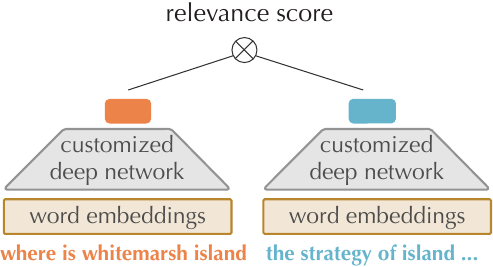}
        \caption{Representation-based neural reranker}
        \label{fig:architecture:representation}
    \end{subfigure}
    \hfill
    \begin{subfigure}{0.5\linewidth}
        \centering
        \includegraphics[width=\linewidth]{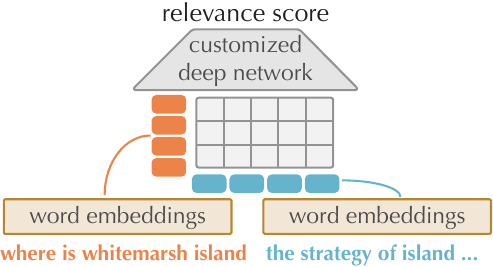}
        \caption{Interaction-based neural reranker}
        \label{fig:architecture:interaction}
    \end{subfigure}

\caption{Illustration on neural ranking models.
Brown boxes indicate uncontextualized word embeddings (e.g., Word2vec).
}
\label{fig:architecture}
\vspace{0pt}
\end{figure*}

Neural ranking models emerged to directly address the key limitations of LTR (\cref{sec:ltr}). By learning semantic representations directly from raw text, they could automatically bridge the lexical gap\,---\,for instance, recognizing that a query for ``computer'' is conceptually related to a document about a ``PC'' without relying on term overlap. This end-to-end approach simultaneously eliminated the laborious process of manual feature engineering, shifting the paradigm from engineering statistical signals to learning semantic patterns from data.\footnote{For structural clarity, this section focuses on neural information retrieval, covering neural network–based retrieval models developed prior to the advent of pre-trained transformers. For a more comprehensive treatment, we refer readers to the dedicated surveys~\citep{onal2018neural,mitra2018introduction,xu2018deep}.}

Depending on how queries interact with documents during network processing, neural ranking models can be roughly divided into \textbf{representation-based models} and \textbf{interaction-based models}~\citep{guo2016deep}. This division reflects a fundamental tradeoff between efficiency and matching depth. Representation-based models pre-encode documents into vectors offline, enabling highly efficient retrieval suitable for first-pass ranking. In contrast, interaction-based models process the query and document together, allowing for deeper and more precise matching at a higher computational cost, making them ideal for reranking a smaller set of candidates.

\subsection{Representation-Based Models} 
This genre of models can be regarded as an extension of Vector Space Models (\cref{sec:traditional}), which independently encode queries and documents into a shared latent vector space, where relevance is determined through simple comparison functions such as cosine similarity or dot product, as illustrated in~\cref{fig:architecture:representation}. This approach maintains clear separation between query and document processing, with no interaction occurring during the encoding procedure.
The core architectural challenge is designing an encoder network that transforms a variable-length sequence of term embeddings into a single, fixed-size semantic vector.

The Deep Structured Semantic Model~\citep[\textsc{DSSM,}][]{huang2013dssm,gao2014modelinginterestingness} is an early example. 
It utilizes word hashing (a technique to manage large vocabularies by grouping words into a smaller number of hash buckets) and multilayer perceptrons (MLPs) to independently encode term vectors of queries and documents, enabling the computation of ranking scores based on the cosine similarity of their embeddings. 
Later works modify \textsc{DSSM}'s encoder network to better capture richer semantic and contextual information. Convolutional \textsc{DSSM}\citep[\textsc{C-DSSM},][]{shen2014cdssm} leverages a convolutional neural network (CNN) architecture. Specifically, it applies 1D convolutions over the sequence of word embeddings, allowing the model to learn representations for n-grams and local phrases. A max-pooling layer then selects the most salient local features to form the final document vector.
Another variant of \textsc{DSSM} replaces MLPs with recurrent layers such as Long Short-Term Memory (LSTM) network~\citep{hochreiter1997long,palangi2016deep,wan2016deep,cohen2016end2end} or tree-structured networks~\citep{tai-etal-2015-improvedsemanticrepresentations}. The LSTM processes the text sequentially, and its recurrent nature allows it to capture word order and long-range dependencies across the entire text, with the final hidden state often used as the comprehensive representation for the query or document. Based on the assumption of documents' hierarchical structure, \cite{yang-etal-2016-hierarchical-attention-networks,Song2018DeepHierarchicalAttentionNetworks,zhu2019hierarchicalattentionretrieval} use attention~\citep{bahdanau2014neural} to model token, phrase and sentence representations for enhanced document/passage representations.

In line with these works, the NLP community has also extensively investigated passage/document representations. \cite{le2014distributedrepresentations} proposed Paragraph Vectors (\textsc{Doc2Vec}), an unsupervised algorithm that learns fixed-length feature representations from variable-length pieces of texts, which is based on single-hidden-layer neural network. \cite{kim-2014-convolutional} studied CNN for sentence representations, while \cite{wieting2016universalparaphrasticsentenceembeddings} proposed to use LSTM network. \cite{arora2017asimplebuttoughtobeatbaseline} reported a weighted average of pre-trained word embeddings fine-tuned with unsupervised random walk algorithm can outperform more complex neural networks on the sentence similarity task. However, weighted averaging word embeddings ignores the word order, which fails to capture the rich, contextual information in longer, more complex documents.

To summarize, representation-based models excel in scenarios requiring global semantic understanding and offer significant computational advantages through their ability to pre-compute document representations offline~\citep{Guo2019ADeepLook}. However, these approaches also face inherent limitations due to their reliance on fixed-size embedding vectors, which can struggle to capture all relevant information from the original text and may not effectively handle precise lexical matching requirements. These limitations are the focus of interaction-based models.

\subsection{Interaction-Based Models} Different from representation-based models, interaction-based models (\cref{fig:architecture:interaction}) process queries and documents jointly through neural networks. 
Instead of compressing each text into a single vector, they first build a detailed, low-level interaction representation between the query and the document, and then use neural networks to learn hierarchical matching patterns from this representation.
The model's output is typically a scalar relevance score of the input query-document pair.
Various network architectures have been proposed under this paradigm. \textsc{MatchPyramid}~\citep{pang2016matchpyramid} employs CNN over the interaction matrix between query and document terms. The interaction matrix is treated as an image, allowing the CNN with its 2D filters to capture local matching patterns (e.g., phrases or bigrams matching) through convolution and pooling operations~\citep{hu2014convolutional}. 

Building upon the concept of interaction-focused models, \cite{guo2016deep} highlighted  the importance of exact term matches in neural ranking models and proposed the Deep Relevance Matching Model (\textsc{DRMM}). Rather than a single interaction matrix, \textsc{DRMM} creates a matching histogram for each query term. This histogram discretizes the similarity scores against all document terms into bins, effectively capturing the distribution of matching signals (e.g., how many terms in the document are an exact match, a strong semantic match, or a weak match to a given query term). An MLP then learns the relevance contribution from these histogram features.

Kernel-Based Neural Ranking Model~\cite[\textsc{K-NRM},][]{xiong2017knrm} further advances interaction-based approaches. It employs radial basis function (RBF) kernels to transform the query-document interaction matrix into a more informative feature representation. Each kernel corresponds to a certain similarity level (e.g., ``exact match'', ``strong match'', ``weak match''). The model uses these kernels to produce ``soft-TF'' counts for each query term\,---\,counting how many document words match the query term at each similarity level. These soft-match features are then aggregated and fed into a simple feed-forward network to compute the final relevance score.
This kernel-based mechanism enables models to capture nuanced matching features, enhancing their ability to model complex query-document interactions.
\textsc{Conv-KNRM}~\citep{dai2018cknrm} later extends it to convolutional kernels to capture n-gram level soft matches, further improving matching granularity.

In line with these works, the interaction matrix-based approach have been explored for short text matching~\citep{lu2013deeparchitectureformatchingshorttexts,yin2016abcnn,yang2016anmm} as well as long document ranking~\cite[\interalia]{mitra2017duet,hui-etal-2017-pacrr,hui2018copacrr}. Multi-Perspective CNN approaches compare sentences via diverse pooling functions and filter widths to capture multiple perspectives between texts~\cite{he-etal-2015-multiperspective}. \textsc{aNMM}~\citep{yang2016anmm}, as an example of attention-based methods, computes passage terms' attention weights over query terms using a query attention network and achieves performance improvement compared to CNN-based baseline~\citep{severyn2015learning}.
Term adjacency and positional information represent another important dimension of interaction modeling. Models such as \textsc{MatchPyramid}, \textsc{PACRR}, and \textsc{ConvKNRM} capture term adjacency patterns and position-dependent interactions~\citep{pang2016matchpyramid,hui-etal-2017-pacrr,dai2018cknrm}.

The interaction functions in these models can be categorized as either non-parametric (using traditional similarity measures like cosine similarity, dot product, or binary indicators) or parametric (learning similarity functions from data through neural networks)~\citep{Dong2022DisentangledGR}. While interaction-focused models require one forward pass through the entire model for each potentially relevant document, making them computationally more expensive than representation-focused approaches, they typically achieve superior ranking quality due to their ability to capture fine-grained matching signals.
We list some representative works in~\cref{tab:neural_ranking_appendix} and direct readers to these works for architectural details. 

\subsection{Hybrid Models} Recognizing the complementary strengths of representation-focused and interaction-focused architectures, researchers have proposed hybrid models that combine the efficiency of representation-based methods with the effectiveness of interaction-based approaches. These models represent a third category in neural ranking architectures, alongside the two primary approaches.

The most notable example is \textsc{DUET}~\citep{mitra2017duet}, which employs two separate deep neural networks operating in parallel. One network performs local interaction-based matching similar to interaction-focused models, while the other learns distributed representations for query and document separately, similar to representation-focused approaches. The term interaction matrix between query and document feeds into the exact matching layers, while term embeddings of the input sequence enter the semantic matching layers. The outputs from both networks are then combined using a fully connected network to produce the final ranking score.

Different from the metric learning theme of representation-based models, a line of works formulates the ranking problem as a classification problem (commonly referred to as \textbf{Extreme Label Classification}, or \textbf{XMC}), where the input is the query, and the output is a probability distribution over the corpus, where each document is a unique ``class'' or ``label''. Instead of optimizing the similarity between query and document representations, XMC models aim to predict the correct subset of relevant document IDs~\citep{prabhu2014fastxml,jain2016extreme,liu2017deeplearningforextrememultilabeltextclassification,jain2019slice}. In the inference time, XMC methods use tree-based hierarchies or cluster-based sampling to quickly narrow down the search path to the likely labels without scanning every candidates~\citep{prabhu2018parabel,you2019attentionxml}. \textsc{AttentionXML}~\citep{you2019attentionxml} uses an attention mechanism to focus on specific parts of the input text that are most relevant to the label's semantic meaning, and thus can be considered a hybrid model. We refer readers to~\citep{dasgupta2023reviewofextrememultilabelclassification} for a comprehensive review of XMC methods.

This hybrid architecture demonstrates that combining distributed representations with traditional local representations is favorable, with the combined approach significantly outperforming either neural network individually. More recent hybrid approaches have focused on reducing computational costs while maintaining effectiveness, with some models incorporating cached token-level representations to enable faster query-document interactions when document representations are pre-computed~\citep{Wrzalik2020CoRTComplementaryRankings}. The success of hybrid models has established that interaction-based and representation-based approaches can be effectively combined for further improvements in ranking performance~\citep{Liu2018EntityDuetNeuralRanking}.

\begin{table*}[t!]
\centering
% \small
\caption{A list of neural ranking models and their model architectures.}
\label{tab:neural_ranking_appendix}

\rowcolors{2}{gray!15}{white} % Alternating row colors
\resizebox{1\textwidth}{!}{
\begin{tabular}{llll}
\toprule
\textbf{Name} & \textbf{Architecture} & \textbf{Backbone}  & \textbf{Embeddings} \\
\midrule
\textsc{DSSM}~\citep{huang2013dssm} & Representation-based & MLP & Semantic Hashing \\
\textsc{CDSSM}~\citep{shen2014cdssm} & Representation-based & CNN &  Semantic Hashing \\
\textsc{CLSM}~\citep{shen2014clsm} & Representation-based & CNN & Semantic Hashing \\
\textsc{ARC-I}~\citep{hu2014convolutional} & Representation-based & CNN &  Word2Vec \\
\cite{tai-etal-2015-improvedsemanticrepresentations} & Representation-based & Tree-structured LSTM & GloVe \\
\textsc{LSTM-RNN}~\citep{palangi2016deep} & Representation-based & LSTM &  Randomly Initialized \\
\textsc{MV-LSTM}~\citep{wan2016deep} & Representation-based & Bi-LSTM &  Word2Vec \\
\textsc{DESM}~\cite{nalisnick2016dualwordembeddings} & Representation-based & MLP & Randomly Initialized \\
\cite{lu2013deeparchitectureformatchingshorttexts} & Representation-based & MLP & Randomly Initialized \\
\textsc{ARC-II}~\citep{hu2014convolutional} & Interaction-based & CNN &  Word2Vec \\
\textsc{MatchPyramid}~\citep{pang2016matchpyramid} & Interaction-based & CNN &  Randomly Initialized \\
\textsc{DRMM}~\citep{guo2016deep} & Interaction-based & MLP & Word2Vec \\
\textsc{ABCNN}~\citep{yin2016abcnn} & Interaction-based & CNN + Attention & Word2Vec \\
\textsc{aNMM}~\citep{yang2016anmm} & Interaction-based & Attention & Word2Vec \\
\textsc{DESM}~\citep{nalisnick2016dualwordembeddings} & Interaction-based & MLP &  Word2Vec \\
\textsc{K-NRM}~\citep{xiong2017knrm} & Interaction-based & MLP + RBF kernels &  Word2Vec \\
\textsc{Conv-KNRM}~\citep{dai2018cknrm} & Interaction-based & CNN & Word2Vec \\
\textsc{PACRR}~\citep{hui-etal-2017-pacrr} & Interaction-based & CNN + RNN & Word2Vec \\
\textsc{Co-PACRR}~\citep{hui2018copacrr} & Interaction-based & CNN & Word2Vec \\
\textsc{TK}~\citep{hofstatter2020interpretable} & Interaction-based & Transformer + Kernel &  GloVe \\
\textsc{TKL}~\citep{hofstatter2020improving} & Interaction-based & Transformer + Kernel &  GloVe \\
\textsc{NDRM}~\citep{mitra2021improving} & Interaction-based & Conformer + Kernel & \textsc{BERT} \\

\bottomrule
\end{tabular}
}

\end{table*}

\subsection{Orthogonal Directions}
In addition to the development of network architecture, pre-trained embeddings~\citep{salakhutdinov2009semantichashing,mikolov2013efficient,pennington-etal-2014-glove,le2014distributedrepresentations} provide semantic-based term representations to enable neural ranking models to focus on learning relevance matching patterns, improving training convergence and retrieval performance on both representation-based and interaction-based models~\citep{levy-etal-2015-improving}. Both GloVe~\citep{pennington-etal-2014-glove} and Word2Vec~\citep{mikolov2013efficient} learn dense vector representations for each vocabulary term from large-scale text corpora. By initializing the embedding layer with these pre-trained vectors, models start with a strong semantic foundation, which proved crucial for performance, especially on smaller training datasets~\citep{guo2016semantic}. Interaction-based models with cross-lingual word embeddings~\citep{joulin2018loss} for cross-lingual reranking have also been explored~\citep{yu2020study}.
\cref{tab:neural_ranking_appendix} shows a list of neural ranking models and backbone architectures. Researchers have explored different backbone neural network architectures in this era, including Convolutional Neural Network~\citep[CNN,][]{lecun1989backpropagation}, Long Short Term Memory~\citep[LSTM,][]{hochreiter1997long} and kernel methods~\citep{vert2004primer,chang2010training,xiong2017knrm}.

Notably, a line of research explores integrating kernel methods with the Transformer architecture~\citep{vaswani2017attention}. 
The main distinction between this line of research and the models discussed in~\cref{sec:transformer} is that the transformer modules here are not pre-trained on large-scale corpora like Wikipedia and C4~\citep{devlin-etal-2019-bert,raffel2020transfer}. We consider this line of research as an intersection between neural ranking models (\cref{sec:neural_ranking}) and retrieval with pre-trained transformers (\cref{sec:transformer}).
\textsc{TK}~\citep{hofstatter2020interpretable} uses a shallow transformer neural network (up to 3 layers) to encode the query $\mathcal{Q}$ and document $\mathcal{D}$ separately. After encoding, the contextualized representations are input to an interaction module inspired by K-NRM, where RBF kernels are used to create soft-match features from the contextualized embeddings. This fusion of a transformer encoder with a kernel-based interaction mechanism allowed the model to achieve better performance-efficiency tradeoff compared to \textsc{BERT}-based reranker~\citep{nogueira2019multi}.
The main bottleneck of applying transformer architectures to long document reranking is $O(n^2)$ time complexity, where $n$ denotes the document length. \textsc{TKL}~\citep{hofstatter2020local} further improves upon $\textsc{TK}$ with a local attention mechanism and leads to performance improvement on long document ranking.

The neural ranking models described above, particularly later developments like \textsc{TK} and \textsc{TKL}, demonstrated the potential of the transformer's attention mechanism for modeling relevance. However, the true paradigm shift occurred when the IR community moved from using these architectures trained from scratch to leveraging massive, pre-trained transformer models like BERT~\citep{devlin-etal-2019-bert} and its variants~\citep{liu2019roberta,sun2019ernie,lan2020albert,beltagy2020longformer}. This marked a fundamental change in approach: instead of designing novel, task-specific network backbones (e.g., CNNs, LSTMs) on top of static word embeddings, research shifted to fine-tuning a single, powerful, and deeply contextualized architecture for IR tasks. This new foundation did not eliminate the core architectural tradeoffs but rather recast them in a more powerful form, leading to the development of cross-encoder rerankers and bi-encoder retrievers, which we explore next.

\section{IR with Pre-Trained Transformers}
\label{sec:transformer}
\begin{figure*}[t]
    \begin{subfigure}{0.48\linewidth}
        \centering
        \includegraphics[width=\linewidth]{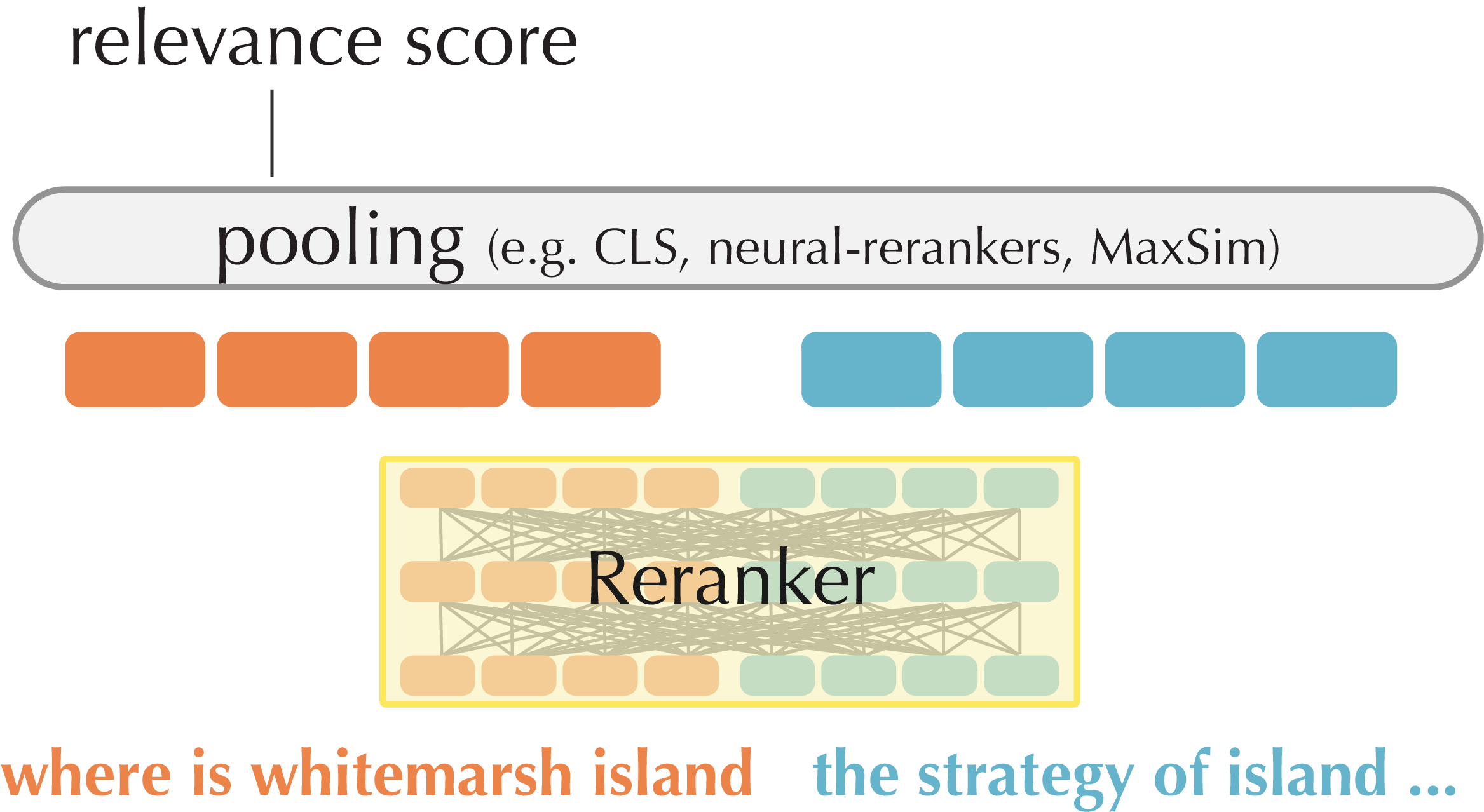}
        \caption{Transformer-based reranker}
        \label{fig:architecture:transformer-reranker}
    \end{subfigure}
    \hfill
    \begin{subfigure}{0.48\linewidth}
        \centering
        \includegraphics[width=\linewidth]{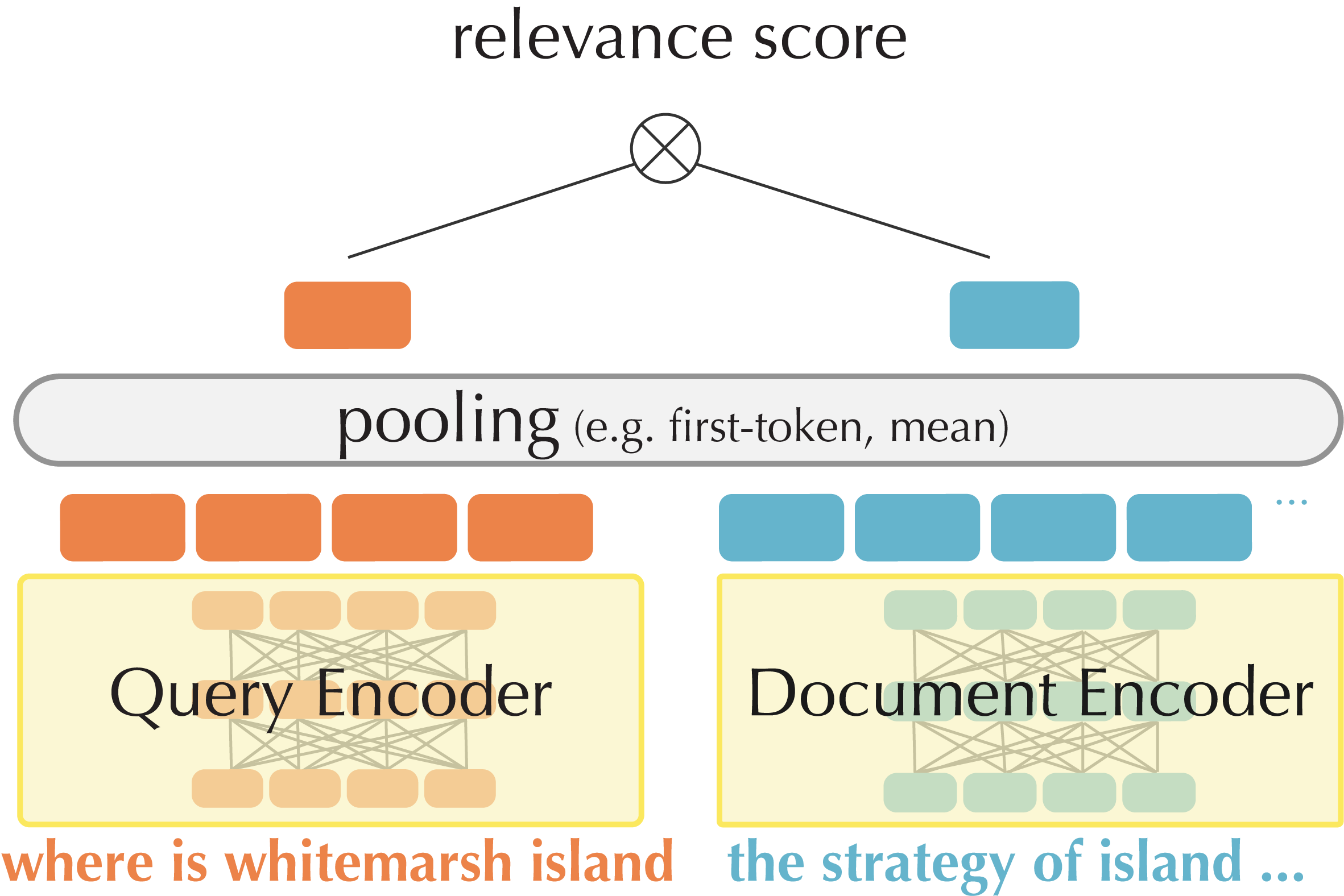}
        \caption{Learned dense retrieval. ($w$ are token-level scalar weights)}
        \label{fig:architecture:single}
    \end{subfigure}
    \hfill
    \begin{subfigure}{0.48\linewidth}
        \centering
        \includegraphics[width=\linewidth]{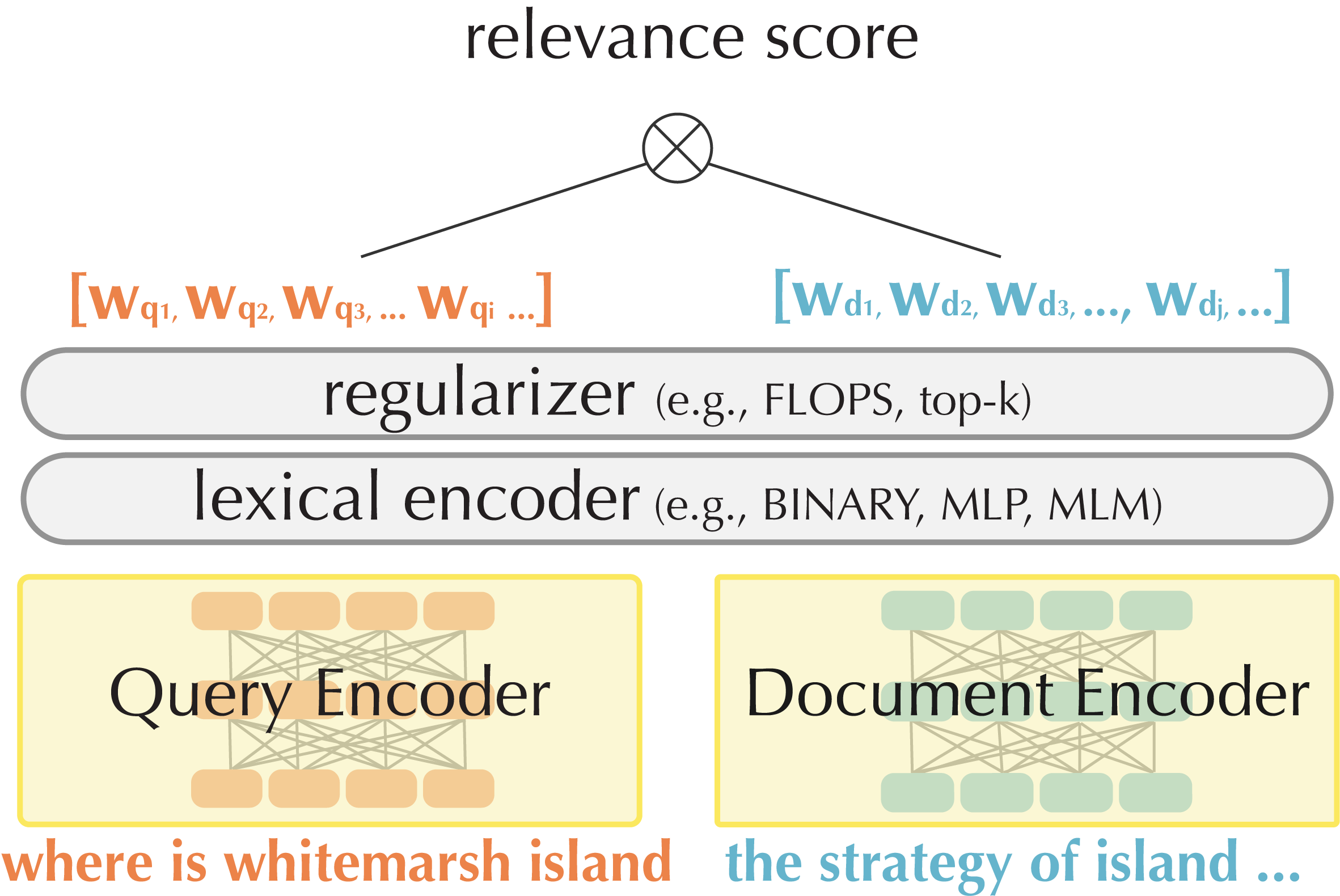}
        \caption{Learned sparse retriever}
        \label{fig:architecture:lsr}
    \end{subfigure}
    \hfill
    \begin{subfigure}{0.48\linewidth}
        \centering
        \includegraphics[width=\linewidth]{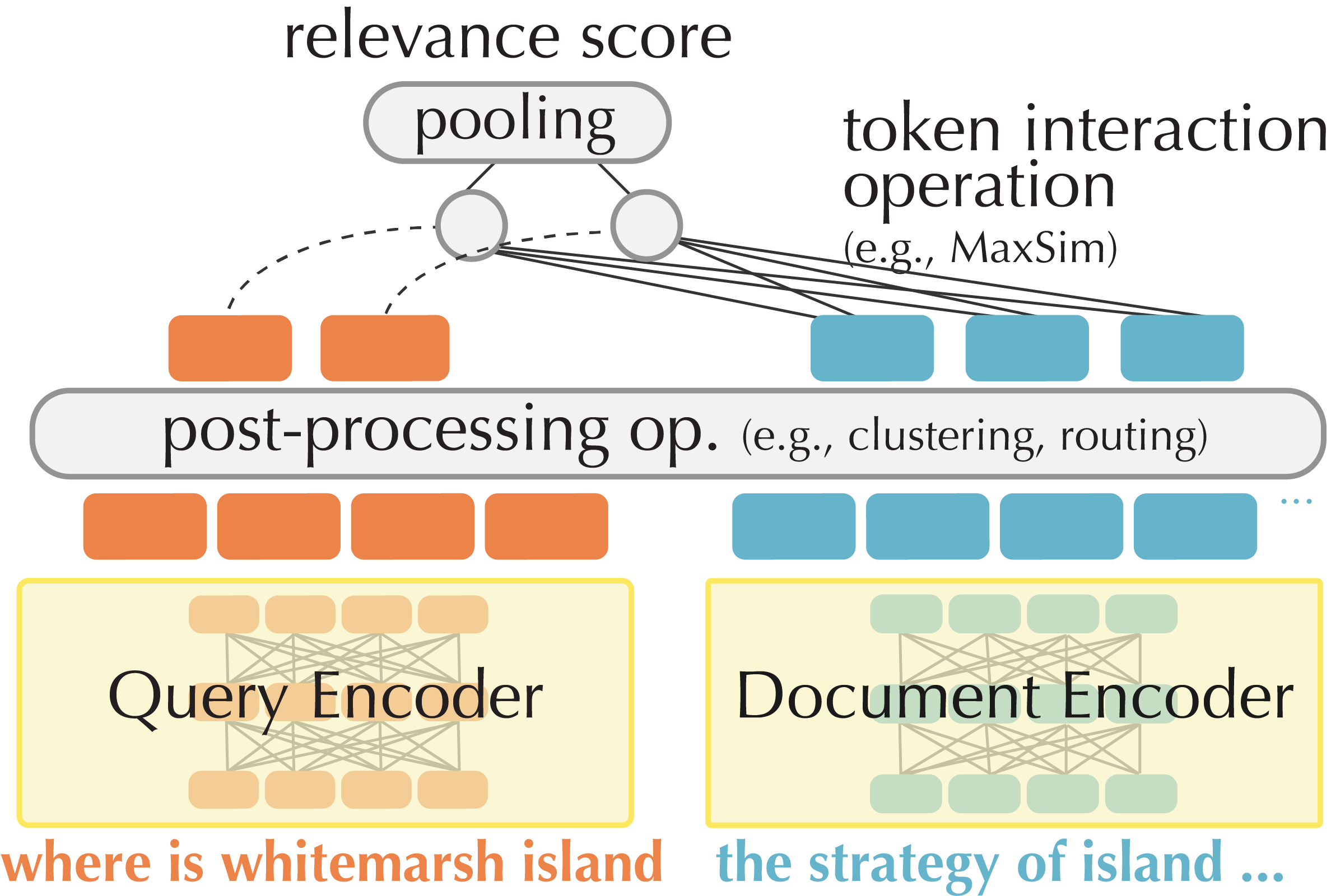}
        \caption{Multi-vector representations}
        \label{fig:architecture:multi}
    \end{subfigure}
\caption{
Illustration on transformer-based retrieval and reranking models.
Yellow boxes indicate pretrained Transformers (e.g., BERT).
Query text, embeddings, and associated weights are color-coded in orange, whereas document representations are color-coded in blue.
}
\label{fig:architecture-transformer}
\vspace{0pt}
\end{figure*}

{\bert}~\citep{devlin-etal-2019-bert} revolutionized research in both natural language processing (NLP) and information retrieval (IR). Its success is largely attributed to two key factors: (1) the Multi-Head Attention (MHA) architecture~\citep{vaswani2017attention}, which enables high-dimensional, contextualized token representations; and (2) large-scale pre-training, which equips \bert~with the ability to capture rich semantics and world knowledge. The expressive power of \bert~has been extensively analyzed in prior work, e.g., \citep{rogers-etal-2020-primer,tenney-etal-2019-bert,clark2019does}.

\paragraph{High-level architectural families.}
Before discussing specific IR modeling architectures, it is useful to understand the architectural families of transformers, as they have distinct implications for IR tasks:
\begin{itemize}
\item \textbf{Encoder-only models} (e.g., \textsc{BERT}~\citep{devlin-etal-2019-bert}, \textsc{RoBERTa}~\citep{liu2019roberta}) use a bidirectional self-attention mechanism, allowing each token's representation to be informed by the entire input sequence (both left and right context). This deep contextual understanding makes them naturally suited for representation-focused tasks. In IR, they have been the workhorses for building powerful bi-encoder retrievers and cross-encoder rerankers.
\item \textbf{Decoder-only models} (e.g., proprietary \textsc{GPT} series~\citep{radford2019language,brown2020language,openai2024gpt4ocard}, open-weight \textsc{Llama}~\citep{touvron2023llama}, \textsc{Mistral}~\citep{Jiang2023Mistral7}) use a unidirectional (causal) self-attention mechanism, where each token can only attend to previous tokens in the sequence. This architecture is optimized for next-token prediction and, by extension, text generation. Their application to representation tasks like retrieval is less direct and often requires architectural adaptations to create meaningful summary vectors from their unidirectional hidden states.
\item \textbf{Encoder-decoder models} (e.g., \textsc{T5}~\citep{raffel2020transfer}, \textsc{BART}~\citep{lewis-etal-2020-bart}) combine both architectures. The encoder processes the input sequence bidirectionally to create a rich representation, which then conditions the decoder to generate an output sequence autoregressively. This ``sequence-to-sequence'' design makes them highly versatile. In IR, they can be framed as rerankers (generating a ``relevant'' or ``irrelevant'' token), or as generative retrievers that directly generate document identifiers.
\end{itemize}

This section discusses IR architectures based on pre-trained transformers, with a focus on {\bert}-type encoder models, which is to be distinct from encoder-decoder models and decoder-only models covered in~\cref{sec:llm4ir}. We structure our review around the fundamental architectural tradeoff between interaction depth and computational efficiency. These constraints necessitate two primary paradigms:
\begin{enumerate}
\item \textbf{Cross-Encoder (Deep Interaction):} A single model processes the concatenated query and document as one sequence, allowing every query token to interact deeply with every document token. This provides state-of-the-art ranking quality but is computationally expensive, making it suitable only for reranking.
\item \textbf{Bi-Encoder (Separable Pre-computation):} Separate encoders process the query and document independently to create fixed-size vectors. Since document vectors can be pre-computed offline, this architecture enables extremely fast similarity search suitable for first-stage retrieval.
\end{enumerate}

We first discuss the crucial training strategies used to optimize these two architectures. We then detail the cross-encoder models that perform deep, full interaction, followed by the separable bi-encoder architectures that prioritize efficiency, exploring their dense, sparse, and multi-vector variants. Finally, we discuss advanced hybrid models and orthogonal improvements such as continual training and interpretability.

\subsection{Training Strategies for Transformer-Based IR}
\label{subsec:transformer_training}

Although we aim to disentangle model architectures from training strategies, the co-evolvement of these two areas is a defining pattern of this era. The architectural dichotomy (cross-encoder vs. bi-encoder) has an impactful influence the training methodology, extending the loss function categories discussed in \cref{sec:ltr} into the deep learning paradigm.

\paragraph{Contrastive and Listwise Objectives.}
The application of contrastive learning builds on the principle of the InfoNCE loss~\citep{oord2018representation}, which is derived from Noise-Contrastive Estimation~\citep{gutmann2012noisecontrastiveestimation}. The general goal is to learn a model that distinguishes a ``positive'' sample from a set of ``negative'' samples.

The InfoNCE framework is primarily used to optimize \textbf{bi-encoders} in the dense retrieval setting. In this context, the relevance score $f$ is a simple similarity function (e.g., dot product) between the query and document vectors, $f(\mathcal{Q}, \mathcal{D}) = \mathrm{sim}(\mathbf{v}_{\mathcal{Q}}, \mathbf{v}_{\mathcal{D}})$. For a query $\mathcal{Q}_i$, a positive document $\mathcal{D}_i^+$, and a set of negative documents $\mathcal{D}_i^-$, the bi-encoder is trained to minimize the negative log probability of correctly classifying the positive document:
$$
    \mathcal{L}_{\mathrm{InfoNCE}}=-\frac{1}{|\mathcal{S}|} \sum\limits_{(\mathcal{Q}_i, \mathcal{D}_i^{+})\in \mathcal{S}} \log  \frac{\exp f_{\theta}(\mathcal{Q}_i, \mathcal{D}_i^{+})}{\exp f_{\theta}(\mathcal{Q}_i, \mathcal{D}_i^{+}) + \sum\limits_{\mathcal{D}_j^{-}\in \mathcal{D}_i^{-}} \exp f_{\theta}(\mathcal{Q}_i, \mathcal{D}_j^{-})}
$$
where $\mathcal{S}$ is the training set and $\mathcal{D}_i^{-}$ is the set of sampled negative documents.

For \textbf{cross-encoders}, which compute relevance over a concatenated list, the objective is also listwise but often simplified to a standard Negative Log-Likelihood (NLL) loss over the final softmax probabilities of the candidates, minimizing the distance between the predicted distribution and the ground truth relevance distribution for the entire list.

\paragraph{Hard Negative Mining.}
A critical challenge in training bi-encoders is generating sufficiently difficult negative examples, as random sampling typically yields easy negatives that do not challenge the model effectively. This is where Hard Negative Mining (HNM) becomes essential. HNM strategies ensure that the model is exposed to challenging cases where positive and negative document features are difficult to distinguish. Key strategies include:
\begin{enumerate}
    \item \textbf{In-Batch Negatives (IBN):} Leveraging other queries' positive documents within the same mini-batch as negative examples for the current query. IBN provides a balance of efficiency and difficulty.
    \item \textbf{Lexical Negatives:} Using documents highly ranked by a traditional sparse model (e.g., BM25) but not labeled as relevant.
    \item \textbf{Iterative Hard Negative Mining:} Employing an existing dense retriever to periodically mine difficult negatives from the collection (i.e., documents that the current model mistakenly ranks highly). Seminal methods like \textsc{ANCE}~\citep{Xiong2020ApproximateNN} and \textsc{ADORE}~\citep{zhan2021optimizingdenseretrievalmodel} use this iterative approach to continually feed the model better training data.
\end{enumerate}

\paragraph{Knowledge Distillation.}
To further narrow the effectiveness gap between efficient bi-encoders and powerful cross-encoders, the community widely adopted Knowledge Distillation (KD)~\citep{bucilua2006modelcompression,hinton2015distillingknowledge}. KD is a technique for training a smaller, efficient ``student'' model by transferring knowledge from a larger, more capable ``teacher'' model~\citep{hinton2015distillingknowledge,gou2021knowledgedistillationasurvey}. In IR, KD is used to create fast rerankers or retrievers that approximate the performance of slower, larger models, which is critical for production systems~\citep{hofstatter2020improving,hofstatter2021efficiently,chen2021simplifiedtinybert,kim2023embeddistillgeometricknowledgedistillation,xu2025distillationvscontrastive,zhang2025qwen3embedding,vera2025embeddinggemmapowerfullightweighttext}.

Let $f_t$ denote the teacher model (typically a cross-encoder) and $f_s$ denote the student model (typically a bi-encoder or a smaller cross-encoder). For a given query $\mathcal{Q}$ and a list of candidate texts $\mathcal{D}=\{\mathcal{D}_1, \mathcal{D}_2, \ldots, \mathcal{D}_k\}$, we first compute relevance scores (logits) from both models:
\begin{align*}
\mathbf{z}_t &= [f_t(\mathcal{Q}, \mathcal{D}_1), f_t(\mathcal{Q}, \mathcal{D}_2), \dots, f_t(\mathcal{Q}, \mathcal{D}_k)] \\
\mathbf{z}_s &= [f_s(\mathcal{Q}, \mathcal{D}_1), f_s(\mathcal{Q}, \mathcal{D}_2), \dots, f_s(\mathcal{Q}, \mathcal{D}_k)]
\end{align*}
The student model $f_s$ is trained to mimic the teacher's output distribution over the candidate texts by minimizing the Kullback-Leibler (KL) divergence between the two softened probability distributions:
$$
\mathcal{L}_{\text{KD}} = D_{\mathrm{KL}}\left( \mathrm{softmax}\left(\frac{\mathbf{z}_t}{T}\right) \bigg| \mathrm{softmax}\left(\frac{\mathbf{z}_s}{T}\right) \right)
$$
where $T$ is the temperature hyperparameter. A higher temperature creates a softer probability distribution, which can help in transferring more nuanced information from the teacher. KD thus provides an essential bridge, allowing the efficiency of bi-encoders to approach the effectiveness ceiling set by cross-encoders.

\subsection{Deep Interaction Models: The Cross-Encoder for Reranking}

The most effective application of transformers in IR involves full, deep interaction between query and document tokens. This architecture, known as a cross-encoder~\citep{Humeau2020Poly-encoders}, takes the concatenated sequence of $(\mathcal{Q}, \mathcal{D})$ as a single input. \cite{nogueira2019multi} first employed this approach in their \textsc{monoBERT} model for reranking candidate passages from a first-stage retriever. The model outputs a relevance score $s$ via a linear layer on top of the final {\bert} representation, typically from a linear layer using the \texttt{[CLS]} token's representation (\cref{fig:architecture:transformer-reranker}). 

Conceptually similar to pre-Transformer interaction-based neural ranking models, this schema has proven effective across various pre-trained encoders~\citep{comparing2021zhang}, as well as other transformer architectures (\cref{sec:llm4ir}). However, this cross-encoder schema faces two primary challenges: (1) the fixed context length of models like \bert~(e.g., 512 tokens) makes processing long documents difficult, and (2) relying on a single token's fixed-dimensional representation (e.g., 768-dimensional representation for \bert) may limit the model's expressive power. 

\paragraph{Handling Long Documents.}
Corresponding mitigations for these two challenges have been extensively investigated in literature. \textbf{Chunk-and-aggregate} approaches represent a practical solution to handle long documents that exceed BERT's input constraints by decomposing the ranking problem into passage-level scoring followed by aggregation~\citep{Gao2022LongDocumentReranking}. The fundamental strategy involves splitting documents into fixed-length passages or sentences, applying \bert-based cross-encoders to score each query-passage pair independently, then combining these scores to produce a final document-level relevance score.
Early work in this direction explored sentence-level aggregation, where \textsc{BERT} scores computed at the sentence level were shown to be effective for document ranking~\citep{MacAvaney2020EfficientDocumentReranking,Yilmaz2019CrossDomainMO}. The BERT-MaxP~\citep{Dai2019DeeperTU} approach became particularly influential, where documents are split into fixed-length passages and the maximum passage score serves as the document score.

Two primary aggregation strategies have emerged: (1) score-pooling and (2) representation aggregation. Score-pooling methods apply simple operations like maximum, sum, or first passage scores to combine passage-level relevance scores~\citep{Dai2019DeeperTU}. In contrast, representation aggregation methods address both the long-document problem and the single-vector expressiveness limitation. Instead of collapsing each passage's signal into a single scalar score, these approaches gather the rich, low-dimensional \texttt{[CLS]} token representations from each passage. This collection of vectors forms a more comprehensive and expressive document-level feature set, which is then processed by additional neural networks~\citep{MacAvaney2020EfficientDocumentReranking}. Notable systems like \textsc{PARADE}~\citep{Li2020PARADEPR} employ CNNs and transformers for aggregation, while \textsc{CEDR}~\citep{macavaney2019cedr} pioneered joint approaches that combine BERT outputs with existing neural IR models through averaging.

While chunk-and-aggregate approaches successfully handle long documents, they fundamentally limit query-document interactions to the passage level, creating information bottlenecks where passage scores or low-dimensional representations constrain the model's ability to capture document-wide relevance patterns. \cite{Hofsttter2021IntraDocumentCascading} argue that this tradeoff between scalability and interaction richness remains a defining characteristic of this approach.

\subsection{Efficient Pre-Computation Models: The Bi-Encoder Architecture}
While cross-encoders offer state-of-the-art effectiveness, their computational cost\,---\,requiring a full transformer pass for every $(\mathcal{Q}, \mathcal{D})$ pair\,---\,makes them infeasible for retrieval over large collections. This limitation motivated the development of bi-encoder architectures, which are conceptually similar to representation-based neural ranking models (\cref{sec:neural_ranking}).\footnote{The term ``bi-encoder'' is also known as a two-tower architecture or an embedding model. We use ``bi-encoder'' to contrast with ``cross-encoder'', which takes concatenated input.} 

A bi-encoder uses a backbone network (typically a transformer) to encode the query $\mathcal{Q}$ and document $\mathcal{D}$ separately. The resulting dense vector representations are then used to compute a relevance score with a simple similarity function like dot product or cosine similarity~\citep{Xiong2020ApproximateNN,Karpukhin2020DensePR,gao-etal-2021-simcse}. The key advantage of this separation is efficiency: the entire document collection can be encoded into a vector index offline. At query time, retrieval becomes a fast approximate nearest neighbor (ANN) search problem~\citep{johnson2019billion,malkov2018hnsw} or search with an inverted index data structure~\citep{zobel1998invertedfiles}, avoiding costly neural network inference. 

A notable insight from~\cite{Lin2021AProposedConceptualFramework} is that the bi-encoder framework provides a unifying lens for understanding diverse retrieval approaches. 
Dense retrieval models, learned sparse retrieval models, and traditional bag-of-words approaches such as BM25 can all be interpreted as parametric variations of this architecture. They differ primarily along two axes: (1) the representational basis (dense semantic embeddings versus sparse lexical vectors), and (2) whether those representations are learned from data or manually engineered.
Existing methods based on this bi-encoder architecture vary primarily in their representation format (dense versus sparse), pooling strategies, and training methodologies.

\subsubsection{Learned Dense Retrieval}
Dense retrieval models typically adopt a standardized dual-encoder architecture built on pre-trained transformer models, most commonly BERT in this context. 
The standard formulation uses separate BERT encoders for queries and documents, with a layer-normalized linear projection applied to the token representation: $\text{Encoder}(\cdot) = \text{Linear}\big(\text{BERT}(\cdot)\big)$. The encoder weights can be separate or shared between query and document sides.

The core architectural principle involves encoding queries and documents into low-dimensional dense vectors, typically 768 dimensions matching \bert's hidden size. Rather than using all hidden representations, most models compress the sequence information using a reduction function, usually the token representation or mean pooling of the final transformer layer outputs. This creates a single dense vector representation per text sequence that captures semantic information beyond simple lexical matching. 

Relevance scoring in dense retrieval is performed through simple similarity functions, most commonly dot product or cosine similarity between query and document vectors. This design enables efficient ANN search over pre-computed document representations~\citep{johnson2019billion}, making dense retrieval practical for large-scale collections while maintaining the semantic understanding capabilities of transformer models. The success of this architecture stems from its ability to learn semantic representations that address the vocabulary mismatch problem inherent in traditional sparse retrieval methods~\citep{Lee2019LatentRF,Karpukhin2020DensePR,Xiong2020ApproximateNN,reimers-gurevych-2019-sentence}. Dense retrieval models have demonstrated notable effectiveness improvements over BM25 baselines across various tasks including open-domain question answering and web search. 

\subsubsection{Learned Sparse Retrieval}
Learned sparse retrieval (LSR, \cref{fig:architecture:lsr}) employs the same bi-encoder architecture as dense retrieval but produces fundamentally different representations.\footnote{We focus on learned sparse retrieval within the bi-encoder formulation, thereby excluding approaches based on learned document expansion (e.g., \cite{nogueira2019doc2query,zbib2019neuralnetworklexicaltranslation}) and document term reweighting (e.g., \textsc{DeepCT}~\citep{dai2019context}). For a unified treatment of learned sparse retrieval that encompasses these lines of work, see \cite{mallia2021learning,basnet2024deeperimpact}.} While sharing the transformer backbone, sparse retrieval models encode queries and documents into high-dimensional sparse vectors whose dimensionality typically matches the vocabulary size of the underlying pre-trained model, often containing tens of thousands of dimensions. Each dimension corresponds to a specific vocabulary term, creating an interpretable representation where non-zero weights indicate term importance~\citep{formal21splade,formal21spladev2,nguyen2023unified}. 

Three key architectural constraints distinguish sparse encoders from their dense counterparts. First, sparsity is enforced through explicit regularization techniques, ensuring most term weights remain zero to maintain efficiency~\citep{formal21splade,xu2025csplade,xu2026laconicdenseleveleffectivenessscalable}. Second, all weights must be non-negative to maintain compatibility with traditional inverted index software designed for lexical search systems like Lucene. Third, the high-dimensional vocabulary-aligned vectors enable integration with existing inverted index infrastructure and optimization algorithms~\citep{turtle1995queryevaluation,broder2003efficientqueryevaluation,bruch2024efficient}. Notably, inference-free learned sparse retrieval methods such as \textsc{SPLADE-Doc}~\citep{formal21spladev2,shen2025exploringl0sparsification} eliminates requirement of specialized accelerators such as GPUs, making them highly efficient for inference on multi-core CPU machines.

At a conceptual level, learned sparse retrieval can be viewed as a sophisticated evolution of traditional term weighting schemes, learning context-aware token importance scores from data rather than relying on heuristic formulas~\citep{zamani2018neural,dai2019context,mallia2021learning,yu2024improved,xu2025csplade,xu2026laconicdenseleveleffectivenessscalable}. This approach inherits desirable properties from bag-of-words models such as exact term matching while leveraging the semantic understanding capabilities of pretrained transformers. Notable implementations include \textsc{SPLADE}~\citep{formal21spladev2}, \textsc{DeepImpact}~\citep{mallia2021learning}, and \textsc{uniCOIL}~\citep{lin2021few}, which demonstrate that transformer-based sparse representations can achieve effectiveness comparable to dense retrieval while maintaining the efficiency benefits of inverted indexes.

\subsection{Bridging the Gap: Advanced Interaction and Hybrid Models}
The standard bi-encoder's lack of term-level interaction is a performance bottleneck compared to cross-encoders. Several lines of research aim to bridge this gap by introducing more granular representations or by combining different retrieval paradigms.

\subsubsection{Multi-Vector Representations}
To re-introduce query-document interaction without the full cost of a cross-encoder, multi-vector models represent queries and documents using multiple vectors. \textsc{Poly-Encoder}~\citep{Humeau2020Poly-encoders} computes a fixed number of vectors per query and aggregates them with softmax attention over document vectors. \textsc{ME-BERT}~\citep{luan2021sparse} represents documents with $m$ vectors and uses the maximum similarity between any query and document vector to estimate relevance. 

In line with this idea, \textsc{ColBERT}~\citep{khattab2020colbert,santhanam-etal-2022-colbertv2,hofstatter2022introducing} represents each token in the query and document as a contextualized vector. It then performs a ``late interaction'' step where each query vector is compared against all document vectors via a MaxSim operator, and the final score is the sum of these maximum similarities. This late interaction scheme (\cref{fig:architecture:multi}) allows \textsc{ColBERT} for end-to-end training to achieve strong performance while still achieving efficient retrieval through a dedicated index structure. On the other hand, it also leads to drastically increased index size, which has been the focus in later studies~\cite[\interalia]{santhanam-etal-2022-colbertv2,hofstatter2022introducing}.

We should also note that multi-vector retrieval can be viewed as a special case of dense retrieval where the learned feature representation is a matrix of size $n \times h$, with $n$ vectors of hidden dimension $h$. This matrix can be conceptually flattened into a single dense vector, showing its connection to the vanilla single-vector retrieval. The key difference lies not in the representation itself but in the richer relevance estimation strategy: instead of applying a simple linear relevance like dot product, models like \textsc{ColBERT} aggregate fine-grained token-level interactions to compute a relevance score.

\subsubsection{Hybrid Retrieval}
Another direction combines the strengths of different retrieval systems. A simple yet effective approach is ranklist fusion~\citep[e.g., Reciprocal Rank Fusion,][]{cormack2009reciprocal}, which merges ranked lists from sparse (e.g., BM25) and dense retrievers post-retrieval without architectural changes. 
More integrated models combine signals at a deeper level. \textsc{COIL}~\citep{gao-etal-2021-coil} enhances traditional bag-of-words retrieval with semantic embeddings from a \bert~encoder. \textsc{uniCOIL}~\citep{lin2021few} simplifies this by reducing the semantic embedding to a single dimension, effectively learning a term weight akin to LSR models like \textsc{SPLADE}~\citep{formal21spladev2,formal21splade}.
A few works fall into the intersection of learned sparse retrieval and multi-vector representations. For example, \textsc{SLIM}~\citep{li2023slim} first maps each contextualized token vector to a sparse, high-dimensional lexical space before performing late interaction between these sparse token embeddings. \textsc{SPLATE}~\citep{formal2024splate} takes an alternative approach to first encode contextualized token vectors, then map these token vectors to a sparse vocabulary space with a partially learned \textsc{SPLADE} module. Both models achieve performance improvement compared to learned sparse retrieval baselines such as \textsc{SPLADE}.

\begin{table*}[t!]
\centering
% \small

\caption{Summary of IR model architecture for passage retrieval and passage ranking based on pre-trained transformers. Dense Retrieval and LSR denote learned dense retrieval and learned sparse retrieval, respectively. \textsc{DeepCT}~\citep{dai2019context} is trained without labeled training set.  Contrastive Learning and in-batch negatives means listwise loss function is used. \textsc{SentenceBERT}~\citep{reimers-gurevych-2019-sentence} is originally designed for the symmetrical sentence similarity tasks, but is quickly expanded to asymmetrical retrieval tasks.}
\label{tab:bert_model_appendix}

\rowcolors{2}{gray!15}{white} % Alternating row colors
\resizebox{1\textwidth}{!}
{
\begin{tabular}{lllll}
\toprule
\textbf{Name} & \textbf{Model} & \textbf{Architecture} & \textbf{Backbone LM} & \textbf{Training strategy} \\
\midrule
\textsc{monoBERT}~\citep{nogueira2019multi} & Reranking & Cross-encoder & \textsc{BERT} & Classification \\
\textsc{CEDR}~\citep{macavaney2019cedr} & Reranking & Cross-encoder & \textsc{BERT} & Contrastive Learning \\
\textsc{BERT-MaxP}~\citep{Dai2019DeeperTU} & Reranking & Cross-encoder & \textsc{BERT} & Pairwise Loss \\
\citet{gao2020understanding} & Reranking & Cross-encoder & \textsc{BERT} & Distillation \\
\textsc{TART-full}~\citep{asai-etal-2023-task} & Reranking & Cross-encoder & \textsc{Flan-T5-Enc} & Instruction Tuning \\
\textsc{ODQA}~\citep{Lee2019LatentRF} & Dense Retrieval & Bi-encoder & \bert & Unsupervised \\
\textsc{SentenceBERT}~\citep{reimers-gurevych-2019-sentence} & Dense Retrieval & Bi-encoder & \bert & Triplet \\
\textsc{DPR}~\citep{Karpukhin2020DensePR} & Dense Retrieval & Bi-encoder & \textsc{BERT} & Contrastive Learning \\
\textsc{ANCE}~\citep{Xiong2020ApproximateNN} & Dense Retrieval & Bi-encoder & \textsc{RoBERTa} & Contrastive Learning \\
\textsc{RepBERT}~\citep{zhan2020repbert} & Dense Retrieval & Bi-encoder & \textsc{BERT} & In-batch negatives \\
\textsc{Margin-MSE}~\citep{hofstatter2020improving} & Dense Retrieval & Bi-encoder & \textsc{DistilBERT} & Distillation \\
\textsc{TAS-B}~\citep{hofstatter2021efficiently} & Dense Retrieval & Bi-encoder & \textsc{BERT} & Distillation \\
\textsc{RocketQA}~\citep{Qu2020RocketQAAO} & Dense Retrieval & Bi-encoder & \textsc{ERNIE} & Contrastive Learning \\
\textsc{RocketQA-v2}~\citep{ren-etal-2021-rocketqav2} & Dense Retrieval & Bi-encoder & \textsc{ERNIE} & Distillation \\
\textsc{GTR}~\citep{ni-etal-2022-large} & Dense Retrieval & Bi-encoder & \textsc{EncT5} & Contrastive Learning \\
\textsc{TART-dual}~\citep{asai-etal-2023-task} & Dense Retrieval & Bi-encoder & \textsc{Contriever} & Instruction Tuning \\
\textsc{E5}~\citep{wang2022e5textembeddings} & Dense Retrieval & Bi-encoder & \textsc{BERT} & Contrastive Learning \\
\textsc{GTE}~\citep{li2023generaltextembedding} & Dense Retrieval & Bi-encoder & \textsc{BERT} & Contrastive Learning \\
\textsc{Poly-encoder}~\citep{Humeau2020Poly-encoders} & Multi-vector & Misc & \textsc{BERT} & In-batch Negatives \\
\textsc{ME-BERT}~\citep{luan2021sparse} & Multi-vector & Bi-encoder & \textsc{BERT} & Contrastive Learning \\
\textsc{ColBERT}~\citep{khattab2020colbert} & Multi-vector & Bi-encoder & \textsc{BERT} & Pairwise Loss \\
\textsc{COIL}~\citep{gao-etal-2021-coil} & Multi-vector & Bi-encoder & \textsc{BERT} & Contrastive Learning\\
\textsc{ColBERT-v2}~\citep{santhanam-etal-2022-colbertv2} & Multi-vector & Bi-encoder & \textsc{BERT} & Distillation \\
\textsc{ColBERTer}~\citep{hofstatter2022introducing} & Multi-vector & Bi-encoder & \textsc{BERT} & Distillation \\
\textsc{DeepCT}~\citep{dai2019context} & LSR & Bi-encoder & \textsc{BERT} & Unsupervised \\
\textsc{SparTerm}~\citep{bai2020sparterm} & LSR & Bi-encoder & \textsc{BERT} & Contrastive Learning \\
\textsc{SPLADE}~\citep{formal21splade} & LSR & Bi-encoder & \textsc{BERT} & Contrastive Learning \\
\textsc{SPLADE-v2}~\citep{formal21spladev2} & LSR & Bi-encoder & \textsc{BERT} & Distillation \\
\textsc{DeepImpact}~\citep{mallia2021learning} & LSR & Bi-encoder & \textsc{BERT} & Contrastive Learning \\
\textsc{uniCOIL}~\cite{lin2021few} & LSR & Bi-encoder & \textsc{BERT} & Contrastive Learning \\
\textsc{SparseEmbed}~\citep{kong2023sparseembed} & LSR & Bi-encoder & \textsc{BERT} & Contrastive Learning \\
\textsc{SLIM}~\citep{li2023slim}  & LSR + Multi-vector & Bi-encoder & \textsc{BERT} & Contrastive Learning \\
\textsc{SLIM++}~\citep{li2023slim}  & LSR + Multi-vector & Bi-encoder & \textsc{BERT} & Distillation \\
\textsc{SPLATE}~\citep{formal2024splate} & LSR + Multi-vector & Bi-encoder & \textsc{BERT} & Distillation \\

\bottomrule
\end{tabular}
}

\end{table*}

\subsection{Orthogonal Improvements and Analysis}
Beyond architectural innovations, performance can be enhanced through improvements to the underlying models and a deeper analysis of their behavior. We show a list of models and their corresponding architectures in~\cref{tab:bert_model_appendix}, a majority of which use \textsc{BERT}~\citep{devlin-etal-2019-bert} as the backbone, with exceptions using \textsc{DistilBERT}~\citep{sanh2019distilbert}, \textsc{RoBERTa}~\citep{liu2019roberta}, and \textsc{T5}~\citep{raffel2020transfer,sanh2022multitask,mo2023convgqr,chung2024scaling}.

\paragraph{Continual Training and Adaptation.}
Instead of modifying the retrieval architecture, this line of work enhances the backbone language model itself through domain adaptation or continued pre-training, a proven strategy in NLP~\citep{howard2018universal,gururangan-etal-2020-dont}. For instance, \citet{Lee2019LatentRF} pre-train \textsc{BERT} with an Inverse-Cloze Task~\citep{Taylor1953ClozePA} for better text representations. \textsc{Condenser}~\citep{gao-callan-2021-condenser} proposes a dedicated pre-training architecture to ``condense'' text representations into the \texttt{[CLS]} token. \textsc{COCO-DR}~\citep{yu-etal-2022-coco} extends \textsc{Condenser} by using a technique named Distributionally Robust Optimization to mitigate distribution shifts in dense retrieval.
A line of works have also explored other pre-training objectives such as masked auto-encoders~\citep{xiao-etal-2022-retromae,wu2023contextualmaskedautoencoder} and bag-of-words prediction~\citep{ma2024dropyourdecoder}.
Recent works~\citep[\interalia]{wang2023improving,nussbaum2025nomicembed,yu2024arctic} have employed a ``middle-stage'' training on large-scale unlabeled text pairs, between pretrained encoder models and supervised fine-tuning on labeled text pairs, and have demonstrated the corresponding performance improvement compared to the traditional two-stage pipeline. 
We refer readers to the original papers for details.

\paragraph{Interpretability and Explainability.}
A few works have attempted to interpret what transformer-based models learn, i.e., mechanistic interpretability~\citep{elhage2021mathematical,saphra2024mechanistic}. \citet{macavaney-etal-2022-abnirml} showed that neural models rely less on exact match signals and instead encode rich semantic information. \citet{ram-etal-2023-token} connected dense and sparse retrieval by projecting a dense retriever's intermediate representations into the vocabulary space. Separately, other work focuses on designing systems that provide model-agnostic explanations~\citep{rahimi2021explaining,yu2022towards,xu2024cfe2} to satisfy desiderata like faithfulness~\citep{jacovi-goldberg-2020-towards,xu2023reusable}. As IR systems become integral to other applied ML domains, we believe it is important to study and design interpretable, truthful, and trustworthy IR models.

The architectural innovations discussed in this section highlight a mature research field dedicated to harnessing pre-trained transformers for information retrieval. The central theme has been the architectural tradeoff between interaction depth and computational cost, giving rise to a spectrum of models from highly effective cross-encoder rerankers to efficient bi-encoder retrievers. 
By developing advanced representations\,---\,whether dense, sparse, or multi-vector\,---\,and hybridizing these approaches, the community has pushed the boundaries of the classic ``retrieve-then-rank'' paradigm.
However, these models still primarily function as specialized components for representation and scoring. 
The next wave of innovation emerged from models capable not only of understanding text but also of generating it, marking the beginning of the era of Large Language Models.

\section{Large Language Models for IR}
\label{sec:llm4ir}
The natural evolution from pre-trained encoders is the recent ascendance of Large Language Models (LLMs).\footnote{The term ``Large Language Model'' lacks a precise, universally accepted definition in the literature. In this survey, we use the term to refer to models with over one billion parameters that are pre-trained with a text generation objective, such as text infilling (e.g., \textsc{T5}) or causal language modeling (e.g., the \textsc{GPT} series), and are often optionally post-trained for instruction following and human preference alignment.} 
While building on the same transformer principles discussed in~\cref{sec:transformer}, the sheer scale and generative capabilities of modern instruction-following LLMs are reshaping the architectural landscape of IR. These models are not just larger backbones for feature extraction; their proficiency in language understanding, generation, and instruction-following allows them to take on entirely new roles. Trained to align with human preferences~\citep{achiam2023gpt,team2023gemini,bai2022constitutional}, LLMs can perform complex tasks such as reasoning~\citep{wei2022chain,openai2024gpt4ocard,guo2025deepseekr1incentivizingreasoningcapabilityinllmsviareinforcementlearning}, tool usage~\citep{schick2023toolformer,patil2024gorilla,qin2024toolllm,patil2025bfcl} and planning~\citep{song2023llm,huang2024understandingtheplanningofllmagents}. In this section, we review how these powerful models\,---\,spanning \textit{decoder-only} and \textit{encoder-decoder} architectures\,---\,are being adapted for IR tasks, moving beyond established paradigms into new frontiers of retrieval, reranking, and direct generation.

\subsection{LLM as Retriever}
A straightforward yet highly effective application of LLMs is to serve as the backbone for bi-encoder retrieval models. We categorize these developments into backbone scaling, architectural adaptation, and unified modeling. We show a shortlist of works in~\cref{tab:llm4retrieval_appendix}.

\paragraph{Scaling Bi-Encoders.}
The dramatic increase in parameter count and training data provides LLMs with richer world knowledge and a more nuanced understanding of semantics compared to their smaller BERT-sized predecessors. This directly translates to performance improvements.
\citet{neelakantan2022text} fine-tuned a series of off-the-shelf \textsc{GPT} models towards text and code representation. 
They empirically verified that the bi-encoder retriever's performance can benefit from increased backbone language model capacity. \cite{muennighoff2022sgpt,ma2024fine} empirically verified the effectiveness of scaling the size of dense retrievers with open-weight models such as \textsc{GPT-J}~\citep{wang2021gpt-j} and \textsc{Llama-2}~\citep{touvron2023llama}.
Today, common text retrieval benchmarks like BEIR~\citep{thakur2021beir} and MTEB~\citep{muennighoff-etal-2023-mteb} are dominated by proprietary and open-source LLM-based embedding models~\cite[\interalia]{wang2023improving,li2023generaltextembedding,lee2024nvembed,zhang2025qwen3embedding,muennighoff2025generative}.

\paragraph{Decoder-Only Adaptation.}
A parallel line of research focuses on adapting the unidirectional architecture of decoder-only LLMs (e.g., \textsc{Llama}) to better suit the needs of bidirectional text representation. Standard decoder-only models are optimized for next-token prediction, which may not be ideal for creating a single summary vector for a whole text. To address this, \textsc{LLM2Vec}~\citep{behnamghader2024llmvec} enables bidirectional attention and further trains \textsc{Llama}-2~\citep{touvron2023llama} with specific unsupervised and supervised adaptive tasks. Similarly, \textsc{NV-Embed}~\citep{lee2024nvembed} introduces a new latent attention mechanism to produce improved representations, leading to improved performance on the MTEB benchmark compared to directly enabling bi-directional attention. 

\paragraph{Unified Modeling.}
\textsc{GritLM}~\citep{muennighoff2025generative} fine-tunes \textsc{Mistral} family models with both dense retrieval task and text generation task with different attention mechanisms and demonstrate the potential of unifying retrieval and generation with one single foundation model.

\subsection{LLM as Reranker}
The reranking task has also been significantly advanced by LLMs, which introduce new capabilities beyond the cross-encoder architecture discussed in \cref{sec:transformer}. This evolution can be categorized into two main architectural approaches: fine-tuning and zero-shot prompting. We show a shortlist of works that use LLM as rerankers in~\cref{tab:llm4reranking_appendix}.

\paragraph{Fine-Tuned Rerankers.}
First, LLMs can be fine-tuned as powerful rerankers. Extending earlier work on \textsc{BERT}-type models, researchers have applied similar techniques to larger encoder-decoder models like \textsc{T5}~\citep{raffel2020transfer} and decoder-only models like \textsc{Llama}~\citep{touvron2023llama}. 
\begin{itemize}
    \item \textbf{Pointwise and Pairwise:} \citet{nogueira2020documentrankingpretrainedsequencetosequence} fine-tuned \textsc{T5} with a classification loss, treating reranking as a binary relevance decision. \textsc{RankT5}~\citep{zhuang2023rankt5} took a more direct approach by fine-tuning \textsc{T5} to output a numerical relevance score, optimizing the model with established ranking losses like \textsc{RankNet}~\citep{burges2010ranknet}. Further, \citet{zhuang2023rankt5} also investigated the impact of language model architectures (\textsc{T5} encoder-decoder versus \textsc{T5} encoder), loss functions (pointwise, pairwise, listwise), and pooling strategies to ranking performance. 
    \item \textbf{Listwise:} Instead of scoring documents individually, \textsc{ListT5}~\citep{yoon-etal-2024-listt5} adopts a Fusion-in-Decoder architecture~\citep{izacard-grave-2021-leveraging} to process and rank an entire list of candidate documents in a single forward pass. More specifically, the architecture consists of an encoder and decoder, where the encoder takes a query and multiple passages as input in parallel, and the decoder outputs a sorted list of input passages in the decreasing order of relevance, achieving better reranking efficiency compared to pointwise methods. \cite{yang2025rankk} fine-tuned \textsc{QwQ}-32B~\citep{alibaba2025qwq32b} for listwise reranking where multiple passages are concatenated together as input and achieved better performance than pointwise reranking, leveraging the reasoning and long context capability of the strong base model. 
    \item \textbf{Long-Context Reranking:} The long-context capabilities of modern LLMs have also enabled a new reranking paradigm. \textsc{RankLlama}~\citep{ma2024fine} demonstrated superior pointwise reranking performance compared to \bert~and \textsc{T5}-based rerankers for long document reranking where the input is truncated at 4,096 tokens. 
\end{itemize}

\paragraph{Zero-Shot / Few-Shot Prompting.}
Second, the instruction-following ability of modern LLMs has unlocked zero-shot and few-shot reranking via prompting. This paradigm requires no task-specific fine-tuning. Instead, the LLM is prompted with a query and a list of candidate documents and asked to identify the most relevant ones~\citep{ma2023zeroshotlistwisedocumentreranking,zhang2023rank,pradeep2023rankvicunazeroshotlistwisedocument,pradeep2023rankzephyreffectiverobustzeroshot,sun2023chatgpt}. This listwise approach is a natural fit for the long context windows of models like GPT-4. To mitigate the high computational cost and context length limitations of processing full documents, \citet{liu2024leveraging} proposed using passage embeddings as compact document representations for the LLM, training a specialized reranker that operates on these embeddings to improve efficiency. 
As this line of research primarily involves prompt engineering rather than architectural changes, we refer readers to a recent survey~\citep{zhu2023large} for further details.

\subsection{Generative Retrieval}
\label{subsec:generative_retrieval}
Perhaps the most radical architectural shift enabled by LLMs is generative retrieval. Traditional IR systems follow the ``retrieve-then-rerank'' paradigm~\citep{schutze2008introduction,asai2024openscholar,xu2025distillationvscontrastive}. Generative retrieval fundamentally challenges this by reframing retrieval as a sequence-to-sequence task. Instead of searching an index, an autoregressive language model is trained to directly generate the unique document identifiers (DocIDs) of relevant documents in response to a query.

\paragraph{Evolution of the Paradigm.}
The foundational work in generative retrieval emerged from the entity linking domain with Generation of ENtity REtrieval~\citep[GENRE,][]{de2020autoregressive}. Rather than treating entity retrieval as a classification problem over atomic labels with dense representations, GENRE reframed it as a generative task where an encoder-decoder model produces entity names autoregressively, token-by-token, conditioned on the input context.
Building on GENRE's success, DSI introduced generative retrieval to the broader document retrieval domain~\citep{tay2022transformer}. The core innovation was fully parameterizing traditional ``retrieve-then-rerank'' pipelines within a single neural model, where all corpus information is encoded in the model parameters rather than external indices~\citep{tay2022transformer,pradeep-etal-2023-generative}. DSI operates through two fundamental sequence-to-sequence tasks that can be trained jointly or sequentially~\citep{He2024ReindexThenAdapt}: the indexing task (Learn to Index) and the retrieval task (Learn to Retrieve).

% --- Table added to address reviewer comment on taxonomy ---
\begin{table}[t]
\centering
\caption{Taxonomy of identifier types in Generative Retrieval. The choice of identifier is a key architectural distinction.}
\label{tab:gen_retrieval_ids}
\small
\begin{tabular}{@{}lp{0.7\linewidth}@{}}
\toprule
\textbf{Identifier Type} & \textbf{Description} \\ \midrule
\textbf{Atomic Identifiers} & Unique integers assigned to documents. Simple but lacks semantic generalization. \\
\textbf{String Identifiers} & Natural language strings (titles, URLs). Leverages pre-trained knowledge but can be ambiguous. \\
\textbf{Semantic Identifiers} & Structured IDs derived from clustering embeddings. Enables semantic generalization in the ID space. \\
\bottomrule
\end{tabular}
\end{table}
% -----------------------------------------------------------

\paragraph{Taxonomy of Identifiers.}
A critical design choice in DSI involves document identifier representation (summarized in \cref{tab:gen_retrieval_ids}).
\begin{itemize}
    \item \textbf{Atomic and String Identifiers:} Early work explored atomic identifiers (unique integers) and simple string identifiers (titles, URLs)~\citep{Chen2023ContinualLearningforGenerativeRetrieval}. DSI can be implemented in two variants: classification-based approaches that use a classification layer to output atomic document identifiers, and generative approaches that autoregressively generate identifier strings~\citep{Mehta2022dsi}. More sophisticated methods have introduced n-gram-based identifiers~\citep{bevilacqua2022autoregressive}.
    \item \textbf{Semantic Identifiers:} Semantically structured identifiers created through clustering algorithms proved most effective~\citep{zhu2023large}. This often involves hierarchical representations using techniques like residual quantization~\citep{zeng2024scalable}, where the model learns to associate document content with corresponding semantically meaningful document identifiers~\citep{kishore2023incdsi}.
\end{itemize}

\paragraph{Inference and Constraints.}
The technical implementation of generative retrieval systems centers on sequence-to-sequence modeling. A critical technical requirement is ensuring only valid document identifiers are generated during inference. This is typically achieved through constrained beam search over a prefix tree (trie) constructed from all valid document identifiers~\citep{Tang2023RecentAdvancesInGenerative}. Alternative approaches include constrained greedy search and FM-index structures~\citep{Tang2023RecentAdvancesInGenerative}. The \textsc{T5} model backbone~\citep{raffel2020transfer} serves as the foundation for most DSI implementations, trained with cross-entropy loss on both indexing and retrieval objectives.

\paragraph{Challenges.}
Generative retrieval faces significant challenges in dynamic environments. The tight coupling between index and retrieval modules makes updating the corpus computationally expensive, typically requiring full model retraining~\citep{Mehta2022dsi}. Scalability poses a major challenge; most research has focused on relatively small collections, as the memory and computational requirements grow substantially as corpus size increases.
Generative retrieval is an active and rapidly evolving research area; we direct interested readers to a dedicated survey~\citep{li2025frommatchingtogenerationasurveyongenerativeinformationretrieval} for a comprehensive review.

\begin{table*}[t!]
\centering
% \small

\caption{Summary of IR model architecture utilizing large language models as retrieval backbone.}
\label{tab:llm4retrieval_appendix}

\rowcolors{2}{gray!15}{white} % Alternating row colors
\resizebox{1\textwidth}{!}
{
\begin{tabular}{llll}
\toprule
\textbf{Name} & \textbf{Architecture} & \textbf{Backbone LM} & \textbf{Training strategy} \\
\midrule
\textsc{CPT-Text}~\citep{neelakantan2022text} & LLM Encoder & \textsc{GPT-3} & Listwise Loss \\
\textsc{SGPT-BE}~\citep{muennighoff2022sgpt} & LLM Encoder & \textsc{GPT-J} \& \textsc{GPT-NeoX} & Listwise Loss \\
\textsc{GTR}~\citep{ni-etal-2022-large} & LLM Encoder & \textsc{T5} & Listwise Loss \\
\textsc{RepLlama}~\citep{ma2024fine} & LLM Encoder & \textsc{Llama} & Listwise Loss \\
\textsc{E5-Mistral}~\citep{wang2023improving} & LLM Encoder & \textsc{Mistral} & Synthetic Data + Listwise Loss \\
\textsc{LLaRA}~\citep{li2023makinglargelanguagemodelsabetterfoundationfordenseretrieval} & LLM Encoder & \textsc{Llama} & Adaptation + Contrastive Training \\
\textsc{MambaRetriever}~\citep{zhang2024mambaretriever} & LLM Encoder & \textsc{Mamba} & Listwise Loss \\
\textsc{LLM2Vec}~\citep{behnamghader2024llmvec} & LLM Encoder & \textsc{Llama} \& \textsc{Mistral} & Adaptation + Contrastive Pre-training \\
\textsc{Grit-LM}~\citep{muennighoff2025generative} & LLM & \textsc{Mistral} \& \textsc{Mixtral 8x7B} & Generative/Embedding Joint Training \\ 
\textsc{NVEmbed}~\citep{lee2024nvembed} & LLM Encoder & \textsc{Mistral} & Adaptation + Synthetic Data + Listwise Loss \\
\textsc{GTE-Qwen2-Instruct}~\citep{li2023generaltextembedding} & LLM Encoder & \textsc{Qwen} & Adaptation + Synthetic Data + Listwise Loss \\
\textsc{Qwen3-Retriever}~\citep{zhang2025qwen3embedding} & LLM Encoder & \textsc{Qwen3} & Synthetic Data + Listwise Loss \\
\bottomrule
\end{tabular}
}

\end{table*}

\begin{table*}[t!]
\centering
% \small

\caption{Summary of IR model architecture utilizing large language models for reranking. \citet{nogueira-dos-santos-etal-2020-beyond} and \citet{zhuang2021deep} revisit the statistic language model problem with modern transformer-based models, including \textsc{BART}~\citep{lewis-etal-2020-bart}, \textsc{T5}~\citep{raffel2020transfer}, and \textsc{GPT-2}~\citep{radford2019language}. We use Seq2Seq LLM to refer to encoder-decoder architecture language models such as \textsc{T5} and \textsc{BART}, and Casual LLM to refer to modern LLMs with causal language model architecture like \textsc{GPT} family models.}
\label{tab:llm4reranking_appendix}

\rowcolors{2}{gray!15}{white} % Alternating row colors
\resizebox{1\textwidth}{!}
{
\begin{tabular}{llll}
\toprule
\textbf{Name} & \textbf{Architecture} & \textbf{Backbone LM} & \textbf{Training / Prompting Strategy} \\
\midrule
\multicolumn{4}{l}{\emph{Fine-tune LLM for Reranking}} \\
\textsc{monoT5}~\citep{nogueira2020documentrankingpretrainedsequencetosequence} & Seq2Seq LM & \textsc{T5} & Classification \\
\citet{nogueira-dos-santos-etal-2020-beyond} & Seq2Seq LLM & \textsc{BART} & Unlikelihood \\
\textsc{QLM-T5}~\citep{zhuang2021deep} & Seq2Seq LLM & \textsc{T5} & Language Modeling \\
\textsc{duoT5}~\citep{pradeep2021expando} & Seq2Seq LLM & \textsc{T5} & Pairwise Loss \\
\textsc{RankT5}~\citep{zhuang2023rankt5} & Seq2Seq LLM Encoder + Prediction Layer & \textsc{T5} & Listwise Loss \\
\textsc{ListT5}~\citep{yoon-etal-2024-listt5} & Fusion-in-decoder & \textsc{T5} & Listwise Loss \\
\textsc{SGPT-CE}~\citep{muennighoff2022sgpt} & Causal LLM & \textsc{GPT-J} \& \textsc{GPT-Neo} & Listwise Loss \\
\textsc{RankLlama}~\citep{ma2024fine} & Casual LLM Encoder + Prediction Layer & \textsc{Llama} & Listwise Loss \\
\textsc{RankMamba}~\citep{xu2024rankmamba} & Causal LLM Encoder + Prediction Layer & \textsc{Mamba} & Listwise Loss \\
\textsc{RankVicuna}~\citep{pradeep2023rankvicunazeroshotlistwisedocument} & Causal LLM & \textsc{Vicuna} & Listwise \\
\textsc{RankZephyr}~\citep{pradeep2023rankzephyreffectiverobustzeroshot} & Causal LLM & \textsc{Zephyr} & Listwise \\
\citet{zhang2023rank} & Causal LLM & \textsc{Code-LLaMA-Instruct} & Listwise \\
\citet{liu2024leveraging} & Embedding + Causal LLM & \textsc{Mistral} & Listwise \\
\textsc{Qwen3-Reranker}~\citep{zhang2025qwen3embedding} & Causal LLM & \textsc{Qwen3} & Synthetic Data + Pairwise Loss \\

\multicolumn{4}{l}{\emph{Prompt LLM for Reranking}} \\
\citet{zhuang-etal-2023-open} & Causal LLM & Multiple & Pointwise Prompting \\
\citet{zhuang-etal-2024-beyond} & Causal LLM & \textsc{Flan-PaLM-S} & Pointwise Prompting \\
\textsc{UPR}~\citep{sachan-etal-2022-improving} & Seq2Seq LM \& Causal LLM & \textsc{T5} \& \textsc{GPT-Neo} & Pointwise Prompting \\
\textsc{PRP}~\citep{qin-etal-2024-large} & Seq2Seq LM & \textsc{Flan-UL2} & Pairwise Prompting \\
\citet{yan-etal-2024-consolidating} & Seq2Seq LM & \textsc{Flan-UL2} & Pairwise Prompting \\
\citet{zhuang2024setwise} & Seq2Seq LM & \textsc{Flan-T5} & Pairwise \& Setwise Prompting \\
\textsc{LRL}~\citep{ma2023zeroshotlistwisedocumentreranking} & Casual LLM & \textsc{GPT-3} & Listwise Prompting \\
\textsc{RankGPT-3.5}~\citep{sun2023chatgpt} & Causal LLM & \textsc{GPT-3.5} & Listwise Prompting \\
\textsc{RankGPT-4}~\citep{sun2023chatgpt} & Causal LLM & \textsc{GPT-4} & Listwise Prompting \\

\bottomrule
\end{tabular}
}

\end{table*}

\subsection{Broader Ecosystem and Concluding Remarks}
Beyond core architectural changes, LLMs are influencing the entire IR ecosystem. Their advanced generative and understanding capabilities are being harnessed for crucial supporting tasks:
\begin{itemize}
    \item \textbf{Data Synthesis:} Modern IR systems require extensive labeled data for training, which is expensive to create. A promising line of work is to use LLMs to synthesize high-quality training data (e.g., queries, relevant passages, and hard negatives)~\citep{bonifacio2022inpars,boytsov2024inpars,dai2023promptagator,lee2024gecko,mo2024chiq,mo2024convsdg,zhang2025qwen3embedding}. 
    \item \textbf{Evaluation:} From an evaluation perspective, LLMs' language understanding has led to research on using them as proxies for human relevance judges, which could dramatically accelerate the evaluation cycle~\citep{faggioli2023perspectives,faggioli2024determines,clarke2024llm}. 
\end{itemize}
We also point readers to a comprehensive survey on conversational information retrieval~\citep{mo2024survey}, another area being reshaped by LLMs.

In conclusion, the adoption of LLMs in IR represents more than a simple increase in model scale. While they certainly serve as more powerful feature extractors within existing bi-encoder and cross-encoder frameworks, their unique generative and instruction-following abilities are forging entirely new architectural paradigms like generative retrieval and zero-shot listwise reranking. However, the advancement of IR architecture is not driven solely by the pursuit of superior semantic matching capabilities. The practical deployment of these systems --- ranging from lightweight encoders to massive LLMs --- necessitates architectures that can withstand rigorous efficiency constraints, handle diverse data modalities, and ensure reliability. We examine these cross-cutting architectural adaptations in the following section.

\section{Architectures for Diverse Scenarios and Constraints}
\label{sec:scenarios_constraints}

The evolution of IR models described in previous sections --- from vector space models to LLMs --- primarily traces the pursuit of better semantic matching for English text. However, deploying these models in real-world environments requires navigating complex scenarios and constraints beyond pure textual relevance. These include handling diverse data modalities and languages, balancing the inherent tradeoff between accuracy and latency, and ensuring model reliability through calibration. In this section, we review how IR architectures are adapted to meet these specific requirements. Across these settings, architectural choices such as representation granularity and modularity, serve as the primary mechanisms to balance task-specific constraints with scalable retrieval.

\subsection{Architectures for Multimodal and Multilingual Data}
\label{subsec:multimodal_multilingual}

\subsubsection{Multilingual and Cross-Lingual Architectures}

\paragraph{Problem Definitions and Retrieval Settings.}
Although often used interchangeably, Multilingual Information Retrieval (MLIR) and Cross-Lingual Information Retrieval (CLIR) correspond to distinct retrieval scenarios. CLIR refers to the setting in which a user issues a query in a source language and retrieves documents written in a different target language~\citep{oard1998evaluatingcrosslingual,fluhr1999multilingualinformationretrieval}. This distinction has direct architectural implications: CLIR requires explicit cross-language alignment at query time, whereas MLIR emphasizes building shared or interoperable representations during indexing. The central architectural challenge in CLIR is bridging the lexical and semantic gap between the query and document language spaces~\citep{goworek2025bridginglanguagegaps}. In contrast, MLIR is a broader paradigm in which a system indexes and searches over document collections containing multiple languages, potentially serving queries in any of them~\citep{oard1996survey,oard1999multilingualinformationdiscoveryandaccess,fluhr1999multilingualinformationretrieval}. Architecturally, CLIR can be viewed as a special case of MLIR that explicitly requires cross-language alignment at retrieval time.

\paragraph{Early Translation-Centric Architectures.}
The architectural foundations of cross-lingual retrieval date back to early work on the Vector Space Model~\citep{salton1975vector}, where cross-language retrieval was framed as a synonymy problem. Early systems relied on bilingual thesauri to map query terms into a shared conceptual space prior to retrieval, treating translation as a distinct pre-processing step~\citep{salton1969interactiveinformationretrieval,salton1970automaticprocessingofforeignlanguagedocuemnts}. Architecturally, these systems isolated linguistic complexity into a separate translation component, leaving the retrieval engine itself unchanged and monolingual. When high-quality thesauri were available, such architectures could approach monolingual retrieval performance; however, they were brittle due to limited vocabulary coverage, poor handling of polysemy, and an inability to model multi-word expressions or contextual meaning.

\paragraph{Statistical and Representation-Based Cross-Lingual Models.}
The availability of large parallel corpora in the 1990s, such as the Canadian Hansards\footnote{\url{https://en.wikipedia.org/wiki/Hansard}}, enabled a shift from static dictionaries to statistically grounded architectures. A major departure from direct translation was introduced by Cross-Language Latent Semantic Indexing (\textsc{CL-LSI}), which projected documents and queries from different languages into a shared, language-independent latent space using Singular Value Decomposition over parallel data~\citep{dumais1997automaticcrosslanguageretrieval}. In parallel, Statistical Machine Translation (SMT) became a dominant architectural component in CLIR systems, with retrieval pipelines adopting either query translation or document translation using probabilistic alignment models such as the IBM Models~\citep{brown1993mathematicsofstatisticalmachinetranslation}. Both CL-LSI and SMT-based pipelines preserved monolingual retrieval backends, differing primarily in whether cross-lingual alignment was achieved through latent semantic projection or probabilistic translation. These systems established the translation-based retrieval paradigm, where the retrieval engine itself remained monolingual and linguistic complexity was isolated within the translation module.

\paragraph{Multilingual Transformers and End-to-End Retrieval.}
The introduction of multilingual pre-trained transformers, including mBERT and XLM-R, marked a fundamental architectural shift away from explicit translation toward shared semantic representation learning~\citep{devlin-etal-2019-bert,conneau2020unsupervised,MacAvaney2019TeachingAN}. Trained on large-scale multilingual corpora, these models enabled a single encoder to represent queries and documents across languages, substantially reducing dependence on parallel data~\citep{conneau-etal-2020-emerging,feng-etal-2022-language,shi-etal-2020-cross,goswami-etal-2021-cross}. However, general-purpose multilingual encoders often underperform in retrieval settings due to insufficient cross-lingual alignment induced during pretraining~\citep{Zhang2022Towardbestpractices,elmahdy2024synergisticapproachsimultaneousoptimization}, which has been the focus of subsequent research works.

\paragraph{Per-Language Modules.} To improve the performance of multilingual pre-trained transformers on low-resource languages, a line of works proposed to add language-specific adapters to enable zero-shot or few-shot cross-lingual transfer. Early explorations in NLP community focus on classical NLP tasks such as dependency parsing and named entity recognition~\citep{ustun-etal-2020-udapter,pfeiffer-etal-2020-mad,artetxe-etal-2020-cross,pfeiffer-etal-2021-adapterfusion}.
In the context of IR, \cite{litschko-etal-2022-parameter} compared the performance of adapter-based approaches and Sparse-Fine-Tuning Masks~\citep{ansell-etal-2022-composable} to NMT-based approaches and reported their efficacy in both performance and training efficiency.

\paragraph{Alignment-Focused and Modern Architectures.}
Tackling the same challenge of per-language modules approaches, specialized multilingual retrieval models such as LaREQA, InfoXLM, and LaBSE enforce tighter alignment between semantically equivalent cross-lingual pairs using parallel corpora and contrastive objectives~\citep{roy-etal-2020-lareqa,chi-etal-2021-infoxlm,feng-etal-2022-language,huang-etal-2024-languageconcepterasure}. Notably, these approaches typically preserve the standard transformer backbone, focusing architectural innovation on representation learning objectives rather than structural redesign. Recent work further refines end-to-end multilingual retrieval through improved data curation, supervision, and knowledge distillation, continuing the shift away from multi-stage translation pipelines toward unified models that directly match meaning across languages~\citep{zhang2023miracl,Yang2024Distillationformultilingual}.

\subsubsection{Multimodal Architectures}

\paragraph{From Shallow Fusion to Deep Joint Embeddings.}
Early multimodal retrieval architectures relied on modality-specific feature extraction followed by shallow fusion, with text typically represented using bag-of-words models and images encoded via hand-crafted descriptors such as scale-invariant feature transform (SIFT)~\citep{lowe1999objectrecognition}. To enable cross-modal comparison, projection-based methods such as canonical correlation analysis (CCA)~\citep{hotelling1992relationsbetweentwosetsofvariates} mapped heterogeneous features into a shared embedding space, establishing the foundational principle that multimodal retrieval requires learning semantic correspondences across modalities rather than treating them independently~\citep{rasiwasia2010newapproachtocrossmodalmultimediaretrieval,sharma2012generalizedmultiviewanalysis,gong2014multiviewembeddingspace,ranjan2015multilabelcrossmodalretrieval}. The transition to deep learning replaced manual feature engineering with end-to-end representation learning, exemplified by early visual-semantic embedding models such as \textsc{DeViSE}~\citep{frome2013devise,wang2016comprehensivesurveycrossmodalretrieval}. Sentence encoders incorporating syntactic structure, including Dependency Tree Recursive Neural Networks (\textsc{DT-RNNs}), further improved alignment by modeling relational semantics~\citep{socher2014groundedcompositionalsemantics}. Fragment-level architectures extended this paradigm by decomposing images and sentences into finer-grained units and aligning them with structured max-margin objectives, enabling both global and local cross-modal reasoning~\citep{karpathy2014deepfragmentembeddings}. Adversarial frameworks such as \textsc{ACMR} subsequently refined joint embeddings by introducing nonlinear projections and modality-invariant regularization~\citep{wang2017adversarialcrossmodalretrieval}. We refer to~\cite{wang2024crossmodalretrievalsystematicreview} for a comprehensive survey.

\paragraph{Dual-Encoder Contrastive Models and Controlled Interaction.}
Large-scale vision--language datasets enabled a paradigm shift toward bi-encoder architectures trained with contrastive objectives. \textsc{CLIP} and \textsc{ALIGN} independently encoded images and text and aligned them through contrastive learning on hundreds of millions to billions of image--text pairs, establishing scalable and efficient retrieval backbones~\citep{Radford2021LearningTransferableVisualModels,Jia2021ScalingUpVisual,wei2025uniir}. This architectural separation facilitated efficient indexing and retrieval and rapidly generalized to additional modalities including video, audio, depth, and sensor data~\citep{Girdhar2023ImageBindOneEmbeddingSpace,Chen2023VAST,Kong2025ModalityCuration}. Subsequent refinements explored the efficiency--expressivity tradeoff within this paradigm, most notably through late-interaction architectures such as \textsc{ColBERT}, which replaced single-vector embeddings with multi-vector representations to enable token-level matching while preserving much of the efficiency of bi-encoders~\citep{khattab2020colbert,faysse2025colpaliefficientdocumentretrieval,Wan2025CLaMRCL}. Together, these models established a dominant architectural family centered on modality-separated encoding with limited but scalable interaction.

\paragraph{Rich Fusion, Hierarchical Interaction, and Unified Architectures.}
Beyond bi-encoders, multimodal retrieval architectures increasingly incorporated sophisticated mechanisms to support complex reasoning, particularly in video--text retrieval~\citep{Chen2020FineGrainedVideoTextRetrieval}. Transformer-based and hierarchical models decomposed alignment into global-to-local or multi-granular stages, combining cross-modal attention with temporal and semantic structure~\citep{Gabeur2020MultimodalTransformerforVideoRetrieval,Liu2021HiTHierarchicalTransformer,Zhang2023DebiasedVideoTextRetrieval,Gorti2022XPoolCrossModelLanguageVideoAttention}. Hybrid designs integrated CLIP-style encoders with cross-modal fusion or temporal alignment modules, balancing pretrained representations with task-specific interaction~\citep{PortilloQuintero2021ASF,Fang2021CLIP2VideoMasteringVideoText}. These architectures substantially increase computational cost, often restricting their use to reranking or small candidate sets. 

Recent architectures further differentiate between multi-encoder designs that preserve modality-specific encoders and single-encoder models employing full cross-attention at higher computational cost~\citep{Li2019UnicoderVLAU,Li2022BLIPBootstrappingLanguageImage,zhai2023sigmoidlossforlanguageimagepretraining,Kim2025GENIUS}. At the same time, multimodal large language models (MLLMs) and multi-agent retrieval frameworks unify retrieval, reasoning, and generation within a single system, while modality-preserving and any-to-any architectures emphasize flexible interaction without collapsing modality structure~\citep{Liu2023VisualInstructionTuning,Xie2024Showo,Liu2025HMRAG,Xu2025OmniEmbedNemotronAU,ju-etal-2025-mire}.

\paragraph{Remarks.}
From a structural perspective, multimodal retrieval architectures can be broadly grouped into: (1) shallow fusion and projection-based models, (2) deep joint-embedding architectures with global or fragment-level alignment, (3) contrastively trained dual-encoder models emphasizing scalability, (4) late-interaction hybrids balancing efficiency and expressivity, and (5) cross-attention and MLLM-centric architectures enabling deep fusion and unified multimodal reasoning. This progression reflects a recurring architectural tension between interaction richness and computational efficiency.

\subsection{Performance-Efficiency Tradeoffs and the Role of Architectural Choices}
\label{subsec:tradeoff}
Retrieval systems face a fundamental tradeoff between effectiveness (e.g., recall, MRR, nDCG) and resource efficiency (latency, throughput, memory footprint, and index/update cost). Architectural choices\,---\,how queries and documents are represented, how similarity is computed, and how candidates are staged and scored\,---\,drive where a method falls on this tradeoff curve. Modern neural retrievers demonstrated that learned dense embeddings can substantially improve effectiveness over classic lexical baselines, but achieving that gain requires extra compute and index engineering compared to lightweight lexical methods.

A common high-performance design is the bi-encoder (single-vector) architecture (\cref{sec:transformer,sec:llm4ir}): queries and documents are encoded independently into compact vectors and approximate nearest-neighbor search (ANN) retrieves candidates quickly. This architecture is attractive for strict latency budgets and large corpora because it enables highly optimized ANN indexes (e.g., HNSW~\citep{malkov2018hnsw}) that deliver millisecond-scale queries at large scale; however, single-vector representations can miss fine-grained token-level matches that matter for some queries. 

To improve effectiveness, researchers have pursued richer interaction patterns. Late-interaction or multi-vector models (e.g., \textsc{ColBERT} and \textsc{ColPali}~\citep{khattab2020colbert,faysse2025colpaliefficientdocumentretrieval}) keep token-granular signals and aggregate local token similarities, which raises effectiveness but increases index size and per-query compute; compression and residual quantization techniques can reduce the space/latency penalty but do not eliminate it entirely. Similarly, learned sparse/lexical models (e.g., \textsc{SPLADE}~\citep{formal21spladev2,formal21splade}) produce high-dimensional but sparse representations that recover lexical signals and interpretability while trading off somewhat higher compute or indexing complexity versus classical BM25. These architectural variants illustrate the spectrum: more expressive interactions often translates to better quality, typically at higher memory and latency cost (unless mitigated by compression).

A pragmatic pattern in production is staged (two- or multi-stage) retrieval: a fast, coarse first-stage (BM25 or compact dense ANN) produces a small candidate set, and a more expensive cross-encoder or interaction-based reranker refines the top results~\citep{huang2020embeddingbasedretrieval,su2025modernizingfacebookscopedsearch}. This cascade yields most of the accuracy of expensive models while preserving throughput, but it requires careful budgeting (how many candidates to pass) and engineering (batching, caching, and efficient GPU/CPU placement). 
Indexing and compression techniques (e.g., product quantization, pruning, residual quantization), along with hybrid lexical–dense pipelines, are commonly employed to navigate the tradeoff curve as operational constraints evolve.

\paragraph{Architectural Design Heuristics.} 
Adopt a single-vector dual-encoder with efficient ANN search when low latency, low cost, and frequent updates are the primary constraints. Favor multi-vector or interaction-intensive models when top-tier ranking quality is required and computational budget permits. To balance these objectives, employ a two-stage pipeline comprising coarse retrieval followed by a specialized reranker. 
In large-scale deployments, the effectiveness-efficiency operating point is chiefly adjusted through index compression techniques, hybrid lexical-dense signals, and careful calibration of candidate set size.

\subsection{Calibration and Confidence Estimation}
\label{subsec:calibration}
Calibration in IR aims to align model scores with true probabilities of relevance, such that confidence faithfully reflects correctness. Unlike traditional deterministic rankers that output single relevance scores, calibrated IR models explicitly represent uncertainty, often as a distribution over scores whose mean encodes relevance belief and whose variance captures uncertainty~\citep{Cohen2021NotAllRelevanceScoresareEqual}. This distinction is architecturally significant: uncertainty-aware scoring exposes information that is otherwise hidden in standard ranking pipelines. The importance of calibration is formalized by the Probability Ranking Principle (PRP), which guarantees optimal ranking only when relevance probabilities are well calibrated and reported with certainty~\citep{penha-hauff-2021-calibration}. However, modern neural rankers frequently violate these assumptions, motivating calibration-aware architectural design.

\paragraph{Neural Architectures and Uncertainty Modeling.}
Calibration properties in IR are strongly dependent on architectural choices, with empirical studies showing mixed calibration behavior across neural model families~\citep{Guo2017OnCalibrationofModern,Minderer2021RevisitingTheCalibrationofModernNeuralNetworks}. Transformer-based rankers, including \textsc{BERT} variants, are often poorly calibrated, with calibration quality varying by model scale and design~\citep{dan-roth-2021-effects-transformer,li-etal-2022-calibration}. Introducing stochasticity at the architectural level\,---\,through stochastic inference or approximate Bayesian formulations\,---\,consistently improves calibration compared to deterministic counterparts, while adding only modest computational overhead~\citep{penha-hauff-2021-calibration,Cohen2021NotAllRelevanceScoresareEqual}. These findings position stochastic and Bayesian architectures as principled mechanisms for embedding uncertainty directly into relevance estimation.

\paragraph{Calibration in Multi-Component Retrieval Systems.}
Calibration challenges are amplified in multi-stage retrieval architectures, such as the ``retrieve-then-rerank'' pipeline, where hard top-$k$ retrieval steps break differentiability and preclude end-to-end calibration. Architectural interventions, including differentiable sampling mechanisms based on Gumbel approximations, restore gradient flow and enable joint calibration of retriever and reader components~\citep{dhuliawala-etal-2022-calibration}. Jointly calibrated systems produce more reliable confidence estimates than calibrating individual modules in isolation, underscoring how calibration requirements can directly shape architectural design in complex IR pipelines.

\paragraph{Modularity and Calibration-Aware Computation.}
Beyond end-to-end design, modular architectures enable calibration to be treated as an attachable component rather than an intrinsic model property. Universal post-hoc calibrators can be applied across heterogeneous retrieval architectures without modifying their internals, offering scalable and architecture-agnostic improvements~\citep{zhang-etal-2021-knowing}. When uncertainty is explicitly modeled, calibration becomes operational rather than merely diagnostic: confidence estimates can guide adaptive computation, such as selective reranking or deferred inference for ambiguous queries, improving both efficiency and robustness~\citep{Cohen2021NotAllRelevanceScoresareEqual,yoon2025acurankuncertaintyawareadaptivecomputation}. Recent evidence further suggests that architectural specialization\,---\,e.g., modeling relevance across multiple criteria instead of a single scalar score\,---\,can inherently reduce calibration error, reinforcing calibration as a first-class architectural consideration in IR system design~\citep{penha-hauff-2021-calibration,javdan2025crestimprovinginterpretabilityeffectiveness}. To summarize, how to incorporate calibration into modern retrieval systems is still an open question in the IR community.

\subsection{Domain-Specific Applications}
\label{subsec:applications}

\paragraph{General Web Search.}
General-purpose web search has historically been dominated by the ``retrieve-then-rerank'' paradigm, built around inverted indexes and multi-stage cascade ranking to ensure efficiency at web scale~\citep{shen2014cdssm,mitra2016dual,Qin2022RankFlowJointOptimizatio,zhang2025agenticinformationretrieval}. These architectures decompose retrieval into heterogeneous components for query understanding, candidate generation, ranking, and re-ranking, each optimized independently~\citep{Wang2023RethinkingEcommerceSearch}. As web content diversified, search engines incorporated additional signals such as anchor text, hyperlinks, and layout-based features to enrich document representations beyond plain text~\citep{Oliveira2023from10bluelinks}. While highly effective, these pipelines impose rigid, predefined information flows that limit adaptability. The recent generative retrieval (\cref{subsec:generative_retrieval}) thus explores replacing this modular pipeline with model-centric architectures, where a single LLM performs indexing, retrieval, and ranking end-to-end.

\paragraph{Domain-Specific and Thematic Retrieval.}
Domain-specific search engines target focused corpora such as scientific literature, medicinal chemistry databases, and job postings, leveraging domain knowledge to improve relevance beyond general web search. These applications share architectural challenges arising from specialized terminology, underspecified expert queries, and narrowly scoped user intents~\citep{Wang2023RethinkingEcommerceSearch,Kang2024ImprovingRetrievalInThemeSpecificApplications}. Generic pretrained models often fail to capture domain-specific semantics without architectural support for structured knowledge. Traditional solutions rely on query expansion, lexical analysis, and large task-specific training sets, but suffer from scalability and complexity limitations~\citep{xu2025rethinkingonpolicyoptimizationquery}. In response, modern architectures increasingly integrate entity- and relation-centric representations to better align retrieval with domain semantics~\citep{Dong2022IncorporatingExplicitKnowledge}. As a result, domain-specific IR architectures emphasize knowledge-aware representations and domain adaptation over purely generic pretrained models.

\paragraph{Medical and Legal Information Retrieval.}
Medical and legal search systems represent high-stakes instantiations of domain-specific retrieval, imposing stricter architectural constraints. Medical IR must operate over long temporal records that combine unstructured clinical notes with structured diagnoses, laboratory values, and medications, motivating architectures that integrate information extraction, faceted search, and structured database querying~\citep{Sonntag2018AnArchitectureOf}. These systems must additionally handle specialized medical terminology, clinical documentation standards, and stringent privacy requirements. Legal IR faces analogous challenges due to highly specialized language, complex document structures, and limited labeled relevance data~\citep{Althammer2020CrossdomainRetrieval}. Despite recent efforts to adapt pretrained language models such as \textsc{BERT}, neural approaches often fail to consistently outperform strong lexical baselines like BM25, reinforcing the continued relevance of hybrid and lexically grounded architectures in legally specialized domains~\citep{araujo2014information,Althammer2020CrossdomainRetrieval}.

\paragraph{E-commerce Search.}
E-commerce search systems operate over structured product catalogs rather than unstructured web pages, fundamentally shaping their architectures~\citep{Wang2023RethinkingEcommerceSearch,Ren2024InformationDiscoveryInEcommerce}. The dominant approach employs embedding-based retrieval using the bi-encoder architecture combined with approximate nearest neighbor search~\citep{Nigam2019SemanticProductSearch,Lin2024EnhancingRelevance}. These systems must address severe vocabulary mismatch between customer queries and product descriptions, motivating query rewriting and semantic bridging components, often implemented using LLMs trained with contrastive or instruction-based objectives~\citep{Peng2024LLMBasedLongTailQueryRewriting}.

\paragraph{Ads Ranking.}
Ads ranking is another important industrial IR application scenario. Modern advertising recommendation systems face the fundamental challenge of processing massive candidate sets while maintaining strict latency requirements and ranking quality. The traditional approach employs multi-stage cascading architectures that decompose the Ad ranking problem into sequential stages: retrieval, pre-ranking (or lightweight ranking), ranking (or heavyweight ranking), and auction~\citep{Gallagher2019JointOptimizationOfCascadeRankingModels,Wang2023TowardsTheBetterRankingConsistency,Zheng2024FullStackLearningToRank,Yang2025MTMD}, where the research focus is similar to that of ad hoc retrieval. 

In the context of Ads ranking, the XMC formulation described  in~\cref{sec:neural_ranking} has been shown to improve ranking quality of tail items/Ads~\citep{jain2019slice,dahiya2021siamesexml,yu2022pecos,dahiya2023deepencoderswithauxiliaryparameters,gupta2024dualencoders}, which is often critical for revenue maximization. However, it also comes with higher engineering complexity and often requires retraining of the models when new items/Ads are added. 

The recent breakthrough of LLMs has catalyzed a paradigm shift toward generative recommendation frameworks that can directly generate personalized item sequences from user interaction histories. These approaches index items with meaningful IDs using vector quantization algorithms and generate items from the entire item set for recommendation, conceptualizing online advertising systems as a unified generative process that eliminates inherent goal conflicts between different pipeline stages~\citep{Rajput2023Recommendersystemswithgenerativeretrieval,Zheng2023AdaptingLargeLanguageModelsByIntegratingCollaborativeSemantics,Zhai2024ActionsSpeaksLoaderThanWords}. Given space constraints, we refer readers to the cited works for a more in-depth treatment of the field. 

\paragraph{Cross-Application IR.}
Across applications, several architectural trends are reshaping IR system design. Retrieval-augmented generation (RAG) architectures combine dense vector search with LLMs  to support search, reasoning, and synthesis over structured and unstructured data~\citep{asai2024reliable}. Modular and multi-agent retrieval frameworks further decouple retrieval functions into interoperable components, enabling dynamic adaptation to application-specific requirements~\citep{chang-etal-2025-main,zhang2025agenticinformationretrieval}. Finally, contextual personalization and the convergence of search and recommendation architectures reflect a broader shift toward unified representations of users, documents, and intent~\citep{Zamani2018JointModelingAndOptimization}.

Taken together, the diversity of application-driven architectures underscores that modern IR systems are no longer optimized solely for static relevance ranking, but must increasingly adapt to new roles, constraints, and integration patterns that give rise to emerging architectural directions and open challenges.

\section{Emerging Directions and Challenges}
\label{sec:future_direction}

IR systems have become crucial across diverse domains, from retrieval-augmented language modeling~\citep{Khandelwal2020Generalization,borgeaud2022improving} to applications in agents~\citep{wu2023autogen,wang2024llmagentsurvey,wang2025reinforcementlearningselfimprovingagent}, code generation~\citep{wang2024coderag,zhang-etal-2023-repocoder}, robotics~\citep{anwar2024remembr}, medicine~\citep{jeong2024improving}, and protein research~\citep{jumper2021highly}, \interalia. These developments present new challenges for IR research. Drawing from the evolution of IR architectures (\cref{sec:traditional,sec:ltr,sec:neural_ranking,sec:transformer,sec:llm4ir,sec:scenarios_constraints}), we examine emerging trends, open problems, and potential research directions.
Our discussion is structured around three key areas: advancing the core components of IR models, adapting to the new paradigm of retrieval for AI, and tackling the pragmatic challenges of real-world deployment.

\subsection{Advancing the Core Components of IR Models}
\label{subsec:direction_core_models}
At the heart of any IR system are the models that extract features and estimate relevance. As IR moves toward more compute-intensive practices, we identify key areas for improving these fundamental components.
\paragraph{More Powerful and Efficient Foundation Models.}
Scaling has been a winning recipe for modern neural networks~\citep[\interalia]{kaplan2020scaling,hoffmann2022training,dehghani2023scaling,fang2024scaling,shao2024scaling}. However, for IR to leverage this trend sustainably, several challenges in model design and training must be addressed:
\begin{itemize}
\item \textbf{Data and training efficiency.} Current transformer-based IR models demand extensive training data~\citep{fang2024scaling}, making them impractical for many real-world applications.
Developing architectures that can learn effectively from limited data or in a few-shot/zero-shot setting remains crucial. Additionally, models should support parallel processing and low-precision training to reduce costs and accelerate convergence~\citep{pool2021nvidia,fishman2024scaling,liu2024deepseek}.

\item \textbf{Inference optimization.} Real-time applications like conversational search~\citep{mo2024survey} and agent-based systems~\citep{yao2023react} require efficient handling of variable-length queries, necessitating advanced compression and optimization techniques for both retriever backbones and index structures~\citep[\interalia]{dettmers2023case,kumar2024scaling,xu-etal-2024-beyond-perplexity,warner2024modernbert,bruch2024efficient,xu2025csplade}.

\item \textbf{Better lite foundation models.} IR models need to process queries in real time, and using compact-sized foundation models is often a practical solution. \cite{warner2024modernbert} presented an interesting study on wide-and-shallow versus deep-and-narrow architecture in the context of training a ``modern'' \textsc{BERT} model. Recent works~\citep{gunther2024jinaembeddings28192token,fu2023monarchmixer,portes2023mosaicbert} studied training efficient \textsc{BERT} model that supports longer context length, while \cite{nussbaum2025trainingsparsemixtureexperts} investigated the efficacy of a mixture-of-expert \textsc{BERT}-style encoder model. The best recipe for lite foundational models for IR applications remains an open question.

\item \textbf{Transformer alternatives.} 
While transformers have dominated recent IR research, their quadratic complexity in attention computation remains a significant bottleneck. Recent advances in linear RNNs~\citep{peng-etal-2023-rwkv,peng2024eagle,qin2024hierarchically}, state space models~\citep{gu2024mamba,dao2024mamba2}, and linear attention~\citep{katharopoulos2020transformers,yang2024gated} offer alternatives with theoretical linear complexity. Although preliminary studies~\citep{xu2024rankmamba,zhang2024mambaretriever,xu-etal-2025-state,xu2025distillationvscontrastive} show limited gains compared to optimized transformers, developing efficient alternative architectures for transformers could revolutionize large-scale IR.
\end{itemize}

Ultimately, strong foundation models have proven crucial for IR performance~\citep[\interalia]{neelakantan2022text,ma2024fine}. As IR applications expand, developing foundation models that balance computational efficiency with robust performance across tasks and modalities emerges as a key research priority.

\paragraph{Flexible and Scalable Relevance Estimators.}
As discussed in~\cref{sec:transformer}, cross-encoders provide complex non-linear relevance estimation but are computationally expensive. In contrast, bi-encoder architectures used in dense and sparse retrieval rely on linear similarity functions (e.g., inner product) to enable fast retrieval through nearest neighbor search and inverted indexing. Balancing complex relevance matching and scalable retrieval remains challenging. \textsc{ColBERT}~\citep{khattab2020colbert} addresses this by using document representation matrices with MaxSim operations, while recent work \citep{killingback2025hypencoder} explores Hypernetworks \citep{ha2022hypernetworks} to generate query-specific neural networks for relevance estimation. The design of flexible yet scalable relevance estimators remains an active research direction.

\subsection{The Shifting Paradigm: From Search for Humans to Retrieval for AI}
\label{subsec:direction_paradigm_shift}

The integration of IR systems into other research domains presents new challenges. We discuss key implications for future IR modeling research as the primary ``user'' of retrieval shifts from humans to AI models.

\paragraph{The End ``User'' of Retrieval.}
While traditional IR systems focus on providing search results to humans, retrieval is increasingly used to support ML models, particularly LLMs, in tasks such as generation~\citep{gao2023retrieval}, reasoning~\citep{yao2024mcqg,islam-etal-2024-open}, tool usage~\citep{schick2023toolformer,patil2024gorilla,qin2024toolllm,patil2025bfcl} and planning~\citep{song2023llm}. This shifting paradigm raises fundamental questions about task formulation, evaluation, and system optimization:
\begin{itemize}
\item Current IR research is grounded in human information-seeking behavior~\citep[\interalia]{wilson2000human,marchionini2006exploratorysearch,white2009exploratory}. When the end user becomes another ML model, we must reconsider how to define and assess \textit{relevance}. For example, a document might be irrelevant to a human but contain a key factual nugget that an LLM needs to answer a question. This suggests a need for flexible, data-efficient models that are adaptable to various downstream tasks.
\item Traditional IR metrics, which are designed for human-centric evaluation, may not align with downstream task performance in retrieval-augmented systems~\citep{lewis2020retrievalaugmentedgeneration,petroni2021kilt,asai2024reliable}. Future IR models should support end-to-end system optimization rather than focusing solely on ranking metrics~\citep{deepresearchsystemcard,huang2025deepresearchagents}.
\end{itemize}

\paragraph{Retrieval Augmented Generation.}
Retrieval Augmented Generation (RAG) refers to architectures that combine an external retrieval module with a generative model, enabling the system to access and condition on external knowledge before producing output. Early RAG formulations separate retrieval and generation as distinct components:
a dense or sparse retriever extracts relevant passages for a given input, and a generator (usually an LLM) conditions on those retrieved results to produce text~\citep{lewis2020retrievalaugmentedgeneration,ram2023incontextretrievalaugmented,mallen-etal-2023-trust,asai2024reliable,chen2020abstractivesnippetgeneration,xu2023lightweightconstrainedgeneration}. This architectural decoupling allows retrieval to be optimized independently from generation, which is useful for modularity and scalability.
A typical RAG pipeline comprises:
\begin{itemize}
    \item Corpus embedding and indexing (vector stores or sparse indexes) for fast retrieval;
    \item Retriever model (dense vectors or hybrid techniques) that returns top-$k$ relevant items;
    \item Fusion mechanism that combines retrieved content with the query (e.g., concatenation or cross-attention) before passing it to the generator;
    \item Generator that produces the final answer based on fused context.
\end{itemize}
This simple architecture improves factual accuracy and reduces hallucination by grounding generation on external context. However, it also introduces key engineering and modeling considerations:
\begin{itemize}
    \item \textbf{Retriever-Generator Coupling}: The architectural choice of where and how to combine retrieval results can affect quality and efficiency. Early approaches rely on simple concatenation of text passages into the generator’s context window~\citep{lewis2020retrievalaugmentedgeneration,ram2023incontextretrievalaugmented}, which do not alter the architecture of the generator. More advanced designs use cross-attention layers or rerankers to improve precision~\citep{izacard-grave-2021-leveraging,park2023rink,an2025hyperragenhancingqualityefficiencytradeoffs}.
    \item \textbf{Multi-Modal and Structured Retrieval}: Recent work explores RAG variants that extend beyond plain text. For example, GraphRAG and related graph-enhanced frameworks integrate structured knowledge sources, enabling retrieval of relational paths not easily captured by vector similarity alone~\citep{edge2025localglobalgraphrag,li2025simpleiseffective}. These architectures introduce additional components like graph encoders and reasoning layers, allowing the system to extract multi-hop or entity-centric context before generation.
    \item \textbf{Adaptive Retrieval}: Newer RAG pipelines incorporate reasoning capabilities into the generator, to dynamically decide whether to retrieval or to generate answer~\citep{trivedi-etal-2023-interleavingretrievalwithchainofthoughtreasoning,jiang2023activeretrievalaugmentedgeneration,asai2024selfrag}.
\end{itemize}
These architectural choices have direct implications: strong retrieval can reduce generation errors but may increase pipeline complexity and latency; graph-aware models can support structured reasoning at the cost of larger indexes and additional components; adaptive retrieval strategies can improve effectiveness but require careful task planning and retrieval coordination.

Overall, RAG architectures represent a spectrum from simple retrieval-plus-generation pipelines to complex multi-component systems that integrate heterogeneous knowledge representations and adaptive retrieval planning to improve both relevance and reasoning capability. We refer readers to~\citep{gao2023retrieval,li2025towardsagenticrag,mei2025surveymultimodalretrievalaugmentedgeneration} for more comprehensive reviews of this topic.

\paragraph{The Rise of Autonomous Search Agents.}
Complex tasks often require retrieving long-tail knowledge using lengthy, complex queries~\citep{soudani2024finetuning,su2024bright}, demanding retrieval models capable of instruction following~\citep{weller2024followir,ravfogel2024descriptionbased} and reasoning~\citep{su2024bright,shao2025reasonir,liang2025reasoningrag}. This has spurred the development of autonomous search agents. Existing attempts can be divided into two main directions.
One line of work focuses on training more capable retrievers and rerankers. This involves creating new data pipelines to synthesize instruction-following and reasoning-focused training data~\citep[\interalia]{oh2024instructir,weller2024promptriever,shao2025reasonir} and leveraging strong backbone language models such as the proprietary GPT-4 family models~\citep{achiam2023gpt} and open-weight \textsc{Qwen}~\citep{yang2025qwen3technicalreport} to imbue the retriever with reasoning capabilities.

Another line of work treats the search/retrieval system as a static tool and focuses on improving a large reasoning model's (LRM) ability to use the retrieval tool to improve reasoning and handle in-context memory~\citep[\interalia]{jin2025searchr1trainingllmstoreasonandleveragesearchengines,yu2025memagentreshapinglongcontextllm,gao2025turnsunlockinglonghorizonagentic,xu2025reconreasoningcondensationefficient}. In this setup, the LRM decides when, where, and how to conduct a search, and the results subsequently influence its further reasoning and decision-making~\citep{nakano2021webgpt,he2025pasa}. 
This line of work, commonly referred to as \textbf{Deep Research Agents}, represents a paradigm shift from \textit{retrieval as the end goal} to \textit{retrieval as tools for LLMs}.
Formally, we use the definition from \cite{huang2025deepresearchagents}: 
deep research agents refer to AI agents powered by LLMs, integrating dynamic reasoning, adaptive planning, multi-iteration external data retrieval and tool use, and comprehensive analytical report generation for informational research tasks. 

From the retrieval perspective, the retrieval usage of deep research agents can be broadly categorized into two types: (1) API-based search engines which interact with structured data sources, such as search engine APIs or scientific database APIs; (2) browser-based search engines, which simulate human-like interactions with web pages and enable real-time extraction of dynamic or unstructured content leveraging LLMs' long-context, code understanding and multimodality capabilities. 

Both formulations have strengths and weaknesses. API-based retrieval, in line with traditional IR literature, is a fast, efficient, structured and scalable method to enable deep research agents' access to external knowledge~\citep{schutze2008introduction,Singh2025Ai2scholarqa}. 
Browser-based retrieval~\citep{nakano2021webgpt,deepresearchsystemcard} offers the advantage of simulating real-time, interactive information-seeking behavior\,---\ conceptually similar to that of human users. However, it incurs additional latency and token consumption, and introduces further complexity arising from the web-browsing environment.
Given the current landscape of retrieval integration within deep research agents, a central open question is how to design a hybrid retrieval architecture that combines the strengths of both paradigms to achieve a superior performance-efficiency tradeoff.

From the model training perspective, recent efforts have abstracted away the details of retrieval models, treating retrieval as static tools and instead focusing on improving the capabilities of LLM-based search agents.
For example, many recent efforts aim to train LRMs to use search tools more effectively via reinforcement learning or specialized fine-tuning~\citep[\interalia]{li2025searcho1agenticsearchenhancedlargereasoningmodels,jin2025searchr1trainingllmstoreasonandleveragesearchengines,li2025webthinker,chen2025learningtoreasonwithsearchforllmsviareinforcementlearning,wu-etal-2025-webwalker,wu2025webdancer}. This approach is central to agentic frameworks that orchestrate tool use for complex task completion~\citep{wu2023autogen,shinn2023reflexion,asai2024selfrag}.
Despite this exciting progress, key limitations remain. To enable retrievers' reasoning capability often requires strong backbone models (e.g., 7B scale), which can be infeasible for production systems. Even larger models (e.g., 32B scale) augmented with retrieval and trained via expensive reinforcement learning~\citep{jin2025searchr1trainingllmstoreasonandleveragesearchengines,chen2025learningtoreasonwithsearchforllmsviareinforcementlearning} still sometimes underperform simpler baselines with query decomposition and chain-of-thought prompting~\citep{khot2023decomposedprompting,trivedi-etal-2023-interleavingretrievalwithchainofthoughtreasoning}. A key open question is how to endow retrievers with strong reasoning capabilities using lightweight, scalable models. Another challenge lies in the joint optimization of retrievers and language models within a unified, reasoning-aware framework. Lastly, the human factors of applying such autonomous search agents remain to be studied. We refer readers to~\citep{singh2025agenticragsurvey,liang2025reasoningrag,xi2025surveyofllmbaseddeepsearchagents,lin2025comprehensivesurveyreinforcementlearningbased} for more comprehensive reviews of this topic.

\subsection{Retrieval Beyond Simple Relevance: Instruction-Following and Reasoning-Aware Retrieval}
\label{subsec:if_and_reasoning}

\paragraph{Instruction-Following Retrieval.}
Instruction-following retrieval extends the standard query-document formulation by conditioning the retriever on a natural-language instruction. Architecturally, this is typically implemented by jointly encoding the instruction and the query in a shared bi-encoder, for example through simple concatenation or lightweight attention layers, so that the query representation adapts to task intent~\citep{weller2024followir,oh2024instructir}. Training objectives often use contrastive learning with instruction-conditioned positives and hard negatives, allowing the model to capture fine-grained task semantics without large cross-encoder computations~\citep{weller2024promptriever}. Instruction-aware retrievers benefit from curated or synthetic datasets pairing instructions, queries, and relevant passages, which improves robustness to paraphrasing and query phrasing variations. These design choices preserve scalability while improving performance on instruction-heavy tasks. Hybrid pipelines may further combine instruction-conditioned retrieval with downstream rerankers or fusion mechanisms, ensuring that retrieved evidence aligns with the specific needs of generation or reasoning tasks~\citep{shao2024scaling,weller2025rank1testtimecomputereranking}.

\paragraph{Reasoning-Aware Retrieval.}
Reasoning-aware retrieval aims to retrieve evidence that supports multi-step inference and complex reasoning~\citep{su2024bright}. One line of work uses \textbf{graph-based architectures}, where the retriever constructs and encodes relational graphs or multi-hop chains of passages. These components\,---\,graph construction, encoding, path selection, and ranking\,---\,allow the retriever to return connected evidence that reflects reasoning dependencies~\citep{liu-etal-2025-hoprag,edge2025localglobalgraphrag}. While effective for compositional queries, these approaches increase system complexity, latency, and index size.

A different line of work, exemplified by \textsc{ReasonIR}~\citep{shao2025reasonir} and \textsc{RankR1}~\citep{zhuang2025rankr1enhancingreasoningllmbased}, improves reasoning capability through \textit{data curation and task-aligned training} rather than altering the underlying ``retrieve-then-rerank'' pipeline. By generating multi-step reasoning datasets or instruction-guided examples, these methods teach standard retrievers and rerankers to prioritize evidence that is useful for downstream reasoning, achieving stronger performance without introducing graph-based components. From an evaluation perspective, reasoning-aware retrieval highlights the need for task-centered metrics that measure downstream reasoning success, emphasizing the tight connection between architectural or training choices and reasoning effectiveness.

\subsection{Deployment, Robustness, and Trustworthiness}
\label{subsec:direction_deployment}
As modern IR systems become more powerful and integrated into high-stakes applications, ensuring their practical deployability and reliability is paramount.

\paragraph{The Efficiency-Effectiveness Tradeoff at Scale.}
Traditional retrieval systems face significant challenges when scaling to web-scale document corpus, and deploying such systems requires a blend of science and engineering expertise~\citep{dean2009challenges,huang2020embeddingbasedretrieval,li2021embeddingbasedproductretrievalintaobaosearch}. In recent years,
retrieval-augmented generation, conversational search, and agentic systems with memory have been widely adopted for information access~\citep[\interalia]{guu2020retrievalaugmentedlanguagemodelpretraining,lewis2020retrievalaugmentedgeneration,google2019search,openai2024introducingchatgptsearch,google2024groundingwithgooglesearch}. These applications often require multiple rounds of retrieval and operate on dynamic corpuses, urging for efficient and effective retrieval. Mainstream inference optimization frameworks such as vLLM~\citep{kwon2023vllmpagedattention} and SGLang~\citep{zheng2024sglang} have provided support for embedding models. From the modeling perspective, an open question is to design and pre-train models explicitly for retrieval purposes~\citep{warner2024modernbert,nussbaum2025nomicembed,gunther2024jinaembeddings28192token}.

\paragraph{Ensuring Robustness in a Noisy and Adversarial World.}
We discuss a few challenges in IR models' deployment in noisy environments, especially when used in retrieval-augmented generation systems. We should note that while these challenges have been studied by prior works, it remains an open question on how to mitigate these challenges from the perspective of IR modeling and architectures.
\begin{itemize}
\item \textbf{Robustness to AI-generated content.} With the advent of LLMs, the amount of AI-generated content is also increasing. \citet{dai2024neuralretrieversarebiasedtowardsllmgeneratedcontent} showed that neural retrievers are biased towards AI-generated documents. \citet{xu2024invisiblerelevancebiastextimageretrievalmdoel} showed that similar problems persist in text-image retrieval models. Future IR modeling research should also consider the robustness of models to AI-generated content.
\item \textbf{Robustness to adversarial attacks.} Recent works on ad hoc retrieval and RAG LLM safety have discussed \textsc{BERT}-based models' brittleness to adversarial attacks~\citep{wang2022bertrankersbrittle}, as well as the threat of corpus poisoning where injected harmful documents lead to unsafe RAG system outputs~\citep[\interalia]{zhong-etal-2023-poisoningretrievalcorporabyinjectingadversarianpassages,xiang2024certifiablyrobustragagainstretrievalcorruption,deng2024pandorajailbreakgptsbyretrievalaugmentedgenerationpoisoning}. This topic is also relevant to the safety of LLM agents using tools~\citep{deng2025aiagentsunderthreat,tian2023evil,xiang2024guardagent}, noting the importance of IR models being robust to adversarial attacks for downstream applications.
\item \textbf{Robustness to bias and toxicity.} As noted by a recent work~\citep{an-etal-2025-ragllmsarenotsafer}, documents that contain biases and toxic materials can potentially jailbreak aligned LLMs. This observation highlights the importance for IR models to be robust to bias and toxic content.
\item \textbf{Robustness to imperfect retrieval results.} Different works have pointed out that existing RAG systems show performance degradation when the retrieval results contain irrelevant documents~\citep[\interalia]{yoran2024makingretrievalaugmentedlanguagemodelsrobusttoirrelevantcontext,chang-etal-2025-main,yu-etal-2024-chainofnote}. Therefore, the RAG paradigm demands more precise results from the retrieval models.
\item \textbf{Robustness to out-of-distribution input.} Given the fact that modern neural retrieval models are trained with data-driven approaches, perhaps it is not surprising to find their performance may vary with different linguistic properties of the queries and documents, i.e., out-of-distribution input from the training data. Several works have reported cross-encoder rerankers' performance drops on out-of-domain datasets~\citep{iurii2021systematicevaluation,thakur2021beir}.
In the context of retrieval-augmented generation, \citet{cao2025outofstyleragfragilitytolinguisticvariation} conducted a rigorous benchmarking, and found that formality, readability, politeness and grammatical correctness\,---\,fundamental aspects of real-world user-LLM queries\,---\,can lead to significant performance variances of retrievers and RAG systems. This observation highlights the importance of retrieval models' robustness to out-of-distribution (OOD)  input~\citep{gupta-etal-2024-whispersofdoubt}.
\end{itemize}
We refer readers for more detailed discussions on IR models' robustness to dedicated surveys~\citep{asai2024reliable,liu2025robustinformationretrieval,zhou2024trustworthinessinretrievalaugmentedgenerationsystemssurvey}. Addressing these robustness issues at the model architecture level is a critical and underexplored direction for future research.

\section{Conclusions and Closing Thoughts}
\label{sec:conclusion}

The journey of IR model architectures, as we have charted, is a story of escalating abstraction and semantic depth. Beginning with the foundational principles of term-based matching in Boolean, vector space, and probabilistic models, the field systematically evolved. Learning-to-Rank introduced the power of machine learning to combine diverse statistical features, but it was the advent of neural networks that marked the first major leap toward semantic understanding. These neural ranking models, with their ability to learn representations directly from text, began to bridge the lexical gap that had long constrained traditional methods. The arrival of pre-trained transformers like \textsc{BERT} then catalyzed a paradigm shift, providing a powerful, universal foundation for both highly-effective cross-encoder rerankers and efficient bi-encoder retrievers. Most recently, the ascent of LLMs has not only scaled up these existing architectures but has also introduced entirely new paradigms, such as generative retrieval and zero-shot listwise reranking, fundamentally reshaping what a retrieval system can do.

Throughout this evolution, a core architectural tension has persisted: the tradeoff between effectiveness and efficiency. This is the fundamental conflict between deep, fine-grained interaction models (like interaction-based neural rankers and cross-encoders) that offer high accuracy, and scalable representation-based models (like vector space models, dense retrievers) that enable fast, pre-computable search over massive collections. The enduring ``retrieve-then-rerank'' pipeline is a direct architectural answer to this tradeoff. Innovations like multi-vector models (e.g., \textsc{ColBERT}) and hybrid sparse-dense systems represent sophisticated attempts to find a better balance on this spectrum, pushing the Pareto frontier of what is possible.

Today, we stand at another inflection point. The primary ``user'' of IR is shifting from a human at a search bar to an AI model within a larger system. IR is no longer just a tool for finding documents; it is becoming a critical cognitive component for retrieval-augmented generation, autonomous agents, and complex reasoning systems. This shift, as outlined in \cref{sec:future_direction}, forces us to re-evaluate our core assumptions. Relevance is no longer solely about satisfying human information needs but about providing the precise factual or contextual information an AI needs to complete a downstream task. This demands new model architectures that are not only powerful and efficient but also instruction-aware, contextually flexible, and seamlessly integrable into end-to-end differentiable systems.

Looking ahead, the future of IR model architecture will be defined by its ability to meet these new demands. The grand challenges lie in building foundation models that are multimodal, multilingual, and computationally sustainable; in designing systems that are robust, trustworthy, and resistant to adversarial manipulation; and in developing autonomous search agents that can reason, plan, and interact with the world's information on our behalf. As IR becomes ever more deeply embedded in the fabric of artificial intelligence, its continued evolution will be crucial, not just for the future of search, but for the future of intelligent systems themselves.

\section*{Acknowledgments}
\label{sec:ack}
This work was partially supported by grants from NSF (DMS 2134223 and IIS 2205418). 

\bibliography{references}

\begin{thebibliography}{521}
\providecommand{\natexlab}[1]{#1}
\providecommand{\url}[1]{\texttt{#1}}
\expandafter\ifx\csname urlstyle\endcsname\relax
  \providecommand{\doi}[1]{doi: #1}\else
  \providecommand{\doi}{doi: \begingroup \urlstyle{rm}\Url}\fi

\bibitem[Ai et~al.(2018{\natexlab{a}})Ai, Bi, Guo, and Croft]{ai2018learning}
Qingyao Ai, Keping Bi, Jiafeng Guo, and W.~Bruce Croft.
\newblock Learning a deep listwise context model for ranking refinement.
\newblock In \emph{Proceedings of the 41st International ACM SIGIR Conference on Research \& Development in Information Retrieval}, pp.\  135--144, 2018{\natexlab{a}}.

\bibitem[Ai et~al.(2018{\natexlab{b}})Ai, Mao, Liu, and Croft]{ai2018unbiased}
Qingyao Ai, Jiaxin Mao, Yiqun Liu, and W.~Bruce Croft.
\newblock Unbiased learning to rank: Theory and practice.
\newblock In \emph{Proceedings of the 27th ACM International Conference on Information and Knowledge Management}, CIKM '18, pp.\  2305--2306, New York, NY, USA, 2018{\natexlab{b}}. Association for Computing Machinery.
\newblock \doi{10.1145/3269206.3274274}.

\bibitem[Ai et~al.(2019)Ai, Wang, Bruch, Golbandi, Bendersky, and Najork]{ai2019learning}
Qingyao Ai, Xuanhui Wang, Sebastian Bruch, Nadav Golbandi, Michael Bendersky, and Marc Najork.
\newblock Learning groupwise multivariate scoring functions using deep neural networks.
\newblock In \emph{Proceedings of the 2019 ACM SIGIR International Conference on Theory of Information Retrieval}, pp.\  85--92, 2019.

\bibitem[Althammer et~al.(2020)Althammer, Hofst{\"a}tter, and Hanbury]{Althammer2020CrossdomainRetrieval}
Sophia Althammer, Sebastian Hofst{\"a}tter, and Allan Hanbury.
\newblock Cross-domain retrieval in the legal and patent domains: a reproducability study.
\newblock In \emph{European Conference on Information Retrieval}, 2020.

\bibitem[An et~al.(2025{\natexlab{a}})An, Zhang, and Dredze]{an-etal-2025-ragllmsarenotsafer}
Bang An, Shiyue Zhang, and Mark Dredze.
\newblock {RAG} {LLM}s are not safer: A safety analysis of retrieval-augmented generation for large language models.
\newblock In Luis Chiruzzo, Alan Ritter, and Lu~Wang (eds.), \emph{Proceedings of the 2025 Conference of the Nations of the Americas Chapter of the Association for Computational Linguistics: Human Language Technologies (Volume 1: Long Papers)}, pp.\  5444--5474, Albuquerque, New Mexico, April 2025{\natexlab{a}}. Association for Computational Linguistics.
\newblock URL \url{https://aclanthology.org/2025.naacl-long.281/}.

\bibitem[An et~al.(2025{\natexlab{b}})An, Cheng, Park, and Jiang]{an2025hyperragenhancingqualityefficiencytradeoffs}
Yuwei An, Yihua Cheng, Seo~Jin Park, and Junchen Jiang.
\newblock {HyperRAG}: Enhancing quality-efficiency tradeoffs in retrieval-augmented generation with reranker kv-cache reuse.
\newblock \emph{arXiv preprint arXiv:2504.02921}, 2025{\natexlab{b}}.

\bibitem[Ansell et~al.(2022)Ansell, Ponti, Korhonen, and Vuli{\'c}]{ansell-etal-2022-composable}
Alan Ansell, Edoardo Ponti, Anna Korhonen, and Ivan Vuli{\'c}.
\newblock Composable sparse fine-tuning for cross-lingual transfer.
\newblock In Smaranda Muresan, Preslav Nakov, and Aline Villavicencio (eds.), \emph{Proceedings of the 60th Annual Meeting of the Association for Computational Linguistics (Volume 1: Long Papers)}, pp.\  1778--1796, Dublin, Ireland, May 2022. Association for Computational Linguistics.
\newblock \doi{10.18653/v1/2022.acl-long.125}.
\newblock URL \url{https://aclanthology.org/2022.acl-long.125/}.

\bibitem[Anwar et~al.(2024)Anwar, Welsh, Biswas, Pouya, and Chang]{anwar2024remembr}
Abrar Anwar, John Welsh, Joydeep Biswas, Soha Pouya, and Yan Chang.
\newblock Remembr: Building and reasoning over long-horizon spatio-temporal memory for robot navigation.
\newblock \emph{arXiv preprint arXiv:2409.13682}, 2024.

\bibitem[Arora et~al.(2017)Arora, Liang, and Ma]{arora2017asimplebuttoughtobeatbaseline}
Sanjeev Arora, Yingyu Liang, and Tengyu Ma.
\newblock A simple but tough-to-beat baseline for sentence embeddings.
\newblock In \emph{International Conference on Learning Representations}, 2017.
\newblock URL \url{https://openreview.net/forum?id=SyK00v5xx}.

\bibitem[Artetxe et~al.(2020)Artetxe, Ruder, and Yogatama]{artetxe-etal-2020-cross}
Mikel Artetxe, Sebastian Ruder, and Dani Yogatama.
\newblock On the cross-lingual transferability of monolingual representations.
\newblock In Dan Jurafsky, Joyce Chai, Natalie Schluter, and Joel Tetreault (eds.), \emph{Proceedings of the 58th Annual Meeting of the Association for Computational Linguistics}, pp.\  4623--4637, Online, July 2020. Association for Computational Linguistics.
\newblock \doi{10.18653/v1/2020.acl-main.421}.
\newblock URL \url{https://aclanthology.org/2020.acl-main.421/}.

\bibitem[Asai et~al.(2023)Asai, Schick, Lewis, Chen, Izacard, Riedel, Hajishirzi, and Yih]{asai-etal-2023-task}
Akari Asai, Timo Schick, Patrick Lewis, Xilun Chen, Gautier Izacard, Sebastian Riedel, Hannaneh Hajishirzi, and Wen-tau Yih.
\newblock Task-aware retrieval with instructions.
\newblock In Anna Rogers, Jordan Boyd-Graber, and Naoaki Okazaki (eds.), \emph{Findings of the Association for Computational Linguistics: ACL 2023}, pp.\  3650--3675, Toronto, Canada, July 2023. Association for Computational Linguistics.
\newblock \doi{10.18653/v1/2023.findings-acl.225}.
\newblock URL \url{https://aclanthology.org/2023.findings-acl.225/}.

\bibitem[Asai et~al.(2024{\natexlab{a}})Asai, He, Shao, Shi, Singh, Chang, Lo, Soldaini, Feldman, D'arcy, et~al.]{asai2024openscholar}
Akari Asai, Jacqueline He, Rulin Shao, Weijia Shi, Amanpreet Singh, Joseph~Chee Chang, Kyle Lo, Luca Soldaini, Sergey Feldman, Mike D'arcy, et~al.
\newblock {OpenScholar}: Synthesizing scientific literature with retrieval-augmented {LMs}.
\newblock \emph{arXiv preprint arXiv:2411.14199}, 2024{\natexlab{a}}.

\bibitem[Asai et~al.(2024{\natexlab{b}})Asai, Wu, Wang, Sil, and Hajishirzi]{asai2024selfrag}
Akari Asai, Zeqiu Wu, Yizhong Wang, Avirup Sil, and Hannaneh Hajishirzi.
\newblock Self-{RAG}: Learning to retrieve, generate, and critique through self-reflection.
\newblock In \emph{The 12th International Conference on Learning Representations}, 2024{\natexlab{b}}.
\newblock URL \url{https://openreview.net/forum?id=hSyW5go0v8}.

\bibitem[Asai et~al.(2024{\natexlab{c}})Asai, Zhong, Chen, Koh, Zettlemoyer, Hajishirzi, and Yih]{asai2024reliable}
Akari Asai, Zexuan Zhong, Danqi Chen, Pang~Wei Koh, Luke Zettlemoyer, Hannaneh Hajishirzi, and Wen-tau Yih.
\newblock Reliable, adaptable, and attributable language models with retrieval.
\newblock \emph{arXiv preprint arXiv:2403.03187}, 2024{\natexlab{c}}.

\bibitem[Baeza-Yates \& Ribeiro-Neto(1999)Baeza-Yates and Ribeiro-Neto]{baeza1999modern}
Ricardo Baeza-Yates and Berthier Ribeiro-Neto.
\newblock \emph{Modern Information Retrieval}.
\newblock ACM Press; Addison-Wesley, 1st edition, 1999.

\bibitem[Bahdanau et~al.(2014)Bahdanau, Cho, and Bengio]{bahdanau2014neural}
Dzmitry Bahdanau, Kyunghyun Cho, and Yoshua Bengio.
\newblock Neural machine translation by jointly learning to align and translate.
\newblock \emph{arXiv preprint arXiv:1409.0473}, 2014.

\bibitem[Bai et~al.(2020)Bai, Li, Wang, Zhang, Shang, Xu, Wang, Wang, and Liu]{bai2020sparterm}
Yang Bai, Xiaoguang Li, Gang Wang, Chaoliang Zhang, Lifeng Shang, Jun Xu, Zhaowei Wang, Fangshan Wang, and Qun Liu.
\newblock {SparTerm}: Learning term-based sparse representation for fast text retrieval.
\newblock \emph{arXiv preprint arXiv:2010.00768}, 2020.

\bibitem[Bai et~al.(2022)Bai, Kadavath, Kundu, Askell, Kernion, Jones, Chen, Goldie, Mirhoseini, McKinnon, et~al.]{bai2022constitutional}
Yuntao Bai, Saurav Kadavath, Sandipan Kundu, Amanda Askell, Jackson Kernion, Andy Jones, Anna Chen, Anna Goldie, Azalia Mirhoseini, Cameron McKinnon, et~al.
\newblock Constitutional {AI}: Harmlessness from {AI} feedback.
\newblock \emph{arXiv preprint arXiv:2212.08073}, 2022.

\bibitem[Basnet et~al.(2024)Basnet, Gou, Mallia, and Suel]{basnet2024deeperimpact}
Soyuj Basnet, Jerry Gou, Antonio Mallia, and Torsten Suel.
\newblock Deeperimpact: Optimizing sparse learned index structures.
\newblock \emph{arXiv preprint arXiv:2405.17093}, 2024.

\bibitem[BehnamGhader et~al.(2024)BehnamGhader, Adlakha, Mosbach, Bahdanau, Chapados, and Reddy]{behnamghader2024llmvec}
Parishad BehnamGhader, Vaibhav Adlakha, Marius Mosbach, Dzmitry Bahdanau, Nicolas Chapados, and Siva Reddy.
\newblock {LLM2Vec}: Large language models are secretly powerful text encoders.
\newblock In \emph{1st Conference on Language Modeling}, 2024.
\newblock URL \url{https://openreview.net/forum?id=IW1PR7vEBf}.

\bibitem[Beltagy et~al.(2020)Beltagy, Peters, and Cohan]{beltagy2020longformer}
Iz~Beltagy, Matthew~E. Peters, and Arman Cohan.
\newblock Longformer: The long-document transformer.
\newblock \emph{arXiv preprint arXiv:2004.05150}, 2020.

\bibitem[Berger \& Lafferty(1999)Berger and Lafferty]{berger1999information}
Adam Berger and John Lafferty.
\newblock Information retrieval as statistical translation.
\newblock In \emph{Proceedings of the 22nd Annual International ACM SIGIR Conference on Research and Development in Information Retrieval}, pp.\  222--229, 1999.

\bibitem[Bevilacqua et~al.(2022)Bevilacqua, Ottaviano, Lewis, Yih, Riedel, and Petroni]{bevilacqua2022autoregressive}
Michele Bevilacqua, Giuseppe Ottaviano, Patrick Lewis, Scott Yih, Sebastian Riedel, and Fabio Petroni.
\newblock Autoregressive search engines: Generating substrings as document identifiers.
\newblock In Alice~H. Oh, Alekh Agarwal, Danielle Belgrave, and Kyunghyun Cho (eds.), \emph{Advances in Neural Information Processing Systems}, 2022.
\newblock URL \url{https://openreview.net/forum?id=Z4kZxAjg8Y}.

\bibitem[Biega et~al.(2018)Biega, Gummadi, and Weikum]{biega2018equity}
Asia~J. Biega, Krishna~P. Gummadi, and Gerhard Weikum.
\newblock Equity of attention: Amortizing individual fairness in rankings.
\newblock In \emph{The 41st International ACM SIGIR Conference on Research \& Development in Information Retrieval}, pp.\  405--414, 2018.

\bibitem[Bonifacio et~al.(2022)Bonifacio, Abonizio, Fadaee, and Nogueira]{bonifacio2022inpars}
Luiz Bonifacio, Hugo Abonizio, Marzieh Fadaee, and Rodrigo Nogueira.
\newblock {InPars}: Unsupervised dataset generation for information retrieval.
\newblock In \emph{Proceedings of the 45th International ACM SIGIR Conference on Research and Development in Information Retrieval}, pp.\  2387--2392, 2022.

\bibitem[Borgeaud et~al.(2022)Borgeaud, Mensch, Hoffmann, Cai, Rutherford, Millican, Van Den~Driessche, Lespiau, Damoc, Clark, et~al.]{borgeaud2022improving}
Sebastian Borgeaud, Arthur Mensch, Jordan Hoffmann, Trevor Cai, Eliza Rutherford, Katie Millican, George~Bm Van Den~Driessche, Jean-Baptiste Lespiau, Bogdan Damoc, Aidan Clark, et~al.
\newblock Improving language models by retrieving from trillions of tokens.
\newblock In \emph{International Conference on Machine Learning}, pp.\  2206--2240. PMLR, 2022.

\bibitem[Boytsov et~al.(2024)Boytsov, Patel, Sourabh, Nisar, Kundu, Ramanathan, and Nyberg]{boytsov2024inpars}
Leonid Boytsov, Preksha Patel, Vivek Sourabh, Riddhi Nisar, Sayani Kundu, Ramya Ramanathan, and Eric Nyberg.
\newblock {InPars-Light}: Cost-effective unsupervised training of efficient rankers.
\newblock \emph{Transactions on Machine Learning Research}, 2024.

\bibitem[Brin \& Page(1998)Brin and Page]{brin1998anatomy}
Sergey Brin and Lawrence Page.
\newblock The anatomy of a large-scale hypertextual web search engine.
\newblock \emph{Computer Networks}, 30:\penalty0 107--117, 1998.
\newblock URL \url{http://www-db.stanford.edu/~backrub/google.html}.

\bibitem[Broder et~al.(2003)Broder, Carmel, Herscovici, Soffer, and Zien]{broder2003efficientqueryevaluation}
Andrei~Z. Broder, David Carmel, Michael Herscovici, Aya Soffer, and Jason Zien.
\newblock Efficient query evaluation using a two-level retrieval process.
\newblock In \emph{Proceedings of the 12th International Conference on Information and Knowledge Management}, pp.\  426--434, 2003.

\bibitem[Brown et~al.(1993)Brown, Della~Pietra, Della~Pietra, and Mercer]{brown1993mathematicsofstatisticalmachinetranslation}
Peter~F. Brown, Stephen~A. Della~Pietra, Vincent~J. Della~Pietra, and Robert~L. Mercer.
\newblock The mathematics of statistical machine translation: Parameter estimation.
\newblock \emph{Computational linguistics}, 19\penalty0 (2):\penalty0 263--311, 1993.

\bibitem[Brown et~al.(2020)Brown, Mann, Ryder, Subbiah, Kaplan, Dhariwal, Neelakantan, Shyam, Sastry, Askell, et~al.]{brown2020language}
Tom Brown, Benjamin Mann, Nick Ryder, Melanie Subbiah, Jared~D. Kaplan, Prafulla Dhariwal, Arvind Neelakantan, Pranav Shyam, Girish Sastry, Amanda Askell, et~al.
\newblock Language models are few-shot learners.
\newblock \emph{Advances in neural information processing systems}, 33:\penalty0 1877--1901, 2020.

\bibitem[Bruch et~al.(2019)Bruch, Zoghi, Bendersky, and Najork]{bruch2019revisiting}
Sebastian Bruch, Masrour Zoghi, Michael Bendersky, and Marc Najork.
\newblock Revisiting approximate metric optimization in the age of deep neural networks.
\newblock In \emph{Proceedings of the 42nd International ACM SIGIR Conference on Research and Development in Information Retrieval}, pp.\  1241--1244, 2019.

\bibitem[Bruch et~al.(2024)Bruch, Nardini, Rulli, and Venturini]{bruch2024efficient}
Sebastian Bruch, Franco~Maria Nardini, Cosimo Rulli, and Rossano Venturini.
\newblock Efficient inverted indexes for approximate retrieval over learned sparse representations.
\newblock In \emph{Proceedings of the 47th International ACM SIGIR Conference on Research and Development in Information Retrieval}, pp.\  152--162, 2024.

\bibitem[Bucilu\u{a} et~al.(2006)Bucilu\u{a}, Caruana, and Niculescu-Mizil]{bucilua2006modelcompression}
Cristian Bucilu\u{a}, Rich Caruana, and Alexandru Niculescu-Mizil.
\newblock Model compression.
\newblock In \emph{Proceedings of the 12th ACM SIGKDD International Conference on Knowledge Discovery and Data Mining}, pp.\  535--541, 2006.

\bibitem[Burges et~al.(2005)Burges, Shaked, Renshaw, Lazier, Deeds, Hamilton, and Hullender]{burges2005learning}
Chris Burges, Tal Shaked, Erin Renshaw, Ari Lazier, Matt Deeds, Nicole Hamilton, and Greg Hullender.
\newblock Learning to rank using gradient descent.
\newblock In \emph{Proceedings of the 22nd International Conference on Machine Learning}, pp.\  89--96, 2005.

\bibitem[Burges et~al.(2006)Burges, Ragno, and Le]{burges2006learning}
Christopher Burges, Robert Ragno, and Quoc Le.
\newblock Learning to rank with nonsmooth cost functions.
\newblock \emph{Advances in neural information processing systems}, 19, 2006.

\bibitem[Burges(2010)]{burges2010ranknet}
Christopher~J.C. Burges.
\newblock From ranknet to lambdarank to lambdamart: An overview.
\newblock \emph{Learning}, 11\penalty0 (23-581):\penalty0 81, 2010.

\bibitem[Cao et~al.(2025)Cao, Bhandari, Yerukola, Asai, and Sap]{cao2025outofstyleragfragilitytolinguisticvariation}
Tianyu Cao, Neel Bhandari, Akhila Yerukola, Akari Asai, and Maarten Sap.
\newblock Out of style: {RAG}'s fragility to linguistic variation.
\newblock \emph{arXiv preprint arXiv:2504.08231}, 2025.

\bibitem[Cao et~al.(2007)Cao, Qin, Liu, Tsai, and Li]{cao2007learning}
Zhe Cao, Tao Qin, Tie-Yan Liu, Ming-Feng Tsai, and Hang Li.
\newblock Learning to rank: from pairwise approach to listwise approach.
\newblock In \emph{Proceedings of the 24th International Conference on Machine Learning}, pp.\  129--136, 2007.

\bibitem[Chang et~al.(2025)Chang, Jiang, Rakesh, Pan, Yeh, Wang, Hu, Xu, Zheng, Das, and Zou]{chang-etal-2025-main}
Chia-Yuan Chang, Zhimeng Jiang, Vineeth Rakesh, Menghai Pan, Chin-Chia~Michael Yeh, Guanchu Wang, Mingzhi Hu, Zhichao Xu, Yan Zheng, Mahashweta Das, and Na~Zou.
\newblock {MAIN}-{RAG}: Multi-agent filtering retrieval-augmented generation.
\newblock In Wanxiang Che, Joyce Nabende, Ekaterina Shutova, and Mohammad~Taher Pilehvar (eds.), \emph{Proceedings of the 63rd Annual Meeting of the Association for Computational Linguistics (Volume 1: Long Papers)}, pp.\  2607--2622, Vienna, Austria, July 2025. Association for Computational Linguistics.
\newblock \doi{10.18653/v1/2025.acl-long.131}.
\newblock URL \url{https://aclanthology.org/2025.acl-long.131/}.

\bibitem[Chang et~al.(2010)Chang, Hsieh, Chang, Ringgaard, and Lin]{chang2010training}
Yin-Wen Chang, Cho-Jui Hsieh, Kai-Wei Chang, Michael Ringgaard, and Chih-Jen Lin.
\newblock Training and testing low-degree polynomial data mappings via linear svm.
\newblock \emph{Journal of Machine Learning Research}, 11\penalty0 (48):\penalty0 1471--1490, 2010.
\newblock URL \url{http://jmlr.org/papers/v11/chang10a.html}.

\bibitem[Chen et~al.(2023{\natexlab{a}})Chen, Zhang, Guo, de~Rijke, Chen, Fan, and Cheng]{Chen2023ContinualLearningforGenerativeRetrieval}
Jiangui Chen, Ruqing Zhang, Jiafeng Guo, M.~de~Rijke, Wei Chen, Yixing Fan, and Xueqi Cheng.
\newblock Continual learning for generative retrieval over dynamic corpora.
\newblock \emph{Proceedings of the 32nd ACM International Conference on Information and Knowledge Management}, 2023{\natexlab{a}}.

\bibitem[Chen et~al.(2025)Chen, Li, Sun, Zhou, Zhu, Wang, Pan, Zhang, Chen, Yang, et~al.]{chen2025learningtoreasonwithsearchforllmsviareinforcementlearning}
Mingyang Chen, Tianpeng Li, Haoze Sun, Yijie Zhou, Chenzheng Zhu, Haofen Wang, Jeff~Z. Pan, Wen Zhang, Huajun Chen, Fan Yang, et~al.
\newblock Learning to reason with search for {LLMs} via reinforcement learning.
\newblock \emph{arXiv preprint arXiv:2503.19470}, 2025.

\bibitem[Chen et~al.(2020{\natexlab{a}})Chen, Zhao, Jin, and Wu]{Chen2020FineGrainedVideoTextRetrieval}
Shizhe Chen, Yida Zhao, Qin Jin, and Qi~Wu.
\newblock Fine-grained video-text retrieval with hierarchical graph reasoning.
\newblock \emph{2020 IEEE/CVF Conference on Computer Vision and Pattern Recognition (CVPR)}, pp.\  10635--10644, 2020{\natexlab{a}}.

\bibitem[Chen et~al.(2023{\natexlab{b}})Chen, Li, Wang, Zhao, Sun, Zhu, and Liu]{Chen2023VAST}
Sihan Chen, Handong Li, Qunbo Wang, Zijia Zhao, Ming-Ting Sun, Xinxin Zhu, and J.~Liu.
\newblock {VAST}: A vision-audio-subtitle-text omni-modality foundation model and dataset.
\newblock \emph{ArXiv preprint arXiv:2305.18500}, 2023{\natexlab{b}}.

\bibitem[Chen et~al.(2020{\natexlab{b}})Chen, Syed, Stein, Hagen, and Potthast]{chen2020abstractivesnippetgeneration}
Wei-Fan Chen, Shahbaz Syed, Benno Stein, Matthias Hagen, and Martin Potthast.
\newblock Abstractive snippet generation.
\newblock In \emph{Proceedings of the Web Conference 2020}, WWW '20, pp.\  1309–1319, New York, NY, USA, 2020{\natexlab{b}}. Association for Computing Machinery.
\newblock ISBN 9781450370233.
\newblock \doi{10.1145/3366423.3380206}.
\newblock URL \url{https://doi.org/10.1145/3366423.3380206}.

\bibitem[Chen et~al.(2021)Chen, He, Hui, Sun, and Sun]{chen2021simplifiedtinybert}
Xuanang Chen, Ben He, Kai Hui, Le~Sun, and Yingfei Sun.
\newblock Simplified {TinyBERT}: Knowledge distillation for document retrieval.
\newblock In \emph{European Conference on Information Retrieval}, pp.\  241--248. Springer, 2021.

\bibitem[Chi et~al.(2021)Chi, Dong, Wei, Yang, Singhal, Wang, Song, Mao, Huang, and Zhou]{chi-etal-2021-infoxlm}
Zewen Chi, Li~Dong, Furu Wei, Nan Yang, Saksham Singhal, Wenhui Wang, Xia Song, Xian-Ling Mao, Heyan Huang, and Ming Zhou.
\newblock {I}nfo{XLM}: An information-theoretic framework for cross-lingual language model pre-training.
\newblock In Kristina Toutanova, Anna Rumshisky, Luke Zettlemoyer, Dilek Hakkani-Tur, Iz~Beltagy, Steven Bethard, Ryan Cotterell, Tanmoy Chakraborty, and Yichao Zhou (eds.), \emph{Proceedings of the 2021 Conference of the North American Chapter of the Association for Computational Linguistics: Human Language Technologies}, pp.\  3576--3588, Online, June 2021. Association for Computational Linguistics.
\newblock \doi{10.18653/v1/2021.naacl-main.280}.
\newblock URL \url{https://aclanthology.org/2021.naacl-main.280/}.

\bibitem[Chung et~al.(2024)Chung, Hou, Longpre, Zoph, Tay, Fedus, Li, Wang, Dehghani, Brahma, et~al.]{chung2024scaling}
Hyung~Won Chung, Le~Hou, Shayne Longpre, Barret Zoph, Yi~Tay, William Fedus, Yunxuan Li, Xuezhi Wang, Mostafa Dehghani, Siddhartha Brahma, et~al.
\newblock Scaling instruction-finetuned language models.
\newblock \emph{Journal of Machine Learning Research}, 25\penalty0 (70):\penalty0 1--53, 2024.

\bibitem[Clark et~al.(2019)Clark, Khandelwal, Levy, and Manning]{clark2019does}
Kevin Clark, Urvashi Khandelwal, Omer Levy, and Christopher~D. Manning.
\newblock What does {BERT} look at? {An Analysis of BERT's Attention}.
\newblock In Tal Linzen, Grzegorz Chrupa{\l}a, Yonatan Belinkov, and Dieuwke Hupkes (eds.), \emph{Proceedings of the 2019 ACL Workshop BlackboxNLP: Analyzing and Interpreting Neural Networks for NLP}, pp.\  276--286, Florence, Italy, August 2019. Association for Computational Linguistics.
\newblock \doi{10.18653/v1/W19-4828}.
\newblock URL \url{https://aclanthology.org/W19-4828/}.

\bibitem[Clarke \& Dietz(2024)Clarke and Dietz]{clarke2024llm}
Charles L.~A. Clarke and Laura Dietz.
\newblock {LLM}-based relevance assessment still can't replace human relevance assessment.
\newblock \emph{arXiv preprint arXiv:2412.17156}, 2024.

\bibitem[Cohen \& Croft(2016)Cohen and Croft]{cohen2016end2end}
Daniel Cohen and W.~Bruce Croft.
\newblock End to end long short term memory networks for non-factoid question answering.
\newblock In \emph{Proceedings of the 2016 ACM International Conference on the Theory of Information Retrieval}, ICTIR '16, pp.\  143--146, New York, NY, USA, 2016. Association for Computing Machinery.
\newblock \doi{10.1145/2970398.2970438}.

\bibitem[Cohen et~al.(2021)Cohen, Mitra, Lesota, Rekabsaz, and Eickhoff]{Cohen2021NotAllRelevanceScoresareEqual}
Daniel Cohen, Bhaskar Mitra, Oleg Lesota, Navid Rekabsaz, and Carsten Eickhoff.
\newblock Not all relevance scores are equal: Efficient uncertainty and calibration modeling for deep retrieval models.
\newblock \emph{Proceedings of the 44th International ACM SIGIR Conference on Research and Development in Information Retrieval}, 2021.

\bibitem[Collobert et~al.(2011)Collobert, Weston, Bottou, Karlen, Kavukcuoglu, and Kuksa]{collobert2011natural}
Ronan Collobert, Jason Weston, L{\'e}on Bottou, Michael Karlen, Koray Kavukcuoglu, and Pavel Kuksa.
\newblock Natural language processing (almost) from scratch.
\newblock \emph{Journal of machine learning research}, 12\penalty0 (7), 2011.

\bibitem[Conneau et~al.(2020{\natexlab{a}})Conneau, Khandelwal, Goyal, Chaudhary, Wenzek, Guzm{\'a}n, Grave, Ott, Zettlemoyer, and Stoyanov]{conneau2020unsupervised}
Alexis Conneau, Kartikay Khandelwal, Naman Goyal, Vishrav Chaudhary, Guillaume Wenzek, Francisco Guzm{\'a}n, Edouard Grave, Myle Ott, Luke Zettlemoyer, and Veselin Stoyanov.
\newblock Unsupervised cross-lingual representation learning at scale.
\newblock In \emph{Proceedings of the 58th Annual Meeting of the Association for Computational Linguistics}, pp.\  8440--8451, 2020{\natexlab{a}}.
\newblock \doi{10.18653/v1/2020.acl-main.747}.
\newblock URL \url{https://aclanthology.org/2020.acl-main.747/}.

\bibitem[Conneau et~al.(2020{\natexlab{b}})Conneau, Wu, Li, Zettlemoyer, and Stoyanov]{conneau-etal-2020-emerging}
Alexis Conneau, Shijie Wu, Haoran Li, Luke Zettlemoyer, and Veselin Stoyanov.
\newblock Emerging cross-lingual structure in pretrained language models.
\newblock In Dan Jurafsky, Joyce Chai, Natalie Schluter, and Joel Tetreault (eds.), \emph{Proceedings of the 58th Annual Meeting of the Association for Computational Linguistics}, pp.\  6022--6034, Online, July 2020{\natexlab{b}}. Association for Computational Linguistics.
\newblock \doi{10.18653/v1/2020.acl-main.536}.
\newblock URL \url{https://aclanthology.org/2020.acl-main.536/}.

\bibitem[Cormack et~al.(2009)Cormack, Clarke, and Buettcher]{cormack2009reciprocal}
Gordon~V. Cormack, Charles L.~A. Clarke, and Stefan Buettcher.
\newblock Reciprocal rank fusion outperforms condorcet and individual rank learning methods.
\newblock In \emph{Proceedings of the 32nd International ACM SIGIR Conference on Research and Development in Information Retrieval}, SIGIR '09, pp.\  758--759, New York, NY, USA, 2009. Association for Computing Machinery.
\newblock \doi{10.1145/1571941.1572114}.

\bibitem[Cortes \& Vapnik(1995)Cortes and Vapnik]{cortes1995supportvectornetworks}
Corinna Cortes and Vladimir Vapnik.
\newblock Support-vector networks.
\newblock \emph{Machine learning}, 20\penalty0 (3):\penalty0 273--297, 1995.

\bibitem[Dahiya et~al.(2021)Dahiya, Agarwal, Saini, Jiao, Singh, Agarwal, Kar, Varma, et~al.]{dahiya2021siamesexml}
Kunal Dahiya, Ananye Agarwal, Deepak Saini, Jian Jiao, Amit Singh, Sumeet Agarwal, Purushottam Kar, Manik Varma, et~al.
\newblock Siamesexml: Siamese networks meet extreme classifiers with 100m labels.
\newblock In \emph{International Conference on Machine Learning}, pp.\  2330--2340. PMLR, 2021.

\bibitem[Dahiya et~al.(2023)Dahiya, Yadav, Sondhi, Saini, Mehta, Jiao, Agarwal, Kar, and Varma]{dahiya2023deepencoderswithauxiliaryparameters}
Kunal Dahiya, Sachin Yadav, Sushant Sondhi, Deepak Saini, Sonu Mehta, Jian Jiao, Sumeet Agarwal, Purushottam Kar, and Manik Varma.
\newblock Deep encoders with auxiliary parameters for extreme classification.
\newblock In \emph{Proceedings of the 29th ACM SIGKDD Conference on Knowledge Discovery and Data Mining}, pp.\  358--367, 2023.

\bibitem[Dai et~al.(2024)Dai, Zhou, Pang, Liu, Hu, Liu, Zhang, Wang, and Xu]{dai2024neuralretrieversarebiasedtowardsllmgeneratedcontent}
Sunhao Dai, Yuqi Zhou, Liang Pang, Weihao Liu, Xiaolin Hu, Yong Liu, Xiao Zhang, Gang Wang, and Jun Xu.
\newblock Neural retrievers are biased towards {LLM}-generated content.
\newblock In \emph{Proceedings of the 30th ACM SIGKDD Conference on Knowledge Discovery and Data Mining}, pp.\  526--537, 2024.

\bibitem[Dai \& Callan(2019{\natexlab{a}})Dai and Callan]{Dai2019DeeperTU}
Zhuyun Dai and Jamie Callan.
\newblock Deeper text understanding for ir with contextual neural language modeling.
\newblock \emph{Proceedings of the 42nd International ACM SIGIR Conference on Research and Development in Information Retrieval}, 2019{\natexlab{a}}.

\bibitem[Dai \& Callan(2019{\natexlab{b}})Dai and Callan]{dai2019context}
Zhuyun Dai and Jamie Callan.
\newblock Context-aware sentence/passage term importance estimation for 1st stage retrieval.
\newblock \emph{arXiv preprint arXiv:1910.10687}, 2019{\natexlab{b}}.

\bibitem[Dai et~al.(2018)Dai, Xiong, Callan, and Liu]{dai2018cknrm}
Zhuyun Dai, Chenyan Xiong, Jamie Callan, and Zhiyuan Liu.
\newblock Convolutional neural networks for soft-matching n-grams in ad-hoc search.
\newblock In \emph{Proceedings of the 11th ACM International Conference on Web Search and Data Mining}, WSDM '18, pp.\  126--134, New York, NY, USA, 2018. Association for Computing Machinery.
\newblock \doi{10.1145/3159652.3159659}.

\bibitem[Dai et~al.(2023)Dai, Zhao, Ma, Luan, Ni, Lu, Bakalov, Guu, Hall, and Chang]{dai2023promptagator}
Zhuyun Dai, Vincent~Y. Zhao, Ji~Ma, Yi~Luan, Jianmo Ni, Jing Lu, Anton Bakalov, Kelvin Guu, Keith Hall, and Ming-Wei Chang.
\newblock Promptagator: Few-shot dense retrieval from 8 examples.
\newblock In \emph{The 11th International Conference on Learning Representations}, 2023.

\bibitem[Dan \& Roth(2021)Dan and Roth]{dan-roth-2021-effects-transformer}
Soham Dan and Dan Roth.
\newblock On the effects of transformer size on in- and out-of-domain calibration.
\newblock In Marie-Francine Moens, Xuanjing Huang, Lucia Specia, and Scott Wen-tau Yih (eds.), \emph{Findings of the Association for Computational Linguistics: EMNLP 2021}, pp.\  2096--2101, Punta Cana, Dominican Republic, November 2021. Association for Computational Linguistics.
\newblock \doi{10.18653/v1/2021.findings-emnlp.180}.
\newblock URL \url{https://aclanthology.org/2021.findings-emnlp.180/}.

\bibitem[Dao \& Gu(2024)Dao and Gu]{dao2024mamba2}
Tri Dao and Albert Gu.
\newblock Transformers are {SSMs}: Generalized models and efficient algorithms through structured state space duality.
\newblock \emph{Proceedings of the 41st International Conference on Machine Learning}, pp.\  10041 -- 10071, 2024.

\bibitem[Dasgupta et~al.(2023)Dasgupta, Katyan, Das, and Kumar]{dasgupta2023reviewofextrememultilabelclassification}
Arpan Dasgupta, Siddhant Katyan, Shrutimoy Das, and Pawan Kumar.
\newblock Review of extreme multilabel classification.
\newblock \emph{arXiv preprint arXiv:2302.05971}, 2023.

\bibitem[de~Araujo et~al.(2014)de~Araujo, M{\"u}ller, Chishman, and Rigo]{araujo2014information}
Denis~Andrei de~Araujo, Carolina M{\"u}ller, Rove Chishman, and Sandro~Jos{\'e} Rigo.
\newblock Information extraction for legal knowledge representation--a review of approaches and trends.
\newblock \emph{Revista Brasileira de Computa{\c{c}}{\~a}o Aplicada}, 6\penalty0 (2):\penalty0 2--19, 2014.

\bibitem[De~Cao et~al.(2021)De~Cao, Izacard, Riedel, and Petroni]{de2020autoregressive}
Nicola De~Cao, G.~Izacard, S.~Riedel, and F.~Petroni.
\newblock Autoregressive entity retrieval.
\newblock In \emph{ICLR 2021-9th International Conference on Learning Representations}, volume 2021. ICLR, 2021.

\bibitem[Dean(2009)]{dean2009challenges}
Jeffrey Dean.
\newblock Challenges in building large-scale information retrieval systems: invited talk.
\newblock In \emph{Proceedings of the 2nd ACM International Conference on Web Search and Data Mining}, 2009.
\newblock \doi{10.1145/1498759.1498761}.

\bibitem[Dehghani et~al.(2023)Dehghani, Djolonga, Mustafa, Padlewski, Heek, Gilmer, Steiner, Caron, Geirhos, Alabdulmohsin, et~al.]{dehghani2023scaling}
Mostafa Dehghani, Josip Djolonga, Basil Mustafa, Piotr Padlewski, Jonathan Heek, Justin Gilmer, Andreas~Peter Steiner, Mathilde Caron, Robert Geirhos, Ibrahim Alabdulmohsin, et~al.
\newblock Scaling vision transformers to 22 billion parameters.
\newblock In \emph{International Conference on Machine Learning}, pp.\  7480--7512. PMLR, 2023.

\bibitem[Deng et~al.(2024)Deng, Liu, Wang, Li, Zhang, and Liu]{deng2024pandorajailbreakgptsbyretrievalaugmentedgenerationpoisoning}
Gelei Deng, Yi~Liu, Kailong Wang, Yuekang Li, Tianwei Zhang, and Yang Liu.
\newblock {Pandora}: Jailbreak {GPTs} by retrieval augmented generation poisoning.
\newblock \emph{arXiv preprint arXiv:2402.08416}, 2024.

\bibitem[Deng et~al.(2025)Deng, Guo, Han, Ma, Xiong, Wen, and Xiang]{deng2025aiagentsunderthreat}
Zehang Deng, Yongjian Guo, Changzhou Han, Wanlun Ma, Junwu Xiong, Sheng Wen, and Yang Xiang.
\newblock {AI} agents under threat: A survey of key security challenges and future pathways.
\newblock \emph{ACM Computing Surveys}, 57\penalty0 (7):\penalty0 1--36, 2025.

\bibitem[Dettmers \& Zettlemoyer(2023)Dettmers and Zettlemoyer]{dettmers2023case}
Tim Dettmers and Luke Zettlemoyer.
\newblock The case for 4-bit precision: k-bit inference scaling laws.
\newblock In \emph{International Conference on Machine Learning}, pp.\  7750--7774. PMLR, 2023.

\bibitem[Devlin et~al.(2019)Devlin, Chang, Lee, and Toutanova]{devlin-etal-2019-bert}
Jacob Devlin, Ming-Wei Chang, Kenton Lee, and Kristina Toutanova.
\newblock {BERT}: Pre-training of deep bidirectional transformers for language understanding.
\newblock In Jill Burstein, Christy Doran, and Thamar Solorio (eds.), \emph{Proceedings of the 2019 Conference of the North {A}merican Chapter of the Association for Computational Linguistics: Human Language Technologies, Volume 1 (Long and Short Papers)}, pp.\  4171--4186, Minneapolis, Minnesota, June 2019. Association for Computational Linguistics.
\newblock \doi{10.18653/v1/N19-1423}.
\newblock URL \url{https://aclanthology.org/N19-1423/}.

\bibitem[Dhuliawala et~al.(2022)Dhuliawala, Adolphs, Das, and Sachan]{dhuliawala-etal-2022-calibration}
Shehzaad Dhuliawala, Leonard Adolphs, Rajarshi Das, and Mrinmaya Sachan.
\newblock Calibration of machine reading systems at scale.
\newblock In Smaranda Muresan, Preslav Nakov, and Aline Villavicencio (eds.), \emph{Findings of the Association for Computational Linguistics: ACL 2022}, pp.\  1682--1693, Dublin, Ireland, May 2022. Association for Computational Linguistics.
\newblock \doi{10.18653/v1/2022.findings-acl.133}.
\newblock URL \url{https://aclanthology.org/2022.findings-acl.133/}.

\bibitem[Dong et~al.(2022{\natexlab{a}})Dong, Liu, Cheng, Wang, Cheng, Niu, and Yin]{Dong2022IncorporatingExplicitKnowledge}
Qian Dong, Yiding Liu, Suqi Cheng, Shuaiqiang Wang, Zhicong Cheng, Shuzi Niu, and Dawei Yin.
\newblock Incorporating explicit knowledge in pre-trained language models for passage re-ranking.
\newblock \emph{Proceedings of the 45th International ACM SIGIR Conference on Research and Development in Information Retrieval}, 2022{\natexlab{a}}.

\bibitem[Dong et~al.(2022{\natexlab{b}})Dong, Niu, Yuan, and Li]{Dong2022DisentangledGR}
Qian Dong, Shuzi Niu, Tao Yuan, and Yucheng Li.
\newblock Disentangled graph recurrent network for document ranking.
\newblock \emph{Data Science and Engineering}, 7:\penalty0 30 -- 43, 2022{\natexlab{b}}.

\bibitem[Dumais et~al.(1997)Dumais, Letsche, Littman, and Landauer]{dumais1997automaticcrosslanguageretrieval}
Susan~T. Dumais, Todd~A. Letsche, Michael~L. Littman, and Thomas~K. Landauer.
\newblock Automatic cross-language retrieval using latent semantic indexing.
\newblock In \emph{AAAI Spring Symposium on Cross-Language Text and Speech Retrieval}, volume~15, pp.\ ~21, 1997.

\bibitem[Edge et~al.(2025)Edge, Trinh, Cheng, Bradley, Chao, Mody, Truitt, Metropolitansky, Ness, and Larson]{edge2025localglobalgraphrag}
Darren Edge, Ha~Trinh, Newman Cheng, Joshua Bradley, Alex Chao, Apurva Mody, Steven Truitt, Dasha Metropolitansky, Robert~Osazuwa Ness, and Jonathan Larson.
\newblock From local to global: A graph {RAG} approach to query-focused summarization.
\newblock \emph{arXiv preprint arXiv:2404.16130}, 2025.

\bibitem[Elhage et~al.(2021)Elhage, Nanda, Olsson, Henighan, Joseph, Mann, Askell, Bai, Chen, Conerly, DasSarma, Drain, Ganguli, Hatfield-Dodds, Hernandez, Jones, Kernion, Lovitt, Ndousse, Amodei, Brown, Clark, Kaplan, McCandlish, and Olah]{elhage2021mathematical}
Nelson Elhage, Neel Nanda, Catherine Olsson, Tom Henighan, Nicholas Joseph, Ben Mann, Amanda Askell, Yuntao Bai, Anna Chen, Tom Conerly, Nova DasSarma, Dawn Drain, Deep Ganguli, Zac Hatfield-Dodds, Danny Hernandez, Andy Jones, Jackson Kernion, Liane Lovitt, Kamal Ndousse, Dario Amodei, Tom Brown, Jack Clark, Jared Kaplan, Sam McCandlish, and Chris Olah.
\newblock A mathematical framework for transformer circuits.
\newblock \emph{Transformer Circuits Thread}, 2021.
\newblock URL \url{https://transformer-circuits.pub/2021/framework/index.html}.

\bibitem[Elmahdy et~al.(2024)Elmahdy, Lin, and Ahmad]{elmahdy2024synergisticapproachsimultaneousoptimization}
Adel Elmahdy, Sheng-Chieh Lin, and Amin Ahmad.
\newblock Synergistic approach for simultaneous optimization of monolingual, cross-lingual, and multilingual information retrieval.
\newblock \emph{arXiv preprint arXiv:2408.10536}, 2024.

\bibitem[Elman(1990)]{elman1990findingstructureintime}
Jeffrey~L. Elman.
\newblock Finding structure in time.
\newblock \emph{Cognitive Science}, 14\penalty0 (2):\penalty0 179--211, 1990.
\newblock \doi{10.1016/0364-0213(90)90002-E}.

\bibitem[Faggioli et~al.(2023)Faggioli, Dietz, Clarke, Demartini, Hagen, Hauff, Kando, Kanoulas, Potthast, Stein, et~al.]{faggioli2023perspectives}
Guglielmo Faggioli, Laura Dietz, Charles L.~A. Clarke, Gianluca Demartini, Matthias Hagen, Claudia Hauff, Noriko Kando, Evangelos Kanoulas, Martin Potthast, Benno Stein, et~al.
\newblock Perspectives on large language models for relevance judgment.
\newblock In \emph{Proceedings of the 2023 ACM SIGIR International Conference on Theory of Information Retrieval}, pp.\  39--50, 2023.

\bibitem[Faggioli et~al.(2024)Faggioli, Dietz, Clarke, Demartini, Hagen, Hauff, Kando, Kanoulas, Potthast, Stein, et~al.]{faggioli2024determines}
Guglielmo Faggioli, Laura Dietz, Charles L.~A. Clarke, Gianluca Demartini, Matthias Hagen, Claudia Hauff, Noriko Kando, Evangelos Kanoulas, Martin Potthast, Benno Stein, et~al.
\newblock Who determines what is relevant? humans or {AI}? why not both?
\newblock \emph{Communications of the ACM}, 67\penalty0 (4):\penalty0 31--34, 2024.

\bibitem[Fang et~al.(2021)Fang, Xiong, Xu, and Chen]{Fang2021CLIP2VideoMasteringVideoText}
Han Fang, Pengfei Xiong, Luhui Xu, and Yu~Chen.
\newblock {CLIP2Video}: Mastering video-text retrieval via image {CLIP}.
\newblock \emph{ArXiv preprint arXiv:2106.11097}, 2021.

\bibitem[Fang et~al.(2024)Fang, Zhan, Ai, Mao, Su, Chen, and Liu]{fang2024scaling}
Yan Fang, Jingtao Zhan, Qingyao Ai, Jiaxin Mao, Weihang Su, Jia Chen, and Yiqun Liu.
\newblock Scaling laws for dense retrieval.
\newblock In \emph{Proceedings of the 47th International ACM SIGIR Conference on Research and Development in Information Retrieval}, pp.\  1339--1349, 2024.

\bibitem[Faysse et~al.(2025)Faysse, Sibille, Wu, Omrani, Viaud, Hudelot, and Colombo]{faysse2025colpaliefficientdocumentretrieval}
Manuel Faysse, Hugues Sibille, Tony Wu, Bilel Omrani, Gautier Viaud, C{\'e}line Hudelot, and Pierre Colombo.
\newblock {ColPali}: Efficient document retrieval with vision language models.
\newblock \emph{arXiv preprint arXiv:2407.01449}, 2025.

\bibitem[Feng et~al.(2022)Feng, Yang, Cer, Arivazhagan, and Wang]{feng-etal-2022-language}
Fangxiaoyu Feng, Yinfei Yang, Daniel Cer, Naveen Arivazhagan, and Wei Wang.
\newblock Language-agnostic {BERT} sentence embedding.
\newblock In Smaranda Muresan, Preslav Nakov, and Aline Villavicencio (eds.), \emph{Proceedings of the 60th Annual Meeting of the Association for Computational Linguistics (Volume 1: Long Papers)}, pp.\  878--891, Dublin, Ireland, May 2022. Association for Computational Linguistics.
\newblock \doi{10.18653/v1/2022.acl-long.62}.
\newblock URL \url{https://aclanthology.org/2022.acl-long.62/}.

\bibitem[Fishman et~al.(2024)Fishman, Chmiel, Banner, and Soudry]{fishman2024scaling}
Maxim Fishman, Brian Chmiel, Ron Banner, and Daniel Soudry.
\newblock Scaling fp8 training to trillion-token {LLMs}.
\newblock \emph{arXiv preprint arXiv:2409.12517}, 2024.

\bibitem[Fluhr et~al.(1999)Fluhr, Frederking, Oard, Okumura, Ishikawa, and Satoh]{fluhr1999multilingualinformationretrieval}
Christian Fluhr, Robert~E. Frederking, Doug Oard, Akitoshi Okumura, Kai Ishikawa, and Kenji Satoh.
\newblock Multilingual (or cross-lingual) information retrieval.
\newblock \emph{Proceedings of the Multilingual Information Management: Current Levels and Future Abilities}, pp.\  10--13, 1999.

\bibitem[Formal et~al.(2021{\natexlab{a}})Formal, Lassance, Piwowarski, and Clinchant]{formal21spladev2}
Thibault Formal, Carlos Lassance, Benjamin Piwowarski, and St{\'e}phane Clinchant.
\newblock {SPLADE} v2: Sparse lexical and expansion model for information retrieval.
\newblock \emph{arXiv preprint arXiv:2109.10086}, 2021{\natexlab{a}}.

\bibitem[Formal et~al.(2021{\natexlab{b}})Formal, Piwowarski, and Clinchant]{formal21splade}
Thibault Formal, Benjamin Piwowarski, and St\'{e}phane Clinchant.
\newblock {SPLADE}: Sparse lexical and expansion model for 1st stage ranking.
\newblock In \emph{Proceedings of the 44th International ACM SIGIR Conference on Research and Development in Information Retrieval}, pp.\  2288--2292, New York, NY, USA, 2021{\natexlab{b}}. Association for Computing Machinery.
\newblock \doi{10.1145/3404835.3463098}.

\bibitem[Formal et~al.(2024)Formal, Clinchant, D{\'e}jean, and Lassance]{formal2024splate}
Thibault Formal, St{\'e}phane Clinchant, Herv{\'e} D{\'e}jean, and Carlos Lassance.
\newblock {SPLATE}: Sparse late interaction retrieval.
\newblock In \emph{Proceedings of the 47th International ACM SIGIR Conference on Research and Development in Information Retrieval}, pp.\  2635--2640, 2024.
\newblock \doi{10.1145/3626772.3657968}.

\bibitem[Freund \& Schapire(1995)Freund and Schapire]{freund1995desiciontheoreticgeneralization}
Yoav Freund and Robert~E Schapire.
\newblock A desicion-theoretic generalization of on-line learning and an application to boosting.
\newblock In \emph{European Conference on Computational Learning Theory}, pp.\  23--37. Springer, 1995.

\bibitem[Freund et~al.(2003)Freund, Iyer, Schapire, and Singer]{freund2003efficient}
Yoav Freund, Raj Iyer, Robert~E. Schapire, and Yoram Singer.
\newblock An efficient boosting algorithm for combining preferences.
\newblock \emph{Journal of machine learning research}, 4:\penalty0 933--969, 2003.

\bibitem[Friedman(2001)]{friedman2001greedy}
Jerome~H. Friedman.
\newblock Greedy function approximation: a gradient boosting machine.
\newblock \emph{Annals of statistics}, pp.\  1189--1232, 2001.

\bibitem[Frome et~al.(2013)Frome, Corrado, Shlens, Bengio, Dean, Ranzato, and Mikolov]{frome2013devise}
Andrea Frome, Greg~S Corrado, Jon Shlens, Samy Bengio, Jeff Dean, Marc\textquotesingle~Aurelio Ranzato, and Tomas Mikolov.
\newblock {DeViSE}: A deep visual-semantic embedding model.
\newblock In C.J. Burges, L.~Bottou, M.~Welling, Z.~Ghahramani, and K.Q. Weinberger (eds.), \emph{Advances in Neural Information Processing Systems}, volume~26. Curran Associates, Inc., 2013.

\bibitem[Fu et~al.(2023)Fu, Arora, Grogan, Johnson, Eyuboglu, Thomas, Spector, Poli, Rudra, and R{\'e}]{fu2023monarchmixer}
Dan Fu, Simran Arora, Jessica Grogan, Isys Johnson, Evan~Sabri Eyuboglu, Armin Thomas, Benjamin Spector, Michael Poli, Atri Rudra, and Christopher R{\'e}.
\newblock {Monarch Mixer}: A simple sub-quadratic {GEMM}-based architecture.
\newblock \emph{Advances in Neural Information Processing Systems}, 36:\penalty0 77546--77603, 2023.

\bibitem[Fuhr(1992)]{fuhr1992probabilistic}
Norbert Fuhr.
\newblock Probabilistic models in information retrieval.
\newblock \emph{The computer journal}, 35\penalty0 (3):\penalty0 243--255, 1992.

\bibitem[Gabeur et~al.(2020)Gabeur, Sun, Karteek, and Schmid]{Gabeur2020MultimodalTransformerforVideoRetrieval}
Valentin Gabeur, Chen Sun, Alahari Karteek, and Cordelia Schmid.
\newblock Multi-modal transformer for video retrieval.
\newblock \emph{arXiv preprint arXiv: 2007.10639}, 2020.

\bibitem[Gallagher et~al.(2019)Gallagher, Chen, Blanco, and Culpepper]{Gallagher2019JointOptimizationOfCascadeRankingModels}
Luke Gallagher, Ruey-Cheng Chen, Roi Blanco, and J.~Shane Culpepper.
\newblock Joint optimization of cascade ranking models.
\newblock \emph{Proceedings of the 12th ACM International Conference on Web Search and Data Mining}, 2019.

\bibitem[Gao et~al.(2014)Gao, Pantel, Gamon, He, and Deng]{gao2014modelinginterestingness}
Jianfeng Gao, Patrick Pantel, Michael Gamon, Xiaodong He, and Li~Deng.
\newblock Modeling interestingness with deep neural networks.
\newblock In \emph{Proceedings of the 2014 Conference on Empirical Methods in Natural Language Processing (EMNLP)}, pp.\  2--13, 2014.

\bibitem[Gao et~al.(2025)Gao, Fu, Xie, Xu, He, Mei, Zhu, and Wu]{gao2025turnsunlockinglonghorizonagentic}
Jiaxuan Gao, Wei Fu, Minyang Xie, Shusheng Xu, Chuyi He, Zhiyu Mei, Banghua Zhu, and Yi~Wu.
\newblock Beyond ten turns: Unlocking long-horizon agentic search with large-scale asynchronous rl, 2025.
\newblock URL \url{https://arxiv.org/abs/2508.07976}.

\bibitem[Gao \& Callan(2021)Gao and Callan]{gao-callan-2021-condenser}
Luyu Gao and Jamie Callan.
\newblock Condenser: a pre-training architecture for dense retrieval.
\newblock In Marie-Francine Moens, Xuanjing Huang, Lucia Specia, and Scott Wen-tau Yih (eds.), \emph{Proceedings of the 2021 Conference on Empirical Methods in Natural Language Processing}, pp.\  981--993, Online and Punta Cana, Dominican Republic, November 2021. Association for Computational Linguistics.
\newblock \doi{10.18653/v1/2021.emnlp-main.75}.
\newblock URL \url{https://aclanthology.org/2021.emnlp-main.75/}.

\bibitem[Gao \& Callan(2022)Gao and Callan]{Gao2022LongDocumentReranking}
Luyu Gao and Jamie Callan.
\newblock Long document re-ranking with modular re-ranker.
\newblock \emph{Proceedings of the 45th International ACM SIGIR Conference on Research and Development in Information Retrieval}, 2022.

\bibitem[Gao et~al.(2020)Gao, Dai, and Callan]{gao2020understanding}
Luyu Gao, Zhuyun Dai, and Jamie Callan.
\newblock Understanding {BERT} rankers under distillation.
\newblock In \emph{Proceedings of the 2020 ACM SIGIR on International Conference on Theory of Information Retrieval}, pp.\  149--152, 2020.

\bibitem[Gao et~al.(2021{\natexlab{a}})Gao, Dai, and Callan]{gao-etal-2021-coil}
Luyu Gao, Zhuyun Dai, and Jamie Callan.
\newblock {COIL}: Revisit exact lexical match in information retrieval with contextualized inverted list.
\newblock In Kristina Toutanova, Anna Rumshisky, Luke Zettlemoyer, Dilek Hakkani-Tur, Iz~Beltagy, Steven Bethard, Ryan Cotterell, Tanmoy Chakraborty, and Yichao Zhou (eds.), \emph{Proceedings of the 2021 Conference of the North American Chapter of the Association for Computational Linguistics: Human Language Technologies}, pp.\  3030--3042, Online, June 2021{\natexlab{a}}. Association for Computational Linguistics.
\newblock \doi{10.18653/v1/2021.naacl-main.241}.
\newblock URL \url{https://aclanthology.org/2021.naacl-main.241/}.

\bibitem[Gao et~al.(2021{\natexlab{b}})Gao, Yao, and Chen]{gao-etal-2021-simcse}
Tianyu Gao, Xingcheng Yao, and Danqi Chen.
\newblock {S}im{CSE}: Simple contrastive learning of sentence embeddings.
\newblock In Marie-Francine Moens, Xuanjing Huang, Lucia Specia, and Scott Wen-tau Yih (eds.), \emph{Proceedings of the 2021 Conference on Empirical Methods in Natural Language Processing}, pp.\  6894--6910, Online and Punta Cana, Dominican Republic, November 2021{\natexlab{b}}. Association for Computational Linguistics.
\newblock \doi{10.18653/v1/2021.emnlp-main.552}.
\newblock URL \url{https://aclanthology.org/2021.emnlp-main.552/}.

\bibitem[Gao et~al.(2023)Gao, Xiong, Gao, Jia, Pan, Bi, Dai, Sun, and Wang]{gao2023retrieval}
Yunfan Gao, Yun Xiong, Xinyu Gao, Kangxiang Jia, Jinliu Pan, Yuxi Bi, Yi~Dai, Jiawei Sun, and Haofen Wang.
\newblock Retrieval-augmented generation for large language models: A survey.
\newblock \emph{arXiv preprint arXiv:2312.10997}, 2023.

\bibitem[{Gemini Team Google} et~al.(2023){Gemini Team Google}, Anil, Borgeaud, Alayrac, Yu, Soricut, Schalkwyk, Dai, Hauth, Millican, et~al.]{team2023gemini}
{Gemini Team Google}, Rohan Anil, Sebastian Borgeaud, Jean-Baptiste Alayrac, Jiahui Yu, Radu Soricut, Johan Schalkwyk, Andrew~M. Dai, Anja Hauth, Katie Millican, et~al.
\newblock {Gemini}: A family of highly capable multimodal models.
\newblock \emph{arXiv preprint arXiv:2312.11805}, 2023.

\bibitem[Girdhar et~al.(2023)Girdhar, El-Nouby, Liu, Singh, Alwala, Joulin, and Misra]{Girdhar2023ImageBindOneEmbeddingSpace}
Rohit Girdhar, Alaaeldin El-Nouby, Zhuang Liu, Mannat Singh, Kalyan~Vasudev Alwala, Armand Joulin, and Ishan Misra.
\newblock {ImageBind} one embedding space to bind them all.
\newblock \emph{2023 IEEE/CVF Conference on Computer Vision and Pattern Recognition (CVPR)}, pp.\  15180--15190, 2023.

\bibitem[Gong et~al.(2014)Gong, Ke, Isard, and Lazebnik]{gong2014multiviewembeddingspace}
Yunchao Gong, Qifa Ke, Michael Isard, and Svetlana Lazebnik.
\newblock A multi-view embedding space for modeling internet images, tags, and their semantics.
\newblock \emph{International journal of computer vision}, 106\penalty0 (2):\penalty0 210--233, 2014.

\bibitem[Google(2024)]{google2024groundingwithgooglesearch}
Google.
\newblock Grounding with google search.
\newblock \url{https://ai.google.dev/gemini-api/docs/grounding?lang=python}, 2024.

\bibitem[Gorti et~al.(2022)Gorti, Vouitsis, Ma, Golestan, Volkovs, Garg, and Yu]{Gorti2022XPoolCrossModelLanguageVideoAttention}
Satya~Krishna Gorti, No{\"e}l Vouitsis, Junwei Ma, Keyvan Golestan, Maksims Volkovs, Animesh Garg, and Guangwei Yu.
\newblock {X-Pool}: Cross-modal language-video attention for text-video retrieval.
\newblock \emph{2022 IEEE/CVF Conference on Computer Vision and Pattern Recognition (CVPR)}, pp.\  4996--5005, 2022.

\bibitem[Goswami et~al.(2021)Goswami, Dutta, Assem, Fransen, and McCrae]{goswami-etal-2021-cross}
Koustava Goswami, Sourav Dutta, Haytham Assem, Theodorus Fransen, and John~P. McCrae.
\newblock Cross-lingual sentence embedding using multi-task learning.
\newblock In Marie-Francine Moens, Xuanjing Huang, Lucia Specia, and Scott Wen-tau Yih (eds.), \emph{Proceedings of the 2021 Conference on Empirical Methods in Natural Language Processing}, pp.\  9099--9113, Online and Punta Cana, Dominican Republic, November 2021. Association for Computational Linguistics.
\newblock \doi{10.18653/v1/2021.emnlp-main.716}.
\newblock URL \url{https://aclanthology.org/2021.emnlp-main.716/}.

\bibitem[Gou et~al.(2021)Gou, Yu, Maybank, and Tao]{gou2021knowledgedistillationasurvey}
Jianping Gou, Baosheng Yu, Stephen~J. Maybank, and Dacheng Tao.
\newblock Knowledge distillation: A survey.
\newblock \emph{International journal of computer vision}, 129\penalty0 (6):\penalty0 1789--1819, 2021.

\bibitem[Goworek et~al.(2025)Goworek, Macmillan-Scott, and {\"O}zyi{\u{g}}it]{goworek2025bridginglanguagegaps}
Roksana Goworek, Olivia Macmillan-Scott, and Eda~B. {\"O}zyi{\u{g}}it.
\newblock Bridging language gaps: Advances in cross-lingual information retrieval with multilingual {LLMs}.
\newblock \emph{arXiv preprint arXiv:2510.00908}, 2025.

\bibitem[Gu \& Dao(2024)Gu and Dao]{gu2024mamba}
Albert Gu and Tri Dao.
\newblock {Mamba}: Linear-time sequence modeling with selective state spaces.
\newblock In \emph{1st Conference on Language Modeling}, 2024.

\bibitem[G{\"u}nther et~al.(2024)G{\"u}nther, Ong, Mohr, Abdessalem, Abel, Akram, Guzman, Mastrapas, Sturua, Wang, Werk, Wang, and Xiao]{gunther2024jinaembeddings28192token}
Michael G{\"u}nther, Jackmin Ong, Isabelle Mohr, Alaeddine Abdessalem, Tanguy Abel, Mohammad~Kalim Akram, Susana Guzman, Georgios Mastrapas, Saba Sturua, Bo~Wang, Maximilian Werk, Nan Wang, and Han Xiao.
\newblock {Jina} {Embeddings} 2: 8192-token general-purpose text embeddings for long documents.
\newblock arXiv preprint arXiv:2310.19923, 2024.

\bibitem[Guo et~al.(2017)Guo, Pleiss, Sun, and Weinberger]{Guo2017OnCalibrationofModern}
Chuan Guo, Geoff Pleiss, Yu~Sun, and Kilian~Q. Weinberger.
\newblock On calibration of modern neural networks.
\newblock \emph{arXiv preprint arXiv:1706.04599}, 2017.

\bibitem[Guo et~al.(2025)Guo, Yang, Zhang, Song, Zhang, Xu, Zhu, Ma, Wang, Bi, et~al.]{guo2025deepseekr1incentivizingreasoningcapabilityinllmsviareinforcementlearning}
Daya Guo, Dejian Yang, Haowei Zhang, Junxiao Song, Ruoyu Zhang, Runxin Xu, Qihao Zhu, Shirong Ma, Peiyi Wang, Xiao Bi, et~al.
\newblock {DeepSeek-R1}: Incentivizing reasoning capability in {LLMs} via reinforcement learning.
\newblock \emph{arXiv preprint arXiv:2501.12948}, 2025.

\bibitem[Guo et~al.(2016{\natexlab{a}})Guo, Fan, Ai, and Croft]{guo2016deep}
Jiafeng Guo, Yixing Fan, Qingyao Ai, and W.~Bruce Croft.
\newblock A deep relevance matching model for ad-hoc retrieval.
\newblock In \emph{Proceedings of the 25th ACM International on Conference on Information and Knowledge Management}, CIKM '16, pp.\  55--64, New York, NY, USA, 2016{\natexlab{a}}. Association for Computing Machinery.
\newblock \doi{10.1145/2983323.2983769}.

\bibitem[Guo et~al.(2016{\natexlab{b}})Guo, Fan, Ai, and Croft]{guo2016semantic}
Jiafeng Guo, Yixing Fan, Qingyao Ai, and W.~Bruce Croft.
\newblock Semantic matching by non-linear word transportation for information retrieval.
\newblock In \emph{Proceedings of the 25th ACM International on Conference on Information and Knowledge Management}, CIKM '16, pp.\  701--710, New York, NY, USA, 2016{\natexlab{b}}. Association for Computing Machinery.
\newblock \doi{10.1145/2983323.2983768}.

\bibitem[Guo et~al.(2020)Guo, Fan, Pang, Yang, Ai, Zamani, Wu, Croft, and Cheng]{Guo2019ADeepLook}
Jiafeng Guo, Yixing Fan, Liang Pang, Liu Yang, Qingyao Ai, Hamed Zamani, Chen Wu, W~Bruce Croft, and Xueqi Cheng.
\newblock A deep look into neural ranking models for information retrieval.
\newblock \emph{Information Processing \& Management}, 57\penalty0 (6):\penalty0 102067, 2020.

\bibitem[Gupta et~al.(2024{\natexlab{a}})Gupta, Rajendhran, Stringham, Srikumar, and Marasovic]{gupta-etal-2024-whispersofdoubt}
Ashim Gupta, Rishanth Rajendhran, Nathan Stringham, Vivek Srikumar, and Ana Marasovic.
\newblock Whispers of doubt amidst echoes of triumph in {NLP} robustness.
\newblock In Kevin Duh, Helena Gomez, and Steven Bethard (eds.), \emph{Proceedings of the 2024 Conference of the North American Chapter of the Association for Computational Linguistics: Human Language Technologies (Volume 1: Long Papers)}, pp.\  5533--5590, Mexico City, Mexico, June 2024{\natexlab{a}}. Association for Computational Linguistics.
\newblock \doi{10.18653/v1/2024.naacl-long.310}.
\newblock URL \url{https://aclanthology.org/2024.naacl-long.310/}.

\bibitem[Gupta et~al.(2024{\natexlab{b}})Gupta, Devvrit, Rawat, Bhojanapalli, Jain, and Dhillon]{gupta2024dualencoders}
Nilesh Gupta, Fnu Devvrit, Ankit~Singh Rawat, Srinadh Bhojanapalli, Prateek Jain, and Inderjit~S. Dhillon.
\newblock Dual-encoders for extreme multi-label classification.
\newblock In \emph{The 12th International Conference on Learning Representations}, 2024{\natexlab{b}}.
\newblock URL \url{https://openreview.net/forum?id=dNe1T0Ahby}.

\bibitem[Gururangan et~al.(2020)Gururangan, Marasovi{\'c}, Swayamdipta, Lo, Beltagy, Downey, and Smith]{gururangan-etal-2020-dont}
Suchin Gururangan, Ana Marasovi{\'c}, Swabha Swayamdipta, Kyle Lo, Iz~Beltagy, Doug Downey, and Noah~A. Smith.
\newblock Don`t stop pretraining: Adapt language models to domains and tasks.
\newblock In Dan Jurafsky, Joyce Chai, Natalie Schluter, and Joel Tetreault (eds.), \emph{Proceedings of the 58th Annual Meeting of the Association for Computational Linguistics}, pp.\  8342--8360, Online, July 2020. Association for Computational Linguistics.
\newblock \doi{10.18653/v1/2020.acl-main.740}.
\newblock URL \url{https://aclanthology.org/2020.acl-main.740/}.

\bibitem[Gutmann \& Hyv{\"a}rinen(2012)Gutmann and Hyv{\"a}rinen]{gutmann2012noisecontrastiveestimation}
Michael~U. Gutmann and Aapo Hyv{\"a}rinen.
\newblock Noise-contrastive estimation of unnormalized statistical models, with applications to natural image statistics.
\newblock \emph{The journal of machine learning research}, 13\penalty0 (1):\penalty0 307--361, 2012.

\bibitem[Guu et~al.(2020)Guu, Lee, Tung, Pasupat, and Chang]{guu2020retrievalaugmentedlanguagemodelpretraining}
Kelvin Guu, Kenton Lee, Zora Tung, Panupong Pasupat, and Mingwei Chang.
\newblock Retrieval augmented language model pre-training.
\newblock In \emph{International Conference on Machine Learning}, pp.\  3929--3938. PMLR, 2020.

\bibitem[Ha et~al.(2022)Ha, Dai, and Le]{ha2022hypernetworks}
David Ha, Andrew~M. Dai, and Quoc~V. Le.
\newblock Hypernetworks.
\newblock In \emph{International Conference on Learning Representations}, 2022.

\bibitem[He et~al.(2015)He, Gimpel, and Lin]{he-etal-2015-multiperspective}
Hua He, Kevin Gimpel, and Jimmy Lin.
\newblock Multi-perspective sentence similarity modeling with convolutional neural networks.
\newblock In Llu{\'\i}s M{\`a}rquez, Chris Callison-Burch, and Jian Su (eds.), \emph{Proceedings of the 2015 Conference on Empirical Methods in Natural Language Processing}, pp.\  1576--1586, Lisbon, Portugal, September 2015. Association for Computational Linguistics.
\newblock \doi{10.18653/v1/D15-1181}.
\newblock URL \url{https://aclanthology.org/D15-1181/}.

\bibitem[He et~al.(2025)He, Huang, Feng, Lin, Zhang, Li, et~al.]{he2025pasa}
Yichen He, Guanhua Huang, Peiyuan Feng, Yuan Lin, Yuchen Zhang, Hang Li, et~al.
\newblock {PaSa}: An {LLM} agent for comprehensive academic paper search.
\newblock \emph{arXiv preprint arXiv:2501.10120}, 2025.

\bibitem[He et~al.(2024)He, Xie, Steck, Liang, Jha, Kallus, and McAuley]{He2024ReindexThenAdapt}
Zhankui He, Zhouhang Xie, Harald Steck, Dawen Liang, Rahul Jha, Nathan Kallus, and Julian McAuley.
\newblock Reindex-then-adapt: Improving large language models for conversational recommendation.
\newblock \emph{Proceedings of the 18th ACM International Conference on Web Search and Data Mining}, 2024.

\bibitem[Hiemstra \& Kraaij(1999)Hiemstra and Kraaij]{hiemstra1999twentyonetrec7}
Djoerd Hiemstra and Wessel Kraaij.
\newblock Twenty-one at {TREC}-7: Ad-hoc and cross-language track.
\newblock In \emph{7th Text REtrieval Conference, TREC-7 1998}, pp.\  227--238. National Institute of Standards and Technology, 1999.

\bibitem[Hinton et~al.(2015)Hinton, Vinyals, and Dean]{hinton2015distillingknowledge}
Geoffrey Hinton, Oriol Vinyals, and Jeff Dean.
\newblock Distilling the knowledge in a neural network.
\newblock \emph{stat}, 1050:\penalty0 9, 2015.

\bibitem[Hochreiter \& Schmidhuber(1997)Hochreiter and Schmidhuber]{hochreiter1997long}
S.~Hochreiter and J.~Schmidhuber.
\newblock Long short-term memory.
\newblock \emph{Neural Computation MIT-Press}, 1997.

\bibitem[Hoffmann et~al.(2022)Hoffmann, Borgeaud, Mensch, Buchatskaya, Cai, Rutherford, Casas, Hendricks, Welbl, Clark, et~al.]{hoffmann2022training}
Jordan Hoffmann, Sebastian Borgeaud, Arthur Mensch, Elena Buchatskaya, Trevor Cai, Eliza Rutherford, Diego de~Las Casas, Lisa~Anne Hendricks, Johannes Welbl, Aidan Clark, et~al.
\newblock Training compute-optimal large language models.
\newblock \emph{arXiv preprint arXiv:2203.15556}, 2022.

\bibitem[Hofst{\"a}tter et~al.(2020{\natexlab{a}})Hofst{\"a}tter, Althammer, Schr{\"o}der, Sertkan, and Hanbury]{hofstatter2020improving}
Sebastian Hofst{\"a}tter, Sophia Althammer, Michael Schr{\"o}der, Mete Sertkan, and Allan Hanbury.
\newblock Improving efficient neural ranking models with cross-architecture knowledge distillation.
\newblock \emph{arXiv preprint arXiv:2010.02666}, 2020{\natexlab{a}}.

\bibitem[Hofst{\"a}tter et~al.(2020{\natexlab{b}})Hofst{\"a}tter, Zamani, Mitra, Craswell, and Hanbury]{hofstatter2020local}
Sebastian Hofst{\"a}tter, Hamed Zamani, Bhaskar Mitra, Nick Craswell, and Allan Hanbury.
\newblock Local self-attention over long text for efficient document retrieval.
\newblock In \emph{Proceedings of the 43rd International ACM SIGIR Conference on Research and Development in Information Retrieval}, pp.\  2021--2024, 2020{\natexlab{b}}.

\bibitem[Hofst{\"a}tter et~al.(2020{\natexlab{c}})Hofst{\"a}tter, Zlabinger, and Hanbury]{hofstatter2020interpretable}
Sebastian Hofst{\"a}tter, Markus Zlabinger, and Allan Hanbury.
\newblock Interpretable \& time-budget-constrained contextualization for re-ranking.
\newblock In \emph{ECAI 2020}, pp.\  513--520. IOS Press, 2020{\natexlab{c}}.

\bibitem[Hofst{\"a}tter et~al.(2021{\natexlab{a}})Hofst{\"a}tter, Lin, Yang, Lin, and Hanbury]{hofstatter2021efficiently}
Sebastian Hofst{\"a}tter, Sheng-Chieh Lin, Jheng-Hong Yang, Jimmy Lin, and Allan Hanbury.
\newblock Efficiently teaching an effective dense retriever with balanced topic aware sampling.
\newblock In \emph{Proceedings of the 44th International ACM SIGIR Conference on Research and Development in Information Retrieval}, pp.\  113--122, 2021{\natexlab{a}}.

\bibitem[Hofst{\"a}tter et~al.(2021{\natexlab{b}})Hofst{\"a}tter, Mitra, Zamani, Craswell, and Hanbury]{Hofsttter2021IntraDocumentCascading}
Sebastian Hofst{\"a}tter, Bhaskar Mitra, Hamed Zamani, Nick Craswell, and Allan Hanbury.
\newblock Intra-document cascading: Learning to select passages for neural document ranking.
\newblock \emph{Proceedings of the 44th International ACM SIGIR Conference on Research and Development in Information Retrieval}, 2021{\natexlab{b}}.

\bibitem[Hofst{\"a}tter et~al.(2022)Hofst{\"a}tter, Khattab, Althammer, Sertkan, and Hanbury]{hofstatter2022introducing}
Sebastian Hofst{\"a}tter, Omar Khattab, Sophia Althammer, Mete Sertkan, and Allan Hanbury.
\newblock Introducing neural bag of whole-words with {ColBERT}er: Contextualized late interactions using enhanced reduction.
\newblock In \emph{Proceedings of the 31st ACM International Conference on Information \& Knowledge Management}, pp.\  737--747, 2022.

\bibitem[Hotelling(1992)]{hotelling1992relationsbetweentwosetsofvariates}
Harold Hotelling.
\newblock Relations between two sets of variates.
\newblock In \emph{Breakthroughs in Statistics: Methodology and Distribution}, pp.\  162--190. Springer, 1992.

\bibitem[Howard \& Ruder(2018)Howard and Ruder]{howard2018universal}
Jeremy Howard and Sebastian Ruder.
\newblock Universal language model fine-tuning for text classification.
\newblock \emph{arXiv preprint arXiv:1801.06146}, 2018.

\bibitem[Hu et~al.(2014)Hu, Lu, Li, and Chen]{hu2014convolutional}
Baotian Hu, Zhengdong Lu, Hang Li, and Qingcai Chen.
\newblock Convolutional neural network architectures for matching natural language sentences.
\newblock In \emph{Proceedings of the 28th International Conference on Neural Information Processing Systems - Volume 2}, NIPS'14, pp.\  2042--2050, Cambridge, MA, USA, 2014. MIT Press.

\bibitem[Hu et~al.(2025)Hu, Zhou, Fan, Nie, Xia, Sun, Ye, Jin, Li, Chen, et~al.]{hu2025owloptimizedworkforcelearning}
Mengkang Hu, Yuhang Zhou, Wendong Fan, Yuzhou Nie, Bowei Xia, Tao Sun, Ziyu Ye, Zhaoxuan Jin, Yingru Li, Qiguang Chen, et~al.
\newblock {OWL}: Optimized workforce learning for general multi-agent assistance in real-world task automation.
\newblock \emph{arXiv preprint arXiv:2505.23885}, 2025.

\bibitem[Hu et~al.(2019)Hu, Wang, Peng, and Li]{hu2019unbiased}
Ziniu Hu, Yang Wang, Qu~Peng, and Hang Li.
\newblock Unbiased {LambdaMART}: An unbiased pairwise learning-to-rank algorithm.
\newblock In \emph{The World Wide Web Conference}, pp.\  2830--2836, 2019.

\bibitem[Huang et~al.(2020)Huang, Sharma, Sun, Xia, Zhang, Pronin, Padmanabhan, Ottaviano, and Yang]{huang2020embeddingbasedretrieval}
Jui-Ting Huang, Ashish Sharma, Shuying Sun, Li~Xia, David Zhang, Philip Pronin, Janani Padmanabhan, Giuseppe Ottaviano, and Linjun Yang.
\newblock Embedding-based retrieval in facebook search.
\newblock In \emph{Proceedings of the 26th ACM SIGKDD International Conference on Knowledge Discovery \& Data Mining}, pp.\  2553--2561, 2020.

\bibitem[Huang et~al.(2013)Huang, He, Gao, Deng, Acero, and Heck]{huang2013dssm}
Po-Sen Huang, Xiaodong He, Jianfeng Gao, Li~Deng, Alex Acero, and Larry Heck.
\newblock Learning deep structured semantic models for web search using clickthrough data.
\newblock In \emph{Proceedings of the 22nd ACM International Conference on Information \& Knowledge Management}, CIKM '13, pp.\  2333--2338, New York, NY, USA, 2013. Association for Computing Machinery.
\newblock \doi{10.1145/2505515.2505665}.

\bibitem[Huang et~al.(2024{\natexlab{a}})Huang, Liu, Chen, Wang, Wang, Lian, Wang, Tang, and Chen]{huang2024understandingtheplanningofllmagents}
Xu~Huang, Weiwen Liu, Xiaolong Chen, Xingmei Wang, Hao Wang, Defu Lian, Yasheng Wang, Ruiming Tang, and Enhong Chen.
\newblock Understanding the planning of {LLM} agents: A survey.
\newblock \emph{arXiv preprint arXiv:2402.02716}, 2024{\natexlab{a}}.

\bibitem[Huang et~al.(2025)Huang, Chen, Zhang, Li, Fang, Yang, Li, Shang, Xu, Hao, et~al.]{huang2025deepresearchagents}
Yuxuan Huang, Yihang Chen, Haozheng Zhang, Kang Li, Meng Fang, Linyi Yang, Xiaoguang Li, Lifeng Shang, Songcen Xu, Jianye Hao, et~al.
\newblock Deep research agents: A systematic examination and roadmap.
\newblock \emph{arXiv preprint arXiv:2506.18096}, 2025.

\bibitem[Huang et~al.(2024{\natexlab{b}})Huang, Yu, Ravfogel, and Allan]{huang-etal-2024-languageconcepterasure}
Zhiqi Huang, Puxuan Yu, Shauli Ravfogel, and James Allan.
\newblock Language concept erasure for language-invariant dense retrieval.
\newblock In Yaser Al-Onaizan, Mohit Bansal, and Yun-Nung Chen (eds.), \emph{Proceedings of the 2024 Conference on Empirical Methods in Natural Language Processing}, pp.\  13261--13273, Miami, Florida, USA, November 2024{\natexlab{b}}. Association for Computational Linguistics.
\newblock \doi{10.18653/v1/2024.emnlp-main.736}.
\newblock URL \url{https://aclanthology.org/2024.emnlp-main.736/}.

\bibitem[Hui et~al.(2017)Hui, Yates, Berberich, and de~Melo]{hui-etal-2017-pacrr}
Kai Hui, Andrew Yates, Klaus Berberich, and Gerard de~Melo.
\newblock {PACRR}: A position-aware neural {IR} model for relevance matching.
\newblock In Martha Palmer, Rebecca Hwa, and Sebastian Riedel (eds.), \emph{Proceedings of the 2017 Conference on Empirical Methods in Natural Language Processing}, pp.\  1049--1058, Copenhagen, Denmark, September 2017. Association for Computational Linguistics.
\newblock \doi{10.18653/v1/D17-1110}.
\newblock URL \url{https://aclanthology.org/D17-1110/}.

\bibitem[Hui et~al.(2018)Hui, Yates, Berberich, and de~Melo]{hui2018copacrr}
Kai Hui, Andrew Yates, Klaus Berberich, and Gerard de~Melo.
\newblock {Co-PACRR}: A context-aware neural ir model for ad-hoc retrieval.
\newblock In \emph{Proceedings of the 11th ACM International Conference on Web Search and Data Mining}, WSDM '18, pp.\  279--287, New York, NY, USA, 2018. Association for Computing Machinery.
\newblock \doi{10.1145/3159652.3159689}.

\bibitem[Humeau et~al.(2020)Humeau, Shuster, Lachaux, and Weston]{Humeau2020Poly-encoders}
Samuel Humeau, Kurt Shuster, Marie-Anne Lachaux, and Jason Weston.
\newblock Poly-encoders: Architectures and pre-training strategies for fast and accurate multi-sentence scoring.
\newblock In \emph{International Conference on Learning Representations}, 2020.
\newblock URL \url{https://openreview.net/forum?id=SkxgnnNFvH}.

\bibitem[Islam et~al.(2024)Islam, Rahman, Hossain, Hoque, Joty, and Parvez]{islam-etal-2024-open}
Shayekh~Bin Islam, Md~Asib Rahman, K.~S. M.~Tozammel Hossain, Enamul Hoque, Shafiq Joty, and Md~Rizwan Parvez.
\newblock Open-{RAG}: Enhanced retrieval augmented reasoning with open-source large language models.
\newblock In Yaser Al-Onaizan, Mohit Bansal, and Yun-Nung Chen (eds.), \emph{Findings of the Association for Computational Linguistics: EMNLP 2024}, pp.\  14231--14244, Miami, Florida, USA, November 2024. Association for Computational Linguistics.
\newblock \doi{10.18653/v1/2024.findings-emnlp.831}.
\newblock URL \url{https://aclanthology.org/2024.findings-emnlp.831/}.

\bibitem[Izacard \& Grave(2021)Izacard and Grave]{izacard-grave-2021-leveraging}
Gautier Izacard and Edouard Grave.
\newblock Leveraging passage retrieval with generative models for open domain question answering.
\newblock In Paola Merlo, Jorg Tiedemann, and Reut Tsarfaty (eds.), \emph{Proceedings of the 16th Conference of the European Chapter of the Association for Computational Linguistics: Main Volume}, pp.\  874--880, Online, April 2021. Association for Computational Linguistics.
\newblock \doi{10.18653/v1/2021.eacl-main.74}.
\newblock URL \url{https://aclanthology.org/2021.eacl-main.74/}.

\bibitem[Jacovi \& Goldberg(2020)Jacovi and Goldberg]{jacovi-goldberg-2020-towards}
Alon Jacovi and Yoav Goldberg.
\newblock Towards faithfully interpretable {NLP} systems: How should we define and evaluate faithfulness?
\newblock In Dan Jurafsky, Joyce Chai, Natalie Schluter, and Joel Tetreault (eds.), \emph{Proceedings of the 58th Annual Meeting of the Association for Computational Linguistics}, pp.\  4198--4205, Online, July 2020. Association for Computational Linguistics.
\newblock \doi{10.18653/v1/2020.acl-main.386}.
\newblock URL \url{https://aclanthology.org/2020.acl-main.386/}.

\bibitem[Jain et~al.(2016)Jain, Prabhu, and Varma]{jain2016extreme}
Himanshu Jain, Yashoteja Prabhu, and Manik Varma.
\newblock Extreme multi-label loss functions for recommendation, tagging, ranking \& other missing label applications.
\newblock In \emph{Proceedings of the 22nd ACM SIGKDD International Conference on Knowledge Discovery and Data Mining}, pp.\  935--944, 2016.

\bibitem[Jain et~al.(2019)Jain, Balasubramanian, Chunduri, and Varma]{jain2019slice}
Himanshu Jain, Venkatesh Balasubramanian, Bhanu Chunduri, and Manik Varma.
\newblock Slice: Scalable linear extreme classifiers trained on 100 million labels for related searches.
\newblock In \emph{Proceedings of the 12th {ACM} International Conference on Web Search and Data Mining}, pp.\  528--536, 2019.
\newblock \doi{10.1145/3289600.3290979}.

\bibitem[Javdan et~al.(2025)Javdan, Krishnamoorthy, and Baysal]{javdan2025crestimprovinginterpretabilityeffectiveness}
Soroush Javdan, Pragash Krishnamoorthy, and Olga Baysal.
\newblock {CREST}: Improving interpretability and effectiveness of troubleshooting at ericsson through criterion-specific trouble report retrieval.
\newblock \emph{arXiv preprint arXiv:2511.17417}, 2025.

\bibitem[Jelinek(1980)]{jelinek1980interpolatedestimation}
Frederick Jelinek.
\newblock Interpolated estimation of {Markov} source parameters from sparse data.
\newblock In \emph{Proceedings of Workshop on Pattern Recognition in Practice}, 1980.

\bibitem[Jeong et~al.(2024)Jeong, Sohn, Sung, and Kang]{jeong2024improving}
Minbyul Jeong, Jiwoong Sohn, Mujeen Sung, and Jaewoo Kang.
\newblock Improving medical reasoning through retrieval and self-reflection with retrieval-augmented large language models.
\newblock \emph{Bioinformatics}, 40\penalty0 (Supplement\_1):\penalty0 i119--i129, 2024.
\newblock \doi{10.1093/bioinformatics/btae238}.

\bibitem[Jia et~al.(2021)Jia, Yang, Xia, Chen, Parekh, Pham, Le, Sung, Li, and Duerig]{Jia2021ScalingUpVisual}
Chao Jia, Yinfei Yang, Ye~Xia, Yi-Ting Chen, Zarana Parekh, Hieu Pham, Quoc~V. Le, Yun-Hsuan Sung, Zhen Li, and Tom Duerig.
\newblock Scaling up visual and vision-language representation learning with noisy text supervision.
\newblock In \emph{International Conference on Machine Learning}, 2021.

\bibitem[Jiang et~al.(2023{\natexlab{a}})Jiang, Sablayrolles, Mensch, Bamford, Chaplot, de~Las~Casas, Bressand, Lengyel, Lample, Saulnier, Lavaud, Lachaux, Stock, Scao, Lavril, Wang, Lacroix, and Sayed]{Jiang2023Mistral7}
Albert~Qiaochu Jiang, Alexandre Sablayrolles, Arthur Mensch, Chris Bamford, Devendra~Singh Chaplot, Diego de~Las~Casas, Florian Bressand, Gianna Lengyel, Guillaume Lample, Lucile Saulnier, L'elio~Renard Lavaud, Marie-Anne Lachaux, Pierre Stock, Teven~Le Scao, Thibaut Lavril, Thomas Wang, Timoth{\'e}e Lacroix, and William~El Sayed.
\newblock {Mistral 7B}.
\newblock \emph{arXiv preprint arXiv:2310.06825}, 2023{\natexlab{a}}.

\bibitem[Jiang et~al.(2023{\natexlab{b}})Jiang, Xu, Gao, Sun, Liu, Dwivedi-Yu, Yang, Callan, and Neubig]{jiang2023activeretrievalaugmentedgeneration}
Zhengbao Jiang, Frank~F. Xu, Luyu Gao, Zhiqing Sun, Qian Liu, Jane Dwivedi-Yu, Yiming Yang, Jamie Callan, and Graham Neubig.
\newblock Active retrieval augmented generation.
\newblock In \emph{Proceedings of the 2023 Conference on Empirical Methods in Natural Language Processing}, pp.\  7969--7992, 2023{\natexlab{b}}.

\bibitem[Jin et~al.(2025)Jin, Zeng, Yue, Yoon, Arik, Wang, Zamani, and Han]{jin2025searchr1trainingllmstoreasonandleveragesearchengines}
Bowen Jin, Hansi Zeng, Zhenrui Yue, Jinsung Yoon, Sercan Arik, Dong Wang, Hamed Zamani, and Jiawei Han.
\newblock {Search-R1}: Training {LLMs} to reason and leverage search engines with reinforcement learning.
\newblock \emph{arXiv preprint arXiv:2503.09516}, 2025.

\bibitem[Joachims(2006)]{joachims2006training}
Thorsten Joachims.
\newblock Training linear {SVMs} in linear time.
\newblock In \emph{Proceedings of the 12th ACM SIGKDD International Conference on Knowledge Discovery and Data Mining}, KDD '06, pp.\  217--226, New York, NY, USA, 2006. Association for Computing Machinery.
\newblock \doi{10.1145/1150402.1150429}.

\bibitem[Joachims et~al.(2017)Joachims, Swaminathan, and Schnabel]{joachims2017unbiased}
Thorsten Joachims, Adith Swaminathan, and Tobias Schnabel.
\newblock Unbiased learning-to-rank with biased feedback.
\newblock In \emph{Proceedings of the 10th ACM International Conference on Web Search and Data Mining}, pp.\  781--789, 2017.

\bibitem[Johnson et~al.(2019)Johnson, Douze, and J{\'e}gou]{johnson2019billion}
Jeff Johnson, Matthijs Douze, and Herv{\'e} J{\'e}gou.
\newblock Billion-scale similarity search with gpus.
\newblock \emph{IEEE Transactions on Big Data}, 7\penalty0 (3):\penalty0 535--547, 2019.

\bibitem[Joulin et~al.(2018)Joulin, Bojanowski, Mikolov, J{\'e}gou, and Grave]{joulin2018loss}
Armand Joulin, Piotr Bojanowski, Tom{\'a}{\v{s}} Mikolov, Herv{\'e} J{\'e}gou, and {\'E}douard Grave.
\newblock Loss in translation: Learning bilingual word mapping with a retrieval criterion.
\newblock In \emph{Proceedings of the 2018 Conference on Empirical Methods in Natural Language Processing}, pp.\  2979--2984, 2018.

\bibitem[Ju et~al.(2025)Ju, Kim, and Lee]{ju-etal-2025-mire}
Yeong-Joon Ju, Ho-Joong Kim, and Seong-Whan Lee.
\newblock {MIRe}: Enhancing multimodal queries representation via fusion-free modality interaction for multimodal retrieval.
\newblock In Wanxiang Che, Joyce Nabende, Ekaterina Shutova, and Mohammad~Taher Pilehvar (eds.), \emph{Findings of the Association for Computational Linguistics: ACL 2025}, pp.\  5350--5363, Vienna, Austria, July 2025. Association for Computational Linguistics.
\newblock \doi{10.18653/v1/2025.findings-acl.279}.
\newblock URL \url{https://aclanthology.org/2025.findings-acl.279/}.

\bibitem[Jumper et~al.(2021)Jumper, Evans, Pritzel, Green, Figurnov, Ronneberger, Tunyasuvunakool, Bates, {\v{Z}}{\'\i}dek, Potapenko, et~al.]{jumper2021highly}
John Jumper, Richard Evans, Alexander Pritzel, Tim Green, Michael Figurnov, Olaf Ronneberger, Kathryn Tunyasuvunakool, Russ Bates, Augustin {\v{Z}}{\'\i}dek, Anna Potapenko, et~al.
\newblock Highly accurate protein structure prediction with {AlphaFold}.
\newblock \emph{Nature}, 596\penalty0 (7873):\penalty0 583--589, 2021.

\bibitem[Kang et~al.(2024)Kang, Agarwal, Jin, Lee, Yu, and Han]{Kang2024ImprovingRetrievalInThemeSpecificApplications}
SeongKu Kang, Shivam Agarwal, Bowen Jin, Dongha Lee, Hwanjo Yu, and Jiawei Han.
\newblock Improving retrieval in theme-specific applications using a corpus topical taxonomy.
\newblock \emph{Proceedings of the ACM Web Conference 2024}, 2024.

\bibitem[Kaplan et~al.(2020)Kaplan, McCandlish, Henighan, Brown, Chess, Child, Gray, Radford, Wu, and Amodei]{kaplan2020scaling}
Jared Kaplan, Sam McCandlish, Tom Henighan, Tom~B. Brown, Benjamin Chess, Rewon Child, Scott Gray, Alec Radford, Jeffrey Wu, and Dario Amodei.
\newblock Scaling laws for neural language models.
\newblock \emph{arXiv preprint arXiv:2001.08361}, 2020.

\bibitem[Karpathy et~al.(2014)Karpathy, Joulin, and Fei-Fei]{karpathy2014deepfragmentembeddings}
Andrej Karpathy, Armand Joulin, and Li~Fei-Fei.
\newblock Deep fragment embeddings for bidirectional image sentence mapping.
\newblock \emph{Advances in neural information processing systems}, 27, 2014.

\bibitem[Karpukhin et~al.(2020)Karpukhin, O{\u g}uz, Min, Lewis, Wu, Edunov, Chen, and tau Yih]{Karpukhin2020DensePR}
Vladimir Karpukhin, Barlas O{\u g}uz, Sewon Min, Patrick Lewis, Ledell~Yu Wu, Sergey Edunov, Danqi Chen, and Wen tau Yih.
\newblock Dense passage retrieval for open-domain question answering.
\newblock \emph{arXiv preprint arXiv:2004.04906}, 2020.

\bibitem[Katharopoulos et~al.(2020)Katharopoulos, Vyas, Pappas, and Fleuret]{katharopoulos2020transformers}
Angelos Katharopoulos, Apoorv Vyas, Nikolaos Pappas, and Fran{\c{c}}ois Fleuret.
\newblock Transformers are {RNNs}: Fast autoregressive transformers with linear attention.
\newblock In \emph{International Conference on Machine Learning}, pp.\  5156--5165. PMLR, 2020.

\bibitem[Ke et~al.(2017)Ke, Meng, Finley, Wang, Chen, Ma, Ye, and Liu]{ke2017lightgbm}
Guolin Ke, Qi~Meng, Thomas Finley, Taifeng Wang, Wei Chen, Weidong Ma, Qiwei Ye, and Tie-Yan Liu.
\newblock Lightgbm: A highly efficient gradient boosting decision tree.
\newblock \emph{Advances in neural information processing systems}, 30, 2017.

\bibitem[Kearns \& Valiant(1994)Kearns and Valiant]{kearns1994cryptographiclimitations}
Michael Kearns and Leslie Valiant.
\newblock Cryptographic limitations on learning boolean formulae and finite automata.
\newblock \emph{Journal of the ACM (JACM)}, 41\penalty0 (1):\penalty0 67--95, 1994.

\bibitem[Khandelwal et~al.(2020)Khandelwal, Levy, Jurafsky, Zettlemoyer, and Lewis]{Khandelwal2020Generalization}
Urvashi Khandelwal, Omer Levy, Dan Jurafsky, Luke Zettlemoyer, and Mike Lewis.
\newblock Generalization through memorization: Nearest neighbor language models.
\newblock In \emph{International Conference on Learning Representations}, 2020.
\newblock URL \url{https://openreview.net/forum?id=HklBjCEKvH}.

\bibitem[Khattab \& Zaharia(2020)Khattab and Zaharia]{khattab2020colbert}
Omar Khattab and Matei Zaharia.
\newblock {ColBERT}: Efficient and effective passage search via contextualized late interaction over {BERT}.
\newblock In \emph{Proceedings of the 43rd International ACM SIGIR Conference on Research and Development in Information Retrieval}, pp.\  39--48, 2020.

\bibitem[Khot et~al.(2023)Khot, Trivedi, Finlayson, Fu, Richardson, Clark, and Sabharwal]{khot2023decomposedprompting}
Tushar Khot, Harsh Trivedi, Matthew Finlayson, Yao Fu, Kyle Richardson, Peter Clark, and Ashish Sabharwal.
\newblock Decomposed prompting: A modular approach for solving complex tasks.
\newblock In \emph{The 11th International Conference on Learning Representations}, 2023.
\newblock URL \url{https://openreview.net/forum?id=_nGgzQjzaRy}.

\bibitem[Killingback et~al.(2025)Killingback, Zeng, and Zamani]{killingback2025hypencoder}
Julian Killingback, Hansi Zeng, and Hamed Zamani.
\newblock Hypencoder: Hypernetworks for information retrieval.
\newblock \emph{arXiv preprint arXiv:2502.05364}, 2025.

\bibitem[Kim et~al.(2023)Kim, Rawat, Zaheer, Jayasumana, Sadhanala, Jitkrittum, Menon, Fergus, and Kumar]{kim2023embeddistillgeometricknowledgedistillation}
Seungyeon Kim, Ankit~Singh Rawat, Manzil Zaheer, Sadeep Jayasumana, Veeranjaneyulu Sadhanala, Wittawat Jitkrittum, Aditya~Krishna Menon, Rob Fergus, and Sanjiv Kumar.
\newblock {EmbedDistill}: A geometric knowledge distillation for information retrieval.
\newblock \emph{arXiv preprint arXiv:2301.12005}, 2023.

\bibitem[Kim et~al.(2025)Kim, Zhu, Lin, Bastan, Gray, and Kwak]{Kim2025GENIUS}
Sungyeon Kim, Xinliang Zhu, Xiaofan Lin, Muhammet Bastan, Douglas Gray, and Suha Kwak.
\newblock {GENIUS}: A generative framework for universal multimodal search.
\newblock \emph{2025 IEEE/CVF Conference on Computer Vision and Pattern Recognition (CVPR)}, pp.\  19659--19669, 2025.

\bibitem[Kim(2014)]{kim-2014-convolutional}
Yoon Kim.
\newblock Convolutional neural networks for sentence classification.
\newblock In Alessandro Moschitti, Bo~Pang, and Walter Daelemans (eds.), \emph{Proceedings of the 2014 Conference on Empirical Methods in Natural Language Processing ({EMNLP})}, pp.\  1746--1751, Doha, Qatar, October 2014. Association for Computational Linguistics.
\newblock \doi{10.3115/v1/D14-1181}.
\newblock URL \url{https://aclanthology.org/D14-1181/}.

\bibitem[Kishore et~al.(2023)Kishore, Wan, Lovelace, Artzi, and Weinberger]{kishore2023incdsi}
Varsha Kishore, Chao Wan, Justin Lovelace, Yoav Artzi, and Kilian~Q. Weinberger.
\newblock Incdsi: Incrementally updatable document retrieval.
\newblock In \emph{International Conference on Machine Learning}, pp.\  17122--17134. PMLR, 2023.

\bibitem[Kong et~al.(2025)Kong, Zhang, Liu, Zhang, Feng, Yang, Wang, Tian, Wang, Zhang, and Zhou]{Kong2025ModalityCuration}
Fanheng Kong, Jingyuan Zhang, Yahui Liu, Hongzhi Zhang, Shi Feng, Xiaocui Yang, Daling Wang, Yu~Tian, Qi~Wang, Fuzheng Zhang, and Guorui Zhou.
\newblock Modality curation: Building universal embeddings for advanced multimodal information retrieval.
\newblock \emph{arXiv preprint arXiv:2505.19650}, 2025.

\bibitem[Kong et~al.(2023)Kong, Dudek, Li, Zhang, and Bendersky]{kong2023sparseembed}
Weize Kong, Jeffrey~M. Dudek, Cheng Li, Mingyang Zhang, and Michael Bendersky.
\newblock Sparseembed: Learning sparse lexical representations with contextual embeddings for retrieval.
\newblock In \emph{Proceedings of the 46th International ACM SIGIR Conference on Research and Development in Information Retrieval}, pp.\  2399--2403, 2023.

\bibitem[Kraft \& Buell(1983)Kraft and Buell]{kraft1983fuzzy}
Donald~H. Kraft and Duncan~A. Buell.
\newblock Fuzzy sets and generalized boolean retrieval systems.
\newblock \emph{International journal of man-machine studies}, 19\penalty0 (1):\penalty0 45--56, 1983.

\bibitem[Kumar et~al.(2024)Kumar, Ankner, Spector, Bordelon, Muennighoff, Paul, Pehlevan, R{\'e}, and Raghunathan]{kumar2024scaling}
Tanishq Kumar, Zachary Ankner, Benjamin~F. Spector, Blake Bordelon, Niklas Muennighoff, Mansheej Paul, Cengiz Pehlevan, Christopher R{\'e}, and Aditi Raghunathan.
\newblock Scaling laws for precision.
\newblock \emph{arXiv preprint arXiv:2411.04330}, 2024.

\bibitem[Kwon et~al.(2023)Kwon, Li, Zhuang, Sheng, Zheng, Yu, Gonzalez, Zhang, and Stoica]{kwon2023vllmpagedattention}
Woosuk Kwon, Zhuohan Li, Siyuan Zhuang, Ying Sheng, Lianmin Zheng, Cody~Hao Yu, Joseph Gonzalez, Hao Zhang, and Ion Stoica.
\newblock Efficient memory management for large language model serving with pagedattention.
\newblock In \emph{Proceedings of the 29th Symposium on Operating Systems Principles}, pp.\  611--626, 2023.

\bibitem[Lan et~al.(2020)Lan, Chen, Goodman, Gimpel, Sharma, and Soricut]{lan2020albert}
Zhenzhong Lan, Mingda Chen, Sebastian Goodman, Kevin Gimpel, Piyush Sharma, and Radu Soricut.
\newblock {ALBERT}: A lite {BERT} for self-supervised learning of language representations.
\newblock In \emph{International Conference on Learning Representations}, 2020.
\newblock URL \url{https://openreview.net/forum?id=H1eA7AEtvS}.

\bibitem[Le \& Mikolov(2014)Le and Mikolov]{le2014distributedrepresentations}
Quoc Le and Tomas Mikolov.
\newblock Distributed representations of sentences and documents.
\newblock \emph{Proceedings of the 31st International Conference on Machine Learning, PMLR}, 32\penalty0 (2):\penalty0 1188--1196, 22--24 Jun 2014.
\newblock URL \url{https://proceedings.mlr.press/v32/le14.html}.

\bibitem[LeCun et~al.(1989)LeCun, Boser, Denker, Henderson, Howard, Hubbard, and Jackel]{lecun1989backpropagation}
Yann LeCun, Bernhard Boser, John~S. Denker, Donnie Henderson, Richard~E. Howard, Wayne Hubbard, and Lawrence~D. Jackel.
\newblock Backpropagation applied to handwritten zip code recognition.
\newblock \emph{Neural computation}, 1\penalty0 (4):\penalty0 541--551, 1989.

\bibitem[Lee et~al.(2025)Lee, Roy, Xu, Raiman, Shoeybi, Catanzaro, and Ping]{lee2024nvembed}
Chankyu Lee, Rajarshi Roy, Mengyao Xu, Jonathan Raiman, Mohammad Shoeybi, Bryan Catanzaro, and Wei Ping.
\newblock {NV}-embed: Improved techniques for training {LLM}s as generalist embedding models.
\newblock In \emph{The 13th International Conference on Learning Representations}, 2025.
\newblock URL \url{https://openreview.net/forum?id=lgsyLSsDRe}.

\bibitem[Lee et~al.(2024)Lee, Dai, Ren, Chen, Cer, Cole, Hui, Boratko, Kapadia, Ding, et~al.]{lee2024gecko}
Jinhyuk Lee, Zhuyun Dai, Xiaoqi Ren, Blair Chen, Daniel Cer, Jeremy~R. Cole, Kai Hui, Michael Boratko, Rajvi Kapadia, Wen Ding, et~al.
\newblock Gecko: Versatile text embeddings distilled from large language models.
\newblock \emph{arXiv preprint arXiv:2403.20327}, 2024.

\bibitem[Lee et~al.(2019)Lee, Chang, and Toutanova]{Lee2019LatentRF}
Kenton Lee, Ming-Wei Chang, and Kristina Toutanova.
\newblock Latent retrieval for weakly supervised open domain question answering.
\newblock \emph{arXiv preprint arXiv:1906.00300}, 2019.

\bibitem[Levy et~al.(2015)Levy, Goldberg, and Dagan]{levy-etal-2015-improving}
Omer Levy, Yoav Goldberg, and Ido Dagan.
\newblock Improving distributional similarity with lessons learned from word embeddings.
\newblock \emph{Transactions of the Association for Computational Linguistics}, 3:\penalty0 211--225, 2015.
\newblock \doi{10.1162/tacl_a_00134}.
\newblock URL \url{https://aclanthology.org/Q15-1016/}.

\bibitem[Lewis et~al.(2020{\natexlab{a}})Lewis, Liu, Goyal, Ghazvininejad, Mohamed, Levy, Stoyanov, and Zettlemoyer]{lewis-etal-2020-bart}
Mike Lewis, Yinhan Liu, Naman Goyal, Marjan Ghazvininejad, Abdelrahman Mohamed, Omer Levy, Veselin Stoyanov, and Luke Zettlemoyer.
\newblock {BART}: Denoising sequence-to-sequence pre-training for natural language generation, translation, and comprehension.
\newblock In Dan Jurafsky, Joyce Chai, Natalie Schluter, and Joel Tetreault (eds.), \emph{Proceedings of the 58th Annual Meeting of the Association for Computational Linguistics}, pp.\  7871--7880, Online, July 2020{\natexlab{a}}. Association for Computational Linguistics.
\newblock \doi{10.18653/v1/2020.acl-main.703}.
\newblock URL \url{https://aclanthology.org/2020.acl-main.703/}.

\bibitem[Lewis et~al.(2020{\natexlab{b}})Lewis, Perez, Piktus, Petroni, Karpukhin, Goyal, K{\"u}ttler, Lewis, Yih, Rockt{\"a}schel, et~al.]{lewis2020retrievalaugmentedgeneration}
Patrick Lewis, Ethan Perez, Aleksandra Piktus, Fabio Petroni, Vladimir Karpukhin, Naman Goyal, Heinrich K{\"u}ttler, Mike Lewis, Wen-tau Yih, Tim Rockt{\"a}schel, et~al.
\newblock Retrieval-augmented generation for knowledge-intensive nlp tasks.
\newblock \emph{Advances in neural information processing systems}, 33:\penalty0 9459--9474, 2020{\natexlab{b}}.

\bibitem[Li et~al.(2023{\natexlab{a}})Li, Yates, MacAvaney, He, and Sun]{Li2020PARADEPR}
Canjia Li, Andrew Yates, Sean MacAvaney, Ben He, and Yingfei Sun.
\newblock {PARADE}: Passage representation aggregation fordocument reranking.
\newblock \emph{ACM Transactions on Information Systems}, 42\penalty0 (2), September 2023{\natexlab{a}}.
\newblock \doi{10.1145/3600088}.

\bibitem[Li et~al.(2023{\natexlab{b}})Li, Liu, Xiao, and Shao]{li2023makinglargelanguagemodelsabetterfoundationfordenseretrieval}
Chaofan Li, Zheng Liu, Shitao Xiao, and Yingxia Shao.
\newblock Making large language models a better foundation for dense retrieval.
\newblock \emph{arXiv preprint arXiv:2312.15503}, 2023{\natexlab{b}}.

\bibitem[Li et~al.(2022{\natexlab{a}})Li, Hu, and Chen]{li-etal-2022-calibration}
Dongfang Li, Baotian Hu, and Qingcai Chen.
\newblock Calibration meets explanation: A simple and effective approach for model confidence estimates.
\newblock In Yoav Goldberg, Zornitsa Kozareva, and Yue Zhang (eds.), \emph{Proceedings of the 2022 Conference on Empirical Methods in Natural Language Processing}, pp.\  2775--2784, Abu Dhabi, United Arab Emirates, December 2022{\natexlab{a}}. Association for Computational Linguistics.
\newblock \doi{10.18653/v1/2022.emnlp-main.178}.
\newblock URL \url{https://aclanthology.org/2022.emnlp-main.178/}.

\bibitem[Li et~al.(2019)Li, Duan, Fang, Jiang, and Zhou]{Li2019UnicoderVLAU}
Gen Li, Nan Duan, Yuejian Fang, Daxin Jiang, and Ming Zhou.
\newblock {Unicoder-VL}: A universal encoder for vision and language by cross-modal pre-training.
\newblock In \emph{AAAI Conference on Artificial Intelligence}, 2019.

\bibitem[Li(2011)]{li2011learning}
Hang Li.
\newblock Learning for ranking aggregation.
\newblock In \emph{Learning to Rank for Information Retrieval and Natural Language Processing}, pp.\  33--35. Springer, 2011.

\bibitem[Li et~al.(2022{\natexlab{b}})Li, Li, Xiong, and Hoi]{Li2022BLIPBootstrappingLanguageImage}
Junnan Li, Dongxu Li, Caiming Xiong, and Steven Hoi.
\newblock {BLIP}: Bootstrapping language-image pre-training for unified vision-language understanding and generation.
\newblock In \emph{International Conference on Machine Learning}, pp.\  12888--12900. PMLR, 2022{\natexlab{b}}.

\bibitem[Li et~al.(2023{\natexlab{c}})Li, Lin, Ma, and Lin]{li2023slim}
Minghan Li, Sheng-Chieh Lin, Xueguang Ma, and Jimmy Lin.
\newblock {SLIM}: Sparsified late interaction for multi-vector retrieval with inverted indexes.
\newblock In \emph{Proceedings of the 46th International ACM SIGIR Conference on Research and Development in Information Retrieval}, SIGIR '23, pp.\  1954--1959, New York, NY, USA, 2023{\natexlab{c}}. Association for Computing Machinery.
\newblock \doi{10.1145/3539618.3591977}.

\bibitem[Li et~al.(2025{\natexlab{a}})Li, Miao, and Li]{li2025simpleiseffective}
Mufei Li, Siqi Miao, and Pan Li.
\newblock Simple is effective: The roles of graphs and large language models in knowledge-graph-based retrieval-augmented generation.
\newblock In \emph{The 13th International Conference on Learning Representations}, 2025{\natexlab{a}}.
\newblock URL \url{https://openreview.net/forum?id=JvkuZZ04O7}.

\bibitem[Li et~al.(2021)Li, Lv, Jin, Lin, Yang, Zeng, Wu, and Ma]{li2021embeddingbasedproductretrievalintaobaosearch}
Sen Li, Fuyu Lv, Taiwei Jin, Guli Lin, Keping Yang, Xiaoyi Zeng, Xiao-Ming Wu, and Qianli Ma.
\newblock Embedding-based product retrieval in taobao search.
\newblock In \emph{Proceedings of the 27th ACM SIGKDD Conference on Knowledge Discovery \& Data Mining}, pp.\  3181--3189, 2021.

\bibitem[Li et~al.(2025{\natexlab{b}})Li, Dong, Jin, Zhang, Zhou, Zhu, Zhang, and Dou]{li2025searcho1agenticsearchenhancedlargereasoningmodels}
Xiaoxi Li, Guanting Dong, Jiajie Jin, Yuyao Zhang, Yujia Zhou, Yutao Zhu, Peitian Zhang, and Zhicheng Dou.
\newblock {Search-o1}: Agentic search-enhanced large reasoning models.
\newblock \emph{arXiv preprint arXiv:2501.05366}, 2025{\natexlab{b}}.

\bibitem[Li et~al.(2025{\natexlab{c}})Li, Jin, Dong, Qian, Zhu, Wu, Wen, and Dou]{li2025webthinker}
Xiaoxi Li, Jiajie Jin, Guanting Dong, Hongjin Qian, Yutao Zhu, Yongkang Wu, Ji-Rong Wen, and Zhicheng Dou.
\newblock Webthinker: Empowering large reasoning models with deep research capability.
\newblock \emph{arXiv preprint arXiv:2504.21776}, 2025{\natexlab{c}}.

\bibitem[Li et~al.(2025{\natexlab{d}})Li, Jin, Zhou, Zhang, Zhang, Zhu, and Dou]{li2025frommatchingtogenerationasurveyongenerativeinformationretrieval}
Xiaoxi Li, Jiajie Jin, Yujia Zhou, Yuyao Zhang, Peitian Zhang, Yutao Zhu, and Zhicheng Dou.
\newblock From matching to generation: A survey on generative information retrieval.
\newblock \emph{ACM Transactions on Information Systems}, 43\penalty0 (3), May 2025{\natexlab{d}}.
\newblock \doi{10.1145/3722552}.

\bibitem[Li et~al.(2025{\natexlab{e}})Li, Zhang, Yang, Huang, Wu, Luo, Bei, Zou, Luo, Zhao, et~al.]{li2025towardsagenticrag}
Yangning Li, Weizhi Zhang, Yuyao Yang, Wei-Chieh Huang, Yaozu Wu, Junyu Luo, Yuanchen Bei, Henry~Peng Zou, Xiao Luo, Yusheng Zhao, et~al.
\newblock Towards agentic {RAG} with deep reasoning: A survey of {RAG}-reasoning systems in {LLMs}.
\newblock \emph{arXiv preprint arXiv:2507.09477}, 2025{\natexlab{e}}.

\bibitem[Li et~al.(2023{\natexlab{d}})Li, Zhang, Zhang, Long, Xie, and Zhang]{li2023generaltextembedding}
Zehan Li, Xin Zhang, Yanzhao Zhang, Dingkun Long, Pengjun Xie, and Meishan Zhang.
\newblock Towards general text embeddings with multi-stage contrastive learning.
\newblock \emph{arXiv preprint arXiv:2308.03281}, 2023{\natexlab{d}}.

\bibitem[Liang et~al.(2025)Liang, Su, Lin, Wu, Zhao, and Li]{liang2025reasoningrag}
Jintao Liang, Gang Su, Huifeng Lin, You Wu, Rui Zhao, and Ziyue Li.
\newblock Reasoning {RAG} via system 1 or system 2: A survey on reasoning agentic retrieval-augmented generation for industry challenges.
\newblock \emph{arXiv preprint arXiv:2506.10408}, 2025.

\bibitem[Lin(2021)]{Lin2021AProposedConceptualFramework}
Jimmy Lin.
\newblock A proposed conceptual framework for a representational approach to information retrieval.
\newblock \emph{ACM SIGIR Forum}, 55:\penalty0 1--29, 2021.

\bibitem[Lin \& Ma(2021)Lin and Ma]{lin2021few}
Jimmy Lin and Xueguang Ma.
\newblock A few brief notes on {DeepImpact}, {COIL}, and a conceptual framework for information retrieval techniques.
\newblock \emph{arXiv preprint arXiv:2106.14807}, 2021.

\bibitem[Lin et~al.(2022)Lin, Nogueira, and Yates]{lin2022pretrained}
Jimmy Lin, Rodrigo Nogueira, and Andrew Yates.
\newblock \emph{Pretrained transformers for text ranking: {BERT} and beyond}.
\newblock Springer Nature, 2022.

\bibitem[Lin et~al.(2024)Lin, Yadav, Liu, Rossi, Suram, Chembolu, Chandran, Mohapatra, Lee, Magnani, and Liao]{Lin2024EnhancingRelevance}
Juexin Lin, Sachin Yadav, Feng Liu, Nicholas Rossi, Praveen~Reddy Suram, Satya Chembolu, Prijith Chandran, Hrushikesh Mohapatra, Tony Lee, Alessandro Magnani, and Ciya Liao.
\newblock Enhancing relevance of embedding-based retrieval at {Walmart}.
\newblock \emph{Proceedings of the 33rd ACM International Conference on Information and Knowledge Management}, 2024.

\bibitem[Lin et~al.(2025)Lin, Wu, Xu, Liu, Tang, He, Aggarwal, Liu, Zhang, and Wang]{lin2025comprehensivesurveyreinforcementlearningbased}
Minhua Lin, Zongyu Wu, Zhichao Xu, Hui Liu, Xianfeng Tang, Qi~He, Charu Aggarwal, Hui Liu, Xiang Zhang, and Suhang Wang.
\newblock A comprehensive survey on reinforcement learning-based agentic search: Foundations, roles, optimizations, evaluations, and applications.
\newblock \emph{arXiv preprint arXiv:2510.16724}, 2025.

\bibitem[Litschko et~al.(2022)Litschko, Vuli{\'c}, and Glava{\v{s}}]{litschko-etal-2022-parameter}
Robert Litschko, Ivan Vuli{\'c}, and Goran Glava{\v{s}}.
\newblock Parameter-efficient neural reranking for cross-lingual and multilingual retrieval.
\newblock In Nicoletta Calzolari, Chu-Ren Huang, Hansaem Kim, James Pustejovsky, Leo Wanner, Key-Sun Choi, Pum-Mo Ryu, Hsin-Hsi Chen, Lucia Donatelli, Heng Ji, Sadao Kurohashi, Patrizia Paggio, Nianwen Xue, Seokhwan Kim, Younggyun Hahm, Zhong He, Tony~Kyungil Lee, Enrico Santus, Francis Bond, and Seung-Hoon Na (eds.), \emph{Proceedings of the 29th International Conference on Computational Linguistics}, pp.\  1071--1082, Gyeongju, Republic of Korea, October 2022. International Committee on Computational Linguistics.
\newblock URL \url{https://aclanthology.org/2022.coling-1.90/}.

\bibitem[Liu et~al.(2024{\natexlab{a}})Liu, Feng, Xue, Wang, Wu, Lu, Zhao, Deng, Zhang, Ruan, et~al.]{liu2024deepseek}
Aixin Liu, Bei Feng, Bing Xue, Bingxuan Wang, Bochao Wu, Chengda Lu, Chenggang Zhao, Chengqi Deng, Chenyu Zhang, Chong Ruan, et~al.
\newblock {DeepSeek}-v3 technical report.
\newblock \emph{arXiv preprint arXiv:2412.19437}, 2024{\natexlab{a}}.

\bibitem[Liu et~al.(2025{\natexlab{a}})Liu, Wang, Chen, Li, Xiong, Yu, and Zhang]{liu-etal-2025-hoprag}
Hao Liu, Zhengren Wang, Xi~Chen, Zhiyu Li, Feiyu Xiong, Qinhan Yu, and Wentao Zhang.
\newblock {H}op{RAG}: Multi-hop reasoning for logic-aware retrieval-augmented generation.
\newblock In Wanxiang Che, Joyce Nabende, Ekaterina Shutova, and Mohammad~Taher Pilehvar (eds.), \emph{Findings of the Association for Computational Linguistics: ACL 2025}, pp.\  1897--1913, Vienna, Austria, July 2025{\natexlab{a}}. Association for Computational Linguistics.
\newblock \doi{10.18653/v1/2025.findings-acl.97}.
\newblock URL \url{https://aclanthology.org/2025.findings-acl.97/}.

\bibitem[Liu et~al.(2023)Liu, Li, Wu, and Lee]{Liu2023VisualInstructionTuning}
Haotian Liu, Chunyuan Li, Qingyang Wu, and Yong~Jae Lee.
\newblock Visual instruction tuning.
\newblock \emph{arXiv preprint arXiv:2304.08485}, 2023.

\bibitem[Liu et~al.(2017)Liu, Chang, Wu, and Yang]{liu2017deeplearningforextrememultilabeltextclassification}
Jingzhou Liu, Wei-Cheng Chang, Yuexin Wu, and Yiming Yang.
\newblock Deep learning for extreme multi-label text classification.
\newblock In \emph{Proceedings of the 40th International ACM SIGIR Conference on Research and Development in Information Retrieval}, pp.\  115--124, 2017.

\bibitem[Liu et~al.(2025{\natexlab{b}})Liu, Liu, Yao, Liu, Meng, Wang, and Ma]{Liu2025HMRAG}
Pei Liu, Xin Liu, Ruoyu Yao, Junming Liu, Siyuan Meng, Ding Wang, and Jun Ma.
\newblock Hm-{RAG}: Hierarchical multi-agent multimodal retrieval augmented generation.
\newblock \emph{Proceedings of the 33rd ACM International Conference on Multimedia}, 2025{\natexlab{b}}.

\bibitem[Liu et~al.(2024{\natexlab{b}})Liu, Wang, Wang, and Mao]{liu2024leveraging}
Qi~Liu, Bo~Wang, Nan Wang, and Jiaxin Mao.
\newblock Leveraging passage embeddings for efficient listwise reranking with large language models.
\newblock In \emph{Proceedings of the ACM on Web Conference 2025}, 2024{\natexlab{b}}.
\newblock \doi{10.1145/3696410.3714554}.

\bibitem[Liu et~al.(2021)Liu, Fan, Qian, Chen, Ding, and Wang]{Liu2021HiTHierarchicalTransformer}
Song Liu, Haoqi Fan, Shengsheng Qian, Yiru Chen, Wenkui Ding, and Zhongyuan Wang.
\newblock {HiT}: Hierarchical transformer with momentum contrast for video-text retrieval.
\newblock \emph{2021 IEEE/CVF International Conference on Computer Vision (ICCV)}, pp.\  11895--11905, 2021.

\bibitem[Liu(2009)]{liu2009ltr}
Tie-Yan Liu.
\newblock Learning to rank for information retrieval.
\newblock \emph{Foundations and Trends in Information Retrieval}, 3\penalty0 (3):\penalty0 225--331, March 2009.
\newblock \doi{10.1561/1500000016}.

\bibitem[Liu et~al.(2019)Liu, Ott, Goyal, Du, Joshi, Chen, Levy, Lewis, Zettlemoyer, and Stoyanov]{liu2019roberta}
Yinhan Liu, Myle Ott, Naman Goyal, Jingfei Du, Mandar Joshi, Danqi Chen, Omer Levy, Mike Lewis, Luke Zettlemoyer, and Veselin Stoyanov.
\newblock {RoBERTa}: A robustly optimized {BERT} pretraining approach.
\newblock \emph{arXiv preprint arXiv:1907.11692}, 2019.

\bibitem[Liu et~al.(2025{\natexlab{c}})Liu, Zhang, Guo, and de~Rijke]{liu2025robustinformationretrieval}
Yu-An Liu, Ruqing Zhang, Jiafeng Guo, and Maarten de~Rijke.
\newblock Robust information retrieval.
\newblock In \emph{Proceedings of the 18th ACM International Conference on Web Search and Data Mining}, pp.\  1008--1011, 2025{\natexlab{c}}.

\bibitem[Liu et~al.(2018)Liu, Xiong, Sun, and Liu]{Liu2018EntityDuetNeuralRanking}
Zhenghao Liu, Chenyan Xiong, Maosong Sun, and Zhiyuan Liu.
\newblock Entity-duet neural ranking: Understanding the role of knowledge graph semantics in neural information retrieval.
\newblock In \emph{Proceedings of the 56th Annual Meeting of the Association for Computational Linguistics (Volume 1: Long Papers)}, pp.\  2395--2405, 2018.

\bibitem[Logeswaran et~al.(2019)Logeswaran, Chang, Lee, Toutanova, Devlin, and Lee]{logeswaran-etal-2019-zero}
Lajanugen Logeswaran, Ming-Wei Chang, Kenton Lee, Kristina Toutanova, Jacob Devlin, and Honglak Lee.
\newblock Zero-shot entity linking by reading entity descriptions.
\newblock In Anna Korhonen, David Traum, and Llu{\'\i}s M{\`a}rquez (eds.), \emph{Proceedings of the 57th Annual Meeting of the Association for Computational Linguistics}, pp.\  3449--3460, Florence, Italy, July 2019. Association for Computational Linguistics.
\newblock \doi{10.18653/v1/P19-1335}.
\newblock URL \url{https://aclanthology.org/P19-1335/}.

\bibitem[Lowe(1999)]{lowe1999objectrecognition}
David~G. Lowe.
\newblock Object recognition from local scale-invariant features.
\newblock In \emph{Proceedings of the 7th IEEE International Conference on Computer Vision}, volume~2, pp.\  1150--1157, 1999.

\bibitem[Lu \& Li(2013)Lu and Li]{lu2013deeparchitectureformatchingshorttexts}
Zhengdong Lu and Hang Li.
\newblock A deep architecture for matching short texts.
\newblock \emph{Advances in neural information processing systems}, 26, 2013.

\bibitem[Luan et~al.(2021)Luan, Eisenstein, Toutanova, and Collins]{luan2021sparse}
Yi~Luan, Jacob Eisenstein, Kristina Toutanova, and Michael Collins.
\newblock Sparse, dense, and attentional representations for text retrieval.
\newblock \emph{Transactions of the Association for Computational Linguistics}, 9:\penalty0 329--345, 2021.

\bibitem[Ma et~al.(2024{\natexlab{a}})Ma, Wu, Lin, and Hu]{ma2024dropyourdecoder}
Guangyuan Ma, Xing Wu, Zijia Lin, and Songlin Hu.
\newblock Drop your decoder: Pre-training with bag-of-word prediction for dense passage retrieval.
\newblock In \emph{Proceedings of the 47th International ACM SIGIR Conference on Research and Development in Information Retrieval}, pp.\  1818--1827, 2024{\natexlab{a}}.

\bibitem[Ma et~al.(2023)Ma, Zhang, Pradeep, and Lin]{ma2023zeroshotlistwisedocumentreranking}
Xueguang Ma, Xinyu Zhang, Ronak Pradeep, and Jimmy Lin.
\newblock Zero-shot listwise document reranking with a large language model.
\newblock \emph{arXiv preprint arXiv: 2305.02156}, 2023.

\bibitem[Ma et~al.(2024{\natexlab{b}})Ma, Wang, Yang, Wei, and Lin]{ma2024fine}
Xueguang Ma, Liang Wang, Nan Yang, Furu Wei, and Jimmy Lin.
\newblock Fine-tuning {LLaMA} for multi-stage text retrieval.
\newblock In \emph{Proceedings of the 47th International ACM SIGIR Conference on Research and Development in Information Retrieval}, pp.\  2421--2425, 2024{\natexlab{b}}.

\bibitem[MacAvaney et~al.(2019{\natexlab{a}})MacAvaney, Soldaini, and Goharian]{MacAvaney2019TeachingAN}
Sean MacAvaney, Luca Soldaini, and Nazli Goharian.
\newblock Teaching a new dog old tricks: Resurrecting multilingual retrieval using zero-shot learning.
\newblock \emph{Advances in Information Retrieval}, 12036:\penalty0 246 -- 254, 2019{\natexlab{a}}.

\bibitem[MacAvaney et~al.(2019{\natexlab{b}})MacAvaney, Yates, Cohan, and Goharian]{macavaney2019cedr}
Sean MacAvaney, Andrew Yates, Arman Cohan, and Nazli Goharian.
\newblock {CEDR}: Contextualized embeddings for document ranking.
\newblock In \emph{Proceedings of the 42nd International ACM SIGIR Conference on Research and Development in Information Retrieval}, pp.\  1101--1104, 2019{\natexlab{b}}.

\bibitem[MacAvaney et~al.(2020)MacAvaney, Nardini, Perego, Tonellotto, Goharian, and Frieder]{MacAvaney2020EfficientDocumentReranking}
Sean MacAvaney, Franco~Maria Nardini, R.~Perego, Nicola Tonellotto, Nazli Goharian, and Ophir Frieder.
\newblock Efficient document re-ranking for transformers by precomputing term representations.
\newblock \emph{Proceedings of the 43rd International ACM SIGIR Conference on Research and Development in Information Retrieval}, 2020.

\bibitem[MacAvaney et~al.(2022)MacAvaney, Feldman, Goharian, Downey, and Cohan]{macavaney-etal-2022-abnirml}
Sean MacAvaney, Sergey Feldman, Nazli Goharian, Doug Downey, and Arman Cohan.
\newblock {ABNIRML}: Analyzing the behavior of neural {IR} models.
\newblock \emph{Transactions of the Association for Computational Linguistics}, 10:\penalty0 224--239, 2022.
\newblock \doi{10.1162/tacl_a_00457}.
\newblock URL \url{https://aclanthology.org/2022.tacl-1.13/}.

\bibitem[MacKay \& Peto(1995)MacKay and Peto]{mackay1995hierarchicaldirichlet}
David J.~C. MacKay and Linda C.~Bauman Peto.
\newblock A hierarchical {Dirichlet} language model.
\newblock \emph{Natural language engineering}, 1\penalty0 (3):\penalty0 289--308, 1995.

\bibitem[Malkov \& Yashunin(2018)Malkov and Yashunin]{malkov2018hnsw}
Yu~A. Malkov and Dmitry~A. Yashunin.
\newblock Efficient and robust approximate nearest neighbor search using hierarchical navigable small world graphs.
\newblock \emph{IEEE transactions on pattern analysis and machine intelligence}, 42\penalty0 (4):\penalty0 824--836, 2018.

\bibitem[Mallen et~al.(2023)Mallen, Asai, Zhong, Das, Khashabi, and Hajishirzi]{mallen-etal-2023-trust}
Alex Mallen, Akari Asai, Victor Zhong, Rajarshi Das, Daniel Khashabi, and Hannaneh Hajishirzi.
\newblock When not to trust language models: Investigating effectiveness of parametric and non-parametric memories.
\newblock In Anna Rogers, Jordan Boyd-Graber, and Naoaki Okazaki (eds.), \emph{Proceedings of the 61st Annual Meeting of the Association for Computational Linguistics (Volume 1: Long Papers)}, pp.\  9802--9822, Toronto, Canada, July 2023. Association for Computational Linguistics.
\newblock \doi{10.18653/v1/2023.acl-long.546}.
\newblock URL \url{https://aclanthology.org/2023.acl-long.546/}.

\bibitem[Mallia et~al.(2021)Mallia, Khattab, Suel, and Tonellotto]{mallia2021learning}
Antonio Mallia, Omar Khattab, Torsten Suel, and Nicola Tonellotto.
\newblock Learning passage impacts for inverted indexes.
\newblock In \emph{Proceedings of the 44th International ACM SIGIR Conference on Research and Development in Information Retrieval}, pp.\  1723--1727, 2021.

\bibitem[Marchionini(2006)]{marchionini2006exploratorysearch}
Gary Marchionini.
\newblock Exploratory search: from finding to understanding.
\newblock \emph{Communications of the ACM}, 49\penalty0 (4):\penalty0 41--46, April 2006.
\newblock \doi{10.1145/1121949.1121979}.

\bibitem[Mehta et~al.(2023)Mehta, Gupta, Tay, Dehghani, Tran, Rao, Najork, Strubell, and Metzler]{Mehta2022dsi}
Sanket~Vaibhav Mehta, Jai Gupta, Yi~Tay, Mostafa Dehghani, Vinh~Q Tran, Jinfeng Rao, Marc Najork, Emma Strubell, and Donald Metzler.
\newblock {DSI++}: Updating transformer memory with new documents.
\newblock In \emph{Proceedings of the 2023 Conference on Empirical Methods in Natural Language Processing}, pp.\  8198--8213, 2023.

\bibitem[Mei et~al.(2025)Mei, Mo, Yang, and Chen]{mei2025surveymultimodalretrievalaugmentedgeneration}
Lang Mei, Siyu Mo, Zhihan Yang, and Chong Chen.
\newblock A survey of {Multimodal Retrieval-Augmented Generation}.
\newblock arXiv preprint arXiv:2504.08748, 2025.

\bibitem[Mikolov et~al.(2013)Mikolov, Chen, Corrado, and Dean]{mikolov2013efficient}
Tomas Mikolov, Kai Chen, Greg Corrado, and Jeffrey Dean.
\newblock Efficient estimation of word representations in vector space.
\newblock \emph{arXiv preprint arXiv:1301.3781}, 2013.

\bibitem[Miller et~al.(1999)Miller, Leek, and Schwartz]{miller1999hidden}
David R.~H. Miller, Tim Leek, and Richard~M. Schwartz.
\newblock A hidden markov model information retrieval system.
\newblock In \emph{Proceedings of the 22nd Annual International ACM SIGIR Conference on Research and Development in Information Retrieval}, pp.\  214--221, 1999.

\bibitem[Minderer et~al.(2021)Minderer, Djolonga, Romijnders, Hubis, Zhai, Houlsby, Tran, and Lucic]{Minderer2021RevisitingTheCalibrationofModernNeuralNetworks}
Matthias Minderer, Josip Djolonga, Rob Romijnders, Frances~Ann Hubis, Xiaohua Zhai, Neil Houlsby, Dustin Tran, and Mario Lucic.
\newblock Revisiting the calibration of modern neural networks.
\newblock \emph{arXiv preprint arXiv:2106.07998}, 2021.

\bibitem[Mitra et~al.(2016)Mitra, Nalisnick, Craswell, and Caruana]{mitra2016dual}
Bhaskar Mitra, Eric Nalisnick, Nick Craswell, and Rich Caruana.
\newblock A dual embedding space model for document ranking.
\newblock \emph{arXiv preprint arXiv:1602.01137}, 2016.

\bibitem[Mitra et~al.(2017)Mitra, Diaz, and Craswell]{mitra2017duet}
Bhaskar Mitra, Fernando Diaz, and Nick Craswell.
\newblock Learning to match using local and distributed representations of text for web search.
\newblock In \emph{Proceedings of the 26th International Conference on World Wide Web}, WWW '17, pp.\  1291--1299, Republic and Canton of Geneva, CHE, 2017. International World Wide Web Conferences Steering Committee.
\newblock \doi{10.1145/3038912.3052579}.

\bibitem[Mitra et~al.(2018)Mitra, Craswell, et~al.]{mitra2018introduction}
Bhaskar Mitra, Nick Craswell, et~al.
\newblock An introduction to neural information retrieval.
\newblock \emph{Foundations and Trends in Information Retrieval}, 13\penalty0 (1):\penalty0 1--126, 2018.

\bibitem[Mitra et~al.(2021)Mitra, Hofst{\"a}tter, Zamani, and Craswell]{mitra2021improving}
Bhaskar Mitra, Sebastian Hofst{\"a}tter, Hamed Zamani, and Nick Craswell.
\newblock Improving transformer-kernel ranking model using conformer and query term independence.
\newblock In \emph{Proceedings of the 44th International ACM SIGIR Conference on Research and Development in Information Retrieval}, pp.\  1697--1702, 2021.

\bibitem[Mo et~al.(2023)Mo, Mao, Zhu, Wu, Huang, and Nie]{mo2023convgqr}
Fengran Mo, Kelong Mao, Yutao Zhu, Yihong Wu, Kaiyu Huang, and Jian-Yun Nie.
\newblock {ConvGQR}: Generative query reformulation for conversational search.
\newblock In \emph{Proceedings of the 61st Annual Meeting of the Association for Computational Linguistics (Volume 1: Long Papers)}, pp.\  4998--5012, 2023.

\bibitem[Mo et~al.(2024{\natexlab{a}})Mo, Ghaddar, Mao, Rezagholizadeh, Chen, Liu, and Nie]{mo2024chiq}
Fengran Mo, Abbas Ghaddar, Kelong Mao, Mehdi Rezagholizadeh, Boxing Chen, Qun Liu, and Jian-Yun Nie.
\newblock {CHIQ}: Contextual history enhancement for improving query rewriting in conversational search.
\newblock \emph{arXiv preprint arXiv:2406.05013}, 2024{\natexlab{a}}.

\bibitem[Mo et~al.(2024{\natexlab{b}})Mo, Mao, Zhao, Qian, Chen, Cheng, Li, Zhu, Dou, and Nie]{mo2024survey}
Fengran Mo, Kelong Mao, Ziliang Zhao, Hongjin Qian, Haonan Chen, Yiruo Cheng, Xiaoxi Li, Yutao Zhu, Zhicheng Dou, and Jian-Yun Nie.
\newblock A survey of conversational search.
\newblock \emph{arXiv preprint arXiv:2410.15576}, 2024{\natexlab{b}}.

\bibitem[Mo et~al.(2024{\natexlab{c}})Mo, Yi, Mao, Qu, Huang, and Nie]{mo2024convsdg}
Fengran Mo, Bole Yi, Kelong Mao, Chen Qu, Kaiyu Huang, and Jian-Yun Nie.
\newblock {ConvSDG}: Session data generation for conversational search.
\newblock In \emph{Companion Proceedings of the ACM on Web Conference 2024}, pp.\  1634--1642, 2024{\natexlab{c}}.

\bibitem[Mokrii et~al.(2021)Mokrii, Boytsov, and Braslavski]{iurii2021systematicevaluation}
Iurii Mokrii, Leonid Boytsov, and Pavel Braslavski.
\newblock A systematic evaluation of transfer learning and pseudo-labeling with {BERT}-based ranking models.
\newblock In \emph{Proceedings of the 44th International ACM SIGIR Conference on Research and Development in Information Retrieval}, SIGIR '21, pp.\  2081--2085, New York, NY, USA, 2021. Association for Computing Machinery.
\newblock \doi{10.1145/3404835.3463093}.

\bibitem[Muennighoff(2022)]{muennighoff2022sgpt}
Niklas Muennighoff.
\newblock {SGPT}: {GPT} sentence embeddings for semantic search.
\newblock \emph{arXiv preprint arXiv:2202.08904}, 2022.

\bibitem[Muennighoff et~al.(2023)Muennighoff, Tazi, Magne, and Reimers]{muennighoff-etal-2023-mteb}
Niklas Muennighoff, Nouamane Tazi, Loic Magne, and Nils Reimers.
\newblock {MTEB}: Massive text embedding benchmark.
\newblock In Andreas Vlachos and Isabelle Augenstein (eds.), \emph{Proceedings of the 17th Conference of the European Chapter of the Association for Computational Linguistics}, pp.\  2014--2037, Dubrovnik, Croatia, May 2023. Association for Computational Linguistics.
\newblock \doi{10.18653/v1/2023.eacl-main.148}.
\newblock URL \url{https://aclanthology.org/2023.eacl-main.148/}.

\bibitem[Muennighoff et~al.(2025)Muennighoff, Su, Wang, Yang, Wei, Yu, Singh, and Kiela]{muennighoff2025generative}
Niklas Muennighoff, Hongjin Su, Liang Wang, Nan Yang, Furu Wei, Tao Yu, Amanpreet Singh, and Douwe Kiela.
\newblock Generative representational instruction tuning.
\newblock In \emph{The 13th International Conference on Learning Representations}, 2025.
\newblock URL \url{https://openreview.net/forum?id=BC4lIvfSzv}.

\bibitem[Nakano et~al.(2021)Nakano, Hilton, Balaji, Wu, Ouyang, Kim, Hesse, Jain, Kosaraju, Saunders, et~al.]{nakano2021webgpt}
Reiichiro Nakano, Jacob Hilton, Suchir Balaji, Jeff Wu, Long Ouyang, Christina Kim, Christopher Hesse, Shantanu Jain, Vineet Kosaraju, William Saunders, et~al.
\newblock {WebGPT}: Browser-assisted question-answering with human feedback.
\newblock \emph{arXiv preprint arXiv:2112.09332}, 2021.

\bibitem[Nalisnick et~al.(2016)Nalisnick, Mitra, Craswell, and Caruana]{nalisnick2016dualwordembeddings}
Eric Nalisnick, Bhaskar Mitra, Nick Craswell, and Rich Caruana.
\newblock Improving document ranking with dual word embeddings.
\newblock In \emph{Proceedings of the 25th International Conference Companion on World Wide Web}, WWW '16 Companion, pp.\  83--84, Republic and Canton of Geneva, CHE, 2016. International World Wide Web Conferences Steering Committee.
\newblock \doi{10.1145/2872518.2889361}.

\bibitem[Nayak(2019)]{google2019search}
Pandu Nayak.
\newblock Understanding searches better than ever before.
\newblock \url{https://blog.google/products/search/search-language-understanding-bert/}, 2019.

\bibitem[Neelakantan et~al.(2022)Neelakantan, Xu, Puri, Radford, Han, Tworek, Yuan, Tezak, Kim, Hallacy, et~al.]{neelakantan2022text}
Arvind Neelakantan, Tao Xu, Raul Puri, Alec Radford, Jesse~Michael Han, Jerry Tworek, Qiming Yuan, Nikolas Tezak, Jong~Wook Kim, Chris Hallacy, et~al.
\newblock Text and code embeddings by contrastive pre-training.
\newblock \emph{arXiv preprint arXiv:2201.10005}, 2022.

\bibitem[Nguyen et~al.(2023)Nguyen, MacAvaney, and Yates]{nguyen2023unified}
Thong Nguyen, Sean MacAvaney, and Andrew Yates.
\newblock A unified framework for learned sparse retrieval.
\newblock In \emph{European Conference on Information Retrieval}, pp.\  101--116. Springer, 2023.

\bibitem[Ni et~al.(2022{\natexlab{a}})Ni, Hernandez~Abrego, Constant, Ma, Hall, Cer, and Yang]{ni-etal-2022-sentence}
Jianmo Ni, Gustavo Hernandez~Abrego, Noah Constant, Ji~Ma, Keith Hall, Daniel Cer, and Yinfei Yang.
\newblock Sentence-{T5}: Scalable sentence encoders from pre-trained text-to-text models.
\newblock In Smaranda Muresan, Preslav Nakov, and Aline Villavicencio (eds.), \emph{Findings of the Association for Computational Linguistics: ACL 2022}, pp.\  1864--1874, Dublin, Ireland, May 2022{\natexlab{a}}. Association for Computational Linguistics.
\newblock \doi{10.18653/v1/2022.findings-acl.146}.
\newblock URL \url{https://aclanthology.org/2022.findings-acl.146/}.

\bibitem[Ni et~al.(2022{\natexlab{b}})Ni, Qu, Lu, Dai, Hernandez~Abrego, Ma, Zhao, Luan, Hall, Chang, and Yang]{ni-etal-2022-large}
Jianmo Ni, Chen Qu, Jing Lu, Zhuyun Dai, Gustavo Hernandez~Abrego, Ji~Ma, Vincent Zhao, Yi~Luan, Keith Hall, Ming-Wei Chang, and Yinfei Yang.
\newblock Large dual encoders are generalizable retrievers.
\newblock In Yoav Goldberg, Zornitsa Kozareva, and Yue Zhang (eds.), \emph{Proceedings of the 2022 Conference on Empirical Methods in Natural Language Processing}, pp.\  9844--9855, Abu Dhabi, United Arab Emirates, December 2022{\natexlab{b}}. Association for Computational Linguistics.
\newblock \doi{10.18653/v1/2022.emnlp-main.669}.
\newblock URL \url{https://aclanthology.org/2022.emnlp-main.669/}.

\bibitem[Nie(2010)]{nie2010cross}
Jian-Yun Nie.
\newblock \emph{Cross-language information retrieval}.
\newblock Morgan \& Claypool Publishers, 2010.

\bibitem[Nigam et~al.(2019)Nigam, Song, Mohan, Lakshman, Ding, Shingavi, Teo, Gu, and Yin]{Nigam2019SemanticProductSearch}
Priyank Nigam, Yiwei Song, Vijai Mohan, Vihan Lakshman, Weitian Ding, Ankit Shingavi, Choon~Hui Teo, Hao Gu, and Bing Yin.
\newblock Semantic product search.
\newblock \emph{Proceedings of the 25th ACM SIGKDD International Conference on Knowledge Discovery \& Data Mining}, 2019.

\bibitem[Nogueira et~al.(2019{\natexlab{a}})Nogueira, Lin, and Epistemic]{nogueira2019doc2query}
Rodrigo Nogueira, Jimmy Lin, and A.~I. Epistemic.
\newblock From doc2query to {docTTTTTquery}.
\newblock \emph{Online preprint}, 6\penalty0 (2), 2019{\natexlab{a}}.

\bibitem[Nogueira et~al.(2019{\natexlab{b}})Nogueira, Yang, Cho, and Lin]{nogueira2019multi}
Rodrigo Nogueira, Wei Yang, Kyunghyun Cho, and Jimmy Lin.
\newblock Multi-stage document ranking with {BERT}.
\newblock \emph{arXiv preprint arXiv:1910.14424}, 2019{\natexlab{b}}.

\bibitem[Nogueira et~al.(2020)Nogueira, Jiang, and Lin]{nogueira2020documentrankingpretrainedsequencetosequence}
Rodrigo Nogueira, Zhiying Jiang, and Jimmy Lin.
\newblock Document ranking with a pretrained sequence-to-sequence model.
\newblock arXiv preprint arXiv:2003.06713, 2020.

\bibitem[Nogueira~dos Santos et~al.(2020)Nogueira~dos Santos, Ma, Nallapati, Huang, and Xiang]{nogueira-dos-santos-etal-2020-beyond}
Cicero Nogueira~dos Santos, Xiaofei Ma, Ramesh Nallapati, Zhiheng Huang, and Bing Xiang.
\newblock Beyond [{CLS}] through ranking by generation.
\newblock In Bonnie Webber, Trevor Cohn, Yulan He, and Yang Liu (eds.), \emph{Proceedings of the 2020 Conference on Empirical Methods in Natural Language Processing (EMNLP)}, pp.\  1722--1727, Online, November 2020. Association for Computational Linguistics.
\newblock \doi{10.18653/v1/2020.emnlp-main.134}.
\newblock URL \url{https://aclanthology.org/2020.emnlp-main.134/}.

\bibitem[Nussbaum \& Duderstadt(2025)Nussbaum and Duderstadt]{nussbaum2025trainingsparsemixtureexperts}
Zach Nussbaum and Brandon Duderstadt.
\newblock Training sparse mixture of experts text embedding models.
\newblock arXiv preprint arXiv:2502.07972, 2025.

\bibitem[Nussbaum et~al.(2025)Nussbaum, Morris, Mulyar, and Duderstadt]{nussbaum2025nomicembed}
Zach Nussbaum, John~Xavier Morris, Andriy Mulyar, and Brandon Duderstadt.
\newblock {Nomic Embed}: Training a reproducible long context text embedder.
\newblock \emph{Transactions on Machine Learning Research}, 2025.
\newblock URL \url{https://openreview.net/forum?id=IPmzyQSiQE}.

\bibitem[Oard et~al.(1999)Oard, Peters, Ruiz, Frederking, Klavans, and Sheridan]{oard1999multilingualinformationdiscoveryandaccess}
Douglas Oard, Carol Peters, Miguel Ruiz, Robert Frederking, Judith Klavans, and Paraic Sheridan.
\newblock {Multilingual Information Discovery and AccesS} ({MIDAS}).
\newblock \emph{{D-Lib Magazine}}, 5\penalty0 (10), 1999.
\newblock \doi{10.1045/october99-oard}.

\bibitem[Oard \& Dorr(1996)Oard and Dorr]{oard1996survey}
Douglas~W. Oard and Bonnie~J. Dorr.
\newblock A survey of multilingual text retrieval.
\newblock Technical Report UMIACS-TR-96-19 CS-TR-3615, University of Maryland Institute for Advanced Computer Studies, 1996.

\bibitem[Oard \& Dorr(1998)Oard and Dorr]{oard1998evaluatingcrosslingual}
Douglas~W. Oard and Bonnie~J. Dorr.
\newblock Evaluating cross-language text filtering effectiveness.
\newblock In \emph{Cross-Language Information Retrieval}, pp.\  151--161. Springer, 1998.

\bibitem[Oh et~al.(2024)Oh, Lee, Ye, Shin, Jang, Jun, and Seo]{oh2024instructir}
Hanseok Oh, Hyunji Lee, Seonghyeon Ye, Haebin Shin, Hansol Jang, Changwook Jun, and Minjoon Seo.
\newblock {INSTRUCTIR}: A benchmark for instruction following of information retrieval models.
\newblock \emph{arXiv preprint arXiv:2402.14334}, 2024.

\bibitem[Oliveira \& Teixeira~Lopes(2023)Oliveira and Teixeira~Lopes]{Oliveira2023from10bluelinks}
Bruno Oliveira and Carla Teixeira~Lopes.
\newblock From 10 blue links pages to feature-full search engine results pages - analysis of the temporal evolution of {SERP} features.
\newblock In \emph{Proceedings of the 2023 Conference on Human Information Interaction and Retrieval}, pp.\  338--345. ACM, March 2023.
\newblock \doi{10.1145/3576840.3578307}.

\bibitem[Onal et~al.(2018)Onal, Zhang, Altingovde, Rahman, Karagoz, Braylan, Dang, Chang, Kim, McNamara, et~al.]{onal2018neural}
Kezban~Dilek Onal, Ye~Zhang, Ismail~Sengor Altingovde, Md~Mustafizur Rahman, Pinar Karagoz, Alex Braylan, Brandon Dang, Heng-Lu Chang, Henna Kim, Quinten McNamara, et~al.
\newblock Neural information retrieval: At the end of the early years.
\newblock \emph{Information Retrieval Journal}, 21:\penalty0 111--182, 2018.

\bibitem[Oord et~al.(2018)Oord, Li, and Vinyals]{oord2018representation}
Aaron van~den Oord, Yazhe Li, and Oriol Vinyals.
\newblock Representation learning with contrastive predictive coding.
\newblock \emph{arXiv preprint arXiv:1807.03748}, 2018.

\bibitem[OpenAI(2022)]{openai2022chatgpt}
OpenAI.
\newblock Introducing {ChatGPT}.
\newblock \url{https://openai.com/index/chatgpt/}, 2022.

\bibitem[OpenAI(2024)]{openai2024introducingchatgptsearch}
OpenAI.
\newblock Introducing {ChatGPT} search.
\newblock \url{https://openai.com/index/introducing-chatgpt-search/}, 2024.

\bibitem[{OpenAI}(2025)]{deepresearchsystemcard}
{OpenAI}.
\newblock Introducing deep research.
\newblock \url{https://openai.com/index/introducing-deep-research/}, 2025.

\bibitem[{OpenAI} et~al.(2023){OpenAI}, Achiam, Adler, Agarwal, Ahmad, Akkaya, et~al.]{achiam2023gpt}
{OpenAI}, Josh Achiam, Steven Adler, Sandhini Agarwal, Lama Ahmad, Ilge Akkaya, et~al.
\newblock {GPT-4} technical report.
\newblock \emph{arXiv preprint arXiv:2303.08774}, 2023.

\bibitem[{OpenAI} et~al.(2024){OpenAI}, Hurst, Lerer, Goucher, Perelman, Ramesh, et~al.]{openai2024gpt4ocard}
{OpenAI}, Aaron Hurst, Adam Lerer, Adam~P. Goucher, Adam Perelman, Aditya Ramesh, et~al.
\newblock {GPT-4o} system card.
\newblock arXiv preprint arXiv:2410.21276, 2024.

\bibitem[Palangi et~al.(2016)Palangi, Deng, Shen, Gao, He, Chen, Song, and Ward]{palangi2016deep}
Hamid Palangi, Li~Deng, Yelong Shen, Jianfeng Gao, Xiaodong He, Jianshu Chen, Xinying Song, and Rabab Ward.
\newblock Deep sentence embedding using long short-term memory networks: analysis and application to information retrieval.
\newblock \emph{IEEE Transactions on Audio, Speech and Language Processing}, 24\penalty0 (4):\penalty0 694--707, 2016.
\newblock \doi{10.1109/TASLP.2016.2520371}.

\bibitem[Pang et~al.(2016)Pang, Lan, Guo, Xu, Wan, and Cheng]{pang2016matchpyramid}
Liang Pang, Yanyan Lan, Jiafeng Guo, Jun Xu, Shengxian Wan, and Xueqi Cheng.
\newblock Text matching as image recognition.
\newblock In \emph{Proceedings of the 30th AAAI Conference on Artificial Intelligence}, AAAI'16, pp.\  2793--2799. AAAI Press, 2016.

\bibitem[Pang et~al.(2020)Pang, Xu, Ai, Lan, Cheng, and Wen]{pang2020setrank}
Liang Pang, Jun Xu, Qingyao Ai, Yanyan Lan, Xueqi Cheng, and Ji-Rong Wen.
\newblock {SetRank}: Learning a permutation-invariant ranking model for information retrieval.
\newblock In \emph{Proceedings of the 43rd International ACM SIGIR Conference on Research and Development in Information Retrieval}, pp.\  499--508, 2020.

\bibitem[Park et~al.(2023)Park, Lee, Seo, Kim, Kang, and Na]{park2023rink}
Eunhwan Park, Sung-Min Lee, Dearyong Seo, Seonhoon Kim, Inho Kang, and Seung-Hoon Na.
\newblock {RINK}: Reader-inherited evidence reranker for table-and-text open domain question answering authors.
\newblock In \emph{Proceedings of the AAAI Conference on Artificial Intelligence}, volume~37, pp.\  13446--13456, 2023.

\bibitem[Patil et~al.(2024{\natexlab{a}})Patil, Mao, Cheng-Jie~Ji, Yan, Suresh, Stoica, and E.~Gonzalez]{patil2025bfcl}
Shishir~G. Patil, Huanzhi Mao, Charlie Cheng-Jie~Ji, Fanjia Yan, Vishnu Suresh, Ion Stoica, and Joseph E.~Gonzalez.
\newblock The berkeley function calling leaderboard ({BFCL}): From tool use to agentic evaluation of large language models.
\newblock In \emph{Advances in Neural Information Processing Systems}, 2024{\natexlab{a}}.

\bibitem[Patil et~al.(2024{\natexlab{b}})Patil, Zhang, Wang, and Gonzalez]{patil2024gorilla}
Shishir~G. Patil, Tianjun Zhang, Xin Wang, and Joseph~E. Gonzalez.
\newblock {Gorilla}: {Large Language Model} connected with massive {APIs}.
\newblock In \emph{The 38th Annual Conference on Neural Information Processing Systems}, 2024{\natexlab{b}}.
\newblock URL \url{https://openreview.net/forum?id=tBRNC6YemY}.

\bibitem[Peng et~al.(2023)Peng, Alcaide, Anthony, Albalak, Arcadinho, Biderman, Cao, Cheng, Chung, Derczynski, Du, Grella, Gv, He, Hou, Kazienko, Kocon, Kong, Koptyra, Lau, Lin, Mantri, Mom, Saito, Song, Tang, Wind, Wo{\'z}niak, Zhang, Zhou, Zhu, and Zhu]{peng-etal-2023-rwkv}
Bo~Peng, Eric Alcaide, Quentin Anthony, Alon Albalak, Samuel Arcadinho, Stella Biderman, Huanqi Cao, Xin Cheng, Michael Chung, Leon Derczynski, Xingjian Du, Matteo Grella, Kranthi Gv, Xuzheng He, Haowen Hou, Przemyslaw Kazienko, Jan Kocon, Jiaming Kong, Bart{\l}omiej Koptyra, Hayden Lau, Jiaju Lin, Krishna Sri~Ipsit Mantri, Ferdinand Mom, Atsushi Saito, Guangyu Song, Xiangru Tang, Johan Wind, Stanis{\l}aw Wo{\'z}niak, Zhenyuan Zhang, Qinghua Zhou, Jian Zhu, and Rui-Jie Zhu.
\newblock {RWKV}: Reinventing {RNN}s for the transformer era.
\newblock In Houda Bouamor, Juan Pino, and Kalika Bali (eds.), \emph{Findings of the Association for Computational Linguistics: EMNLP 2023}, pp.\  14048--14077, Singapore, 2023. Association for Computational Linguistics.
\newblock \doi{10.18653/v1/2023.findings-emnlp.936}.
\newblock URL \url{https://aclanthology.org/2023.findings-emnlp.936/}.

\bibitem[Peng et~al.(2024{\natexlab{a}})Peng, Goldstein, Anthony, Albalak, Alcaide, Biderman, Cheah, Du, Ferdinan, Hou, et~al.]{peng2024eagle}
Bo~Peng, Daniel Goldstein, Quentin Anthony, Alon Albalak, Eric Alcaide, Stella Biderman, Eugene Cheah, Xingjian Du, Teddy Ferdinan, Haowen Hou, et~al.
\newblock Eagle and finch: {RWKV} with matrix-valued states and dynamic recurrence.
\newblock \emph{arXiv preprint arXiv:2404.05892}, 2024{\natexlab{a}}.

\bibitem[Peng et~al.(2024{\natexlab{b}})Peng, Li, Jiang, Wang, Ou, Zeng, Xu, Xu, and Chen]{Peng2024LLMBasedLongTailQueryRewriting}
Wenjun Peng, Guiyang Li, Yue Jiang, Zilong Wang, Dan Ou, Xiaoyi Zeng, Derong Xu, Tong Xu, and Enhong Chen.
\newblock Large language model based long-tail query rewriting in {Taobao} search.
\newblock In \emph{Companion Proceedings of the ACM Web Conference 2024}, WWW '24, pp.\  20--28, New York, NY, USA, 2024{\natexlab{b}}. Association for Computing Machinery.
\newblock \doi{10.1145/3589335.3648298}.

\bibitem[Penha \& Hauff(2021)Penha and Hauff]{penha-hauff-2021-calibration}
Gustavo Penha and Claudia Hauff.
\newblock On the calibration and uncertainty of neural learning to rank models for conversational search.
\newblock In Paola Merlo, Jorg Tiedemann, and Reut Tsarfaty (eds.), \emph{Proceedings of the 16th Conference of the European Chapter of the Association for Computational Linguistics: Main Volume}, pp.\  160--170, Online, April 2021. Association for Computational Linguistics.
\newblock \doi{10.18653/v1/2021.eacl-main.12}.
\newblock URL \url{https://aclanthology.org/2021.eacl-main.12/}.

\bibitem[Pennington et~al.(2014)Pennington, Socher, and Manning]{pennington-etal-2014-glove}
Jeffrey Pennington, Richard Socher, and Christopher Manning.
\newblock {G}lo{V}e: Global vectors for word representation.
\newblock In Alessandro Moschitti, Bo~Pang, and Walter Daelemans (eds.), \emph{Proceedings of the 2014 Conference on Empirical Methods in Natural Language Processing ({EMNLP})}, pp.\  1532--1543, Doha, Qatar, October 2014. Association for Computational Linguistics.
\newblock \doi{10.3115/v1/D14-1162}.
\newblock URL \url{https://aclanthology.org/D14-1162/}.

\bibitem[Petroni et~al.(2021)Petroni, Piktus, Fan, Lewis, Yazdani, De~Cao, Thorne, Jernite, Karpukhin, Maillard, et~al.]{petroni2021kilt}
Fabio Petroni, Aleksandra Piktus, Angela Fan, Patrick Lewis, Majid Yazdani, Nicola De~Cao, James Thorne, Yacine Jernite, Vladimir Karpukhin, Jean Maillard, et~al.
\newblock {KILT}: a benchmark for knowledge intensive language tasks.
\newblock In \emph{Proceedings of the 2021 Conference of the North American Chapter of the Association for Computational Linguistics: Human Language Technologies}, pp.\  2523--2544, 2021.

\bibitem[Pfeiffer et~al.(2020)Pfeiffer, Vuli{\'c}, Gurevych, and Ruder]{pfeiffer-etal-2020-mad}
Jonas Pfeiffer, Ivan Vuli{\'c}, Iryna Gurevych, and Sebastian Ruder.
\newblock {MAD-X}: {A}n {A}dapter-{B}ased {F}ramework for {M}ulti-{T}ask {C}ross-{L}ingual {T}ransfer.
\newblock In Bonnie Webber, Trevor Cohn, Yulan He, and Yang Liu (eds.), \emph{Proceedings of the 2020 Conference on Empirical Methods in Natural Language Processing (EMNLP)}, pp.\  7654--7673, Online, November 2020. Association for Computational Linguistics.
\newblock \doi{10.18653/v1/2020.emnlp-main.617}.
\newblock URL \url{https://aclanthology.org/2020.emnlp-main.617/}.

\bibitem[Pfeiffer et~al.(2021)Pfeiffer, Kamath, R{\"u}ckl{\'e}, Cho, and Gurevych]{pfeiffer-etal-2021-adapterfusion}
Jonas Pfeiffer, Aishwarya Kamath, Andreas R{\"u}ckl{\'e}, Kyunghyun Cho, and Iryna Gurevych.
\newblock {A}dapter{F}usion: Non-destructive task composition for transfer learning.
\newblock In Paola Merlo, Jorg Tiedemann, and Reut Tsarfaty (eds.), \emph{Proceedings of the 16th Conference of the European Chapter of the Association for Computational Linguistics: Main Volume}, pp.\  487--503, Online, April 2021. Association for Computational Linguistics.
\newblock \doi{10.18653/v1/2021.eacl-main.39}.
\newblock URL \url{https://aclanthology.org/2021.eacl-main.39/}.

\bibitem[Ponte \& Croft(1998)Ponte and Croft]{ponte1998language}
Jay~M. Ponte and W.~Bruce Croft.
\newblock A language modeling approach to information retrieval.
\newblock In \emph{Proceedings of the 21st Annual International ACM SIGIR Conference on Research and Development in Information Retrieval}, pp.\  275--281, 1998.

\bibitem[Pool et~al.(2021)Pool, Sawarkar, and Rodge]{pool2021nvidia}
Jeff Pool, Abhishek Sawarkar, and Jay Rodge.
\newblock Accelerating inference with sparsity using the {NVIDIA} {Ampere} architecture and {NVIDIA TensorRT}.
\newblock NVIDIA Technical Blog, July 2021.
\newblock URL \url{https://developer.nvidia.com/blog/accelerating-inference-with-sparsity-using-ampere-and-tensorrt/}.

\bibitem[Portes et~al.(2023)Portes, Trott, Havens, King, Venigalla, Nadeem, Sardana, Khudia, and Frankle]{portes2023mosaicbert}
Jacob Portes, Alexander Trott, Sam Havens, Daniel King, Abhinav Venigalla, Moin Nadeem, Nikhil Sardana, Daya Khudia, and Jonathan Frankle.
\newblock {MosaicBERT}: A bidirectional encoder optimized for fast pretraining.
\newblock \emph{Advances in Neural Information Processing Systems}, 36:\penalty0 3106--3130, 2023.

\bibitem[Portillo-Quintero et~al.(2021)Portillo-Quintero, carlos Ort{\'\i}z-Bayliss, and Terashima-Mar'in]{PortilloQuintero2021ASF}
Jes'us~Andr'es Portillo-Quintero, Jos{\'e} carlos Ort{\'\i}z-Bayliss, and Hugo Terashima-Mar'in.
\newblock A straightforward framework for video retrieval using {CLIP}.
\newblock In \emph{Mexican Conference on Pattern Recognition}, 2021.

\bibitem[Prabhu \& Varma(2014)Prabhu and Varma]{prabhu2014fastxml}
Yashoteja Prabhu and Manik Varma.
\newblock {FastXML}: A fast, accurate and stable tree-classifier for extreme multi-label learning.
\newblock In \emph{Proceedings of the 20th ACM SIGKDD International Conference on Knowledge Discovery and Data Mining}, pp.\  263--272, 2014.

\bibitem[Prabhu et~al.(2018)Prabhu, Kag, Harsola, Agrawal, and Varma]{prabhu2018parabel}
Yashoteja Prabhu, Anil Kag, Shrutendra Harsola, Rahul Agrawal, and Manik Varma.
\newblock Parabel: Partitioned label trees for extreme classification with application to dynamic search advertising.
\newblock In \emph{Proceedings of the 2018 World Wide Web Conference}, pp.\  993--1002, 2018.

\bibitem[Pradeep et~al.(2021)Pradeep, Nogueira, and Lin]{pradeep2021expando}
Ronak Pradeep, Rodrigo Nogueira, and Jimmy Lin.
\newblock The {Expando-Mono-Duo} design pattern for text ranking with pretrained sequence-to-sequence models.
\newblock \emph{arXiv preprint arXiv:2101.05667}, 2021.

\bibitem[Pradeep et~al.(2023{\natexlab{a}})Pradeep, Hui, Gupta, Lelkes, Zhuang, Lin, Metzler, and Tran]{pradeep-etal-2023-generative}
Ronak Pradeep, Kai Hui, Jai Gupta, Adam Lelkes, Honglei Zhuang, Jimmy Lin, Donald Metzler, and Vinh Tran.
\newblock How does generative retrieval scale to millions of passages?
\newblock In Houda Bouamor, Juan Pino, and Kalika Bali (eds.), \emph{Proceedings of the 2023 Conference on Empirical Methods in Natural Language Processing}, pp.\  1305--1321, Singapore, December 2023{\natexlab{a}}. Association for Computational Linguistics.
\newblock \doi{10.18653/v1/2023.emnlp-main.83}.
\newblock URL \url{https://aclanthology.org/2023.emnlp-main.83/}.

\bibitem[Pradeep et~al.(2023{\natexlab{b}})Pradeep, Sharifymoghaddam, and Lin]{pradeep2023rankvicunazeroshotlistwisedocument}
Ronak Pradeep, Sahel Sharifymoghaddam, and Jimmy Lin.
\newblock {RankVicuna}: Zero-shot listwise document reranking with open-source large language models.
\newblock \emph{arXiv preprint arXiv:2309.15088}, 2023{\natexlab{b}}.

\bibitem[Pradeep et~al.(2023{\natexlab{c}})Pradeep, Sharifymoghaddam, and Lin]{pradeep2023rankzephyreffectiverobustzeroshot}
Ronak Pradeep, Sahel Sharifymoghaddam, and Jimmy Lin.
\newblock {RankZephyr}: Effective and robust zero-shot listwise reranking is a breeze!
\newblock \emph{arXiv preprint arXiv: 2312.02724}, 2023{\natexlab{c}}.

\bibitem[Qin et~al.(2022)Qin, Zhu, Chen, Liu, Liu, Tang, Zhang, Yu, and Zhang]{Qin2022RankFlowJointOptimizatio}
Jiarui Qin, Jiachen Zhu, Bo~Chen, Zhirong Liu, Weiwen Liu, Ruiming Tang, Rui Zhang, Yong Yu, and Weinan Zhang.
\newblock {RankFlow}: Joint optimization of multi-stage cascade ranking systems as flows.
\newblock \emph{Proceedings of the 45th International ACM SIGIR Conference on Research and Development in Information Retrieval}, 2022.

\bibitem[Qin \& Liu(2013)Qin and Liu]{qin2013mslr}
Tao Qin and Tie{-}Yan Liu.
\newblock Introducing {LETOR} 4.0 datasets.
\newblock \emph{arXiv preprint arXiv:1306.2597}, 2013.

\bibitem[Qin et~al.(2010)Qin, Liu, and Li]{qin2010general}
Tao Qin, Tie-Yan Liu, and Hang Li.
\newblock A general approximation framework for direct optimization of information retrieval measures.
\newblock \emph{Information retrieval}, 13:\penalty0 375--397, 2010.

\bibitem[Qin et~al.(2024{\natexlab{a}})Qin, Liang, Ye, Zhu, Yan, Lu, Lin, Cong, Tang, Qian, Zhao, Hong, Tian, Xie, Zhou, Gerstein, dahai li, Liu, and Sun]{qin2024toolllm}
Yujia Qin, Shihao Liang, Yining Ye, Kunlun Zhu, Lan Yan, Yaxi Lu, Yankai Lin, Xin Cong, Xiangru Tang, Bill Qian, Sihan Zhao, Lauren Hong, Runchu Tian, Ruobing Xie, Jie Zhou, Mark Gerstein, dahai li, Zhiyuan Liu, and Maosong Sun.
\newblock Tool{LLM}: Facilitating large language models to master 16000+ real-world {API}s.
\newblock In \emph{The 12th International Conference on Learning Representations}, 2024{\natexlab{a}}.
\newblock URL \url{https://openreview.net/forum?id=dHng2O0Jjr}.

\bibitem[Qin et~al.(2021)Qin, Yan, Zhuang, Tay, Pasumarthi, Wang, Bendersky, and Najork]{qin2021neural}
Zhen Qin, Le~Yan, Honglei Zhuang, Yi~Tay, Rama~Kumar Pasumarthi, Xuanhui Wang, Michael Bendersky, and Marc Najork.
\newblock Are neural rankers still outperformed by gradient boosted decision trees?
\newblock In \emph{International Conference on Learning Representations}, 2021.

\bibitem[Qin et~al.(2024{\natexlab{b}})Qin, Jagerman, Hui, Zhuang, Wu, Yan, Shen, Liu, Liu, Metzler, Wang, and Bendersky]{qin-etal-2024-large}
Zhen Qin, Rolf Jagerman, Kai Hui, Honglei Zhuang, Junru Wu, Le~Yan, Jiaming Shen, Tianqi Liu, Jialu Liu, Donald Metzler, Xuanhui Wang, and Michael Bendersky.
\newblock Large language models are effective text rankers with pairwise ranking prompting.
\newblock In Kevin Duh, Helena Gomez, and Steven Bethard (eds.), \emph{Findings of the Association for Computational Linguistics: NAACL 2024}, pp.\  1504--1518, Mexico City, Mexico, June 2024{\natexlab{b}}. Association for Computational Linguistics.
\newblock \doi{10.18653/v1/2024.findings-naacl.97}.
\newblock URL \url{https://aclanthology.org/2024.findings-naacl.97/}.

\bibitem[Qin et~al.(2024{\natexlab{c}})Qin, Yang, and Zhong]{qin2024hierarchically}
Zhen Qin, Songlin Yang, and Yiran Zhong.
\newblock Hierarchically gated recurrent neural network for sequence modeling.
\newblock \emph{Advances in Neural Information Processing Systems}, 36, 2024{\natexlab{c}}.

\bibitem[Qu et~al.(2020)Qu, Ding, Liu, Liu, Ren, Zhao, Dong, Wu, and Wang]{Qu2020RocketQAAO}
Yingqi Qu, Yuchen Ding, Jing Liu, Kai Liu, Ruiyang Ren, Xin Zhao, Daxiang Dong, Hua Wu, and Haifeng Wang.
\newblock {RocketQA}: An optimized training approach to dense passage retrieval for open-domain question answering.
\newblock In \emph{North American Chapter of the Association for Computational Linguistics}, 2020.

\bibitem[{Qwen Team}(2025)]{alibaba2025qwq32b}
{Qwen Team}.
\newblock {QwQ-32B}: Embracing the power of reinforcement learning.
\newblock \url{https://qwenlm.github.io/blog/qwq-32b/}, March 2025.

\bibitem[Radecki(1979)]{radecki1979fuzzy}
Tadeusz Radecki.
\newblock Fuzzy set theoretical approach to document retrieval.
\newblock \emph{Information Processing \& Management}, 15\penalty0 (5):\penalty0 247--259, 1979.

\bibitem[Radford et~al.(2019)Radford, Wu, Child, Luan, Amodei, and Sutskever]{radford2019language}
Alec Radford, Jeffrey Wu, Rewon Child, David Luan, Dario Amodei, and Ilya Sutskever.
\newblock Language models are unsupervised multitask learners.
\newblock \url{https://cdn.openai.com/better-language-models/language-models.pdf}, 2019.

\bibitem[Radford et~al.(2021)Radford, Kim, Hallacy, Ramesh, Goh, Agarwal, Sastry, Askell, Mishkin, Clark, Krueger, and Sutskever]{Radford2021LearningTransferableVisualModels}
Alec Radford, Jong~Wook Kim, Chris Hallacy, Aditya Ramesh, Gabriel Goh, Sandhini Agarwal, Girish Sastry, Amanda Askell, Pamela Mishkin, Jack Clark, Gretchen Krueger, and Ilya Sutskever.
\newblock Learning transferable visual models from natural language supervision.
\newblock In \emph{International Conference on Machine Learning}, 2021.

\bibitem[Raffel et~al.(2020)Raffel, Shazeer, Roberts, Lee, Narang, Matena, Zhou, Li, and Liu]{raffel2020transfer}
Colin Raffel, Noam Shazeer, Adam Roberts, Katherine Lee, Sharan Narang, Michael Matena, Yanqi Zhou, Wei Li, and Peter~J. Liu.
\newblock Exploring the limits of transfer learning with a unified text-to-text transformer.
\newblock \emph{Journal of Machine Learning Research}, 21\penalty0 (140):\penalty0 1--67, 2020.
\newblock URL \url{http://jmlr.org/papers/v21/20-074.html}.

\bibitem[Rahimi et~al.(2021)Rahimi, Kim, Zamani, and Allan]{rahimi2021explaining}
Razieh Rahimi, Youngwoo Kim, Hamed Zamani, and James Allan.
\newblock Explaining documents' relevance to search queries.
\newblock \emph{arXiv preprint arXiv:2111.01314}, 2021.

\bibitem[Rajput et~al.(2023)Rajput, Mehta, Singh, Keshavan, Vu, Heldt, Hong, Tay, Tran, Samost, Kula, Chi, and Sathiamoorthy]{Rajput2023Recommendersystemswithgenerativeretrieval}
Shashank Rajput, Nikhil Mehta, Anima Singh, Raghunandan~H. Keshavan, Trung~Hieu Vu, Lukasz Heldt, Lichan Hong, Yi~Tay, Vinh~Q. Tran, Jonah Samost, Maciej Kula, Ed~H. Chi, and Maheswaran Sathiamoorthy.
\newblock Recommender systems with generative retrieval.
\newblock \emph{arXiv preprint arXiv:2305.05065}, 2023.

\bibitem[Ram et~al.(2023{\natexlab{a}})Ram, Bezalel, Zicher, Belinkov, Berant, and Globerson]{ram-etal-2023-token}
Ori Ram, Liat Bezalel, Adi Zicher, Yonatan Belinkov, Jonathan Berant, and Amir Globerson.
\newblock What are you token about? {Dense} retrieval as distributions over the vocabulary.
\newblock In Anna Rogers, Jordan Boyd-Graber, and Naoaki Okazaki (eds.), \emph{Proceedings of the 61st Annual Meeting of the Association for Computational Linguistics (Volume 1: Long Papers)}, pp.\  2481--2498, Toronto, Canada, July 2023{\natexlab{a}}. Association for Computational Linguistics.
\newblock \doi{10.18653/v1/2023.acl-long.140}.
\newblock URL \url{https://aclanthology.org/2023.acl-long.140/}.

\bibitem[Ram et~al.(2023{\natexlab{b}})Ram, Levine, Dalmedigos, Muhlgay, Shashua, Leyton-Brown, and Shoham]{ram2023incontextretrievalaugmented}
Ori Ram, Yoav Levine, Itay Dalmedigos, Dor Muhlgay, Amnon Shashua, Kevin Leyton-Brown, and Yoav Shoham.
\newblock In-context retrieval-augmented language models.
\newblock \emph{Transactions of the Association for Computational Linguistics}, 11:\penalty0 1316--1331, 2023{\natexlab{b}}.

\bibitem[Ranjan et~al.(2015)Ranjan, Rasiwasia, and Jawahar]{ranjan2015multilabelcrossmodalretrieval}
Viresh Ranjan, Nikhil Rasiwasia, and C.~V. Jawahar.
\newblock Multi-label cross-modal retrieval.
\newblock In \emph{Proceedings of the IEEE International Conference on Computer Vision}, pp.\  4094--4102, 2015.

\bibitem[Rasiwasia et~al.(2010)Rasiwasia, Costa~Pereira, Coviello, Doyle, Lanckriet, Levy, and Vasconcelos]{rasiwasia2010newapproachtocrossmodalmultimediaretrieval}
Nikhil Rasiwasia, Jose Costa~Pereira, Emanuele Coviello, Gabriel Doyle, Gert R.~G. Lanckriet, Roger Levy, and Nuno Vasconcelos.
\newblock A new approach to cross-modal multimedia retrieval.
\newblock In \emph{Proceedings of the 18th ACM International Conference on Multimedia}, pp.\  251--260, 2010.

\bibitem[Ravfogel et~al.(2024)Ravfogel, Pyatkin, Cohen, Manevich, and Goldberg]{ravfogel2024descriptionbased}
Shauli Ravfogel, Valentina Pyatkin, Amir David~Nissan Cohen, Avshalom Manevich, and Yoav Goldberg.
\newblock Description-based text similarity.
\newblock In \emph{1st Conference on Language Modeling}, 2024.
\newblock URL \url{https://openreview.net/forum?id=W8Rv1jVycX}.

\bibitem[Reimers \& Gurevych(2019)Reimers and Gurevych]{reimers-gurevych-2019-sentence}
Nils Reimers and Iryna Gurevych.
\newblock Sentence-{BERT}: Sentence embeddings using {S}iamese {BERT}-networks.
\newblock In Kentaro Inui, Jing Jiang, Vincent Ng, and Xiaojun Wan (eds.), \emph{Proceedings of the 2019 Conference on Empirical Methods in Natural Language Processing and the 9th International Joint Conference on Natural Language Processing (EMNLP-IJCNLP)}, pp.\  3982--3992, Hong Kong, China, November 2019. Association for Computational Linguistics.
\newblock \doi{10.18653/v1/D19-1410}.
\newblock URL \url{https://aclanthology.org/D19-1410/}.

\bibitem[Ren et~al.(2021)Ren, Qu, Liu, Zhao, She, Wu, Wang, and Wen]{ren-etal-2021-rocketqav2}
Ruiyang Ren, Yingqi Qu, Jing Liu, Wayne~Xin Zhao, QiaoQiao She, Hua Wu, Haifeng Wang, and Ji-Rong Wen.
\newblock {R}ocket{QA}v2: A joint training method for dense passage retrieval and passage re-ranking.
\newblock In Marie-Francine Moens, Xuanjing Huang, Lucia Specia, and Scott Wen-tau Yih (eds.), \emph{Proceedings of the 2021 Conference on Empirical Methods in Natural Language Processing}, pp.\  2825--2835, Online and Punta Cana, Dominican Republic, November 2021. Association for Computational Linguistics.
\newblock \doi{10.18653/v1/2021.emnlp-main.224}.
\newblock URL \url{https://aclanthology.org/2021.emnlp-main.224/}.

\bibitem[Ren et~al.(2024)Ren, He, Yin, and de~Rijke]{Ren2024InformationDiscoveryInEcommerce}
Zhaochun Ren, Xiangnan He, Dawei Yin, and M.~de~Rijke.
\newblock Information discovery in e-commerce.
\newblock \emph{Foundations and Trends in Information Retrieval}, 18:\penalty0 417--690, 2024.

\bibitem[Robertson \& Zaragoza(2009)Robertson and Zaragoza]{robertson2009probabilistic}
Stephen Robertson and Hugo Zaragoza.
\newblock The probabilistic relevance framework: {BM25} and beyond.
\newblock \emph{Foundations and Trends in Information Retrieval}, 3\penalty0 (4):\penalty0 333--389, April 2009.
\newblock \doi{10.1561/1500000019}.

\bibitem[Robertson \& Jones(1976)Robertson and Jones]{robertson1976relevance}
Stephen~E. Robertson and K.~Sparck Jones.
\newblock Relevance weighting of search terms.
\newblock \emph{Journal of the American Society for Information science}, 27\penalty0 (3):\penalty0 129--146, 1976.

\bibitem[Robertson et~al.(1995)Robertson, Walker, Jones, Hancock-Beaulieu, and Gatford]{robertson1995okapi}
Stephen~E. Robertson, Steve Walker, Susan Jones, Micheline~M. Hancock-Beaulieu, and Mike Gatford.
\newblock Okapi at {TREC}-3.
\newblock In \emph{Proceedings of the 3rd Text REtrieval Conference (TREC-3), NIST Special Publication}, pp.\  109--126. National Institute of Standards \& Technology, 1995.

\bibitem[Rogers et~al.(2020)Rogers, Kovaleva, and Rumshisky]{rogers-etal-2020-primer}
Anna Rogers, Olga Kovaleva, and Anna Rumshisky.
\newblock A primer in {BERT}ology: What we know about how {BERT} works.
\newblock \emph{Transactions of the Association for Computational Linguistics}, 8:\penalty0 842--866, 2020.
\newblock \doi{10.1162/tacl_a_00349}.
\newblock URL \url{https://aclanthology.org/2020.tacl-1.54/}.

\bibitem[Roy et~al.(2020)Roy, Constant, Al-Rfou, Barua, Phillips, and Yang]{roy-etal-2020-lareqa}
Uma Roy, Noah Constant, Rami Al-Rfou, Aditya Barua, Aaron Phillips, and Yinfei Yang.
\newblock {LAR}e{QA}: Language-agnostic answer retrieval from a multilingual pool.
\newblock In Bonnie Webber, Trevor Cohn, Yulan He, and Yang Liu (eds.), \emph{Proceedings of the 2020 Conference on Empirical Methods in Natural Language Processing (EMNLP)}, pp.\  5919--5930, Online, November 2020. Association for Computational Linguistics.
\newblock \doi{10.18653/v1/2020.emnlp-main.477}.
\newblock URL \url{https://aclanthology.org/2020.emnlp-main.477/}.

\bibitem[Sachan et~al.(2022)Sachan, Lewis, Joshi, Aghajanyan, Yih, Pineau, and Zettlemoyer]{sachan-etal-2022-improving}
Devendra Sachan, Mike Lewis, Mandar Joshi, Armen Aghajanyan, Wen-tau Yih, Joelle Pineau, and Luke Zettlemoyer.
\newblock Improving passage retrieval with zero-shot question generation.
\newblock In Yoav Goldberg, Zornitsa Kozareva, and Yue Zhang (eds.), \emph{Proceedings of the 2022 Conference on Empirical Methods in Natural Language Processing}, pp.\  3781--3797, Abu Dhabi, United Arab Emirates, December 2022. Association for Computational Linguistics.
\newblock \doi{10.18653/v1/2022.emnlp-main.249}.
\newblock URL \url{https://aclanthology.org/2022.emnlp-main.249/}.

\bibitem[Salakhutdinov \& Hinton(2009)Salakhutdinov and Hinton]{salakhutdinov2009semantichashing}
Ruslan Salakhutdinov and Geoffrey Hinton.
\newblock Semantic hashing.
\newblock \emph{International Journal of Approximate Reasoning}, 50\penalty0 (7):\penalty0 969--978, 2009.

\bibitem[Salton(1969)]{salton1969interactiveinformationretrieval}
Gerard Salton.
\newblock Interactive information retrieval.
\newblock Technical report, Cornell University, 1969.

\bibitem[Salton(1970)]{salton1970automaticprocessingofforeignlanguagedocuemnts}
Gerard Salton.
\newblock Automatic processing of foreign language documents.
\newblock \emph{Journal of the American Society for Information Science}, 21\penalty0 (3):\penalty0 187--194, 1970.

\bibitem[Salton \& Buckley(1988)Salton and Buckley]{salton1988termweighting}
Gerard Salton and Christopher Buckley.
\newblock Term-weighting approaches in automatic text retrieval.
\newblock \emph{Information processing \& management}, 24\penalty0 (5):\penalty0 513--523, 1988.

\bibitem[Salton et~al.(1975)Salton, Wong, and Yang]{salton1975vector}
Gerard Salton, Anita Wong, and Chung-Shu Yang.
\newblock A vector space model for automatic indexing.
\newblock \emph{Communications of the ACM}, 18\penalty0 (11):\penalty0 613--620, 1975.

\bibitem[Salton et~al.(1983)Salton, Fox, and Wu]{salton1983extended}
Gerard Salton, Edward~A. Fox, and Harry Wu.
\newblock Extended {Boolean} information retrieval.
\newblock \emph{Communications of the ACM}, 26\penalty0 (11):\penalty0 1022--1036, 1983.

\bibitem[Sanh et~al.(2019)Sanh, Debut, Chaumond, and Wolf]{sanh2019distilbert}
Victor Sanh, Lysandre Debut, Julien Chaumond, and Thomas Wolf.
\newblock {DistilBERT}, a distilled version of {BERT}: smaller, faster, cheaper and lighter.
\newblock \emph{arXiv preprint arXiv:1910.01108}, 2019.

\bibitem[Sanh et~al.(2022)Sanh, Webson, Raffel, Bach, Sutawika, Alyafeai, Chaffin, Stiegler, Raja, Dey, et~al.]{sanh2022multitask}
Victor Sanh, Albert Webson, Colin Raffel, Stephen Bach, Lintang Sutawika, Zaid Alyafeai, Antoine Chaffin, Arnaud Stiegler, Arun Raja, Manan Dey, et~al.
\newblock Multitask prompted training enables zero-shot task generalization.
\newblock In \emph{International Conference on Learning Representations}, 2022.

\bibitem[Santhanam et~al.(2022)Santhanam, Khattab, Saad-Falcon, Potts, and Zaharia]{santhanam-etal-2022-colbertv2}
Keshav Santhanam, Omar Khattab, Jon Saad-Falcon, Christopher Potts, and Matei Zaharia.
\newblock {C}ol{BERT}v2: Effective and efficient retrieval via lightweight late interaction.
\newblock In Marine Carpuat, Marie-Catherine de~Marneffe, and Ivan~Vladimir Meza~Ruiz (eds.), \emph{Proceedings of the 2022 Conference of the North American Chapter of the Association for Computational Linguistics: Human Language Technologies}, pp.\  3715--3734, Seattle, United States, July 2022. Association for Computational Linguistics.
\newblock \doi{10.18653/v1/2022.naacl-main.272}.
\newblock URL \url{https://aclanthology.org/2022.naacl-main.272/}.

\bibitem[Saphra \& Wiegreffe(2024)Saphra and Wiegreffe]{saphra2024mechanistic}
Naomi Saphra and Sarah Wiegreffe.
\newblock Mechanistic?
\newblock \emph{arXiv preprint arXiv:2410.09087}, 2024.

\bibitem[Schick et~al.(2023)Schick, Dwivedi-Yu, Dess{\`\i}, Raileanu, Lomeli, Hambro, Zettlemoyer, Cancedda, and Scialom]{schick2023toolformer}
Timo Schick, Jane Dwivedi-Yu, Roberto Dess{\`\i}, Roberta Raileanu, Maria Lomeli, Eric Hambro, Luke Zettlemoyer, Nicola Cancedda, and Thomas Scialom.
\newblock Toolformer: Language models can teach themselves to use tools.
\newblock \emph{Advances in Neural Information Processing Systems}, 36:\penalty0 68539--68551, 2023.

\bibitem[Sch{\"u}tze et~al.(2008)Sch{\"u}tze, Manning, and Raghavan]{schutze2008introduction}
Hinrich Sch{\"u}tze, Christopher~D. Manning, and Prabhakar Raghavan.
\newblock \emph{Introduction to information retrieval}, volume~39.
\newblock Cambridge University Press Cambridge, 2008.

\bibitem[Severyn \& Moschitti(2015)Severyn and Moschitti]{severyn2015learning}
Aliaksei Severyn and Alessandro Moschitti.
\newblock Learning to rank short text pairs with convolutional deep neural networks.
\newblock In \emph{Proceedings of the 38th International ACM SIGIR Conference on Research and Development in Information Retrieval}, pp.\  373--382, 2015.

\bibitem[Shao et~al.(2024)Shao, He, Asai, Shi, Dettmers, Min, Zettlemoyer, and Koh]{shao2024scaling}
Rulin Shao, Jacqueline He, Akari Asai, Weijia Shi, Tim Dettmers, Sewon Min, Luke Zettlemoyer, and Pang~Wei Koh.
\newblock Scaling retrieval-based language models with a trillion-token datastore.
\newblock In \emph{The 38th Annual Conference on Neural Information Processing Systems}, 2024.

\bibitem[Shao et~al.(2025)Shao, Qiao, Kishore, Muennighoff, Lin, Rus, Low, Min, tau Yih, Koh, and Zettlemoyer]{shao2025reasonir}
Rulin Shao, Rui Qiao, Varsha Kishore, Niklas Muennighoff, Xi~Victoria Lin, Daniela Rus, Bryan Kian~Hsiang Low, Sewon Min, Wen tau Yih, Pang~Wei Koh, and Luke Zettlemoyer.
\newblock Reason{IR}: Training retrievers for reasoning tasks.
\newblock In \emph{2nd Conference on Language Modeling}, 2025.
\newblock URL \url{https://openreview.net/forum?id=kkBCNLMbGj}.

\bibitem[Sharma et~al.(2012)Sharma, Kumar, Daume, and Jacobs]{sharma2012generalizedmultiviewanalysis}
Abhishek Sharma, Abhishek Kumar, Hal Daume, and David~W. Jacobs.
\newblock Generalized multiview analysis: A discriminative latent space.
\newblock In \emph{2012 IEEE Conference on Computer Vision and Pattern Recognition}, pp.\  2160--2167. IEEE, 2012.

\bibitem[Shen et~al.(2025)Shen, Geng, and Yang]{shen2025exploringl0sparsification}
Xinjie Shen, Zhichao Geng, and Yang Yang.
\newblock Exploring l0 sparsification for inference-free sparse retrievers.
\newblock In \emph{Proceedings of the 48th International ACM SIGIR Conference on Research and Development in Information Retrieval}, SIGIR '25, pp.\  2572--2576, New York, NY, USA, 2025. Association for Computing Machinery.
\newblock \doi{10.1145/3726302.3730192}.

\bibitem[Shen et~al.(2014{\natexlab{a}})Shen, He, Gao, Deng, and Mesnil]{shen2014cdssm}
Yelong Shen, Xiaodong He, Jianfeng Gao, Li~Deng, and Gr\'{e}goire Mesnil.
\newblock Learning semantic representations using convolutional neural networks for web search.
\newblock In \emph{Proceedings of the 23rd International Conference on World Wide Web}, WWW '14 Companion, pp.\  373--374, New York, NY, USA, 2014{\natexlab{a}}. Association for Computing Machinery.
\newblock \doi{10.1145/2567948.2577348}.

\bibitem[Shen et~al.(2014{\natexlab{b}})Shen, He, Gao, Deng, and Mesnil]{shen2014clsm}
Yelong Shen, Xiaodong He, Jianfeng Gao, Li~Deng, and Gr\'{e}goire Mesnil.
\newblock A latent semantic model with convolutional-pooling structure for information retrieval.
\newblock In \emph{Proceedings of the 23rd ACM International Conference on Conference on Information and Knowledge Management}, CIKM '14, pp.\  101--110, New York, NY, USA, 2014{\natexlab{b}}. Association for Computing Machinery.
\newblock \doi{10.1145/2661829.2661935}.

\bibitem[Shi et~al.(2020)Shi, Bai, and Lin]{shi-etal-2020-cross}
Peng Shi, He~Bai, and Jimmy Lin.
\newblock Cross-lingual training of neural models for document ranking.
\newblock In Trevor Cohn, Yulan He, and Yang Liu (eds.), \emph{Findings of the Association for Computational Linguistics: EMNLP 2020}, pp.\  2768--2773, Online, November 2020. Association for Computational Linguistics.
\newblock \doi{10.18653/v1/2020.findings-emnlp.249}.
\newblock URL \url{https://aclanthology.org/2020.findings-emnlp.249/}.

\bibitem[Shinn et~al.(2023)Shinn, Cassano, Gopinath, Narasimhan, and Yao]{shinn2023reflexion}
Noah Shinn, Federico Cassano, Ashwin Gopinath, Karthik Narasimhan, and Shunyu Yao.
\newblock Reflexion: Language agents with verbal reinforcement learning.
\newblock \emph{Advances in Neural Information Processing Systems}, 36:\penalty0 8634--8652, 2023.

\bibitem[Singh et~al.(2025{\natexlab{a}})Singh, Ehtesham, Kumar, and Khoei]{singh2025agenticragsurvey}
Aditi Singh, Abul Ehtesham, Saket Kumar, and Tala~Talaei Khoei.
\newblock Agentic retrieval-augmented generation: A survey on agentic {RAG}.
\newblock \emph{arXiv preprint arXiv:2501.09136}, 2025{\natexlab{a}}.

\bibitem[Singh et~al.(2025{\natexlab{b}})Singh, Chang, Haddad, Naik, Hwang, Kinney, Weld, Downey, and Feldman]{Singh2025Ai2scholarqa}
Amanpreet Singh, Joseph~Chee Chang, Dany Haddad, Aakanksha Naik, Jena~D. Hwang, Rodney Kinney, Daniel~S Weld, Doug Downey, and Sergey Feldman.
\newblock {Ai2} scholar {QA}: Organized literature synthesis with attribution.
\newblock In Pushkar Mishra, Smaranda Muresan, and Tao Yu (eds.), \emph{Proceedings of the 63rd Annual Meeting of the Association for Computational Linguistics (Volume 3: System Demonstrations)}, pp.\  513--523, Vienna, Austria, July 2025{\natexlab{b}}. Association for Computational Linguistics.
\newblock \doi{10.18653/v1/2025.acl-demo.49}.
\newblock URL \url{https://aclanthology.org/2025.acl-demo.49/}.

\bibitem[Singh \& Joachims(2018)Singh and Joachims]{singh2018fairness}
Ashudeep Singh and Thorsten Joachims.
\newblock Fairness of exposure in rankings.
\newblock In \emph{Proceedings of the 24th ACM SIGKDD International Conference on Knowledge Discovery \& Data Mining}, pp.\  2219--2228, 2018.

\bibitem[Socher et~al.(2014)Socher, Karpathy, Le, Manning, and Ng]{socher2014groundedcompositionalsemantics}
Richard Socher, Andrej Karpathy, Quoc Le, Christopher~D. Manning, and Andrew~Y. Ng.
\newblock Grounded compositional semantics for finding and describing images with sentences.
\newblock \emph{Transactions of the association for computational linguistics}, 2:\penalty0 207--218, 2014.

\bibitem[Song et~al.(2023)Song, Wu, Washington, Sadler, Chao, and Su]{song2023llm}
Chan~Hee Song, Jiaman Wu, Clayton Washington, Brian~M. Sadler, Wei-Lun Chao, and Yu~Su.
\newblock {LLM}-planner: Few-shot grounded planning for embodied agents with large language models.
\newblock In \emph{Proceedings of the IEEE/CVF International Conference on Computer Vision}, pp.\  2998--3009, 2023.

\bibitem[Song \& Croft(1999)Song and Croft]{song1999general}
Fei Song and W.~Bruce Croft.
\newblock A general language model for information retrieval.
\newblock In \emph{Proceedings of the 8th International Conference on Information and Knowledge Management}, pp.\  316--321, 1999.

\bibitem[Song et~al.(2018)Song, Liu, and Haihong]{Song2018DeepHierarchicalAttentionNetworks}
Meina Song, Qing Liu, and E.~Haihong.
\newblock Deep hierarchical attention networks for text matching in information retrieval.
\newblock \emph{2018 International Conference on Information Systems and Computer Aided Education (ICISCAE)}, pp.\  476--481, 2018.

\bibitem[Sonntag \& Profitlich(2018)Sonntag and Profitlich]{Sonntag2018AnArchitectureOf}
Daniel Sonntag and Hans-J{\"u}rgen Profitlich.
\newblock An architecture of open-source tools to combine textual information extraction, faceted search and information visualisation.
\newblock \emph{Artificial intelligence in Medicine}, 93:\penalty0 13--28, 2018.

\bibitem[Soudani et~al.(2024)Soudani, Kanoulas, and Hasibi]{soudani2024finetuning}
Heydar Soudani, Evangelos Kanoulas, and Faegheh Hasibi.
\newblock Fine tuning vs. retrieval augmented generation for less popular knowledge.
\newblock In \emph{Proceedings of the 2024 Annual International ACM SIGIR Conference on Research and Development in Information Retrieval in the Asia Pacific Region}, SIGIR-AP 2024, pp.\  12--22, New York, NY, USA, 2024. Association for Computing Machinery.
\newblock \doi{10.1145/3673791.3698415}.

\bibitem[Sparck~Jones(1972)]{sparck1972statistical}
Karen Sparck~Jones.
\newblock A statistical interpretation of term specificity and its application in retrieval.
\newblock \emph{Journal of documentation}, 28\penalty0 (1):\penalty0 11--21, 1972.

\bibitem[Su et~al.(2024)Su, Yen, Xia, Shi, Muennighoff, Wang, Liu, Shi, Siegel, Tang, et~al.]{su2024bright}
Hongjin Su, Howard Yen, Mengzhou Xia, Weijia Shi, Niklas Muennighoff, Han-yu Wang, Haisu Liu, Quan Shi, Zachary~S. Siegel, Michael Tang, et~al.
\newblock {BRIGHT}: A realistic and challenging benchmark for reasoning-intensive retrieval.
\newblock \emph{arXiv preprint arXiv:2407.12883}, 2024.

\bibitem[Su et~al.(2025)Su, Zhang, Kou, Ju, Sarkar, Wang, Liu, and Guo]{su2025modernizingfacebookscopedsearch}
Yongye Su, Zeya Zhang, Jane Kou, Cheng Ju, Shubhojeet Sarkar, Yamin Wang, Ji~Liu, and Shengbo Guo.
\newblock Modernizing facebook scoped search: Keyword and embedding hybrid retrieval with {LLM} evaluation.
\newblock arXiv preprint arXiv:2509.13603, 2025.

\bibitem[Sun et~al.(2023)Sun, Yan, Ma, Wang, Ren, Chen, Yin, and Ren]{sun2023chatgpt}
Weiwei Sun, Lingyong Yan, Xinyu Ma, Shuaiqiang Wang, Pengjie Ren, Zhumin Chen, Dawei Yin, and Zhaochun Ren.
\newblock Is {ChatGPT} good at search? investigating large language models as re-ranking agents.
\newblock In \emph{Proceedings of the 2023 Conference on Empirical Methods in Natural Language Processing}, pp.\  14918--14937, 2023.

\bibitem[Sun et~al.(2019)Sun, Wang, Li, Feng, Chen, Zhang, Tian, Zhu, Tian, and Wu]{sun2019ernie}
Yu~Sun, Shuohuan Wang, Yukun Li, Shikun Feng, Xuyi Chen, Han Zhang, Xin Tian, Danxiang Zhu, Hao Tian, and Hua Wu.
\newblock {ERNIE}: Enhanced representation through knowledge integration.
\newblock \emph{arXiv preprint arXiv:1904.09223}, 2019.

\bibitem[Tai et~al.(2015)Tai, Socher, and Manning]{tai-etal-2015-improvedsemanticrepresentations}
Kai~Sheng Tai, Richard Socher, and Christopher~D. Manning.
\newblock Improved semantic representations from tree-structured long short-term memory networks.
\newblock In Chengqing Zong and Michael Strube (eds.), \emph{Proceedings of the 53rd Annual Meeting of the Association for Computational Linguistics and the 7th International Joint Conference on Natural Language Processing (Volume 1: Long Papers)}, pp.\  1556--1566, Beijing, China, July 2015. Association for Computational Linguistics.
\newblock \doi{10.3115/v1/P15-1150}.
\newblock URL \url{https://aclanthology.org/P15-1150/}.

\bibitem[Tang et~al.(2023)Tang, Zhang, Guo, and de~Rijke]{Tang2023RecentAdvancesInGenerative}
Yubao Tang, Ruqing Zhang, Jiafeng Guo, and Maarten de~Rijke.
\newblock Recent advances in generative information retrieval.
\newblock \emph{Companion Proceedings of the ACM Web Conference 2024}, 2023.

\bibitem[Tay et~al.(2022)Tay, Tran, Dehghani, Ni, Bahri, Mehta, Qin, Hui, Zhao, Gupta, Schuster, Cohen, and Metzler]{tay2022transformer}
Yi~Tay, Vinh~Q. Tran, Mostafa Dehghani, Jianmo Ni, Dara Bahri, Harsh Mehta, Zhen Qin, Kai Hui, Zhe Zhao, Jai Gupta, Tal Schuster, William~W. Cohen, and Donald Metzler.
\newblock Transformer memory as a differentiable search index.
\newblock In Alice~H. Oh, Alekh Agarwal, Danielle Belgrave, and Kyunghyun Cho (eds.), \emph{Advances in Neural Information Processing Systems}, 2022.
\newblock URL \url{https://openreview.net/forum?id=Vu-B0clPfq}.

\bibitem[Taylor et~al.(2008)Taylor, Guiver, Robertson, and Minka]{taylor2008softrank}
Michael Taylor, John Guiver, Stephen Robertson, and Tom Minka.
\newblock Softrank: optimizing non-smooth rank metrics.
\newblock In \emph{Proceedings of the 2008 International Conference on Web Search and Data Mining}, pp.\  77--86, 2008.

\bibitem[Taylor(1953)]{Taylor1953ClozePA}
Wilson~L. Taylor.
\newblock {``Cloze Procedure''}: A new tool for measuring readability.
\newblock \emph{Journalism \& Mass Communication Quarterly}, 30:\penalty0 415 -- 433, 1953.

\bibitem[Tenney et~al.(2019)Tenney, Das, and Pavlick]{tenney-etal-2019-bert}
Ian Tenney, Dipanjan Das, and Ellie Pavlick.
\newblock {BERT} rediscovers the classical {NLP} pipeline.
\newblock In Anna Korhonen, David Traum, and Llu{\'\i}s M{\`a}rquez (eds.), \emph{Proceedings of the 57th Annual Meeting of the Association for Computational Linguistics}, pp.\  4593--4601, Florence, Italy, July 2019. Association for Computational Linguistics.
\newblock \doi{10.18653/v1/P19-1452}.
\newblock URL \url{https://aclanthology.org/P19-1452/}.

\bibitem[Thakur et~al.(2021)Thakur, Reimers, R{\"u}ckl{\'e}, Srivastava, and Gurevych]{thakur2021beir}
Nandan Thakur, Nils Reimers, Andreas R{\"u}ckl{\'e}, Abhishek Srivastava, and Iryna Gurevych.
\newblock {BEIR}: A heterogeneous benchmark for zero-shot evaluation of information retrieval models.
\newblock In \emph{35th Conference on Neural Information Processing Systems Datasets and Benchmarks Track (Round 2)}, 2021.

\bibitem[Tian et~al.(2023)Tian, Yang, Zhang, Dong, and Su]{tian2023evil}
Yu~Tian, Xiao Yang, Jingyuan Zhang, Yinpeng Dong, and Hang Su.
\newblock Evil geniuses: Delving into the safety of {LLM}-based agents.
\newblock \emph{arXiv preprint arXiv:2311.11855}, 2023.

\bibitem[Touvron et~al.(2023)Touvron, Martin, Stone, Albert, Almahairi, Babaei, Bashlykov, Batra, Bhargava, Bhosale, et~al.]{touvron2023llama}
Hugo Touvron, Louis Martin, Kevin Stone, Peter Albert, Amjad Almahairi, Yasmine Babaei, Nikolay Bashlykov, Soumya Batra, Prajjwal Bhargava, Shruti Bhosale, et~al.
\newblock {LLaMA} 2: Open foundation and fine-tuned chat models.
\newblock \emph{arXiv preprint arXiv:2307.09288}, 2023.

\bibitem[Trivedi et~al.(2023)Trivedi, Balasubramanian, Khot, and Sabharwal]{trivedi-etal-2023-interleavingretrievalwithchainofthoughtreasoning}
Harsh Trivedi, Niranjan Balasubramanian, Tushar Khot, and Ashish Sabharwal.
\newblock Interleaving retrieval with chain-of-thought reasoning for knowledge-intensive multi-step questions.
\newblock In Anna Rogers, Jordan Boyd-Graber, and Naoaki Okazaki (eds.), \emph{Proceedings of the 61st Annual Meeting of the Association for Computational Linguistics (Volume 1: Long Papers)}, pp.\  10014--10037, Toronto, Canada, July 2023. Association for Computational Linguistics.
\newblock \doi{10.18653/v1/2023.acl-long.557}.
\newblock URL \url{https://aclanthology.org/2023.acl-long.557/}.

\bibitem[Turtle \& Flood(1995)Turtle and Flood]{turtle1995queryevaluation}
Howard Turtle and James Flood.
\newblock Query evaluation: strategies and optimizations.
\newblock \emph{Information Processing \& Management}, 31\penalty0 (6):\penalty0 831--850, 1995.

\bibitem[{\"U}st{\"u}n et~al.(2020){\"U}st{\"u}n, Bisazza, Bouma, and van Noord]{ustun-etal-2020-udapter}
Ahmet {\"U}st{\"u}n, Arianna Bisazza, Gosse Bouma, and Gertjan van Noord.
\newblock {UD}apter: Language adaptation for truly {U}niversal {D}ependency parsing.
\newblock In Bonnie Webber, Trevor Cohn, Yulan He, and Yang Liu (eds.), \emph{Proceedings of the 2020 Conference on Empirical Methods in Natural Language Processing (EMNLP)}, pp.\  2302--2315, Online, November 2020. Association for Computational Linguistics.
\newblock \doi{10.18653/v1/2020.emnlp-main.180}.
\newblock URL \url{https://aclanthology.org/2020.emnlp-main.180/}.

\bibitem[Van~Rijsbergen(1979)]{van1979information}
C.~Van~Rijsbergen.
\newblock Information retrieval: theory and practice.
\newblock In \emph{Proceedings of the Joint IBM/University of Newcastle Upon Tyne Seminar on Data Base Systems}, volume~79, pp.\  1--14, 1979.

\bibitem[Vaswani et~al.(2017)Vaswani, Shazeer, Parmar, Uszkoreit, Jones, Gomez, Kaiser, and Polosukhin]{vaswani2017attention}
Ashish Vaswani, Noam Shazeer, Niki Parmar, Jakob Uszkoreit, Llion Jones, Aidan~N. Gomez, \L~ukasz Kaiser, and Illia Polosukhin.
\newblock Attention is all you need.
\newblock In I.~Guyon, U.~Von Luxburg, S.~Bengio, H.~Wallach, R.~Fergus, S.~Vishwanathan, and R.~Garnett (eds.), \emph{Advances in Neural Information Processing Systems}, volume~30. Curran Associates, Inc., 2017.
\newblock URL \url{https://proceedings.neurips.cc/paper_files/paper/2017/file/3f5ee243547dee91fbd053c1c4a845aa-Paper.pdf}.

\bibitem[Vera et~al.(2025)Vera, Dua, Zhang, Salz, Mullins, Panyam, et~al.]{vera2025embeddinggemmapowerfullightweighttext}
Henrique~Schechter Vera, Sahil Dua, Biao Zhang, Daniel Salz, Ryan Mullins, Sindhu~Raghuram Panyam, et~al.
\newblock {EmbeddingGemma}: Powerful and lightweight text representations.
\newblock \emph{arXiv preprint arXiv:2509.20354}, 2025.

\bibitem[Vert et~al.(2004)Vert, Tsuda, and Sch{\"o}lkopf]{vert2004primer}
Jean-Philippe Vert, Koji Tsuda, and Bernhard Sch{\"o}lkopf.
\newblock A primer on kernel methods.
\newblock In \emph{Kernel Methods in Computational Biology}, pp.\  35--70. MIT press, 2004.

\bibitem[Wan et~al.(2025)Wan, Wang, Stengel-Eskin, Cho, and Bansal]{Wan2025CLaMRCL}
David Wan, Han Wang, Elias Stengel-Eskin, Jaemin Cho, and Mohit Bansal.
\newblock {CLaMR}: Contextualized late-interaction for multimodal content retrieval.
\newblock \emph{arXiv preprint arXiv:2506.06144}, 2025.

\bibitem[Wan et~al.(2016)Wan, Lan, Guo, Xu, Pang, and Cheng]{wan2016deep}
Shengxian Wan, Yanyan Lan, Jiafeng Guo, Jun Xu, Liang Pang, and Xueqi Cheng.
\newblock A deep architecture for semantic matching with multiple positional sentence representations.
\newblock In \emph{Proceedings of the 30th AAAI Conference on Artificial Intelligence}, AAAI'16, pp.\  2835--2841. AAAI Press, 2016.

\bibitem[Wang \& Komatsuzaki(2021)Wang and Komatsuzaki]{wang2021gpt-j}
Ben Wang and Aran Komatsuzaki.
\newblock {GPT-J-6B: A 6 Billion Parameter Autoregressive Language Model}.
\newblock \url{https://github.com/kingoflolz/mesh-transformer-jax}, May 2021.

\bibitem[Wang et~al.(2017)Wang, Yang, Xu, Hanjalic, and Shen]{wang2017adversarialcrossmodalretrieval}
Bokun Wang, Yang Yang, Xing Xu, Alan Hanjalic, and Heng~Tao Shen.
\newblock Adversarial cross-modal retrieval.
\newblock In \emph{Proceedings of the 25th ACM International Conference on Multimedia}, pp.\  154--162, 2017.

\bibitem[Wang \& Na(2023)Wang and Na]{Wang2023RethinkingEcommerceSearch}
Haixun Wang and Taesik Na.
\newblock Rethinking e-commerce search.
\newblock \emph{ACM SIGIR Forum}, 57:\penalty0 1--19, 2023.

\bibitem[Wang et~al.(2025{\natexlab{a}})Wang, Yan, Wang, Tian, Mishra, Xu, Gandhi, Xu, and Cheong]{wang2025reinforcementlearningselfimprovingagent}
Jiongxiao Wang, Qiaojing Yan, Yawei Wang, Yijun Tian, Soumya~Smruti Mishra, Zhichao Xu, Megha Gandhi, Panpan Xu, and Lin~Lee Cheong.
\newblock Reinforcement learning for self-improving agent with skill library, 2025{\natexlab{a}}.
\newblock URL \url{https://arxiv.org/abs/2512.17102}.

\bibitem[Wang et~al.(2016)Wang, Yin, Wang, Wu, and Wang]{wang2016comprehensivesurveycrossmodalretrieval}
Kaiye Wang, Qiyue Yin, Wei Wang, Shu Wu, and Liang Wang.
\newblock A comprehensive survey on cross-modal retrieval.
\newblock \emph{arXiv preprint arXiv:1607.06215}, 2016.

\bibitem[Wang et~al.(2024{\natexlab{a}})Wang, Ma, Feng, Zhang, Yang, Zhang, Chen, Tang, Chen, Lin, et~al.]{wang2024llmagentsurvey}
Lei Wang, Chen Ma, Xueyang Feng, Zeyu Zhang, Hao Yang, Jingsen Zhang, Zhiyuan Chen, Jiakai Tang, Xu~Chen, Yankai Lin, et~al.
\newblock A survey on large language model based autonomous agents.
\newblock \emph{Frontiers of Computer Science}, 18\penalty0 (6):\penalty0 186345, 2024{\natexlab{a}}.

\bibitem[Wang et~al.(2022{\natexlab{a}})Wang, Yang, Huang, Jiao, Yang, Jiang, Majumder, and Wei]{wang2022e5textembeddings}
Liang Wang, Nan Yang, Xiaolong Huang, Binxing Jiao, Linjun Yang, Daxin Jiang, Rangan Majumder, and Furu Wei.
\newblock Text embeddings by weakly-supervised contrastive pre-training.
\newblock \emph{arXiv preprint arXiv:2212.03533}, 2022{\natexlab{a}}.

\bibitem[Wang et~al.(2023{\natexlab{a}})Wang, Yang, Huang, Yang, Majumder, and Wei]{wang2023improving}
Liang Wang, Nan Yang, Xiaolong Huang, Linjun Yang, Rangan Majumder, and Furu Wei.
\newblock Improving text embeddings with large language models.
\newblock \emph{arXiv preprint arXiv:2401.00368}, 2023{\natexlab{a}}.

\bibitem[Wang et~al.(2024{\natexlab{b}})Wang, Li, Zhu, Li, Zhang, and Shen]{wang2024crossmodalretrievalsystematicreview}
Tianshi Wang, Fengling Li, Lei Zhu, Jingjing Li, Zheng Zhang, and Heng~Tao Shen.
\newblock Cross-modal retrieval: A systematic review of methods and future directions.
\newblock \emph{arXiv preprint arXiv:2308.14263}, 2024{\natexlab{b}}.

\bibitem[Wang et~al.(2018)Wang, Golbandi, Bendersky, Metzler, and Najork]{wang2018position}
Xuanhui Wang, Nadav Golbandi, Michael Bendersky, Donald Metzler, and Marc Najork.
\newblock Position bias estimation for unbiased learning to rank in personal search.
\newblock In \emph{Proceedings of the 11th ACM International Conference on Web Search and Data Mining}, pp.\  610--618, 2018.

\bibitem[Wang et~al.(2023{\natexlab{b}})Wang, Jin, Huang, Zhang, Liu, Zhao, Chen, Zhang, Yang, Wen, Chordia, Chen, and Huang]{Wang2023TowardsTheBetterRankingConsistency}
Xuewei Wang, Qiang Jin, Shengyu Huang, Min Zhang, Xi~Liu, Zhengli Zhao, Yukun Chen, Zhengyu Zhang, Jiyan Yang, Ellie Wen, Sagar Chordia, Wenlin Chen, and Qin Huang.
\newblock Towards the better ranking consistency: A multi-task learning framework for early stage ads ranking.
\newblock \emph{arXiv preprint arXiv:2307.11096}, 2023{\natexlab{b}}.

\bibitem[Wang et~al.(2022{\natexlab{b}})Wang, Lyu, and Anand]{wang2022bertrankersbrittle}
Yumeng Wang, Lijun Lyu, and Avishek Anand.
\newblock {BERT} rankers are brittle: A study using adversarial document perturbations.
\newblock In \emph{Proceedings of the 2022 ACM SIGIR International Conference on Theory of Information Retrieval}, ICTIR '22, pp.\  115--120, New York, NY, USA, 2022{\natexlab{b}}. Association for Computing Machinery.
\newblock \doi{10.1145/3539813.3545122}.

\bibitem[Wang et~al.(2024{\natexlab{c}})Wang, Xu, Srikumar, and Ai]{wang2024depth}
Zhenduo Wang, Zhichao Xu, Vivek Srikumar, and Qingyao Ai.
\newblock An in-depth investigation of user response simulation for conversational search.
\newblock In \emph{Proceedings of the ACM on Web Conference 2024}, pp.\  1407--1418, 2024{\natexlab{c}}.

\bibitem[Wang et~al.(2025{\natexlab{b}})Wang, Asai, Yu, Xu, Xie, Neubig, and Fried]{wang2024coderag}
Zora~Zhiruo Wang, Akari Asai, Xinyan~Velocity Yu, Frank~F. Xu, Yiqing Xie, Graham Neubig, and Daniel Fried.
\newblock {C}ode{RAG}-bench: Can retrieval augment code generation?
\newblock In Luis Chiruzzo, Alan Ritter, and Lu~Wang (eds.), \emph{Findings of the Association for Computational Linguistics: NAACL 2025}, pp.\  3199--3214, Albuquerque, New Mexico, April 2025{\natexlab{b}}. Association for Computational Linguistics.
\newblock \doi{10.18653/v1/2025.findings-naacl.176}.
\newblock URL \url{https://aclanthology.org/2025.findings-naacl.176/}.

\bibitem[Warner et~al.(2025)Warner, Chaffin, Clavi{\'e}, Weller, Hallstr{\"o}m, Taghadouini, Gallagher, Biswas, Ladhak, Aarsen, Adams, Howard, and Poli]{warner2024modernbert}
Benjamin Warner, Antoine Chaffin, Benjamin Clavi{\'e}, Orion Weller, Oskar Hallstr{\"o}m, Said Taghadouini, Alexis Gallagher, Raja Biswas, Faisal Ladhak, Tom Aarsen, Griffin~Thomas Adams, Jeremy Howard, and Iacopo Poli.
\newblock Smarter, better, faster, longer: A modern bidirectional encoder for fast, memory efficient, and long context finetuning and inference.
\newblock In Wanxiang Che, Joyce Nabende, Ekaterina Shutova, and Mohammad~Taher Pilehvar (eds.), \emph{Proceedings of the 63rd Annual Meeting of the Association for Computational Linguistics (Volume 1: Long Papers)}, pp.\  2526--2547, Vienna, Austria, July 2025. Association for Computational Linguistics.
\newblock \doi{10.18653/v1/2025.acl-long.127}.
\newblock URL \url{https://aclanthology.org/2025.acl-long.127/}.

\bibitem[Wei et~al.(2025)Wei, Chen, Chen, Hu, Zhang, Fu, Ritter, and Chen]{wei2025uniir}
Cong Wei, Yang Chen, Haonan Chen, Hexiang Hu, Ge~Zhang, Jie Fu, Alan Ritter, and Wenhu Chen.
\newblock {UniIR}: Training and benchmarking universal multimodal information retrievers.
\newblock In \emph{Computer Vision -- ECCV 2024}, volume 15145 of \emph{Lecture Notes in Computer Science}, pp.\  387--404. Springer, Cham, 2025.
\newblock \doi{10.1007/978-3-031-73021-4_23}.

\bibitem[Wei et~al.(2022)Wei, Wang, Schuurmans, Bosma, Xia, Chi, Le, Zhou, et~al.]{wei2022chain}
Jason Wei, Xuezhi Wang, Dale Schuurmans, Maarten Bosma, Fei Xia, Ed~Chi, Quoc~V. Le, Denny Zhou, et~al.
\newblock Chain-of-thought prompting elicits reasoning in large language models.
\newblock \emph{Advances in neural information processing systems}, 35:\penalty0 24824--24837, 2022.

\bibitem[Weller et~al.(2024)Weller, Van~Durme, Lawrie, Paranjape, Zhang, and Hessel]{weller2024promptriever}
Orion Weller, Benjamin Van~Durme, Dawn Lawrie, Ashwin Paranjape, Yuhao Zhang, and Jack Hessel.
\newblock Promptriever: Instruction-trained retrievers can be prompted like language models.
\newblock \emph{arXiv preprint arXiv:2409.11136}, 2024.

\bibitem[Weller et~al.(2025{\natexlab{a}})Weller, Chang, MacAvaney, Lo, Cohan, Van~Durme, Lawrie, and Soldaini]{weller2024followir}
Orion Weller, Benjamin Chang, Sean MacAvaney, Kyle Lo, Arman Cohan, Benjamin Van~Durme, Dawn Lawrie, and Luca Soldaini.
\newblock {F}ollow{IR}: Evaluating and teaching information retrieval models to follow instructions.
\newblock In Luis Chiruzzo, Alan Ritter, and Lu~Wang (eds.), \emph{Proceedings of the 2025 Conference of the Nations of the Americas Chapter of the Association for Computational Linguistics: Human Language Technologies (Volume 1: Long Papers)}, pp.\  11926--11942, Albuquerque, New Mexico, April 2025{\natexlab{a}}. Association for Computational Linguistics.
\newblock \doi{10.18653/v1/2025.naacl-long.597}.
\newblock URL \url{https://aclanthology.org/2025.naacl-long.597/}.

\bibitem[Weller et~al.(2025{\natexlab{b}})Weller, Ricci, Yang, Yates, Lawrie, and Durme]{weller2025rank1testtimecomputereranking}
Orion Weller, Kathryn Ricci, Eugene Yang, Andrew Yates, Dawn Lawrie, and Benjamin~Van Durme.
\newblock {Rank1}: Test-time compute for reranking in information retrieval.
\newblock In \emph{2nd Conference on Language Modeling}, 2025{\natexlab{b}}.
\newblock URL \url{https://openreview.net/forum?id=Pg0PAvbhGv}.

\bibitem[White \& Roth(2009)White and Roth]{white2009exploratory}
Ryen~W. White and Resa~A. Roth.
\newblock \emph{Exploratory search: Beyond the query-response paradigm}.
\newblock Morgan \& Claypool Publishers, 2009.

\bibitem[Wieting et~al.(2016)Wieting, Bansal, Gimpel, and Livescu]{wieting2016universalparaphrasticsentenceembeddings}
John Wieting, Mohit Bansal, Kevin Gimpel, and Karen Livescu.
\newblock Towards universal paraphrastic sentence embeddings.
\newblock \emph{arXiv preprint arXiv:1511.08198}, 2016.

\bibitem[Wilson(2000)]{wilson2000human}
Thomas~D. Wilson.
\newblock Human information behavior.
\newblock \emph{Informing science}, 3:\penalty0 49, 2000.

\bibitem[Wong \& Yao(1989)Wong and Yao]{wong1989probability}
S.~K.~Michael Wong and Y.~Y. Yao.
\newblock A probability distribution model for information retrieval.
\newblock \emph{Information Processing \& Management}, 25\penalty0 (1):\penalty0 39--53, 1989.

\bibitem[Wrzalik \& Krechel(2021)Wrzalik and Krechel]{Wrzalik2020CoRTComplementaryRankings}
Marco Wrzalik and Dirk Krechel.
\newblock {CoRT}: Complementary rankings from transformers.
\newblock In \emph{Proceedings of the 2021 Conference of the North American Chapter of the Association for Computational Linguistics: Human Language Technologies}, pp.\  4194--4204, 2021.

\bibitem[Wu et~al.(2025{\natexlab{a}})Wu, Li, Fang, Yin, Zhang, Tao, Zhang, Xi, Jiang, Xie, Huang, and Zhou]{wu2025webdancer}
Jialong Wu, Baixuan Li, Runnan Fang, Wenbiao Yin, Liwen Zhang, Zhengwei Tao, Dingchu Zhang, Zekun Xi, Yong Jiang, Pengjun Xie, Fei Huang, and Jingren Zhou.
\newblock {WebDancer}: Towards autonomous information seeking agency.
\newblock \emph{arXiv preprint arXiv:2505.22648}, 2025{\natexlab{a}}.

\bibitem[Wu et~al.(2025{\natexlab{b}})Wu, Yin, Jiang, Wang, Xi, Fang, Zhang, He, Zhou, Xie, and Huang]{wu-etal-2025-webwalker}
Jialong Wu, Wenbiao Yin, Yong Jiang, Zhenglin Wang, Zekun Xi, Runnan Fang, Linhai Zhang, Yulan He, Deyu Zhou, Pengjun Xie, and Fei Huang.
\newblock {W}eb{W}alker: Benchmarking {LLM}s in web traversal.
\newblock In Wanxiang Che, Joyce Nabende, Ekaterina Shutova, and Mohammad~Taher Pilehvar (eds.), \emph{Proceedings of the 63rd Annual Meeting of the Association for Computational Linguistics (Volume 1: Long Papers)}, pp.\  10290--10305, Vienna, Austria, July 2025{\natexlab{b}}. Association for Computational Linguistics.
\newblock URL \url{https://aclanthology.org/2025.acl-long.508/}.

\bibitem[Wu et~al.(2010)Wu, Burges, Svore, and Gao]{wu2010adapting}
Qiang Wu, Christopher~J.C. Burges, Krysta~M. Svore, and Jianfeng Gao.
\newblock Adapting boosting for information retrieval measures.
\newblock \emph{Information Retrieval}, 13:\penalty0 254--270, 2010.

\bibitem[Wu et~al.(2023{\natexlab{a}})Wu, Bansal, Zhang, Wu, Zhang, Zhu, Li, Jiang, Zhang, and Wang]{wu2023autogen}
Qingyun Wu, Gagan Bansal, Jieyu Zhang, Yiran Wu, Shaokun Zhang, Erkang Zhu, Beibin Li, Li~Jiang, Xiaoyun Zhang, and Chi Wang.
\newblock {AutoGen}: Enabling next-gen {LLM} applications via multi-agent conversation.
\newblock \emph{arXiv preprint arXiv:2308.08155}, 2023{\natexlab{a}}.

\bibitem[Wu et~al.(2023{\natexlab{b}})Wu, Ma, Lin, Lin, Wang, and Hu]{wu2023contextualmaskedautoencoder}
Xing Wu, Guangyuan Ma, Meng Lin, Zijia Lin, Zhongyuan Wang, and Songlin Hu.
\newblock Contextual masked auto-encoder for dense passage retrieval.
\newblock \emph{Proceedings of the AAAI Conference on Artificial Intelligence}, 37\penalty0 (4):\penalty0 4738--4746, 2023{\natexlab{b}}.

\bibitem[Xi et~al.(2025)Xi, Lin, Xiao, Zhou, Shan, Gao, Zhu, Liu, Yu, and Zhang]{xi2025surveyofllmbaseddeepsearchagents}
Yunjia Xi, Jianghao Lin, Yongzhao Xiao, Zheli Zhou, Rong Shan, Te~Gao, Jiachen Zhu, Weiwen Liu, Yong Yu, and Weinan Zhang.
\newblock A survey of {LLM}-based deep search agents: Paradigm, optimization, evaluation, and challenges.
\newblock \emph{arXiv preprint arXiv:2508.05668}, 2025.

\bibitem[Xia et~al.(2008)Xia, Liu, Wang, Zhang, and Li]{xia2008listwise}
Fen Xia, Tie-Yan Liu, Jue Wang, Wensheng Zhang, and Hang Li.
\newblock Listwise approach to learning to rank: theory and algorithm.
\newblock In \emph{Proceedings of the 25th International Conference on Machine Learning}, pp.\  1192--1199, 2008.

\bibitem[Xiang et~al.(2024{\natexlab{a}})Xiang, Wu, Zhong, Wagner, Chen, and Mittal]{xiang2024certifiablyrobustragagainstretrievalcorruption}
Chong Xiang, Tong Wu, Zexuan Zhong, David Wagner, Danqi Chen, and Prateek Mittal.
\newblock Certifiably robust {RAG} against retrieval corruption.
\newblock \emph{arXiv preprint arXiv:2405.15556}, 2024{\natexlab{a}}.

\bibitem[Xiang et~al.(2024{\natexlab{b}})Xiang, Zheng, Li, Hong, Li, Xie, Zhang, Xiong, Xie, Yang, et~al.]{xiang2024guardagent}
Zhen Xiang, Linzhi Zheng, Yanjie Li, Junyuan Hong, Qinbin Li, Han Xie, Jiawei Zhang, Zidi Xiong, Chulin Xie, Carl Yang, et~al.
\newblock {GuardAgent}: Safeguard {LLM} agents by a guard agent via knowledge-enabled reasoning.
\newblock \emph{arXiv preprint arXiv:2406.09187}, 2024{\natexlab{b}}.

\bibitem[Xiao et~al.(2022)Xiao, Liu, Shao, and Cao]{xiao-etal-2022-retromae}
Shitao Xiao, Zheng Liu, Yingxia Shao, and Zhao Cao.
\newblock {R}etro{MAE}: Pre-training retrieval-oriented language models via masked auto-encoder.
\newblock In Yoav Goldberg, Zornitsa Kozareva, and Yue Zhang (eds.), \emph{Proceedings of the 2022 Conference on Empirical Methods in Natural Language Processing}, pp.\  538--548, Abu Dhabi, United Arab Emirates, December 2022. Association for Computational Linguistics.
\newblock \doi{10.18653/v1/2022.emnlp-main.35}.
\newblock URL \url{https://aclanthology.org/2022.emnlp-main.35/}.

\bibitem[Xie et~al.(2024{\natexlab{a}})Xie, Mao, Bai, Zhang, Wang, Lin, Gu, Chen, Yang, and Shou]{Xie2024Showo}
Jinheng Xie, Weijia Mao, Zechen Bai, David~Junhao Zhang, Weihao Wang, Kevin~Qinghong Lin, Yuchao Gu, Zhijie Chen, Zhenheng Yang, and Mike~Zheng Shou.
\newblock Show-o: One single transformer to unify multimodal understanding and generation.
\newblock \emph{arXiv preprint arXiv:2408.12528}, 2024{\natexlab{a}}.

\bibitem[Xie et~al.(2024{\natexlab{b}})Xie, Min, Zhang, Xu, Bajaj, Salakhutdinov, Johnson-Roberson, and Bisk]{xie2024embodiedrag}
Quanting Xie, So~Yeon Min, Tianyi Zhang, Kedi Xu, Aarav Bajaj, Russ Salakhutdinov, Matthew Johnson-Roberson, and Yonatan Bisk.
\newblock Embodied-{RAG}: General non-parametric embodied memory for retrieval and generation.
\newblock In \emph{Language Gamification - NeurIPS 2024 Workshop}, 2024{\natexlab{b}}.
\newblock URL \url{https://openreview.net/forum?id=U8p8zpK3jL}.

\bibitem[Xiong et~al.(2017)Xiong, Dai, Callan, Liu, and Power]{xiong2017knrm}
Chenyan Xiong, Zhuyun Dai, Jamie Callan, Zhiyuan Liu, and Russell Power.
\newblock End-to-end neural ad-hoc ranking with kernel pooling.
\newblock In \emph{Proceedings of the 40th International ACM SIGIR Conference on Research and Development in Information Retrieval}, SIGIR '17, pp.\  55--64, New York, NY, USA, 2017. Association for Computing Machinery.
\newblock \doi{10.1145/3077136.3080809}.

\bibitem[Xiong et~al.(2020)Xiong, Xiong, Li, Tang, Liu, Bennett, Ahmed, and Overwijk]{Xiong2020ApproximateNN}
Lee Xiong, Chenyan Xiong, Ye~Li, Kwok-Fung Tang, Jialin Liu, Paul~N. Bennett, Junaid Ahmed, and Arnold Overwijk.
\newblock Approximate nearest neighbor negative contrastive learning for dense text retrieval.
\newblock \emph{arXiv preprint arXiv:2007.00808}, 2020.

\bibitem[Xu et~al.(2018)Xu, He, and Li]{xu2018deep}
Jun Xu, Xiangnan He, and Hang Li.
\newblock Deep learning for matching in search and recommendation.
\newblock In \emph{The 41st International ACM SIGIR Conference on Research \& Development in Information Retrieval}, pp.\  1365--1368, 2018.

\bibitem[Xu et~al.(2025{\natexlab{a}})Xu, Zhou, Babakhin, Moreira, Ak, Osmulski, Liu, Oldridge, and Schifferer]{Xu2025OmniEmbedNemotronAU}
Mengyao Xu, Wenfei Zhou, Yauhen Babakhin, Gabriel De Souza~Pereira Moreira, Ronay Ak, Radek Osmulski, Bo~Liu, Even Oldridge, and Benedikt Schifferer.
\newblock {Omni-Embed-Nemotron}: A unified multimodal retrieval model for text, image, audio, and video.
\newblock \emph{arXiv preprint arXiv:2510.03458}, 2025{\natexlab{a}}.

\bibitem[Xu et~al.(2024{\natexlab{a}})Xu, Hou, Pang, Deng, Xu, Shen, and Cheng]{xu2024invisiblerelevancebiastextimageretrievalmdoel}
Shicheng Xu, Danyang Hou, Liang Pang, Jingcheng Deng, Jun Xu, Huawei Shen, and Xueqi Cheng.
\newblock Invisible relevance bias: Text-image retrieval models prefer {AI}-generated images.
\newblock In \emph{Proceedings of the 47th International ACM SIGIR Conference on Research and Development in Information Retrieval}, SIGIR '24, pp.\  208--217, New York, NY, USA, 2024{\natexlab{a}}. Association for Computing Machinery.
\newblock \doi{10.1145/3626772.3657750}.

\bibitem[Xu(2023)]{xu2026contextawaredecodingreduceshallucination}
Zhichao Xu.
\newblock Context-aware decoding reduces hallucination in query-focused summarization, 2023.
\newblock URL \url{https://arxiv.org/abs/2312.14335}.

\bibitem[Xu(2024)]{xu2024rankmamba}
Zhichao Xu.
\newblock {RankMamba}: Benchmarking {Mamba}'s document ranking performance in the era of transformers.
\newblock \emph{arXiv preprint arXiv:2403.18276}, 2024.

\bibitem[Xu \& Cohen(2023)Xu and Cohen]{xu2023lightweightconstrainedgeneration}
Zhichao Xu and Daniel Cohen.
\newblock A lightweight constrained generation alternative for query-focused summarization.
\newblock In \emph{Proceedings of the 46th International ACM SIGIR Conference on Research and Development in Information Retrieval}, SIGIR '23, pp.\  1745–1749, New York, NY, USA, 2023. Association for Computing Machinery.
\newblock ISBN 9781450394086.
\newblock \doi{10.1145/3539618.3591936}.
\newblock URL \url{https://doi.org/10.1145/3539618.3591936}.

\bibitem[Xu \& Jiang(2024)Xu and Jiang]{xu-jiang-2024-multi}
Zhichao Xu and Jiepu Jiang.
\newblock Multi-dimensional evaluation of empathetic dialogue responses.
\newblock In Yaser Al-Onaizan, Mohit Bansal, and Yun-Nung Chen (eds.), \emph{Findings of the Association for Computational Linguistics: EMNLP 2024}, pp.\  2066--2087, Miami, Florida, USA, November 2024. Association for Computational Linguistics.
\newblock \doi{10.18653/v1/2024.findings-emnlp.113}.
\newblock URL \url{https://aclanthology.org/2024.findings-emnlp.113/}.

\bibitem[Xu et~al.(2022{\natexlab{a}})Xu, Han, Yang, Tran, and Ai]{xu2022learningrankrationalesexplainable}
Zhichao Xu, Yi~Han, Tao Yang, Anh Tran, and Qingyao Ai.
\newblock Learning to rank rationales for explainable recommendation, 2022{\natexlab{a}}.
\newblock URL \url{https://arxiv.org/abs/2206.05368}.

\bibitem[Xu et~al.(2022{\natexlab{b}})Xu, Tran, Yang, and Ai]{xu2022reinforcement}
Zhichao Xu, Anh Tran, Tao Yang, and Qingyao Ai.
\newblock Reinforcement learning to rank with coarse-grained labels.
\newblock \emph{arXiv preprint arXiv:2208.07563}, 2022{\natexlab{b}}.

\bibitem[Xu et~al.(2023)Xu, Zeng, Tan, Fu, Zhang, and Ai]{xu2023reusable}
Zhichao Xu, Hansi Zeng, Juntao Tan, Zuohui Fu, Yongfeng Zhang, and Qingyao Ai.
\newblock A reusable model-agnostic framework for faithfully explainable recommendation and system scrutability.
\newblock \emph{ACM Transactions on Information Systems}, 42\penalty0 (1):\penalty0 1--29, 2023.

\bibitem[Xu et~al.(2024{\natexlab{b}})Xu, Gupta, Li, Bentham, and Srikumar]{xu-etal-2024-beyond-perplexity}
Zhichao Xu, Ashim Gupta, Tao Li, Oliver Bentham, and Vivek Srikumar.
\newblock Beyond perplexity: Multi-dimensional safety evaluation of {LLM} compression.
\newblock In Yaser Al-Onaizan, Mohit Bansal, and Yun-Nung Chen (eds.), \emph{Findings of the Association for Computational Linguistics: EMNLP 2024}, pp.\  15359--15396, Miami, Florida, USA, November 2024{\natexlab{b}}. Association for Computational Linguistics.
\newblock \doi{10.18653/v1/2024.findings-emnlp.901}.
\newblock URL \url{https://aclanthology.org/2024.findings-emnlp.901/}.

\bibitem[Xu et~al.(2024{\natexlab{c}})Xu, Lamba, Ai, Tetreault, and Jaimes]{xu2024cfe2}
Zhichao Xu, Hemank Lamba, Qingyao Ai, Joel Tetreault, and Alex Jaimes.
\newblock {CFE2}: Counterfactual editing for search result explanation.
\newblock In \emph{Proceedings of the 2024 ACM SIGIR International Conference on Theory of Information Retrieval}, pp.\  145--155, 2024{\natexlab{c}}.

\bibitem[Xu et~al.(2025{\natexlab{b}})Xu, Feng, Tian, Ding, and Cheong]{xu2025csplade}
Zhichao Xu, Aosong Feng, Yijun Tian, Haibo Ding, and Lin~Lee Cheong.
\newblock {CSPLADE}: Learned sparse retrieval with causal language models.
\newblock In Kentaro Inui, Sakriani Sakti, Haofen Wang, Derek~F. Wong, Pushpak Bhattacharyya, Biplab Banerjee, Asif Ekbal, Tanmoy Chakraborty, and Dhirendra~Pratap Singh (eds.), \emph{Proceedings of the 14th International Joint Conference on Natural Language Processing and the 4th Conference of the Asia-Pacific Chapter of the Association for Computational Linguistics}, pp.\  99--114, Mumbai, India, December 2025{\natexlab{b}}. The Asian Federation of Natural Language Processing and The Association for Computational Linguistics.
\newblock URL \url{https://aclanthology.org/2025.ijcnlp-long.7/}.

\bibitem[Xu et~al.(2025{\natexlab{c}})Xu, Huang, Zhuang, and Srikumar]{xu2025distillationvscontrastive}
Zhichao Xu, Zhiqi Huang, Shengyao Zhuang, and Vivek Srikumar.
\newblock Distillation versus contrastive learning: How to train your rerankers.
\newblock In Kentaro Inui, Sakriani Sakti, Haofen Wang, Derek~F. Wong, Pushpak Bhattacharyya, Biplab Banerjee, Asif Ekbal, Tanmoy Chakraborty, and Dhirendra~Pratap Singh (eds.), \emph{Proceedings of the 14th International Joint Conference on Natural Language Processing and the 4th Conference of the Asia-Pacific Chapter of the Association for Computational Linguistics}, pp.\  564--578, Mumbai, India, December 2025{\natexlab{c}}. The Asian Federation of Natural Language Processing and The Association for Computational Linguistics.
\newblock URL \url{https://aclanthology.org/2025.findings-ijcnlp.33/}.

\bibitem[Xu et~al.(2025{\natexlab{d}})Xu, Wang, Wang, Ye, Du, Ma, and Tian]{xu2025reconreasoningcondensationefficient}
Zhichao Xu, Minheng Wang, Yawei Wang, Wenqian Ye, Yuntao Du, Yunpu Ma, and Yijun Tian.
\newblock Recon: Reasoning with condensation for efficient retrieval-augmented generation, 2025{\natexlab{d}}.
\newblock URL \url{https://arxiv.org/abs/2510.10448}.

\bibitem[Xu et~al.(2025{\natexlab{e}})Xu, Yan, Gupta, and Srikumar]{xu-etal-2025-state}
Zhichao Xu, Jinghua Yan, Ashim Gupta, and Vivek Srikumar.
\newblock State space models are strong text rerankers.
\newblock In Vaibhav Adlakha, Alexandra Chronopoulou, Xiang~Lorraine Li, Bodhisattwa~Prasad Majumder, Freda Shi, and Giorgos Vernikos (eds.), \emph{Proceedings of the 10th Workshop on Representation Learning for NLP (RepL4NLP-2025)}, pp.\  152--169, Albuquerque, NM, May 2025{\natexlab{e}}. Association for Computational Linguistics.
\newblock URL \url{https://aclanthology.org/2025.repl4nlp-1.12/}.

\bibitem[Xu et~al.(2025{\natexlab{f}})Xu, Zhuang, Ma, Chen, Tian, Mo, Cao, and Srikumar]{xu2025rethinkingonpolicyoptimizationquery}
Zhichao Xu, Shengyao Zhuang, Xueguang Ma, Bingsen Chen, Yijun Tian, Fengran Mo, Jie Cao, and Vivek Srikumar.
\newblock Rethinking on-policy optimization for query augmentation.
\newblock \emph{arXiv preprint arXiv:2510.17139}, 2025{\natexlab{f}}.

\bibitem[Xu et~al.(2026)Xu, Zhuang, Zhang, Ma, Tian, Mehta, Lin, and Srikumar]{xu2026laconicdenseleveleffectivenessscalable}
Zhichao Xu, Shengyao Zhuang, Crystina Zhang, Xueguang Ma, Yijun Tian, Maitrey Mehta, Jimmy Lin, and Vivek Srikumar.
\newblock {LACONIC}: Dense-level effectiveness for scalable sparse retrieval via a two-phase training curriculum.
\newblock \emph{arXiv preprint arXiv:2601.01684}, 2026.

\bibitem[Yan et~al.(2024)Yan, Qin, Zhuang, Jagerman, Wang, Bendersky, and Oosterhuis]{yan-etal-2024-consolidating}
Le~Yan, Zhen Qin, Honglei Zhuang, Rolf Jagerman, Xuanhui Wang, Michael Bendersky, and Harrie Oosterhuis.
\newblock Consolidating ranking and relevance predictions of large language models through post-processing.
\newblock In Yaser Al-Onaizan, Mohit Bansal, and Yun-Nung Chen (eds.), \emph{Proceedings of the 2024 Conference on Empirical Methods in Natural Language Processing}, pp.\  410--423, Miami, Florida, USA, November 2024. Association for Computational Linguistics.
\newblock \doi{10.18653/v1/2024.emnlp-main.25}.
\newblock URL \url{https://aclanthology.org/2024.emnlp-main.25/}.

\bibitem[Yang et~al.(2025{\natexlab{a}})Yang, Li, Yang, Zhang, Hui, Zheng, Yu, Gao, Huang, Lv, et~al.]{yang2025qwen3technicalreport}
An~Yang, Anfeng Li, Baosong Yang, Beichen Zhang, Binyuan Hui, Bo~Zheng, Bowen Yu, Chang Gao, Chengen Huang, Chenxu Lv, et~al.
\newblock {Qwen3} technical report.
\newblock \emph{arXiv preprint arXiv:2505.09388}, 2025{\natexlab{a}}.

\bibitem[Yang et~al.(2024)Yang, Lawrie, and Mayfield]{Yang2024Distillationformultilingual}
Eugene Yang, Dawn~J. Lawrie, and James Mayfield.
\newblock Distillation for multilingual information retrieval.
\newblock \emph{Proceedings of the 47th International ACM SIGIR Conference on Research and Development in Information Retrieval}, 2024.

\bibitem[Yang et~al.(2025{\natexlab{b}})Yang, Yates, Ricci, Weller, Chari, Van~Durme, and Lawrie]{yang2025rankk}
Eugene Yang, Andrew Yates, Kathryn Ricci, Orion Weller, Vivek Chari, Benjamin Van~Durme, and Dawn Lawrie.
\newblock {Rank-K}: Test-time reasoning for listwise reranking.
\newblock \emph{arXiv preprint arXiv:2505.14432}, 2025{\natexlab{b}}.

\bibitem[Yang et~al.(2016{\natexlab{a}})Yang, Ai, Guo, and Croft]{yang2016anmm}
Liu Yang, Qingyao Ai, Jiafeng Guo, and W.~Bruce Croft.
\newblock {aNMM}: Ranking short answer texts with attention-based neural matching model.
\newblock In \emph{Proceedings of the 25th ACM International on Conference on Information and Knowledge Management}, CIKM '16, pp.\  287--296, New York, NY, USA, 2016{\natexlab{a}}. Association for Computing Machinery.
\newblock \doi{10.1145/2983323.2983818}.

\bibitem[Yang et~al.(2025{\natexlab{c}})Yang, Kautz, and Hatamizadeh]{yang2024gated}
Songlin Yang, Jan Kautz, and Ali Hatamizadeh.
\newblock Gated {Delta} networks: Improving {Mamba2} with {Delta} rule.
\newblock In \emph{The 13th International Conference on Learning Representations}, 2025{\natexlab{c}}.
\newblock URL \url{https://openreview.net/forum?id=r8H7xhYPwz}.

\bibitem[Yang et~al.(2023{\natexlab{a}})Yang, Xu, and Ai]{yang2023vertical}
Tao Yang, Zhichao Xu, and Qingyao Ai.
\newblock Vertical allocation-based fair exposure amortizing in ranking.
\newblock In \emph{Proceedings of the Annual International ACM SIGIR Conference on Research and Development in Information Retrieval in the Asia Pacific Region}, pp.\  234--244, 2023{\natexlab{a}}.

\bibitem[Yang et~al.(2023{\natexlab{b}})Yang, Xu, Wang, and Ai]{yang2023fara}
Tao Yang, Zhichao Xu, Zhenduo Wang, and Qingyao Ai.
\newblock {FARA}: Future-aware ranking algorithm for fairness optimization.
\newblock In \emph{Proceedings of the 32nd ACM International Conference on Information and Knowledge Management}, pp.\  2906--2916, 2023{\natexlab{b}}.

\bibitem[Yang et~al.(2023{\natexlab{c}})Yang, Xu, Wang, Tran, and Ai]{yang2023marginal}
Tao Yang, Zhichao Xu, Zhenduo Wang, Anh Tran, and Qingyao Ai.
\newblock Marginal-certainty-aware fair ranking algorithm.
\newblock In \emph{Proceedings of the 16th ACM International Conference on Web Search and Data Mining}, pp.\  24--32, 2023{\natexlab{c}}.

\bibitem[Yang et~al.(2025{\natexlab{d}})Yang, Yin, Engle, Zhuang, and Leng]{Yang2025MTMD}
Xiao Yang, Peifeng Yin, Abe Engle, Jinfeng Zhuang, and Ling Leng.
\newblock {MTMD}: A multi-task multi-domain framework for unified ad lightweight ranking at pinterest.
\newblock \emph{arXiv preprint arXiv:2510.09857}, 2025{\natexlab{d}}.

\bibitem[Yang et~al.(2016{\natexlab{b}})Yang, Yang, Dyer, He, Smola, and Hovy]{yang-etal-2016-hierarchical-attention-networks}
Zichao Yang, Diyi Yang, Chris Dyer, Xiaodong He, Alex Smola, and Eduard Hovy.
\newblock Hierarchical attention networks for document classification.
\newblock In Kevin Knight, Ani Nenkova, and Owen Rambow (eds.), \emph{Proceedings of the 2016 Conference of the North {A}merican Chapter of the Association for Computational Linguistics: Human Language Technologies}, pp.\  1480--1489, San Diego, California, June 2016{\natexlab{b}}. Association for Computational Linguistics.
\newblock \doi{10.18653/v1/N16-1174}.
\newblock URL \url{https://aclanthology.org/N16-1174/}.

\bibitem[Yao et~al.(2023)Yao, Zhao, Yu, Du, Shafran, Narasimhan, and Cao]{yao2023react}
Shunyu Yao, Jeffrey Zhao, Dian Yu, Nan Du, Izhak Shafran, Karthik~R. Narasimhan, and Yuan Cao.
\newblock {ReAct}: Synergizing reasoning and acting in language models.
\newblock In \emph{The 11th International Conference on Learning Representations}, 2023.
\newblock URL \url{https://openreview.net/forum?id=WE_vluYUL-X}.

\bibitem[Yao et~al.(2024)Yao, Parashar, Zhou, Jang, Ouyang, Yang, and Yu]{yao2024mcqg}
Zonghai Yao, Aditya Parashar, Huixue Zhou, Won~Seok Jang, Feiyun Ouyang, Zhichao Yang, and Hong Yu.
\newblock {MCQG-SRefine}: Multiple choice question generation and evaluation with iterative self-critique, correction, and comparison feedback.
\newblock \emph{arXiv preprint arXiv:2410.13191}, 2024.

\bibitem[Yilmaz et~al.(2019)Yilmaz, Yang, Zhang, and Lin]{Yilmaz2019CrossDomainMO}
Zeynep~Akkalyoncu Yilmaz, Wei Yang, Haotian Zhang, and Jimmy Lin.
\newblock Cross-domain modeling of sentence-level evidence for document retrieval.
\newblock In \emph{Conference on Empirical Methods in Natural Language Processing}, 2019.

\bibitem[Yin et~al.(2016)Yin, Sch{\"u}tze, Xiang, and Zhou]{yin2016abcnn}
Wenpeng Yin, Hinrich Sch{\"u}tze, Bing Xiang, and Bowen Zhou.
\newblock {ABCNN}: Attention-based convolutional neural network for modeling sentence pairs.
\newblock \emph{Transactions of the Association for computational linguistics}, 4:\penalty0 259--272, 2016.

\bibitem[Yoon et~al.(2024)Yoon, Choi, Kim, Yun, Kim, and Hwang]{yoon-etal-2024-listt5}
Soyoung Yoon, Eunbi Choi, Jiyeon Kim, Hyeongu Yun, Yireun Kim, and Seung-won Hwang.
\newblock {L}ist{T}5: Listwise reranking with fusion-in-decoder improves zero-shot retrieval.
\newblock In Lun-Wei Ku, Andre Martins, and Vivek Srikumar (eds.), \emph{Proceedings of the 62nd Annual Meeting of the Association for Computational Linguistics (Volume 1: Long Papers)}, pp.\  2287--2308, Bangkok, Thailand, 2024. Association for Computational Linguistics.
\newblock \doi{10.18653/v1/2024.acl-long.125}.
\newblock URL \url{https://aclanthology.org/2024.acl-long.125/}.

\bibitem[Yoon et~al.(2025)Yoon, Kim, Cho, and won Hwang]{yoon2025acurankuncertaintyawareadaptivecomputation}
Soyoung Yoon, Gyuwan Kim, Gyu-Hwung Cho, and Seung won Hwang.
\newblock {AcuRank}: Uncertainty-aware adaptive computation for listwise reranking.
\newblock \emph{arXiv preprint arXiv:2505.18512}, 2025.

\bibitem[Yoran et~al.(2024)Yoran, Wolfson, Ram, and Berant]{yoran2024makingretrievalaugmentedlanguagemodelsrobusttoirrelevantcontext}
Ori Yoran, Tomer Wolfson, Ori Ram, and Jonathan Berant.
\newblock Making retrieval-augmented language models robust to irrelevant context.
\newblock In \emph{ICLR 2024 Workshop on Large Language Model (LLM) Agents}, 2024.

\bibitem[You et~al.(2019)You, Zhang, Wang, Dai, Mamitsuka, and Zhu]{you2019attentionxml}
Ronghui You, Zihan Zhang, Ziye Wang, Suyang Dai, Hiroshi Mamitsuka, and Shanfeng Zhu.
\newblock {AttentionXML}: Label tree-based attention-aware deep model for high-performance extreme multi-label text classification.
\newblock \emph{Advances in neural information processing systems}, 32, 2019.

\bibitem[Yu et~al.(2025)Yu, Chen, Feng, Chen, Dai, Yu, Zhang, Ma, Liu, Wang, and Zhou]{yu2025memagentreshapinglongcontextllm}
Hongli Yu, Tinghong Chen, Jiangtao Feng, Jiangjie Chen, Weinan Dai, Qiying Yu, Ya-Qin Zhang, Wei-Ying Ma, Jingjing Liu, Mingxuan Wang, and Hao Zhou.
\newblock Memagent: Reshaping long-context llm with multi-conv rl-based memory agent, 2025.
\newblock URL \url{https://arxiv.org/abs/2507.02259}.

\bibitem[Yu et~al.(2022{\natexlab{a}})Yu, Zhong, Zhang, Chang, and Dhillon]{yu2022pecos}
Hsiang-Fu Yu, Kai Zhong, Jiong Zhang, Wei-Cheng Chang, and Inderjit~S. Dhillon.
\newblock {PECOS}: Prediction for enormous and correlated output spaces.
\newblock \emph{Journal of Machine Learning Research}, 23\penalty0 (98):\penalty0 1--32, 2022{\natexlab{a}}.

\bibitem[Yu \& Allan(2020)Yu and Allan]{yu2020study}
Puxuan Yu and James Allan.
\newblock A study of neural matching models for cross-lingual {IR}.
\newblock In \emph{Proceedings of the 43rd International ACM SIGIR Conference on Research and Development in Information Retrieval}, pp.\  1637--1640, 2020.

\bibitem[Yu et~al.(2022{\natexlab{b}})Yu, Rahimi, and Allan]{yu2022towards}
Puxuan Yu, Razieh Rahimi, and James Allan.
\newblock Towards explainable search results: a listwise explanation generator.
\newblock In \emph{Proceedings of the 45th International ACM SIGIR Conference on Research and Development in Information Retrieval}, pp.\  669--680, 2022{\natexlab{b}}.

\bibitem[Yu et~al.(2024{\natexlab{a}})Yu, Mallia, and Petri]{yu2024improved}
Puxuan Yu, Antonio Mallia, and Matthias Petri.
\newblock Improved learned sparse retrieval with corpus-specific vocabularies.
\newblock In \emph{European Conference on Information Retrieval}, pp.\  181--194. Springer, 2024{\natexlab{a}}.

\bibitem[Yu et~al.(2024{\natexlab{b}})Yu, Merrick, Nuti, and Campos]{yu2024arctic}
Puxuan Yu, Luke Merrick, Gaurav Nuti, and Daniel Campos.
\newblock {Arctic-Embed 2.0}: Multilingual retrieval without compromise.
\newblock \emph{arXiv preprint arXiv:2412.04506}, 2024{\natexlab{b}}.

\bibitem[Yu et~al.(2024{\natexlab{c}})Yu, Zhang, Pan, Cao, Ma, Li, Wang, and Yu]{yu-etal-2024-chainofnote}
Wenhao Yu, Hongming Zhang, Xiaoman Pan, Peixin Cao, Kaixin Ma, Jian Li, Hongwei Wang, and Dong Yu.
\newblock {Chain-of-Note}: Enhancing robustness in retrieval-augmented language models.
\newblock In Yaser Al-Onaizan, Mohit Bansal, and Yun-Nung Chen (eds.), \emph{Proceedings of the 2024 Conference on Empirical Methods in Natural Language Processing}, pp.\  14672--14685, Miami, Florida, USA, November 2024{\natexlab{c}}. Association for Computational Linguistics.
\newblock \doi{10.18653/v1/2024.emnlp-main.813}.
\newblock URL \url{https://aclanthology.org/2024.emnlp-main.813/}.

\bibitem[Yu et~al.(2022{\natexlab{c}})Yu, Xiong, Sun, Zhang, and Overwijk]{yu-etal-2022-coco}
Yue Yu, Chenyan Xiong, Si~Sun, Chao Zhang, and Arnold Overwijk.
\newblock {COCO}-{DR}: Combating the distribution shift in zero-shot dense retrieval with contrastive and distributionally robust learning.
\newblock In Yoav Goldberg, Zornitsa Kozareva, and Yue Zhang (eds.), \emph{Proceedings of the 2022 Conference on Empirical Methods in Natural Language Processing}, pp.\  1462--1479, Abu Dhabi, United Arab Emirates, December 2022{\natexlab{c}}. Association for Computational Linguistics.
\newblock \doi{10.18653/v1/2022.emnlp-main.95}.
\newblock URL \url{https://aclanthology.org/2022.emnlp-main.95/}.

\bibitem[Zamani \& Croft(2018)Zamani and Croft]{Zamani2018JointModelingAndOptimization}
Hamed Zamani and W.~Bruce Croft.
\newblock Joint modeling and optimization of search and recommendation.
\newblock In \emph{Biennial Conference on Design of Experimental Search \& Information Retrieval Systems}, 2018.

\bibitem[Zamani et~al.(2018)Zamani, Dehghani, Croft, Learned-Miller, and Kamps]{zamani2018neural}
Hamed Zamani, Mostafa Dehghani, W.~Bruce Croft, Erik Learned-Miller, and Jaap Kamps.
\newblock From neural re-ranking to neural ranking: Learning a sparse representation for inverted indexing.
\newblock In \emph{Proceedings of the 27th ACM International Conference on Information and Knowledge Management}, pp.\  497--506, 2018.

\bibitem[Zbib et~al.(2019)Zbib, Zhao, Karakos, Hartmann, DeYoung, Huang, Jiang, Rivkin, Zhang, Schwartz, et~al.]{zbib2019neuralnetworklexicaltranslation}
Rabih Zbib, Lingjun Zhao, Damianos Karakos, William Hartmann, Jay DeYoung, Zhongqiang Huang, Zhuolin Jiang, Noah Rivkin, Le~Zhang, Richard Schwartz, et~al.
\newblock Neural-network lexical translation for cross-lingual ir from text and speech.
\newblock In \emph{Proceedings of the 42nd International ACM SIGIR Conference on Research and Development in Information Retrieval}, pp.\  645--654, 2019.

\bibitem[Zeng et~al.(2024)Zeng, Luo, Jin, Sarwar, Wei, and Zamani]{zeng2024scalable}
Hansi Zeng, Chen Luo, Bowen Jin, Sheikh~Muhammad Sarwar, Tianxin Wei, and Hamed Zamani.
\newblock Scalable and effective generative information retrieval.
\newblock In \emph{Proceedings of the ACM Web Conference 2024}, WWW '24, pp.\  1441--1452, New York, NY, USA, 2024. Association for Computing Machinery.
\newblock \doi{10.1145/3589334.3645477}.

\bibitem[Zhai \& Lafferty(2004)Zhai and Lafferty]{zhai2004study}
ChengXiang Zhai and John Lafferty.
\newblock A study of smoothing methods for language models applied to information retrieval.
\newblock \emph{ACM Transactions on Information Systems (TOIS)}, 22\penalty0 (2):\penalty0 179--214, 2004.

\bibitem[Zhai et~al.(2008)]{zhai2008statistical}
ChengXiang Zhai et~al.
\newblock Statistical language models for information retrieval a critical review.
\newblock \emph{Foundations and Trends in Information Retrieval}, 2\penalty0 (3):\penalty0 137--213, 2008.

\bibitem[Zhai et~al.(2024)Zhai, Liao, Liu, Wang, Li, Cao, Gao, Gong, Gu, He, Lu, and Shi]{Zhai2024ActionsSpeaksLoaderThanWords}
Jiaqi Zhai, Lucy Liao, Xing Liu, Yueming Wang, Rui Li, Xuan Cao, Leon Gao, Zhaojie Gong, Fangda Gu, Michael He, Yin-Hua Lu, and Yu~Shi.
\newblock Actions speak louder than words: Trillion-parameter sequential transducers for generative recommendations.
\newblock \emph{arXiv preprint arXiv:2402.17152}, 2024.

\bibitem[Zhai et~al.(2023)Zhai, Mustafa, Kolesnikov, and Beyer]{zhai2023sigmoidlossforlanguageimagepretraining}
Xiaohua Zhai, Basil Mustafa, Alexander Kolesnikov, and Lucas Beyer.
\newblock Sigmoid loss for language image pre-training.
\newblock In \emph{Proceedings of the IEEE/CVF International Conference on Computer Vision}, pp.\  11975--11986, 2023.

\bibitem[Zhan et~al.(2020)Zhan, Mao, Liu, Zhang, and Ma]{zhan2020repbert}
Jingtao Zhan, Jiaxin Mao, Yiqun Liu, Min Zhang, and Shaoping Ma.
\newblock {RepBERT}: Contextualized text embeddings for 1st-stage retrieval.
\newblock \emph{arXiv preprint arXiv:2006.15498}, 2020.

\bibitem[Zhan et~al.(2021)Zhan, Mao, Liu, Guo, Zhang, and Ma]{zhan2021optimizingdenseretrievalmodel}
Jingtao Zhan, Jiaxin Mao, Yiqun Liu, Jiafeng Guo, Min Zhang, and Shaoping Ma.
\newblock Optimizing dense retrieval model training with hard negatives.
\newblock In \emph{Proceedings of the 44th International ACM SIGIR Conference on Research and Development in Information Retrieval}, pp.\  1503--1512, 2021.

\bibitem[Zhang et~al.(2023{\natexlab{a}})Zhang, Chen, Zhang, Keung, Liu, Zan, Mao, Lou, and Chen]{zhang-etal-2023-repocoder}
Fengji Zhang, Bei Chen, Yue Zhang, Jacky Keung, Jin Liu, Daoguang Zan, Yi~Mao, Jian-Guang Lou, and Weizhu Chen.
\newblock {R}epo{C}oder: Repository-level code completion through iterative retrieval and generation.
\newblock In Houda Bouamor, Juan Pino, and Kalika Bali (eds.), \emph{Proceedings of the 2023 Conference on Empirical Methods in Natural Language Processing}, pp.\  2471--2484, Singapore, December 2023{\natexlab{a}}. Association for Computational Linguistics.
\newblock \doi{10.18653/v1/2023.emnlp-main.151}.
\newblock URL \url{https://aclanthology.org/2023.emnlp-main.151/}.

\bibitem[Zhang et~al.(2024)Zhang, Chen, Mei, Liu, and Mao]{zhang2024mambaretriever}
Hanqi Zhang, Chong Chen, Lang Mei, Qi~Liu, and Jiaxin Mao.
\newblock {Mamba} retriever: Utilizing {Mamba} for effective and efficient dense retrieval.
\newblock In \emph{Proceedings of the 33rd ACM International Conference on Information and Knowledge Management}, pp.\  4268--4272, 2024.

\bibitem[Zhang et~al.(2023{\natexlab{b}})Zhang, Yang, Qi, Qian, and Xu]{Zhang2023DebiasedVideoTextRetrieval}
Huaiwen Zhang, Yang Yang, Fan Qi, Shengsheng Qian, and Changsheng Xu.
\newblock Debiased video-text retrieval via soft positive sample calibration.
\newblock \emph{IEEE Transactions on Circuits and Systems for Video Technology}, 33:\penalty0 5257--5270, 2023{\natexlab{b}}.

\bibitem[Zhang et~al.(2021{\natexlab{a}})Zhang, Gong, and Choi]{zhang-etal-2021-knowing}
Shujian Zhang, Chengyue Gong, and Eunsol Choi.
\newblock Knowing more about questions can help: Improving calibration in question answering.
\newblock In Chengqing Zong, Fei Xia, Wenjie Li, and Roberto Navigli (eds.), \emph{Findings of the Association for Computational Linguistics: ACL-IJCNLP 2021}, pp.\  1958--1970, Online, August 2021{\natexlab{a}}. Association for Computational Linguistics.
\newblock \doi{10.18653/v1/2021.findings-acl.172}.
\newblock URL \url{https://aclanthology.org/2021.findings-acl.172/}.

\bibitem[Zhang et~al.(2025{\natexlab{a}})Zhang, Liao, Li, Du, and Lin]{zhang2025agenticinformationretrieval}
Weinan Zhang, Junwei Liao, Ning Li, Kounianhua Du, and Jianghao Lin.
\newblock Agentic information retrieval.
\newblock \emph{arXiv preprint arXiv:2410.09713}, 2025{\natexlab{a}}.

\bibitem[Zhang et~al.(2021{\natexlab{b}})Zhang, Yates, and Lin]{comparing2021zhang}
Xinyu Zhang, Andrew Yates, and Jimmy Lin.
\newblock Comparing score aggregation approaches for document retrieval with pretrained transformers.
\newblock In Djoerd Hiemstra, Marie-Francine Moens, Josiane Mothe, Raffaele Perego, Martin Potthast, and Fabrizio Sebastiani (eds.), \emph{Advances in Information Retrieval}, pp.\  150--163, Cham, 2021{\natexlab{b}}. Springer International Publishing.

\bibitem[Zhang et~al.(2022)Zhang, Ogueji, Ma, and Lin]{Zhang2022Towardbestpractices}
Xinyu Zhang, Kelechi Ogueji, Xueguang Ma, and Jimmy Lin.
\newblock Toward best practices for training multilingual dense retrieval models.
\newblock \emph{ACM Transactions on Information Systems}, 42:\penalty0 1--33, 2022.

\bibitem[Zhang et~al.(2023{\natexlab{c}})Zhang, Hofst{\"a}tter, Lewis, Tang, and Lin]{zhang2023rank}
Xinyu Zhang, Sebastian Hofst{\"a}tter, Patrick Lewis, Raphael Tang, and Jimmy Lin.
\newblock Rank-without-{GPT}: Building {GPT}-independent listwise rerankers on open-source large language models.
\newblock \emph{arXiv preprint arXiv:2312.02969}, 2023{\natexlab{c}}.

\bibitem[Zhang et~al.(2023{\natexlab{d}})Zhang, Thakur, Ogundepo, Kamalloo, Alfonso-Hermelo, Li, Liu, Rezagholizadeh, and Lin]{zhang2023miracl}
Xinyu Zhang, Nandan Thakur, Odunayo Ogundepo, Ehsan Kamalloo, David Alfonso-Hermelo, Xiaoguang Li, Qun Liu, Mehdi Rezagholizadeh, and Jimmy Lin.
\newblock {MIRACL}: A multilingual retrieval dataset covering 18 diverse languages.
\newblock \emph{Transactions of the Association for Computational Linguistics}, 11:\penalty0 1114--1131, 2023{\natexlab{d}}.

\bibitem[Zhang et~al.(2025{\natexlab{b}})Zhang, Li, Long, Zhang, Lin, Yang, Xie, Yang, Liu, Lin, et~al.]{zhang2025qwen3embedding}
Yanzhao Zhang, Mingxin Li, Dingkun Long, Xin Zhang, Huan Lin, Baosong Yang, Pengjun Xie, An~Yang, Dayiheng Liu, Junyang Lin, et~al.
\newblock {Qwen3} embedding: Advancing text embedding and reranking through foundation models.
\newblock \emph{arXiv preprint arXiv:2506.05176}, 2025{\natexlab{b}}.

\bibitem[Zheng et~al.(2023)Zheng, Hou, Lu, Chen, Zhao, and rong Wen]{Zheng2023AdaptingLargeLanguageModelsByIntegratingCollaborativeSemantics}
Bowen Zheng, Yupeng Hou, Hongyu Lu, Yu~Chen, Wayne~Xin Zhao, and Ji~rong Wen.
\newblock Adapting large language models by integrating collaborative semantics for recommendation.
\newblock \emph{2024 IEEE 40th International Conference on Data Engineering (ICDE)}, pp.\  1435--1448, 2023.

\bibitem[Zheng et~al.(2024{\natexlab{a}})Zheng, Zhao, Huang, Zhang, Mou, Niu, Song, Wang, and Gai]{Zheng2024FullStackLearningToRank}
Kai Zheng, Haijun Zhao, Rui Huang, Beichuan Zhang, Na~Mou, Yanan Niu, Yang Song, Hongning Wang, and Kun Gai.
\newblock Full stage learning to rank: A unified framework for multi-stage systems.
\newblock \emph{Proceedings of the ACM Web Conference 2024}, 2024{\natexlab{a}}.

\bibitem[Zheng et~al.(2024{\natexlab{b}})Zheng, Yin, Xie, Sun, Huang, Yu, Cao, Kozyrakis, Stoica, Gonzalez, et~al.]{zheng2024sglang}
Lianmin Zheng, Liangsheng Yin, Zhiqiang Xie, Chuyue~Livia Sun, Jeff Huang, Cody~Hao Yu, Shiyi Cao, Christos Kozyrakis, Ion Stoica, Joseph~E. Gonzalez, et~al.
\newblock {SGLang}: Efficient execution of structured language model programs.
\newblock \emph{Advances in Neural Information Processing Systems}, 37:\penalty0 62557--62583, 2024{\natexlab{b}}.

\bibitem[Zhong et~al.(2023)Zhong, Huang, Wettig, and Chen]{zhong-etal-2023-poisoningretrievalcorporabyinjectingadversarianpassages}
Zexuan Zhong, Ziqing Huang, Alexander Wettig, and Danqi Chen.
\newblock Poisoning retrieval corpora by injecting adversarial passages.
\newblock In Houda Bouamor, Juan Pino, and Kalika Bali (eds.), \emph{Proceedings of the 2023 Conference on Empirical Methods in Natural Language Processing}, pp.\  13764--13775, Singapore, December 2023. Association for Computational Linguistics.
\newblock \doi{10.18653/v1/2023.emnlp-main.849}.
\newblock URL \url{https://aclanthology.org/2023.emnlp-main.849/}.

\bibitem[Zhou et~al.(2024)Zhou, Liu, Li, Jin, Qian, Liu, Li, Dou, Ho, and Yu]{zhou2024trustworthinessinretrievalaugmentedgenerationsystemssurvey}
Yujia Zhou, Yan Liu, Xiaoxi Li, Jiajie Jin, Hongjin Qian, Zheng Liu, Chaozhuo Li, Zhicheng Dou, Tsung-Yi Ho, and Philip~S. Yu.
\newblock Trustworthiness in retrieval-augmented generation systems: A survey.
\newblock \emph{arXiv preprint arXiv:2409.10102}, 2024.

\bibitem[Zhu et~al.(2019)Zhu, Ahuja, Wei, and Reddy]{zhu2019hierarchicalattentionretrieval}
Ming Zhu, Aman Ahuja, Wei Wei, and Chandan~K. Reddy.
\newblock A hierarchical attention retrieval model for healthcare question answering.
\newblock In \emph{The World Wide Web Conference}, pp.\  2472--2482, 2019.

\bibitem[Zhu et~al.(2023)Zhu, Yuan, Wang, Liu, Liu, Deng, Chen, Liu, Dou, and Wen]{zhu2023large}
Yutao Zhu, Huaying Yuan, Shuting Wang, Jiongnan Liu, Wenhan Liu, Chenlong Deng, Haonan Chen, Zheng Liu, Zhicheng Dou, and Ji-Rong Wen.
\newblock Large language models for information retrieval: A survey.
\newblock \emph{arXiv preprint arXiv:2308.07107}, 2023.

\bibitem[Zhuang et~al.(2023{\natexlab{a}})Zhuang, Qin, Jagerman, Hui, Ma, Lu, Ni, Wang, and Bendersky]{zhuang2023rankt5}
Honglei Zhuang, Zhen Qin, Rolf Jagerman, Kai Hui, Ji~Ma, Jing Lu, Jianmo Ni, Xuanhui Wang, and Michael Bendersky.
\newblock {RankT5}: Fine-tuning {T5} for text ranking with ranking losses.
\newblock In \emph{Proceedings of the 46th International ACM SIGIR Conference on Research and Development in Information Retrieval}, SIGIR '23, pp.\  2308--2313, New York, NY, USA, 2023{\natexlab{a}}. Association for Computing Machinery.
\newblock \doi{10.1145/3539618.3592047}.

\bibitem[Zhuang et~al.(2024{\natexlab{a}})Zhuang, Qin, Hui, Wu, Yan, Wang, and Bendersky]{zhuang-etal-2024-beyond}
Honglei Zhuang, Zhen Qin, Kai Hui, Junru Wu, Le~Yan, Xuanhui Wang, and Michael Bendersky.
\newblock Beyond yes and no: Improving zero-shot {LLM} rankers via scoring fine-grained relevance labels.
\newblock In Kevin Duh, Helena Gomez, and Steven Bethard (eds.), \emph{Proceedings of the 2024 Conference of the North American Chapter of the Association for Computational Linguistics: Human Language Technologies (Volume 2: Short Papers)}, pp.\  358--370, Mexico City, Mexico, June 2024{\natexlab{a}}. Association for Computational Linguistics.
\newblock \doi{10.18653/v1/2024.naacl-short.31}.
\newblock URL \url{https://aclanthology.org/2024.naacl-short.31/}.

\bibitem[Zhuang et~al.(2021)Zhuang, Li, and Zuccon]{zhuang2021deep}
Shengyao Zhuang, Hang Li, and Guido Zuccon.
\newblock Deep query likelihood model for information retrieval.
\newblock In \emph{Advances in Information Retrieval: 43rd European Conference on IR Research, ECIR 2021, Virtual Event, March 28--April 1, 2021, Proceedings, Part II 43}, pp.\  463--470. Springer, 2021.

\bibitem[Zhuang et~al.(2023{\natexlab{b}})Zhuang, Liu, Koopman, and Zuccon]{zhuang-etal-2023-open}
Shengyao Zhuang, Bing Liu, Bevan Koopman, and Guido Zuccon.
\newblock Open-source large language models are strong zero-shot query likelihood models for document ranking.
\newblock In Houda Bouamor, Juan Pino, and Kalika Bali (eds.), \emph{Findings of the Association for Computational Linguistics: EMNLP 2023}, pp.\  8807--8817, Singapore, December 2023{\natexlab{b}}. Association for Computational Linguistics.
\newblock \doi{10.18653/v1/2023.findings-emnlp.590}.
\newblock URL \url{https://aclanthology.org/2023.findings-emnlp.590/}.

\bibitem[Zhuang et~al.(2024{\natexlab{b}})Zhuang, Zhuang, Koopman, and Zuccon]{zhuang2024setwise}
Shengyao Zhuang, Honglei Zhuang, Bevan Koopman, and Guido Zuccon.
\newblock A setwise approach for effective and highly efficient zero-shot ranking with large language models.
\newblock In \emph{Proceedings of the 47th International ACM SIGIR Conference on Research and Development in Information Retrieval}, SIGIR '24, pp.\  38--47, New York, NY, USA, 2024{\natexlab{b}}. Association for Computing Machinery.
\newblock \doi{10.1145/3626772.3657813}.

\bibitem[Zhuang et~al.(2025)Zhuang, Ma, Koopman, Lin, and Zuccon]{zhuang2025rankr1enhancingreasoningllmbased}
Shengyao Zhuang, Xueguang Ma, Bevan Koopman, Jimmy Lin, and Guido Zuccon.
\newblock {Rank-R1}: Enhancing reasoning in {LLM}-based document rerankers via reinforcement learning.
\newblock \emph{arXiv preprint arXiv:2503.06034}, 2025.

\bibitem[Zobel et~al.(1998)Zobel, Moffat, and Ramamohanarao]{zobel1998invertedfiles}
Justin Zobel, Alistair Moffat, and Kotagiri Ramamohanarao.
\newblock Inverted files versus signature files for text indexing.
\newblock \emph{ACM Transactions on Database Systems (TODS)}, 23\penalty0 (4):\penalty0 453--490, 1998.

\end{thebibliography}
\bibliographystyle{tmlr}

\end{document}